  \documentstyle[11pt]{article}
\begin{document}
  \baselineskip 14.18  pt  \parskip 3 pt  
  \renewcommand{\theequation}{\arabic{subsection}.\arabic{equation}}
                \setcounter{subsection}{1}  \setcounter{equation}{0}
                \begin{center} \vspace*{-1.2 cm}
  { Review for ``Reports on Progress in Physics''.
                 Version of 21 June 1995}\\
                 \end{center}  \vspace*{ 2.0 cm}
  {\Large\bf The Flux-Line Lattice in Superconductors}
                                             \\[1.4 cm]
    Ernst Helmut Brandt \\[0.3 cm]
    {\small  Max-Planck-Institut f\"ur Metallforschung,
    Institut f\"ur  Physik, D-70506 Stuttgart, Germany} \\[0.5 cm]
        \\[0.6 cm]
{\bf  Abstract}
        \\[0.5 cm]
   Magnetic flux can penetrate a type-II superconductor in form of
Abrikosov vortices (also called flux lines, flux tubes, or fluxons)
each carrying a quantum of magnetic flux $\phi_0 = h/2e$.
These tiny vortices of supercurrent tend to arrange in a triangular
flux-line lattice (FLL) which is more or less
perturbed by material inhomogeneities that pin the flux lines,
and in high-$T_c$ superconductors (HTSC's) also by thermal
fluctuations.
Many properties of the FLL are well described by the phenomenological
Ginzburg-Landau theory or by the electromagnetic London
theory, which treats the vortex core as a singularity. In Nb alloys
and HTSC's the FLL is very soft mainly because of the large magnetic
penetration depth $\lambda$: The shear modulus of the FLL
is $c_{66} \sim 1/\lambda^2$, and the tilt modulus $c_{44}(k)
\sim (1+k^2\lambda^2)^{-1}$ is dispersive and becomes very
small for short distortion wavelength $2\pi/k \ll \lambda$.
This softness is enhanced further by the pronounced anisotropy
and layered structure of HTSC's, which strongly increases the
penetration depth for currents along the $c$-axis of these
(nearly uniaxial) crystals and may even cause a decoupling of
two-dimensional vortex lattices in the Cu-O layers.
 Thermal fluctuations and softening may ``melt'' the FLL and
cause thermally activated depinning of the flux lines or of
the two-dimensional ``pancake vortices'' in the layers. Various
phase transitions are predicted for the FLL in layered HTSC's.
Although large pinning forces and high critical currents have
been achieved, the small depinning energy so far prevents the
application of HTSC's as  conductors at high temperatures
except in cases when the applied current and the surrounding
magnetic field are small.
  \newpage
  \noindent  %
{\bf Contents}       \\ [0.15 cm]
{\bf  1.~ Introduction}
         \\*[0.13 cm]
{\it 1.1.~ Abrikosov's prediction of the flux-line lattice}
         \\[0.01 cm]
{\it 1.2.~ Observation of the flux-line lattice}
         \\[0.01 cm]
{\it 1.3.~ Ideal flux-line lattice from Ginzburg-Landau Theory}
         \\[0.01 cm]
{\it 1.4.~ Ideal flux-line lattice from microscopic BCS Theory}
         \\[0.01 cm]
{\it 1.5.~ High-$T_c$ Superconductors}
         \\[0.19 cm]
{\bf  2.~ Arbitrary arrangements of flux lines}
         \\*[0.13cm]
{\it 2.1.~ Ginzburg-Landau and London Theories}
         \\[0.01 cm]
{\it 2.2.~ Straight parallel flux lines}
         \\[0.01 cm]
{\it 2.3.~ Curved flux lines}
         \\[0.19 cm]
{\bf  3.~ Anisotropic and layered superconductors}
         \\*[0.13cm]
{\it 3.1.~ Anisotropic London Theory}
         \\[0.01 cm]
{\it 3.2.~ Arrangements of parallel or curved flux lines}
         \\[0.01 cm]
{\it 3.3.~ Layered superconductors}
         \\[0.01 cm]
{\it 3.4.~ Lawrence-Doniach Model}
         \\[0.01 cm]
{\it 3.5.~ Josephson currents and helical instability}
         \\[0.19 cm]
{\bf  4.~ Elasticity of the flux-line lattice}
         \\*[0.13cm]
{\it 4.1.~ Elastic moduli and elastic matrix}
         \\[0.01 cm]
{\it 4.2.~ Nonlocal elasticity}
         \\[0.01 cm]
{\it 4.3.~ FLL elasticity in anisotropic superconductors}
         \\[0.01 cm]
{\it 4.4.~ Line tension of an isolated flux line}
         \\[0.01 cm]
{\it 4.5.~ Examples and elastic pinning}
         \\[0.19 cm]
{\bf  5.~ Thermal fluctuations and melting of the flux-line lattice}
         \\*[0.13cm]
{\it 5.1.~ Thermal fluctuations of the flux-line positions}
         \\[0.01 cm]
{\it 5.2.~ Time scale of the fluctuations}
         \\[0.01 cm]
{\it 5.3.~ Simple melting criteria}
         \\[0.01 cm]
{\it 5.4.~ Monte-Carlo simulations and properties of line liquids}
         \\[0.01 cm]
{\it 5.5.~ Flux-line liquid}
         \\[0.01 cm]
{\it 5.6.~ First-order melting transition}
         \\[0.01 cm]
{\it 5.7.~ Neutrons, muons, and FLL melting}
         \\[0.19 cm]
{\bf 6.~ Fluctuations and phase transitions in layered superconductors}
          \\*[0.13cm]
{\it 6.1.~ 2D melting and the Kosterlitz-Thouless transition}
         \\[0.01 cm]
{\it 6.2.~ Evaporation of flux lines into point vortices}
         \\[0.01 cm]
{\it 6.3.~ Decoupling of the layers}
         \\[0.01 cm]
{\it 6.4.~ Measurements of decoupling}
         \\[0.01 cm]
{\it 6.5.~ Vortex fluctuations modify the magnetization}
         \\[0.01 cm]
{\it 6.6.~ Quantum fluctuations of vortices}
         \\[0.19 cm]
{\bf  7.~ Flux motion}
         \\*[0.13cm]
{\it 7.1.~ Usual flux flow}
         \\[0.01 cm]
{\it 7.2.~ Longitudinal current and helical instability}
         \\[0.01 cm]
{\it 7.3.~ Cross-flow of two components of magnetic flux}
         \\[0.01 cm]
{\it 7.4.~ Dissipation in layered superconductors}
         \\[0.01 cm]
{\it 7.5.~ Flux-flow Hall effect}
         \\[0.01 cm]
{\it 7.6.~ Thermo-electric and thermo-magnetic effects}
         \\[0.01 cm]
{\it 7.7.~ Flux-flow noise}
         \\[0.01 cm]
{\it 7.8.~ Microscopic theories of vortex motion}
         \\[0.19 cm]
{\bf  8.~ Pinning of flux lines}
         \\*[0.13cm]
{\it 8.1.~ Pinning of flux lines by material inhomogeneities}
         \\[0.01 cm]
{\it 8.2.~ Theories of pinning}
         \\[0.01 cm]
{\it 8.3.~ Bean's critical state model in longitudinal geometry}
         \\[0.01 cm]
{\it 8.4.~ Extension to transverse geometry}
         \\[0.01 cm]
{\it 8.5.~ Surface and edge barriers for flux penetration}
         \\[0.01 cm]
{\it 8.6.~ Flux dynamics in thin strips, disks, and rectangles}
         \\[0.01 cm]
{\it 8.7.~ Statistical summation of pinning forces}
         \\[0.19 cm]
{\bf  9.~ Thermally activated depinning of flux lines}
         \\*[0.13cm]
{\it 9.1.~ Pinning force versus pinning energy}
         \\[0.01 cm]
{\it 9.2.~ Linear defects as pinning centres. The ``Bose glass''}
         \\[0.01 cm]
{\it 9.3.~ The Kim-Anderson model}
         \\[0.01 cm]
{\it 9.4.~ Theory of collective creep}
         \\[0.01 cm]
{\it 9.5.~ Vortex-glass scaling}
         \\[0.01 cm]
{\it 9.6.~ Linear ac response}
         \\[0.01 cm]
{\it 9.7.~ Flux creep and activation energy}
         \\[0.01 cm]
{\it 9.8.~ Universality of flux creep}
         \\[0.01 cm]
{\it 9.9.~ Depinning by tunnelling}
         \\[0.19 cm]
{\bf 10~ Concluding remarks}
         \\[0.19 cm]
{\bf \hspace*{0.65 cm} Acknowledgments}
         \\[0.19 cm]
{\bf \hspace*{0.65 cm} References}
  \newpage  \noindent  
%
{\bf  1.~~Introduction}
         \\*[0.5 cm]
{\it 1.1.~ Abrikosov's prediction of the Flux-Line Lattice}
         \\*[0.2 cm]
Numerous metals, alloys, and compounds become superconducting when they
are cooled below a transition temperature $T_c$. This critical
temperature ranges from
$T_c <1\,$K for Al, Zn, Ti, U, W and $T_c=4.15\,$K for Hg
(the superconductor discovered first, Kamerlingh Onnes 1911), over
$T_c =9.2\,$K for Nb (the elemental metal with the highest $T_c$)
and $T_c \approx 23\,$K for Nb$_3$Ge (the highest value from
1973 to 1986, see the overview by Hulm and Matthias 1980)
to the large $T_c$ values of the high-$T_c$
superconductors (HTSC's) discovered by Bednorz and M\"uller (1986),
e.g.\ YBa$_2$Cu$_3$O$_{7-\delta}$ (YBCO, $\delta \ll 1$,
 Wu et al.\ 1987) with $T_c \approx 92.5\,$K and
 Bi$_2$Sr$_2$Ca$_2$Cu$_3$O$_{10+\delta}$ (BSCCO) with
$T_c\approx 120\,$K, and Hg-compounds which under pressure have
reached $T_c \approx 164\,$K (Gao et al.\ 1994,
Kim, Thompson et al.\ 1995).
The  superconducting state is characterized by the vanishing
electric resistivity $\rho(T)$ of the material and by the
complete expulsion of magnetic flux,
irrespective of whether the magnetic field $B_a$ was applied before
or after cooling the superconductor below $T_c$
(Meissner and Ochsenfeld 1933). The existence of this
Meissner effect proves that the superconducting state is
a thermodynamic state, which uniquely depends on the applied
field and temperature but not on previous history.
As opposed to this, an ideal conductor expels the magnetic
flux of a suddenly switched on field $B_a$ but also ``freezes''
 in its interior the magnetic flux which has been there
before the conductivity became ideal.

   It was soon realized  that some
superconductors do not exhibit  expulsion of flux, but
the applied field  partly penetrates and the magnetization
 of the specimen depends on the magnetic history
in a complicated way. Early theories tried to explain this by
a ``sponge-like'' nature of the material, which could trap flux
in microscopic current loops that may become normal conducting
when the circulating current exceeds some critical value.
The true explanation of partial flux penetration was given in a
pioneering work by Abrikosov (1957). Alexei Abrikosov, then a
student of Lew Landau in Moscow, discovered a periodic solution
of the phenomenological theory of superconductivity
conceived a few years earlier by Ginzburg and Landau (GL) (1950).
Abrikosov interpreted his solution as a lattice of
parallel flux lines, now also called flux tubes, fluxons,
or Abrikosov vortex lines, {\bf figure 1}. These flux lines thread
the specimen, each carrying a quantum of magnetic flux
$\phi_0 = h/2e =2.07 \cdot 10^{-15}\,$Tm$^2$.
At the centre of a flux line the superconducting order parameter
$\psi({\bf r})$ (the complex GL function) vanishes. The line
$\psi=0$ is surrounded by a tube of radius $\approx \xi$,
the vortex core, within which $|\psi|$ is suppressed
from its  superconducting value $|\psi|=1$ that it attains
in the Meissner state, {\bf figure} 2. The vortex core is surrounded
by a circulating supercurrent ${\bf J(r)}$ which generates the
magnetic field ${\bf B(r)}$ of the flux line. In bulk specimens
the vortex current and field are confined to a flux tube of radius
$\lambda$; at large distances
$r \gg \lambda$, current and field of an isolated vortex decay
as $\exp(-r/\lambda)$. In thin films of thickness $d\ll \lambda$,
the current and magnetic field of a vortex extend to the larger
distance $\lambda_{\rm film} = 2\lambda^2 /d$, the circulating current
and the parallel field at large distances $r\gg \lambda_{\rm film}$
decrease only as $1/r^2$ and the perpendicular field as $1/r^3$,
and the vortex core has a larger radius
$\approx (12 \lambda_{\rm film}\xi^2)^{1/3}$ (Pearl 1964, 1966,
Fetter and Hohenberg 1967).
The coherence length $\xi$ and magnetic penetration depth $\lambda$
of the GL theory depend on the temperature $T$ and diverge at
$T_c$  as $(1-T/T_c)^{-1/2}$.

   The ratio $\kappa =\lambda/\xi$ is the GL parameter of the
superconductor.
Within  GL theory, which was conceived for temperatures
close to the transition temperature $T_c$, $\kappa$ is independent
of $T$. The flux-line lattice (FLL) (Abrikosov's periodic solution)
exists only in materials with $\kappa > 1/\sqrt 2$,
called type-II superconductors as opposed to type-I
superconductors, which have $\kappa <1/\sqrt 2$. Type-I
superconductors in a parallel field $B_a < B_c(T)$ are in the
 Meissner state, i.e.  flux penetrates only into a thin
surface layer of depth $\lambda(T)$, and at
$B> B_c(T)$  they become normal conducting. Here $B_c(T)$ is
the thermodynamic critical field. Type-II superconductors in a
parallel applied field  $B_a < B_{c1}(T) \le B_c(T)$ are in the
Meissner state $B=0$; in the field range
$B_{c1}(T) < B_a < B_{c2}(T)$ magnetic flux penetrates partly in
form of flux lines (mixed state or Shubnikov phase $0 <B <B_a$);
and at $B_a >B_{c2}(T) \ge B_c(T)$ the material is
normal conducting and thus $B=B_a$. $B_{c1}$ and $B_{c2}$ are the
lower and upper critical fields. Within GL theory, all three critical
fields vanish for $T \to T_c$ as $T_c-T$.
In this paper the notation $B_a=\mu_0 H_a$, $B_c=\mu_0 H_c$,
$B_{c1}=\mu_0 H_{c1}$, and $B_{c2}=\mu_0 H_{c2}$ will be used.

   When the superconductor is not a long specimen in parallel
field then demagnetization effects come into play. For
ellipsoidal specimens with homogeneous magnetization the
demagnetizing field is accounted for by a demagnetization factor $N$
with $0\le N \le 1$, {\bf figure} 3.  If $N>0$, flux penetration starts
earlier, namely, into type-II superconductors at $B_{c1}'=(1-N)B_{c1}$
in form of a FLL, and into type-I superconductors at
$B_c'=(1-N)B_c$ in form of normal conducting lamellae; this
``intermediate state'' is described by Landau and Lifshitz
(1960), see also Hubert (1967),  Fortini and Paumier (1972),
and the review by Livingston and DeSorbo (1969). From GL theory the
wall energy between normal and superconducting domains
is positive (negative) for type-I (type-II) superconductors.
Therefore, at $B_a =B_c$ the homogeneous Meissner state is unstable
in type-II superconductors and tends to split into normal and
superconducting domains
in the finest possible way; this means a FLL appears with normal cores
of radius $\approx \xi$. The field of first penetration of flux is thus
$(1-N)B_{c1} < (1-N)B_c$  for type-II superconductors and, following
Kronm\"uller and Riedel (1976),
$B_p = [\,(1-N)^2 B_c^2 +K^2\,]^{1/2} >(1-N)B_c$
for type-I superconductors with $K$ proportional to the wall energy.
  Superconductivity disappears when $B_a$
reaches the critical field $B_{c2}$ (type-II) or $B_c$ (type-I),
irrespective of demagnetization effects since the
magnetization vanishes at this transition. Within GL theory
these critical fields are
 \begin{equation}  
 B_{c1} \approx \frac{\phi_0}{4\pi \lambda^2} (\ln\kappa +0.5),~~
 B_c =  \frac{\phi_0}{\sqrt 8\pi \xi\lambda},~~
 B_{c2} = \frac{\phi_0}{2\pi \xi^2} = \sqrt2 \kappa B_c.
 \end{equation}
 The order parameter $|\psi({\bf r})|^2$ and microscopic field
$B({\bf r})$ of an isolated flux line for $\kappa\gg 1$ are
approximately given by (Clem 1975a, Hao et al.\  1991)
\begin{eqnarray}  
  |\psi( r)|^2  \approx 1/(1 + 2\xi^2/ r^2) \,, \\
  B(r) \approx \frac{\phi_0}{2\pi \lambda^2}
               K_0[(r^2 +2\xi^2)^{1/2} /\lambda],
\end{eqnarray}
  with $r=(x^2 +y^2)^{1/2}$ and $B\| z$; $K_0(x)$ is a modified
Bessel function with the limits $-\ln(x)$ ($x\ll 1$) and
$(\pi/2x)^{1/2} \exp(-x)$ ($x\gg 1$).
 The field (1.3) exactly minimizes the GL free energy
if the variational ansatz (1.2) is inserted.
The maximum field occurs in the vortex core, $B_{\rm max} = B(0)
 \approx (\phi_0 /2\pi\lambda^2) \ln\kappa \approx 2B_{c1}$ (still
for $\kappa \gg 1$). From (1.3) one obtains the current density
circulating in the vortex $J(r) = \mu_0^{-1} | B'(r)| = (\phi_0
 /2\pi\lambda^2 \mu_0)(r/\lambda \tilde r) K_1(\tilde r /\lambda)$
with $\tilde r = (r^2 +2\xi^2)^{1/2}$. Inserting for the modified Bessel
function $K_1(x)$ the approximation $K_1(x) \approx 1/x$ valid for
$x\ll 1$, one obtains the maximum current density
$J_{\rm max} = J(r =\sqrt 2 \xi) \approx \phi_0/(4\sqrt 2 \pi
\lambda^2 \xi \mu_0) = (27/32)^{1/2} J_0$  where
$J_0 = \phi_0/(3\sqrt 3 \pi \lambda^2 \xi \mu_0)$ is the
``depairing current density'', i.e.\ the maximum super-current density
which can flow within the GL theory in planar geometry (see, e.g.\
Tinkham 1975). Thus, for $\kappa \gg 1$ the field in the flux-line centre
is twice the lower critical field, and the maximum vortex current is the
depairing current.
         \\[0.4 cm]
 {\it 1.2. Observation of the flux-line lattice}
         \\*[0.2 cm]
  First evidence for the existence of Abrikosov's FLL in Nb came from
a weak Bragg peak seen in small-angle  neutron-scattering (SANS)
experiments by Cribier et al.\ (1964),
which later were improved considerably (Weber et al.\ 1973, 1974,
 Thorel et al.\ 1973, Lynn et al.\ 1994).
High-resolution pictures of the FLL were obtained by
Essmann and Tr\"auble (1967, 1971) by a  decoration
method using ferromagnetic microcrystallites generated by
evaporating a wire of Fe or Ni in a He atmosphere of a few Torr
pressure. This ``magnetic smoke'' settles on the
surface of the superconductor at the points where flux lines
end and where the local magnetic field is thus peaked. The resulting
pattern is then observed in an electron microscope; some such
pictures are shown in {\bf figure} 4.

This Bitter decoration of the FLL works also with the new high-$T_c$
superconductors (Gammel et al.\ 1987, 1992a, b, Vinnikov et al.\ 1988,
Dolan et al.\ 1989a, b, Bolle  et al.\ 1991, Grigorieva et al.\ 1993,
Dai et al.\ 1994a, b), see also the review by Grigorieva (1994).
Yao et al.\ (1994) decorate the FLL on both sides of a BSCCO crystal,
and Marchetti and Nelson (1995) show how to extract bulk properties
of the FLL from such double-sided decoration.
Marchevsky et al.\ (1995) observe droplets of FLL in NbSe$_2$
crystals by decoration.
 Small-angle neutron-scattering (SANS) from the FLL in HTSC's was
performed  with YBCO by Forgan et al.\ (1990) and
Yethiraj et al.\ (1993) and with BSCCO by Cubitt et al.\ (1993b) to
 see if the FLL melts
(section 5.7.). Keimer et al.\ (1994) investigate the
symmetry of the FLL in anisotropic YBCO by SANS.
Kleiman et al.\ (1992) by SANS observe the FLL in the heavy-fermion
superconductor UPt$_3$. The 3D flux distribution inside the
superconductor also can be probed by neutron depolarization
(Weber 1974, Roest and Rekveldt 1993).
A further method which probes the field distribution of
the FLL in the bulk is muon spin rotation ($\mu$SR) (Brandt
and Seeger 1986, P\"umpin et al.\ 1988, 1990, Herlach et al.\ 1990,
Harshman et al.\ 1989, 1991, 1993, Brandt 1988a,b, 1991c,
Lee et al.\ 1993, 1995, Cubitt et al.\ 1993a, Weber et al.\ 1993,
Sonier et al.\ 1994, Alves et al.\ 1994, Song 1995,
Riseman et al.\ 1995), see also section 5.7.

Very high resolution of the FLL in NbSe$_2$ ($T_c = 7.2\,$K) was achieved
by a scanning tunnelling microscope (STM), which measures the spatially
varying density of states of the quasiparticles
(Hess et al.\  1990, 1992, 1994, Renner et al.\ 1991,
Behler et al.\ 1994a, b).
By STM Hasegawa and Kitazawa (1990) performed spectroscopy with spatial
resolution of single atoms in the CuO$_2$ and (BiO)$_2$ layers of BSCCO.
Flux lines were imaged also in a magnetic force microscope by
Hug et al.\ (1994) and Moser et al.\ (1995), see the appropriate
theories by Reittu and Laiho (1992), Wadas et al.\ (1992), and
Coffey (1995). Hyun et al.\ (1989) detected the motion and pinning of
a single flux line in a thin film by measuring the diffraction pattern
from a Josephson junction formed by superconducting cross strips.
Kruithof et al.\ (1991) in NbMo near $B_{c2}$ see flux-line motion
by its coupling to a 2D electron gas, which is formed at the interface
between Si and SiO.
Bending et al.\ (1990) and Stoddart et al.\ (1993) achieved {\it single
fluxon resolution} by microscopic Hall probes. Mannhardt et al.\ (1987)
observe trapped flux quanta by scanning electron microscopy,
see also the reviews on this method by Huebener (1984) and
Gross and Koelle (1994).
Harada et al.\ (1993) observed the FLL time-resolved in thin films of
BSCCO in a field-emission transmission electron microscope by
Lorentz microscopy (outside the focal plane)
following a suggestion by Suzuki and Seeger (1971) and early
experiments by Lischke and Rodewald (1974), see also
Bonevich et al.\ (1994a) and the nice work on electron holography of
flux lines by Hasegawa et al.\ (1991) and Bonevich et al.\ (1994b).

The flux density at the specimen surface also may be observed by a
magneto-optical method which uses the Faraday effect in a thin layer
of EuS/EuF$_2$ (Kirchner 1969, Huebener et al.\ 1970,
Habermeier and Kronm\"uller 1977, Br\"ull et al.\ 1991) and EuSe
(Schuster et al.\ 1991, see also the review paper by
Koblischka and Wijngaarden (1995),
or in a high-contrast magneto-optical garnet plate
(Dorosinskii et al.\ 1992, Indenbom et al.\ 1994c)
covering the surface of the superconductor. Magneto-optics was used,
 e.g., for comparing the penetrating flux pattern
with microscopic pictures of the same polycrystalline surface in order
to identify the pinning defects (Forkl et al.\ 1991a,
Koblischka et al.\ 1993, 1994), for determining
the critical current density from the measured flux density profiles
(Forkl et al.\ 1991b, Theuss et al.\ 1992a),
for visualising flux motion in a YBCO film covering a step edge in the
substrate (a so called step-edge junction used for SQUIDS,
Schuster et al.\ 1993c), for  obtaining
ultra-fast time-resolved pictures of thermal instabilities in the
flux-line lattice in thin films (Leiderer et al.\ 1993), for
investigating the flux penetration into partly irradiated platelets
with a weak pinning central zone (Schuster et al.\ 1994d)
and the anisotropy of perpendicular flux penetration induced by a strong
in-plane magnetic field (Indenbom et al.\ 1994c), and for visualizing
pinning of flux lines at twin boundaries (Vlasko-Vlasov et al.\ 1994b).
The flux distribution and motion at the surface may also be observed,
more quantitatively but with lower spatial resolution,
by scanning Hall probes (e.g.\ Br\"ull et al.\ 1991,
Brawner and Ong 1993, Brawner et al.\ 1993, Xing et al.\ 1994) or
Hall-sensor arrays (e.g.\ Tamegai et al.\ 1992, Zeldov et al.\ 1994b).
         \\[0.4 cm]
 {\it 1.3. Ideal flux-line lattice from Ginzburg-Landau Theory}
         \\*[0.2 cm]
  When the applied field $B_a$ is increased, more and more flux
lines penetrate,  and the average induction
$B = \langle B({\bf r}) \rangle =\phi_0 n$
($n$ = flux-line density) grows.
At $B_a=B=B_{c2}$  the vortex cores overlap completely
such that $\psi$ vanishes  and superconductivity disappears.
At {\it low inductions} $B< 0.2 B_{c2}$ the order parameter is a product
of terms of the form (1.2), or of the more practical approximation
$|\psi({\bf r})|^2 \approx 1 -\exp(-r^2/2\xi^2)$, with $r$
replaced by $|{\bf r-r}_i |$ (Brandt and Evetts 1992).
The magnetic field is then  a linear superposition of terms (1.3)
which are centred at the flux-line positions ${\bf r}_i$
({\bf figure} 2, top).
This useful approximation applies to arbitrary arrangements of
parallel flux lines as long as the vortex cores do not overlap;
this means that all distances $|{\bf r}_i -{\bf r}_j|$
should exceed $\approx 5\xi$.
Far from the centre of an isolated vortex a more accurate
perturbation calculation gives
$1-|\psi(r)| \sim \exp(-\sqrt2 r/\xi)$ for $\kappa >\sqrt 2$ and
$1-|\psi(r)| \sim \exp(-2r/ \lambda)$ for $\kappa <\sqrt 2$
(Kramer 1971). At {\it high inductions} $B> 0.5 B_{c2}$,
the GL theory yields for a periodic FLL with arbitrary symmetry
\begin{eqnarray}  
  |\psi({\bf r})|^2  &=& \frac{1-B/B_{c2}}  {[1-1/(2\kappa^2)]
     \beta_A} \sum_{\bf K} a_{\bf K} \cos{\bf Kr},\\
  B({\bf r}) &=& B_a - (\phi_0/4\pi\lambda^2) |\psi({\bf r})|^2 \,.
\end{eqnarray}
Here the sum is over all reciprocal lattice vectors
$K_{mn} = (2\pi/x_1 y_2)(my_2; ~-mx_2 +nx_1)$ of the periodic FLL
with flux-line positions
$R_{mn} = (mx_1 +nx_2; ~ny_2)$ ($m,\,n$ = integer). The unit-cell
area $x_1 y_2$ yields the mean induction $B=\phi_0/(x_1y_2)$.
The first Brillouin zone has the area $\pi k_{\rm BZ}^2$ with
$k_{\rm BZ}^2 = 4\pi B/\phi_0$.
For general lattice symmetry the Fourier coefficients $a_{\bf K}$
and the Abrikosov parameter $\beta_A$  are (Brandt 1974)
\begin{eqnarray}  
 a_{\bf K}\ = (-1)^{m+mn+n} \exp(-K_{mn}^2 x_1 y_2 /8\pi),~~
 \beta_A = \sum_{\bf K} a_{\bf K}^2\,.
\end{eqnarray}
{}From $\psi(0)=0$ follows $\sum_{\bf K} a_{\bf K} = 0$, or with
$a_{{\bf K}=0} =1$, $\sum_{{\bf K}\ne 0} a_{\bf K} =-1$.
In particular, for the triangular FLL with spacing
$a = (2\phi_0/\sqrt3 B)^{1/2}$ one has $x_1=a$, $x_2=a/2$,
$y_2 =\sqrt3 a/2$, $B= 2\phi_0/\sqrt3 a^2$, $a_{\bf K} =
(-1)^\nu \exp(-\pi \nu /\sqrt3 )$ with $\nu = m^2 +mn +n^2 =
R_{mn}/R_{10} = K_{mn}/K_{10}$. This yields
$\beta_A = 1.1595953$ (the reciprocal-lattice sum converges
very rapidly)  and the useful relationships
$K_{10} =2\pi/y_2$, $K_{10}^2 =16\pi^2 /3a^2 = 8\pi^2 B/\sqrt3 \phi_0$.
For the square lattice one has $\beta_A = 1.1803406$.
The free energy $F(B)$ per unit volume and the negative magnetization
$M=B_a -B \ge 0$, where $B_a = \mu_0 \partial F/\partial B$ is the
applied field which is in equilibrium with the induction $B$, are
(still for $B>0.5 B_{c2}$)
\begin{eqnarray}  
  F(B) &=& \frac{B^2}{2\mu_0} - \frac{(B_{c2} -B)^2}{ 2\mu_0
  (2\kappa^2 -1)\beta_A}\,, \\
  M(B) &=&  \frac{B_{c2} -B}{(2\kappa^2 -1) \beta_A} =
            \frac{\phi_0}{4\pi\lambda^2} \langle|\psi|^2\rangle\,.
\end{eqnarray}
  This yields for the periodic field $B({\bf r}) = \sum_{\bf K}
  B_{\bf K} \cos {\bf Kr}$  the Fourier coefficients
$B_{\bf K\ne 0} = Ma_{\bf K}$, $B_{\bf K =0} =
\langle B({\bf r}) \rangle = B$, and for the averaged order
parameter $\langle|\psi|^2\rangle = (4\pi\lambda^2/\phi_0)M$.
Klein and P\"ottinger (1991) rederive Abrikosov's periodic FLL
solution in a simple way using the virial theorem discovered for
the GL equations by Doria et al.\ (1989).

At fields $B_a > B_{c2}$ the bulk superconductivity vanishes, but a
thin surface layer of thickness $\approx \xi$  remains superconducting
as long as the parallel field is smaller than a third critical field
$B_{c3} = 1.695 B_{c2}$ near $T_c$ (Saint-James and DeGennes 1963,
Abrikosov 1964) or $B_{c3} \approx 2 B_{c2}$ near $T=0$ (Kulik 1968).
For recent work on $B_{c3}$ see Suzuki et al.\ (1995) and
Troy and Dorsey (1995).
  A very interesting situation results when a
field $B_a >B_{c2}$ is applied at an angle to the surface of the
superconductor. Then a lattice of {\it tilted vortices of finite
length} appears, which was confirmed by the classical experiment
of  Monceau et al.\ (1975). The theory of these ``Kulik vortices''
(Kulik 1968) is quite intricate and still not complete, although
``only'' the eigenvalue equation (linearized GL equation)
$[ -i\nabla -(2\pi/\phi_0){\bf A} ]^2 \psi = E \psi$ with boundary
condition $[ -i\nabla -(2\pi /\phi_0){\bf A} ] \psi = 0$
has to be solved, where ${\bf A}$ is the vector potential of the
constant (but oblique!) applied field, see the reviews by
Thompson (1975) and Minenko and Kulik (1979).

   At {\it intermediate inductions} $0.2 < B/B_{c2} < 0.5$
the GL  equations have to be solved
numerically. This has been achieved by  a Ritz variational
method using finite Fourier series as trial functions for the real
and periodic functions $|\psi({\bf r})|^2$ and $B({\bf r})$
 (Brandt 1972), thereby extending early analytical calculations
by Kleiner et al.\ (1964) and Eilenberger (1964) to the entire
range of inductions $0\le B \le B_{c2}$. These computations show
that the above expressions derived for low and high
fields are excellent approximations.
         \\[0.4 cm]
 {\it 1.4. Ideal flux-line lattice from microscopic BCS Theory}
         \\*[0.2 cm]
The same variational method was used to compute the ideal FLL in pure
superconductors in the entire range of $B$ and $T$ from the
{\it microscopic} theory
of superconductivity of Bardeen, Cooper, and Schrieffer (BCS) (1957),
reformulated by Gor'kov (1959a,b) to allow for a spatially
varying order parameter. This computation (Brandt 1976a, b) extended
ideas of Uwe Brandt, Pesch and Tewordt (1967) and Delrieu (1972, 1974)
(originally conceived to calculate the electron Green function near
$B_{c2}$) by introducing as small parameters the Fourier coefficients
of the function $V({\bf r, r'}) = \Delta({\bf r})
 \exp[i(2\pi/\phi_0) \int_{\bf r}^{\bf r'}\! {\bf A(l)}\,{\rm d}{\bf l} ]
 \,\Delta^*({\bf r'})$, where $\Delta$ is the gap function and ${\bf A }$
the vector potential. For a periodic FLL, $V({\bf r, r'})$ is
periodic in the variable ${\bf r+r'}$, and for $B \to B_{c2}$ (i.e.,
when the Abrikosov solution for $\Delta$ applies) it is a simple Gaussian,
$V({\bf r,r'}) = \langle |\Delta|^2 \rangle \exp(-|{\bf r-r'}|^2
\pi B /2\phi_0)$; this means that all higher Fourier coefficients of $V$
vanish at $B_{c2}$ and are very small in the large range $B\ge 0.5 B_{c2}$
(Brandt 1974). However, this elegant numerical method appears to be
restricted to pure superconductors with infinite
electron mean free path $l$.

 For superconductors with arbitrary purity the
FLL was computed from microscopic theory in the quasiclassical
formulation by Eilenberger (1968), which expresses the free energy
in terms of energy-integrated Green functions; see also Usadel (1970)
for the limit of impure superconductors (with numerical solutions
by Watts-Tobin et al.\ 1974 and Rammer 1988) and the carefull discussion
of this quasiclassical approximation by Serene and Rainer (1983).
The computations by Pesch and Kramer (1974), Kramer and Pesch (1974),
Watts-Tobin et al.\ (1974), Rammer et al.\ (1987),
and Rammer (1988) used a circular cell method,
which approximates the hexagonal Wigner-Seitz cell of the triangular
FLL by a circle, and the periodic solutions by rotationally symmetric
ones with zero slope at the circular boundary (Ihle 1971).
Improved computations account for the periodicity of the FLL
(Klein 1987, Rammer 1991a, b), see also the calculation of the
density of states near $B_{c2}$ by Pesch (1975) and recent work by
Golubov and Hartmann (1994).

 Remarkably, the circular cell approximation not only yields good
results at low inductions $B\ll B_{c2}$, but within  GL theory it
even yields the exact value for $B_{c2}$ (Ihle 1971), and from
BSC theory a good approximate $B_{c2}$
(Larkin and Ovchinnikov 1995).
{}From the BCS weak-coupling microscopic theory the upper critical
field $B_{c2}(T)$ was calculated for all temperatures $0 \le T < T_c$
by Helfand and Werthamer (1966) and Eilenberger (1966) and from
strong-coupling theory by Riek et al.\ (1991).
Zwicknagl and Fulde (1981) calculate $B_{c2}(T)$ for antiferromagnetic
superconductors, e.g.\ the Chevrel phase compound
Rare-Earth--Mo$_6$S$_8$. For layered
superconductors  $B_{c2}(T)$ was obtained
by Klemm et al.\ (1975), Takezawa et al.\ (1993), and
recently by Ovchinnikov and Kresin (1995), who showed that in the
presence of magnetic impurities $B_{c2}(T)$ shows upward
curvature at low $T$ and is strongly enhanced; such an enhancement
is observed by Cooper et al.\ (1995).
For non-equivalent layers $B_{c2}(T)$ was calculated
by  Koyama and Tachiki (1993).
Spivak and Zhou (1995) consider $B_{c2}(T)$ in disordered
superconductors; in mesoscopic samples they find multiple reentrant
superconducting transitions, but the $T_c$ of bulk samples remains
finite at any $B$ such that $B_{c2}$ diverges for $T\to 0$.

  Calculations near $B_{c2}$ yield the Maki parameters
$\kappa_1(T/T_c)$ and $\kappa_2(T/T_c)$ (Maki 1964) defined by
$\kappa_1 = B_{c2}/\sqrt2 B_c$ (1.1) and
$\kappa_2 = (|2.32\,{\rm d} M/{\rm d}B|_{B_{c2}}^{-1}\! +0.5)^{1/2}$
(1.8), see Auer and Ullmaier (1973) for a careful experimental analysis.
Both ratios $\kappa_1/\kappa$ and $\kappa_2/\kappa$ equal unity at
$T=T_c$ and become $> 1$ at lower $T$. In the clean limit,
$\kappa_2/\kappa$ diverges at $T=0$ since the magnetization becomes
a non-analytic function at $B_a \to B_{c2}$ and $T\to 0$, namely,
$M \sim x/\ln x$ with $x=(1-B_a/B_{c2}) \approx 1-b$ (Delrieu 1974).
For impure superconductors one has
$\kappa_1 \approx \kappa_2 \approx \kappa$ at all temperatures.

  The full $T$- and $B$-dependences of the vortex structure in
HTSC's were calculated from BCS theory by Rammer (1991a), and the
vortex structure in pure Nb from strong-coupling theory
by Rammer (1991b).
The local density of states of the quasiparticles in the vortex core
(Caroli et al.\ 1964, Bergk and Tewordt 1970), which can be measured
by scanning-tunnelling-microscopy (Hess et al.\ 1990, 1992, 1994),
was calculated by Klein (1990), Ullah et al.\ (1990),
Gygi and Schl\"uter (1991), Zhu et al.\ (1995) (accounting for
the atomic crystal field),  Berkovits and Shapiro (1995)
(influence of impurity scattering), and by Schopohl and Maki (1995)
(for a d-wave superconductor). For more work on the flux line
in d-wave (or combined ``s+id''-wave) superconductors see
Volovik (1993) and Soininen et al.\ (1994).
Electronic transport properties like ultrasonic absorption and thermal
conductivity of the FLL (Vinen et al.\ 1971) were calculated by
Pesch et al.\ (1974) and Klimesch and Pesch (1978), see also
Takayama and Maki (1973) and the review on the properties of
type-II superconductors near $B_{c2}$ by Maki (1969a).

   These computations show that for impure superconductors
(with electron mean free path $l$ smaller than the BCS
coherence length $\xi_0$) the GL result is typically a good
approximation. However, as was shown by Delrieu (1972) in a
brilliant experimental and theoretical paper, for clean
superconductors near
$B_{c2}$ the local relation (1.5) between $B({\bf r})$ and
$|\psi({\bf r})|^2$ does not hold. The correct nonlocal
relation (with a kernel of range $\hbar v_F/\pi k_B T$,
$v_F$ = Fermi velocity) yields more slowly decreasing
Fourier coefficients $B_{\bf K}$ at $T<T_c$
(see also Brandt 1973, 1974, Brandt and Seeger 1986,
Brandt and Essmann 1987).
In particular, at $T\ll T_c$, $B \approx B_{c2}$,
and $l \gg \xi_0 = \hbar v_F/\pi \Delta(T\!\! = \!0)$ one has
$B_{\bf K} \sim (-1)^\nu /K^3$.  This Fourier series yields a rather
exotic magnetic field $B({\bf r})$ with sharp conical maxima at the
flux-line centres and conical minima between two flux lines,
and (for the triangular FLL) saddle points with three-fold
point symmetry in the middle between three flux lines, {\bf figure} 5.
The minima and saddle points interchange their positions
at $T/T_c \approx 0.6$. Explicitly, the magnetic field of clean
superconductors near $B_{c2}(T)$ has the form (Brandt 1974)
  \begin{eqnarray}  
  B({\bf r}) = B_a +|M| \sum_{\bf K} b_{\bf K} \cos{\bf Kr} =
          B +|M| \sum_{{\bf K}\ne 0} b_{\bf K} \cos{\bf Kr} \nonumber \\
  b_{\bf K}(c) =-(-)^\nu I_\nu /I_0, ~~~~~~\nu=m^2 +mn +n^2, ~~~~~~~
  p= (\pi^{1/2} /3^{1/4}) \,\sqrt\nu                         \nonumber \\
  I_\nu = \int_0^\infty \! K(cs)\, s\, e^{-s^2 -p^2}\, \frac{\cosh(2ps)-1}
     {2p^2} \,{\rm d}s, ~~~
  I_0 = \int_0^\infty \! K(cs)\, s^3\, e^{-s^2} \,{\rm d}s   \nonumber \\
  K(r) = \int_0^\infty \! [(r^2 +z^2) \sinh\sqrt{r^2 +z^2} \,]^{-1}
     \, {\rm d}z, ~~~ c(T,B) =\frac{2(\phi_0/2\pi B)^{1/2}}
     {\hbar v_F /2\pi k_BT}
  \end{eqnarray}
with the reciprocal FLL vectors ${\bf K}_{mn}$ from (1.4),
   $p^2=K_{mn}^2 x_1 y_2/8\pi$ for general FLL symmetry,
$|M| = B_a -B$ the negative magnetization, and $v_F$ the Fermi velocity.
   From (1.9) one obtains the limiting expressions
$b_{\bf K} \approx |M| (-1)^\nu (3^{3/4}/2\pi)/\nu^{3/2}$
for $T\to 0$ ($c\to 0$) and
$b_{\bf K} = |M| (-1)^\nu \exp(-\nu \pi/\sqrt3)$
for $T\to T_c$ ($c\to \infty$ since $B\to 0$).
It is interesting to note that the limit ${\bf K}\to 0$ yields the
magnetization.    Though derived for
$B\approx B_{c2}$, the field (1.9) approximately applies also to
lower inductions $B$ as long as the Abrikosov solution for $\psi({\bf r})$
is good. The parameter $c(T,B)$ is explicitly obtained from
  \begin{eqnarray}  
  c(T,B) = c(t)/\sqrt{b} \,, ~~~~
  \int_0^\infty \! K(r)\,r\,[1- e^{\displaystyle -s^2/c(t)^2} ]\,
                                                {\rm d}z = -\ln t
  \end{eqnarray}
with $t=T/T_c$ and $b=B/B_{c2}(T)$. From (1.10) one finds to a relative
accuracy of $0.0012$
  \begin{eqnarray}  
  c(t) = t \,(1-t^2)^{-1/2}\, (1.959 +0.182\, t + 0.224\, t^2).
  \end{eqnarray}
{}From $c(t)$ (1.10) the upper critical field of clean superconductors
for  $0 \le T < T_c$ is obtained as
$B_{c2}(T) = 2\phi_0 t^2 /[\pi \xi^2_c c(t)^2 ]$ with
$\xi_c = \hbar v_F /2\pi k_BT_c$; these $B_{c2}$ values
coincide with the result of Helfand and Werthamer (1966). For arbitrary
purity the $b_{\bf K}$ are derived by Delrieu (1974), Brandt (1974),
and Pesch (1975).

  The strange field profile (1.9) has been observed in pure Nb by
nuclear magnetic resonance (NMR, Delrieu 1972) and by muon spin
rotation ($\mu$SR, Herlach et al.\ 1990).
  It is an interesting question whether this zig-zag profile of
$B({\bf r})$ may  be observed in other superconductors.
The type-II superconducting elemental metals are Nb, V, La and Tc.
Single crystals of V with the required high purity became available
only quite recently and have been studied comprehensively by
Moser et al.\ (1982).
The HTSC's appear to be in the clean limit
since their coherence length $\xi$ is so small, smaller than the
electron mean free path $l$; but by the same token their upper critical
field $B_{c2}\sim 1/\xi^2$ (1.1) is extremely large at $T\ll T_c$ such
that the experimental observation of a possible exotic zig-zag
field profile is difficult.

 Clean Nb has the remarkable property that its theoretical
magnetization curve obtained from the generalized GL theory
of Neumann and Tewordt (1966) (an expansion in powers of $T_c -T$)
by Hubert (1972) and from the BCS theory by
Ovchinnikov and Brandt (1975) and Brandt (1976a, b), is not monotonic
but has an S-shape if\, $T$ is not close to  $T_c$
({\bf figure} 3, middle).
This S-shape means that the penetrating flux lines attract each other.
The resulting equilibrium distance
$a_0$ of the flux lines, and a jump in $B(B_a)$ to a finite
induction $B_0 \approx \phi_0/a_0^2$  obtained by a Maxwell
construction from the theoretical S-shape,  were observed in
clean Nb and in Pb-Tl and Ta-N alloys with $\kappa$ values slightly
above $1/\sqrt2$  by decoration and magnetization experiments, see
Sarma (1968), Essmann (1971), Kr\"ageloh (1970),
Kr\"ageloh et al.\ (1971), Auer and Ullmaier (1973), and the review by
Brandt and Essmann (1987).  Decoration revealed both
Meissner phase islands surrounded by FLL ({\bf figure} 4b),
and FLL islands embedded in Meissner phase ({\bf figure} 4c).
The phase diagram of clean
superconductors in the $\kappa$--$t$-plane ($t=T/T_c$) computed
from the BCS theory by Klein (1987) and discussed by
Weber et al.\ (1987), near $\kappa = 1/\sqrt2$ with increasing
$t$ shows the transition from type-I to usual type-II (``type-II-1'')
superconductors via a region where the flux lines attract each other
(``type-II-2'' superconductors).
  Klein et al.\ (1987) give an elegant and surprisingly accurate
interpolation method which constructs the free energy and
magnetization (also the S-shape) of type-II superconductors
using only properties at the upper and lower critical fields.

  A further rather unexpected result of microscopic theory is that
for pure superconductors at low induction $B$ the vortex core
gets rather narrow; possibly the gap function $|\Delta(r)|$ even
attains a vertical slope at $r=0$ when $l\to \infty$, $T\to 0$,
and $B\to 0$. This {\it shrinking of the vortex core} was seen in
computations (Kramer and Pesch 1974, Brandt 1976a, b) and was
investigated analytically by Ovchinnikov (1977), see also
Rammer et al.\ (1987).
         \\[0.4 cm]
{\it 1.5.~ High-$T_c$ Superconductors}
         \\*[0.2 cm]
  With the discovery of high-$T_c$ superconductivity,
the interest in the FLL rose again. At present it is not clear
whether Delrieu's (1972) exotic field distribution
near $B_{c2}$ might be observed in these rather clean materials;
in HTSC's $\xi_0$ is very small ($20...30\,$\AA) such that
$l \gg \xi_0$ is typical (Harris et al.\ 1994, Matsuda et al.\ 1994,
Guinea and Pogorelov 1995) (section 7.8.),  but by the same token
$B_{c2}(0) =\phi_0/2\pi\xi_0^2$ is very large ($\approx 100\,$T),
rendering such experiments difficult, and possibly
 quantum fluctuations smear the cusps in $B({\bf r})$.
 In Tl-based HTSC's the field
distribution of the FLL and the vortex motion can be observed by NMR of
the $^{205}$Tl nuclear spin (Mehring et al.\ 1990, Mehring 1990,
Brom and Alloul 1991, Song et al.\ 1992, 1993, 1994, Carretta 1992, 1993,
Bulaevskii et al.\ 1993b, 1995b,  Moonen and Brom 1995).
Bontemps et al.\ (1991) measure the electron paramagnetic resonance
(EPR) linewidth of a free-radical layer on YBCO.
Steegmans et al.\ (1993) study the FLL by a NMR decoration
technique putting a silicone layer (thinner than the flux-line
spacing) on grains of Nb$_3$Sn powder.
Suh et al.\ (1993) detect the thermal motion of the
flux lines by nuclear spin echo decay measurements and
De Soto et al.\ (1993) observe that intrinsic pinning of the FLL in
organic superconductors enhances the NMR spin--lattice relaxation.

  Besides the field distribution, there are numerous other
exciting problems concerning the
FLL in HTSC's, e.g. the structure and softness of the FLL in these
highly anisotropic materials, the appearance of weakly interacting
two-dimensional (2D) vortex disks (``pancake vortices'', Clem 1991)
in the superconducting Cu-O layers, and of core-less vortex lines
(``Josephson strings'') running between these layers. Furthermore,
the thermal fluctuations of the 3D or 2D vortices may lead to a melted
(shear-stiffness-free)  FLL in a large region of the $B$-$T$-plane.
 Particularly fascinating are recent ideas concerning the statistics
of pinned vortices at finite temperatures, namely, thermally
activated depinning and collective flux creep,
a vortex-glass transition (Fisher 1989,
Fisher, Fisher and Huse 1991) near which the resistivity and magnetic
response obey scaling laws, and the analogy
with a Bose glass (Nelson and Vinokur 1992, 1993) when linear
pins are introduced by irradiation with heavy ions, section 9.2.
All these points will be discussed in the following sections.
More details on these theoretical concepts may be found in
recent review papers by Blatter et al.\ (1994a)
and K.\ H.\ Fischer (1995a).
%
         \\[0.7 cm]
{\bf  2.~ Arbitrary arrangements of flux lines}
         \setcounter{subsection}{2}  \setcounter{equation}{0}
         \\*[0.5 cm]
{\it 2.1.~ Ginzburg-Landau and London Theories}
         \\*[0.2 cm]
The GL theory (Ginzburg and Landau 1951) extends Landau's theory of
second-order phase transitions to a spatially varying complex order
parameter $\psi({\bf r})$. The resulting gradient term is made
gauge-invariant by combining it with the vector potential ${\bf
 A(r)}$ where ${\bf \nabla \times A(r) = B(r)} =\mu_0 {\bf H(r)}$
is the local magnetic field. The two GL equations are obtained
by minimization of the GL free energy
functional $F\{\psi, {\bf A}\}$ with respect to $\psi$ and
${\bf A}$, i.e.\ from $\delta F/\delta \psi = 0$ and
$\delta F/\delta {\bf A}=0$, where in reduced units
\begin{eqnarray}  
  F\{\psi, {\bf A}\} = \mu_0 H_c^2 \int \!{\rm d}^3r \Big[ -|\psi|^2
 +\frac{1}{2} |\psi|^4 +|(-i\nabla/\kappa -{\bf A})\psi|^2 +
 {\bf B}^2 \Big].
\end{eqnarray}
In (2.1)  the length unit is the penetration depth $\lambda$,
$\kappa =\lambda/\xi$ is the GL parameter, $\psi$ is in units of
the BCS order parameter $\psi_0$ ($\psi_0 \sim T_c-T$) such that
$|\psi|=1$ in the Meissner state and $|\psi|=0$ in the
normal conducting state, ${\bf B}$ is in units $\sqrt 2 B_c$, and
${\bf A}$ in units $\sqrt2B_c\lambda$. If desired, the electric field
${\bf E}= - \partial {\bf A}/\partial t$ and the time dependence
may be introduced by the assumption $\partial \psi/\partial t =
-\gamma \delta F/\delta \psi$ with $\gamma$ = const (Schmid 1966,
Vecris and Pelcovits 1991, Dorsey 1992, Troy and Dorsey 1993).
An extension to anisotropic non-cubic superconductors is possible
by multiplication of the gradient term in (2.1) by an
``effective mass tensor'' (Takanaka 1975, Kogan and Clem 1981).
For the $H_{c2}$-anisotropy in cubic superconductors (like Nb) see
Weber et al.\ (1981) and Sauerzopf et al.\ (1982).

   The phenomenological GL theory is one of the most elegant and
powerful concepts in physics, which was applied not only to
superconductivity (see the textbooks of DeGennes 1966,
Saint-James, Sarma and Thomas 1969, Tinkham 1975, Abrikosov 1988,
Buckel 1990, Klemm 1996) but also to other phase
transitions, to nonlinear dynamics, to dissipative systems
with self-organizing pattern formation, and even to
cosmology (melting of a lattice of ``cosmic strings'', Nielsen
and Olesen 1979;  vortex lattice in electroweak theory,
Ambj{\o}rn and Olesen 1989).

  Introducing the modulus $f({\bf r}) = |\psi|$ and phase
$\phi({\bf r})$ of the GL function $\psi = f e^{i\phi}$ one finds
that only this modulus and the gauge-invariant ``supervelocity''
${\bf Q} = \nabla\phi/\kappa - {\bf A}$,  or
${\bf Q} = (\phi_0 /2\pi) \nabla\phi - {\bf A}$ in physical
units, enter the GL theory, but not $\psi$ and {\bf A}
separately. Thus, completely equivalent to (2.1) is the functional
                              \setcounter{equation}{0}
                { \renewcommand{\theequation}{2.1a}
\begin{eqnarray}             
   F\{ f, {\bf Q} \} =  \mu_0 H_c^2 \int \!{\rm d}^3r \Big[
  -f^2 +\frac{1}{2} f^4 + (\nabla f)^2/\kappa^2 + f^2 {\bf Q}^2
  + (\nabla  \times {\bf Q})^2  \Big].   
\end{eqnarray}  } 
  Since only real and gauge-invariant quantities enter in (2.1a),
any calculation based on the GL theory (2.1) should contain
{\it only gauge-invariant equations}, even when approximations
are made. For a straight vortex with $n$ quanta of flux
centred at $r=(x^2+y^2)^{1/2} = 0$ one finds at $r\ll \xi$
the behaviour $f(r) \sim r^n$ and ${\bf Q}(r) =n/(\kappa r)$.
In a particular gauge one has at $r\ll \xi$ for this vortex
$\psi\sim (x+iy)^n$  and ${\bf A}\approx$ const. In the impractical
gauge with real $\psi$ the vector potential {\bf A} would
diverge as $1/(\kappa |{\bf r-r}_\nu|)$ at all vortex positions
${\bf r}_\nu = (x_\nu; y_\nu)$; this {\it real gauge would thus depend
on the vortex positions}.  For a FLL with $n_\nu$ flux quanta
in the $\nu$th flux line, a convenient gauge is the choice
(Brandt 1969b, 1977a, b)
\begin{eqnarray}  
\psi(x,y) = g(x,y) \exp\Big[ -\frac{\pi B}{2 \phi_0} (x^2 +y^2)\Big]
\prod_\nu [(x-x_\nu) +i(y-y_\nu)]^{n_\nu},
\end{eqnarray}
which allows for smooth non-singular functions $g(x,y)$ and
${\bf A}(x,y)$ as exact solutions or trial functions. In particular,
for periodic positions $(x_\nu, y_\nu)$ with $n_\nu = 1$, the choice
$g(x,y)=$ const  in (2.2) describes the ideal FLL (1.4) near $B_{c2}$.
 Approximate expressions for $\psi(x,y)$ and $B(x,y)$ for arbitrary
arrangements of straight or curved flux lines and valid at all
inductions $B$ are given by Brandt (1977b).

  Using the general GL function (2.2) Brandt (1969b) calculated the
energy of structural defects in the 2D FLL from the linearized
GL theory which is valid near $B_{c2}$. This ``lowest Landau level
method'' [which yields formula (2.2)] recently was extended
by Dodgson and Moore (1995) to obtain the energy of 3D defects
in the FLL, e.g., screw dislocations and twisted flux lines
(two-, three-, and six-line braids). These authors correctly
obtain that the length of such defects diverges near $B_{c2}$ as
$a/(1-b)^{1/2}$ ($b=B/B_{c2}$, $a^2 \approx \phi_0 /B$);
this finding coincides with the result of elasticity theory,
$a (c_{44}/c_{66})^{1/2}$ if one notes that near $B_{c2}$
the shear and tilt moduli of the FLL (section 4.1.) behave as
$c_{66}\sim (1-b)^2$ (4.4) and $c_{44}(k\ne 0) \sim 1-b$  (4.6a).
The 3D structural defects of the FLL are thus {\it stretched along $B$}
when $B$ approaches the upper critical field $B_{c2}$, similarly
as the deformations of the FLL and the superconducting nuclei
 caused by point pins (Brandt 1975) are stretched.

  If the vortex cores do not overlap appreciably, one may put
$|\psi({\bf r})| = 1$ outside the vortex core. The condensation
energy is then constant [$\mu_0 H_c^2 (|\psi|^2 -\frac{1}{2} |\psi|^4)
= \mu_0 H_c^2 /2$] and the magnetic energy density $\sim H^2$
and kinetic energy of the supercurrent $\sim f^2 Q^2$ yield
the free energy functional of the London theory,
with ${\bf B(r)} = \mu_0 {\bf H(r)}$,
\begin{eqnarray}  
  F\{ {\bf B} \} =  \frac{\mu_0}{2} \int \!{\rm d}^3r \,
  [\, {\bf B}^2 + \lambda^2 (\nabla \times {\bf B})^2 \,].
\end{eqnarray}
  Minimizing $F\{ {\bf B}\}$ with respect to the local induction
${\bf B(r)}$ using $\nabla {\bf B(r)} =0$
and adding appropriate singularities along the positions ${\bf r}_i$
of the vortex cores, one obtains the modified
London equation
\begin{eqnarray}  
  -\lambda^2 \nabla^2 {\bf B(r)} + {\bf B(r)} = \phi_0 \sum_i
  \! \int\! {\rm d}{\bf r}_i \,\delta_3({\bf r-r}_i)\,,
\end{eqnarray}
  where the line integral is taken along the $i$th flux line,
see equation (2.8) below. The three-dimensional (3D) delta function
$\delta_3({\bf r})$ in (2.4) is the London approximation to a
peaked function of finite width $\approx2\sqrt2\xi$ which
can be obtained from the isolated-vortex solution (1.2), (1.3)
of the GL theory. The assumption of non-interacting vortex cores
means that the ${\bf B(r)}$-dependence of this peaked
r.h.s.\ of (2.4) is disregarded, resulting in an inhomogeneous
linear differential equation. The energy of vacancies and
interstitials in the FLL was calculated from London theory by
Brandt (1969a, b) and Frey et al.\ (1994), see also the nice
London calculations of the stability of the twisted pair and the
twisted triplet in the FLL by Sch\"onenberger et al.\ (1995).
         \\[0.4 cm]
{\it 2.2.~ Straight parallel flux lines}
         \\*[0.2 cm]
  A good approximation to the GL interaction energy of an
arbitrary arrangement of straight parallel flux lines with flux
$\phi_0$ is per unit length (Brandt 1977a, 1986a)
\begin{eqnarray}  
 F_{\rm int}(\{{\bf r}_i\}) = \frac{\phi_0^2}{2\pi\lambda'^2 \mu_0}
    \sum_i \sum_{i>j} \Big[
    K_0\Big( \frac{|{\bf r}_i -{\bf r}_j|}{\lambda'} \Big) -
    K_0\Big( \frac{|{\bf r}_i -{\bf r}_j|}{\xi'    } \Big) \Big] \\
  \lambda'= \lambda/\langle |\psi|^2 \rangle^{1/2} \approx
  \lambda/(1-b)^{1/2}, ~~ \xi'= \xi/[2(1-b)]^{1/2}, ~~b=B/B_{c2}
\end{eqnarray}
where $K_0(r)$ is the Mac-Donald function, cf.\ equation (1.3), and
$\lambda'$ and $\xi'$ are an effective penetration depth and an
effective coherence length and $B = |\langle {\bf B(r)} \rangle|$
is the induction averaged over a few flux-line spacings.
  The first term in (2.5) is the repulsive magnetic interaction
(overlap of magnetic vortex fields) and the second term the
condensation-energy attraction (overlap of the vortex cores, see
also Kramer 1971). The expression (2.5)  with good accuracy
reproduces the dispersive elastic moduli $c_{11}$ and $c_{66}$
of the FLL (section 4.2) in the entire induction range
$0\le B < B_{c2}$ and entire $k$-space, and it reduces to the
London result for
$B\ll B_{c2}$. At $B\ll B_{c2}$ the vortex self-energy per unit
length $\phi_0 H_{c1}$ (with $H_{c1}=B_{c1}/\mu_0$) has to be added
to $F_{\rm int}$ (2.5) to obtain the total energy $F$.
At $B\approx B_{c2}$ the self-energy contribution is not defined
but it may be chosen such as to reproduce equation (1.7) in the case
of a periodic FLL. The fact that for $B/B_{c2}>0.25$ the {\it total}
energy of a flux-line arrangement cannot be written as a sum over
pair interactions plus a (structure-independent) self energy, is the
reason for the failure of Labusch's (1966) elegant method to
extract the shear modulus $c_{66}$ of the FLL from the magnetization
curve $M(H_a)$, which yields a monotonically increasing $c_{66}(B)$
rather than the correct $c_{66} \sim B(B_{c2} -B)^2$ (4.4).

  The magnetic field of arbitrary vortex arrangements for
$\kappa \gg1$ and $B\ll B_{c2}$ is a linear superposition of isolated
vortex fields  (1.3),
\begin{eqnarray}  
   B({\bf r}) = \frac{\phi_0}{2\pi\lambda^2} \sum_i K_0 \Big[
   \frac{ (|{\bf r-r}_i|^2 +2\xi^2)^{1/2}} {\lambda} \Big] \,.
\end{eqnarray}
         \\[0.4 cm]
{\it 2.3.~ Curved flux lines}
         \\*[0.2 cm]
  The structure-dependent part of the GL free energy of an
arrangement of arbitrarily curved flux lines is to a good
approximation (Brandt 1977b, Miesenb\"ock 1984, Brandt 1986a)
\begin{eqnarray}  
 F\{{\bf r}_i(z) \} = \frac{\phi_0^2}{8\pi \lambda'^2 \mu_0}
  \sum_i \sum_j \Big(\! \int\! {\rm d}{\bf r}_i\! \int\! {\rm d}{\bf r}_j
  \frac{\exp(-r_{ij}/\lambda')} {r_{ij}} - \!
                  \int\! |{\rm d}{\bf r}_i| \! \int\! |{\rm d}{\bf r}_j|
  \frac{\exp(-r_{ij}/\xi')} {r_{ij}} \Big)
\end{eqnarray}
 with $r_{ij} = |{\bf r}_i -{\bf r}_j|$. Parametrizing the vortex
lines as ${\bf r}_i = {\bf r}_i(z) =[\, x_i(z); y_i(z); z \,]$
one has ${\rm d}{\bf r}_i = (x_i'; y_i'; 1)\, {\rm d}z$
and may  write the line integrals in (2.8) as ({\bf figure} 6, left)
\begin{eqnarray}  
 \int\! {\rm d}{\bf r}_i \,... =
 \int\! {\rm d} z\, [\,x_i'(z); y_i'(z); 1\,]\,... \,, ~~~
 \int\! |{\rm d}{\bf r}_i| \,... =
 \int\! {\rm d} z\, [\,1 +x_i'(z)^2 +y_i'(z)^2 \,]^{1/2}... \,.
\end{eqnarray}
 Like in (2.5), the first term in (2.8) yields the magnetic repulsion
and the second term the core attraction of the interacting vortex
segments ${\rm d}{\bf r}_i$.
 For straight flux lines formula (2.8) reduces to (2.5); for small
flux-line displacements (2.8) reproduces the nonlocal linear
elastic energy  of section 4; and for $B\ll B_{c2}$ it yields the
London result derived from (2.4) with a GL core cutoff.

In contrast to (2.5), the general
expression (2.8) contains also the terms $i=j$, which yield the
self-energy of arbitrarily curved flux lines. In these
self-energy terms the distance $r_{ij}$ of the line elements
has to be cut off, e.g.\ by replacing the $r_{ij}$ with
$(r_{ij}^2 + 2\xi^2)^{1/2}$ for $\kappa \gg 1$ and $B \ll B_{c2}$.
 The self-energy of a curved flux line
of total length $L$ is approximately  $L \phi_0 H_{c1} = L
(\phi_0^2/4\pi \lambda^2 \mu_0) (\ln\kappa + 0.5)$, cf.\ equation (1.1).
In the same approximation the local magnetic field of this
vortex arrangement is obtained from (2.4) by Fourier transformation,
\begin{eqnarray}  
  {\bf B(r)} = \frac{\phi_0 }{4\pi\lambda^2} \sum_i \int\!
  {\rm d}{\bf r}_i \frac{\exp(-|{\bf r- r}_i| /\lambda )}
  {[\, ({\bf r-r}_i)^2 + 2\xi^2 \,]\,^{1/2} } \,.
\end{eqnarray}
The general field (2.10)  [and also (3.4) below] satisfies div{\bf B}=0.
The London limits $\xi\to 0$ of (2.8) and (2.10) were given
by Goodmann (1966) and Fetter (1967), see also the review on the FLL
by Fetter and Hohenberg (1969).
%
         \\[0.7 cm]
{\bf  3.~ Anisotropic and layered superconductors}
         \\*[0.5 cm]
         \setcounter{subsection}{3}  \setcounter{equation}{0}
{\it 3.1.~ Anisotropic London Theory}
         \\*[0.2 cm]
 The HTSC's are anisotropic oxides which to a good approximation
exhibit uniaxial crystal symmetry. In YBCO a weak
penetration-depth anisotropy in the $ab$-plane of 1.11 to 1.15 was
observed in decoration experiments by Dolan et al.\ (1989b).
Within the anisotropic extension of London theory, uniaxial
HTSC's are characterized by two magnetic penetration
depths for currents in the $ab$-plane, $\lambda_{ab}$, and
along the $c$-axis, $\lambda_c$, and by two coherence lengths
$\xi_{ab}$ and $\xi_c$. The anisotropy ratio
$\Gamma = \lambda_c/\lambda_{ab} = \xi_{ab}/\xi_c$ is
$\Gamma \approx 5$ for YBa$_2$Cu$_3$O$_{7-\delta}$ (YBCO),
$\Gamma > 150$ for Bi$_2$Sr$_2$CaCu$_2$O$_x$  (Bi2212)
 (Okuda et al.\ 1991, Martinez et al.\ 1992), and
$\Gamma \approx 90$ for Tl$_2$Ba$_2$CaCu$_2$O$_x$ (Gray et al.\ 1990).
One defines the GL parameter $\kappa =\lambda_{ab}/\xi_{ab}$.
The anisotropic London theory is obtained from anisotropic
GL theory in the limit $\xi\to 0$.
These theories yield the anisotropic critical fields $B_{c1}$ and
$B_{c2}$ as functions of the angle $\Theta$ between $B$ and the
$c$-axis, $B_{c1}(\Theta) \approx (\phi_0 /4\pi \lambda_{ab}^2
\Gamma) \epsilon(\Theta) \ln\kappa$ (Balatski\u{\i} et al.\ 1986,
see Klemm 1993, 1995 and Sudb{\o} et al.\ 1993 for improved expressions)
and $B_{c2}(\Theta) = [\phi_0 /2\pi \xi_{ab}^2 \epsilon(\Theta)]$ where
$\epsilon(\Theta) =  (\cos^2\Theta + \Gamma^{-2} \sin^2 \Theta)^{1/2}$
(Bulaevskii 1990).
  For a FLL in thermodynamic equilibrium, at large $B\gg B_{c1}$
the angle $\theta$ between $H_a$ and the $c$-axis practically
coincides with $\Theta$ since ${\bf B} \approx \mu_0 {\bf H}_a$.
In general one has $\theta < \Theta$ (Kogan 1988, Bulaevskii 1991).
For the isolated flux line ($B\to 0$) one has
$\epsilon(\theta) \approx 1/\epsilon(\Theta)$ and thus
$B_{c1}(\Theta) \approx B_{c1}(0)  \epsilon(\Theta)$ and
$B_{c1}(\theta) \approx B_{c1}(0) /\epsilon(\theta)$.

The modified London equation for the magnetic induction ${\bf B(r)}$
in a uniaxial superconductor containing arbitrarily
arranged straight or curved flux lines with one quantum of flux
$\phi_0$ follows by minimizing the free energy $F\{{\bf B}\}$
and adding appropriate singularities,
     \begin{eqnarray}  
    F\{{\bf B}\} ~ =~  \frac{1}{2\mu_0}
    \int \Big[\,{\bf B}^2 + (\nabla\!\times\! {\bf B})\,
   \Lambda\, (\nabla \! \times \!{\bf B}) \,\Big]~{\rm d}^3 r\,, \\
   {\bf B}+ \nabla\times [\,\Lambda \cdot (\nabla \times{\bf B})\,] =
   \phi_0\sum_i \int {\rm d}{\bf r}_i\; \delta_3({\bf r-r}_i) .
     \end{eqnarray}
  The tensor $\Lambda$  has the components
   $\Lambda_{\alpha\beta} = \Lambda_1 \delta_{\alpha\beta}$
  $+\Lambda_2 c_\alpha c_\beta$ where $c_\alpha$ are the
  cartesian components of the unit vector $\hat{\bf c}$ along the
crystaline $c$-axis,  $(\alpha,\beta)$ denote $(x,y,z)$,
  $\Lambda_1=\lambda^2_{ab}$, and $\Lambda_2
  =\lambda^2_c - \lambda^2_{ab}$.
     For  {\it isotropic} superconductors one has
   $\Lambda_1=\lambda^2$, $\Lambda_2=0$,
   $\Lambda_{\alpha\beta} = \lambda^2 \delta_{\alpha\beta}$, and
  the l.h.s.\ of (3.1) becomes ${\bf B} -\lambda^2  \nabla^2 {\bf B}$
  [equation (2.4)] since div${\bf B}=0$.
The integral in (3.2) is along the position of the $i$th
flux line as in  equations (2.4) and (2.8).

   Since the discovery of HTSC's, numerous authors have calculated
properties of the ideal FLL from anisotropic London theory:
The energy and symmetry of the FLL (Campbell et al.\ 1988,
Sudb{\o} 1992, Daemen et al.\ 1992); the
local magnetic field (Barford and Gunn 1988, Thiemann et al.\ 1988,
Gri\-shin et al.\ 1990); the magnetic stray field of a straight
 flux line parallel to the surface (Kogan et al.\ 1992) and
tilted to the surface  (Kogan et al.\ 1993a); the isolated vortex
(Klemm 1988, 1990, 1993, 1995, Sudb{\o} et al.\ 1993);
the elastic shear moduli (Kogan and Campbell 1989, Grishin et al.\ 1992,
Barford and Harrison 1994); magnetization curves (Kogan et al.\ 1988,
Hao et al.\ 1991); and the magnetization
in inclined field (Kogan 1988, Buzdin and Simonov 1991, Hao 1993),
which causes a mechanical torque from which the anisotropy ratio
$\Gamma$ can be obtained (Farrell et al.\ 1988, Farrell 1994,
Martinez et al.\ 1992).
  The modification of the flux-line interaction near a flat
surface was calculated for isotropic superconductors by Brandt (1981a)
and for anisotropic superconductors by Marchetti (1992).
  Hao and Clem (1992, 1993) show that for $H_a\gg H_{c1}$ the
angle-dependent thermodynamic and electromagnetic properties of HTSC's
with a FLL depend only on the reduced field
$h=H_a/H_{c2}(\theta, \phi)$. A further useful scaling approach
from isotropic to anisotropic results is given by
Blatter, Geshkenbein, and Larkin (1992). As clarified in a
Reply by Batter, Geshkenbein, and Larkin (1993), such scaling rules
are not unique, but the relevant issue in constructing a scaling
theory is the {\it consistency} requirement on the set of
scaling rules.

 The FLL in anisotropic HTSC's exhibits some unusual features:
(a) A field reversal and attraction of vortices occurring when $B$ is
at an oblique angle may lead to the appearance of vortex chains
(Grishin et al.\ 1990, Buzdin and Simonov 1990,
Kogan et al.\ 1990, Buzdin et al.\ 1991),which were observed
in YBCO at $B=25$\,Gauss by Gammel et al.\ (1992a).
Vortex chains embedded
in a perturbed FLL were observed by decoration at the surface of
inclined crystals of Bi$_2$Sr$_2$CaCu$_2$O$_{8+\delta}$ at
$B=35$\,Gauss
(Bolle et al.\ 1991, Grigorieva et al.\ 1995) and with
YBa$_2$(Cu$_{1-x}$Al$_x$)$_3$O$_{7-\delta}$
at $B=10$\,Gauss (Grigorieva et al.\ 1993).
These chains may possibly be explained by
the {\it coexistence of two interpenetrating vortex lattices}
parallel and perpendicular to the $ab$-plane, which may exhibit
lower energy than a uniformly tilted FLL. The existence of this
{\it combined lattice} in layered superconductors was discussed
(Theodorakis 1990, Huse 1992, Bulaevskii et al.\ 1992a,
Daemen et al.\ 1993b,
Feinberg and Ettouhami 1993a, Feinberg 1994) and calculated
explicitly (Benkraouda and Ledvij 1995), and it was found also
within anisotropic London theory
(Preosti and Muzikar 1993, Doria and de Oliveira 1994b).
(b) The magnetic forces between straight vortices in general are
no longer central forces (Kogan 1990).
(c) Shear instabilities of the FLL with $B$ in or near the $ab$-planes
    were predicted by Ivlev et al.\  (1990, 1991a).
These are related to the metastable states or frustration-induced
disorder of flux lines in layered superconductors
(Levitov 1991, Watson and Canright 1993).
(d) Ivlev and Campbell (1993) show that flux-line chains between
twin boundaries are unstable and buckle.
(e) Multiple $B_{c1}$ values in oblique field were predicted
    by Sudb{\o} et al.\  (1993).

Flux lines in nonuniaxially anisotropic superconductors were
calculated from London theory by Schopohl and Baratoff (1988)
and by Burlachkov (1989). Properties of the anisotropic FLL
were calculated from the GL theory e.g.\ by Takanaka (1975),
Kogan and Clem (1981), Petzinger and  Warren (1990), and
 Petzinger and Tuttle (1993). Using microscopic Eliashberg theory,
Teichler (1975) and Fischer and Teichler (1976) explained the
square FLL which was observed by decoration in some materials
(Obst 1971, Obst and Brandt 1978), as due to atomic lattice
anisotropy, see also the book by Weber (1977).
       \\[0.4 cm]
{\it 3.2.~ Arrangements of parallel or curved flux lines}
         \\*[0.2 cm]
The general solution of (3.1) and (3.2) for the induction ${\bf B(r)}$
and the free energy $F\{{\bf r}_i\}$ for arbitrary vortex
arrangements explicitly read (Brandt 1990c, 1991a, b, 1992a)
       \begin{eqnarray}  
   F\{{\bf r}_i\} &=& \frac{\phi_0^2}{2\mu_0} \sum_i \sum_j
        \int \!{\rm d}{\bf r}_i^\alpha  \int \!{\rm d}{\bf r}_j^\beta
            ~f_{\alpha \beta}({\bf r}_i - {\bf r}_j)~
         \\
   B_\alpha({\bf r})  &=& \phi_0  \sum_i  \int \!{\rm d}{\bf r}_i^\beta
     ~f_{\alpha \beta}({\bf r} - {\bf r}_i)~
         \\
   f_{\alpha \beta}({\bf r})  &=&  \int \! \exp(i{\bf kr})
       f_{\alpha\beta} ({\bf k}) ~   \frac{{\rm d}^3k}{8\pi^3}~
         \\
   f_{\alpha \beta}({\bf k})  &=&   \frac{\exp[ -2g(k,q)\,]}
     {1 + \Lambda_1 k^2} \left( \delta_{\alpha\beta} -
   \frac{q_\alpha ~q_\beta ~\Lambda_2}{1 +\Lambda_1 k^2
   +\Lambda_2 q^2}  \right)
         \\
   g(k,q) &=& \xi_{ab}^2 q^2 +\xi_c^2(k^2 -q^2)
          ~=~ (\Lambda_1 k^2 + \Lambda_2 q^2)/(\Gamma^2 \kappa^2)
        \end{eqnarray}
   where ${\bf q=k\times \hat{c}}$,  ${\bf\hat{c}}$ is the unit vector
along the $c$-axis,  $\Lambda_1 =\lambda_{ab}^2$, and
$\Lambda_2 =\lambda_c^2 -\lambda_{ab}^2 \ge 0$.
The integrals in (3.3) and (3.4) are along the $i$th and $j$th
flux lines ({\bf figure} 6, left) and the integration in (3.5)
is over the infinite $k$-space.
The tensor function $f_{\alpha\beta}({\bf r})$ (3.5) enters both the
interaction potential between the line elements of the flux lines
in (3.3) and the source field generated by a vortex segment
${\rm d}{\bf r}_i$ in (3.4).
Due to the tensorial character of $f_{\alpha\beta}$ this source
field in general is not parallel to ${\rm d}{\bf r}_i$.

The factor $\exp(-2g)$  in (3.6) provides an
elliptical cutoff at large ${\bf k}$ or small ${\bf r}$
and can be derived from GL theory with
anisotropic  coherence length ($\xi_c = \xi_{ab}/\Gamma$).
This cutoff, also derived by Klemm (1993, 1995), originates from the
finite vortex-core radius in the GL theory and means that the 3D delta
function $\delta_3({\bf r})$ in (3.1) is replaced by a 3D
Gaussian of width $\sqrt2 \xi_{ab}$ along $a$ and $b$ and
width $\sqrt2 \xi_c$ along the $c$-axis.
Note that this GL cutoff exhibits the correct anisotropy; choosing
a cutoff with a different symmetry may lead to artificial helical
or tilt-wave instabilities of the flux lines at very short
wavelengths (Sardella and Moore 1993).  Tachiki et al.\ (1991)
mention another trap which yields an erroneous spiral instability.
 As shown by Carneiro et al.\ (1992), in the energy integral (3.3)
[but not in the magnetic field (3.4)] the tensorial
potential (3.6) formally may be replaced by a diagonal potential
since the r.h.s.\ of (3.1) is divergence-free because
vortex lines cannot end inside the bulk.
Kogan et al.\ (1995) consider vortex--vortex interaction via
material strains.

  For {\it isotropic} superconductors one has
$\lambda_{ab} = \lambda_c =\lambda$, thus $\Lambda_2 =0$  and
$f_{\alpha\beta} = \delta_{\alpha\beta}$ $ \exp(-r/\lambda)
/(4\pi r \lambda^2)$; this means the source field is now
spherically symmetric, and equations (3.3), (3.4) reduce to
equations (2.8), (2.10)  (with different cut-off). In the
{\it isotropic} case the magnetic interaction between two line
elements contains  the scalar product
${\rm d}{\bf r}_i^\alpha {\rm d}{\bf r}_j^\beta \delta_{\alpha\beta}$
$={\rm d}{\bf r}_i \cdot{\rm d}{\bf r}_j=
    {\rm d}r_i\cdot {\rm }{\rm d}r_j\,\cos\phi$
where $\phi$ is the angle between these vortex segments; the
repulsive interaction thus {\it vanishes} at $\phi = \pi/2$ and
becomes even attractive when $\phi >\pi/2$. For {\it anisotropic}
superconductors with $\Gamma \gg 1$ and symmetric tilt of the two
flux lines about the $c$-axis, this magnetic attraction begins
at larger angles $\phi = 2\Theta$ close to $\pi$, i.e.\ when the
flux lines are nearly antiparallel and almost in the $ab$-plane.
The spontaneous tilting of the vortices, which prefer to be in
the $ab$-plane and antiparallel to each other, facilitates that
vortices cut each other (Sudb{\o} and Brandt 1991c).
         \\[0.4 cm]
{\it 3.3.~ Layered superconductors}
         \\*[0.2cm]
The HTSC's oxides consist of superconducting Cu-O layers (or multiple
Cu-O layers) of spacing $s$ and thickness $d\ll s$ which interact
with each other by weak Josephson coupling
(Lawrence and Doniach 1971, Klemm et al.\ 1975,
 Bulaevskii 1972, Deutscher and Entin-Wohlman 1978), see the reviews
by Bulaevskii (1990) and Latyshev and Artemenko (1992) and the
book by Klemm (1996).  This layered structure causes two novel
phenomena: pancake vortices and Josephson strings, {\bf figure} 7.

{\it 3.3.1. Pancake vortices.}~
 When $B_a$ is nearly along the $c$-axis the flux lines formally
consist of a stack of 2D point vortices, or {\it pancake vortices}
(Efetov 1979, Artemenko and Kruglov 1990, Guinea 1990, Clem 1991,
Buzdin and Feinberg 1990, Fischer 1991, Genenko 1992)
which have their singularity
(the zero of $\psi$ and infinity of $\nabla \phi$) only
 in {\it one} layer.  The field of a single point vortex
with centre at ${\bf r}_i =0$ is confined to a layer of thickness
$\approx 2\lambda_{ab}$. Its rotationally symmetric $z$-component
and its in-layer radial component are
       \begin{eqnarray}  
B_z(r) &=& (s\phi_0/4\pi\lambda_{ab}^2 r) \exp(-r/\lambda_{ab})
      \\
B_\perp({\bf r}) &=& (s\phi_0 z / 4\pi\lambda_{ab}^2 r_\perp)
[\exp(-|z|/\lambda_{ab}) /|z| - \exp(-r/\lambda_{ab}) /r ]
       \end{eqnarray}
 where ${\bf r}_{\perp} =(x; y)$ and ${\bf r} = (x;y;z)$.
The field (3.8), (3.9) satisfies div{\bf B}=0 as it should be.
In a vortex {\it line}, the radial components of the point-vortex
fields cancel (if $s\ll \lambda$) and only the $z$-components
survive.  A point vortex in plane $z=z_n$ contributes a flux
$\phi(z_n) =(\phi_0 s/2\lambda_{ab}) \exp(-|z_n|/\lambda_{ab})$
 to the flux through the plane $z=0$.
This means the flux of one point vortex is $\ll \phi_0$, but the
sum of all $\phi(z_n)$ along a stack yields $\phi_0$.
The self-energy of a single point vortex diverges
logarithmically with the layer extension, but this fact does not
disturb since the energy of a stack of such point vortices,
a vortex line, has an energy {\it per unit length} and  also
diverges for an infinitely long vortex. Remarkably,
 the interaction between  point vortices is
exactly logarithmic up to arbitrarily large distances,
in contrast to the interaction of point vortices in an isolated
film (Pearl 1964, 1966), which is logarithmic only up to distances
$\lambda_{\rm film} = 2\lambda_{\rm bulk}^2/d$ where $d$ is the
film thickness. Point vortices in the same layer repel each
other, and those in different layers attract each other. This is
the reason why a regular lattice of straight flux lines has
the lowest energy. The pancake vortex in a superconducting
layer squeezed between two infinite superconducting half spaces
is calculated in detail by Ivanchenko et al.\ (1992).
Pudikov (1993) shows that the logarithmic interaction of the
pancake-vortices applies for any number of layers $N>1$,
and that the interaction is softer near the specimen surface.
Theodorakis and Ettouhami (1995) calculate the undulating or
``sausaging'' vortex along $z$ when $T_c(z)$ varies periodically to
model a layered system with a finite order parameter between
 the layers.

{\it 3.3.2. Josephson strings.}~ When the applied field is nearly
along the $ab$-plane the vortex core prefers to run between the
Cu-O layers (Schimmele et al. 1988, Tachiki and Takahashi 1989, 1991,
Tachiki et al.\ 1991).
When the coupling between the layers is weak, the vortex lines along
the $ab$-plane are so called {\it Josephson vortices} or
{\it Josephson strings}. These have no core in the usual sense of a
 vanishing  order parameter $\psi$ since $\psi$ is assumed to
be zero in the space between the layers anyway.
The circulating vortex current thus has to tunnel across the
isolating space between the layers.
The width of the Josephson core is $\lambda_J=\Gamma s$ (the
Josephson length, Barone and Paterno 1982) and its thickness is $s$.
Its aspect ratio  $\lambda_J /s = \Gamma= \xi_{ab}/\xi_c$ is thus
the same as for the core of a London vortex in the $ab$-plane with
width $\xi_{ab}$ and thickness $\xi_c$. The vortex structure
and anisotropy of the penetration depth in various
superconducting/normal multilayers was calculated by
Krasnov et al.\ (1993).

{\it 3.3.3. Lock-in transition and oscillation.}~
When the angle $\Theta$ between the $c$-axis and the applied field
$H_a$ is close to $\pi/2$, the flux lines form steps or kinks
consisting of point vortices separated by Josephson vortices
(Ivlev et al.\ 1991b), see the reviews by Feinberg (1992, 1994),
the work on kink walls by Koshelev (1993), and computer simulations
by Machida and Kaburaki (1995).
If it is energetically favorable and not impeded
by pinning, the flux lines in this case will spontaneously
switch into the $ab$-plane. This lock-in transition  (Feinberg
and Villard 1990, Theodorakis 1990, Feinberg and Ettouhami 1993b,
Koshelev 1993) was observed in HTSC's by Steinmeyer et al.\ (1994a, b),
 Vulcanescu et al.\ (1994), and Janossy et al.\ (1995)
 in mechanical torque experiments,
and in an organic layered superconductor by Mansky et al.\ (1993)
from ac susceptibility.
It means that the perpendicular component of the applied field
should only penetrate above some critical angle, for
$|\pi/2 -\Theta| \ge (1-N)B_{c1,z} /B_a$, where $N$ is the
demagnetization factor.
The anisotropic magnetic moment of layered HTSC's and the mechanical
torque were calculated by Bulaevskii (1991) and measured, e.g.\
by Farrell et al.\ (1989) and Iye et al.\ (1992).
In a nice experiment Oussena et al.\ (1994b) in untwinned YBCO
single crystals with $B$ in the $ab$-plane observe lock-in
oscillations as a function of $B$ which are caused by matching
of the FLL with the Cu-O layers.  In a nice theory
Koshelev and Vinokur (1995) calculate corrections to the resistivity
when the flux lines form a small angle with the layers,
see also section 8.2.7.

{\it 3.3.4. Motion of Josephson vortices.}~
  The viscous drag force on moving Josephson vortices is calculated by
Clem and Coffey (1990), extending the Bardeen-Stephen (1965) model.
Gurevich (1992a, 1993, 1995a), Aliev and Silin (1993),
and Mints and Snapiro (1994, 1995) find nonlocal electrodynamics
and nonlinear viscous
motion of vortices moving along planar defects: Vortices in long
Josephson contacts (characterized by a coupling strength or critical
current density $J_c$) are usually described by the sine-Gordon
equation for the phase difference $\phi(x,t)$; but when $\phi$ varies
over lengths shorter than the London penetration depth $\lambda$,
then a nonlocal description is required, which has as solutions
Josephson vortices (section 3.3.2.) or Abrikosov vortices with
Josephson core (Gurevich 1995a), depending on the value of $J_c$.
In the static case the equations describing this nonlocal Josephson
electrodynamics turn out to be similar to the equations describing
crystal-lattice dislocations in a Peierls potential (Seeger 1955),
allowing one to use results of dislocation theory to obtain
periodic solutions which describe vortex structures in
high-$J_c$ Josephson contacts (Alfimov and Silin 1994)
       \\[0.4 cm]
{\it 3.4.~ The Lawrence-Doniach Model}
         \\*[0.2 cm]
A very useful phenomenological theory of layered superconductors
is the Lawrence-Doniach (LD) theory (Lawrence and Doniach 1971),
which contains the anisotropic GL and London theories
as limiting cases when the coherence length
$\xi_c(T)$ exceeds the layer spacing $s$. The LD model
defines for each superconducting layer a two-dimensional (2D)
GL parameter $\psi_n(x,y)$  ($n$ = layer index, $z_n = ns$)  and
replaces the gradient along $z \| c$ by a
finite difference. The LD free energy functional reads
  \begin{eqnarray}     
  F\{\psi_n({\bf r}_\perp), {\bf A(r) }\}
     &=& s\! \sum_n \!\int\! \! {\rm d}^2r_\perp \!\Big[
   - \alpha|\psi_n|^2 +\frac{\beta}{2}
    |\psi_n|^4  +\frac{\hbar^2}{2m} |(i\nabla +\frac{2e}{\hbar}
            {\bf A} )\psi_n|^2    \nonumber \\
    ~&~&+~\frac{\hbar^2}{2Ms^2} |\psi_{n+1} -\psi_n \exp(i I_n)|^2
      \Big]
     +\int \frac{{\bf B}^2}{2\mu_0} \, {\rm d}^3r  \\
 I_n({\bf r_\perp}) &=& \frac{2e}{\hbar} \!\int_{ns}^{ns+s}
      \!\!\! A_z({\bf r_\perp},z) \, {\rm d}z \, ~\approx~ \,
      \frac{2e}{\hbar} s A_z ({\bf r_\perp}, ns)
      ~~\mbox{ (if $s \ll\lambda$)}\,.
      \end{eqnarray}
  Here ${\bf r_\perp} = (x;y)$, $\alpha \propto T_c -T$ and
  $\beta$ are the usual GL coefficients,
  $m$ and $M$  are the effective masses of Cooper pairs moving in
the $ab$-plane or along $c$,
  ${\bf B} = {\rm rot} {\bf A}$,
  $2e/\hbar = 2\pi/\phi_0$,  $A_z =\hat{\bf z} {\bf A} $,
  and the integral in (3.11) is along a straight line. One has
  $\xi_{ab}^2 = \hbar^2/2m\alpha$,  $\xi_c^2 =\hbar^2 /2M\alpha$,
  $\lambda_{ab}^2  =m \beta/4\mu_0 e^2 \alpha$,
  and $\lambda_c^2 =M \beta/4\mu_0 e^2 \alpha$, thus
  $\lambda_c^2 /\lambda_{ab}^2 = \xi_{ab}^2/\xi_c^2 =M/m=\Gamma^2
  \gg 1$ for HTSC's. For $\xi_c \gg s$  the difference in (3.10) may be
  replaced by the gradient $ ( \partial / \partial z - 2ieA_z/\hbar)
  \psi(x,y,z)$, and the anisotropic GL theory is recovered.
  In the opposite limit, $\xi_c\ll s$, (3.10) describes weakly
  Josephson coupled superconducting layers.
  An extension of the LD model to non-vanishing order parameter
between the layers (i.e.\ to alternating superconducting and
normal layers coupled by proximity effect) was recently given by
Feinberg et al.\ (1994), see also Abrikosov et al.\ (1989).

The LD theory can be derived from a tight binding formulation of the
BCS theory near $T_c$ (Klemm et al.\ 1975, Bulaevski 1990), but like
many phenomenological theories the LD theory is more general than the
microscopic theory from which it may be  derived.
In particular, if $\psi_n(x,y) =$ const$\times \exp[i \phi_n(x,y)]$
is assumed, the resulting London-type LD theory applies
to all temperatures $0<T<T_c$ (Feigel'man et al.\ 1990),
       \begin{eqnarray} 
  F\{ \phi_n({\bf r_\perp}), {\bf A(r)} \} \!
   &=& \!\frac{s \phi_0^2}{4\pi^2 \mu_0 \lambda_{ab}^2} \sum_n
   \!\int\! \! {\rm d}^2r_\perp \Big[ \frac{1}{2}({\bf \nabla}\phi_n
 + \frac{2\pi}{\phi_0} {\bf A} )^2 + \frac{ 1- \cos\delta_n}
 {\Gamma^2 s^2} \Big] + \!\int \!\frac{{\bf B}^2}{2\mu_0}\,{\rm d}^3r
        \nonumber \\
     \delta_n({\bf r}_\perp)\! &=&\! \phi_n - \phi_{n+1} - I_n .
      \end{eqnarray}
  The magnetic field of an arbitrary arrangement of point
vortices at 3D positions ${\bf r}_\mu$ is in the limit of zero
Josephson coupling, corresponding to the limit $\lambda_c \to\infty$,
   \begin{equation}  
{\bf B(r)} = s\phi_0 \int\frac{{\rm d}^3k}{8\pi^3}\,\,
 \frac{\hat{\bf z} - {\bf k}_\perp k_z /k_\perp^2}
  {1+k^2\lambda_{ab}^2}
 \sum_\mu \exp[\, i{\bf k(r - r}_\mu)  \,
  -2\xi_{ab}^2 k_\perp^2 ]
   \end{equation}
where ${\bf k = (k}_\perp; k_z)$,  ${\bf k}_\perp = (k_x; k_y)$,
${\bf \hat{z} = \hat{c}}$ , and I have introduced the cutoff
from (3.6), (3.7) [note that $ k_\perp^2=q^2$ since
${\bf q=k\times\hat c}$ in (3.6)].
Even for finite $\lambda_c$ the field (3.13) is a good
approximation when $B_a$ is along $z$  or close to it and may
be used to calculate the field variance of arbitrary
arrangements of point vortices (Brandt 1991c).
The  free energy of a system of  point-vortices at positions
${\bf r}_\mu =(x_\mu; y_\mu; z_\mu)$ is
composed of their magnetic interaction energy $F_M$
and the Josephson coupling energy $F_J$ caused by the difference
$\delta_n$ (3.12) of the phases $\phi_n$ of $\psi_n$ in adjacent
 layers, $F = F_M + F_J$, with
   \begin{eqnarray}  
 F_M    &=& \frac{s^2\phi_0^2}{2\mu_0} \int\frac{{\rm d}^3k}
  {8\pi^3}\,\, \frac{k^2 / k_\perp^2}{1+k^2\lambda_{ab}^2}
 \sum_\mu \sum_\nu \exp[\, i{\bf k(r}_\mu - {\bf r}_\nu) \,
  -2\xi_{ab}^2 k_\perp^2 ]
             \\
 F_J    &=& \frac{\phi_0^2}{4\pi^2\mu_0 s \lambda_c^2 }
  \int \!\!{\rm d}x\, \int \!\!{\rm d}y\,\,\sum_n
       ~(\,1 - \cos \delta_n \,) \,.
   \end{eqnarray}
In general, $F_J$ is not easily calculated since the
gauge-invariant phase differences have to be determined
by minimizing $F_M + F_J$.
The vortex lattice with ${\bf B}$ parallel to the layers was
calculated analytically (Bulaevskii and Clem 1991) and numerically
(Ichioka 1995).
Solutions of the LD theory for a single straight flux line and
for a periodic FLL for all orientations of {\bf B} have been obtained
by Bulaevskii et al.\ (1992a), Feinberg (1992, 1994), and
Klemm (1995, 1996). The energy of a distorted vortex line was
calculated by Bulaevskii et al.\ (1992b); the interaction between two
pairs of point vortices (four-point interaction) entering that energy,
corresponds to the interaction of two vortex line elements
${\rm d}{\bf r}$ in the London energy, equations (2.8) and (3.3),
see {\bf figure} 6, left.
Phase transitions caused by vortices and by the layered structure
of HTSC's are discussed in section 6.
       \\[0.4 cm]
{\it 3.5.~ Josephson currents and helical instability}
         \\*[0.2 cm]
Within the LD Model the current flowing perpendicular to the layers
is a Josephson current, which tunnels across the insulating layer
and which is determined by the phase difference $\delta_n$ (3.12)
between neighbouring superconducting layers. From the LD energy
(3.12) follows the Josephson relation  (Josephson 1962) for this
current density,
   \begin{equation}  
 J_n(x,y) = J_{c0} \sin\delta_n(x,y), ~~~~
 J_{c0} =\phi_0 /(2\pi s \lambda_c^2 \mu_0),
   \end{equation}
where $J_{c0}$ is the critical current density along the $c$-axis.
Thermal fluctuations of $\delta_n({\bf r}_\perp)$ reduce the
loss-free current which can flow along $c$ (sections 6.3.\ and 7.4.)
(Glazmann and Koshelev 1990, 1991b, Genenko and Medvedev 1992,
Daemen et al.\ 1993a, Ioffe et al.\ 1993).
Considering the depression of the order parameter $|\psi({\bf r})|^2$
by the current $J_n$, Genenko et al.\ (1993) obtain a generally
non-sinusoidal relation between this current and the
phase difference $\delta_n$.

Intrinsic Josephson effects between the Cu-O planes in BSCCO
single crystals were observed by Kleiner et al.\ (1992, 1994)
and M\"uller (1994) in intricate experiments.
Fraunhofer oscillations as a function of $H_a$ in HTSC's with
a width in the $ab$-plane not exceeding the Josephson length
$\lambda_J = \Gamma s$ were predicted by Bulaevskii et al.\ (1992e).

  Interestingly, a critical current density of the same order as the
decoupling current density $J_{c0}$ (3.16) results from anisotropic
London theory if one considers the helical instability
of a flux line (or a stack of point vortices) in the presence of
a longitudinal current of density $J_{\|}$, cf.\ section 7.2.
By a spontaneous helical distortion
${\bf u}(z) = u_0 ( \hat{\bf x}\cos kz +\hat{\bf y}\sin kz)$
a flux line gains an energy from $J_\|$ which may exceed the
elastic energy of this distortion. The total energy per unit
length of the helix is $(u_0^2/2) [\,k^2 P(k) -kJ_\| \phi_0\,]$
where $P(k)$  is the flux-line tension, equation (4.13) below.
An unpinned flux line (Clem 1977) or FLL (Brandt 1980c, 1981b) is
thus unstable with respect to a helical deformation with long pitch
length $2\pi/k$ at arbitrarily small $J_\| > P(0) k/\phi_0 $, but
even weak pinning may suppress this ``inflation'' of a helix.
However, due to the dispersion of $P(k)$ (nonlocal elasticity)
a second instability occurs at the shortest possible pitch length
$2s$, i.e.\ at $k=\pi/s$, where $s$ is the layer spacing. With
$P(k) \approx \phi_0^2/(4\pi \mu_0 \lambda_c^2) $
[putting $\ln(\Gamma /k\xi_{ab}) = \ln( s/\pi \xi_{c})
 \approx 1$ in (4.14)\,] this helical instability of a stack of
point vortices occurs at a longitudinal current density
$J_{c \|} = \phi_0 /(4 s \lambda_c^2 \mu_0) \approx J_{c0}$.
A longitudinal current of the critical value $J_{c0}$ thus,
in zero magnetic field, suppresses superconductivity; but in the
presence of a longitudinal field or of vortices, this current
``blows up'' the flux line into independent point vortices.
During this instability, the point vortices in adjacent layers are
torn apart by the Josephson string which connects them and which
deforms into a circle with growing radius.
%
         \\[0.7 cm]
{\bf  4.~ Elasticity of the flux-line lattice}
         \setcounter{subsection}{4}  \setcounter{equation}{0}
         \\*[0.5 cm]
{\it 4.1.~ Elastic moduli and elastic matrix}
         \\*[0.2 cm]
  From the general expressions (2.5) and (2.8) for the energy of a
distorted FLL, or from equation (3.14) for the magnetic energy of point
vortex arrangements, in principle one can calculate arbitrarily large
vortex displacements caused by pinning forces, by structural
defects in the FLL, by field gradients or currents, by
temperature gradients, or by thermal fluctuations.
However, in many cases, the distortion is so small that it may be
calculated from linear elasticity theory.
The linear elastic energy $F_{\rm elast}$ of the FLL is expressed
most conveniently in $k$-space. First, the displacements
of the flux lines ${\bf u}_i(z) = {\bf r}_i(z) - {\bf R}_i
= (u_{i,x}; u_{i,y}; 0)$ from their ideal positions
${\bf R}_i = (X_i; Y_i; z)$ are expressed by their Fourier
components
       \begin{eqnarray} 
  {\bf u}_i (z) =\int_{\rm BZ}\! \frac{{\rm d}^3 k}{8\pi^3}\,
  {\bf u(k)} \,e^{i{\bf kR}_i},~~~
  {\bf u(k)} =\frac{\phi_0}{B} \sum_i \int\! {\rm d}z\,
  {\bf u}_i(z) \,e^{-i{\bf kR}_i}
       \end{eqnarray}
and then the most general quadratic form in the
${\bf u(k)} =(u_x; u_y; 0)$ is written down,
       \begin{eqnarray} 
  F_{\rm elast} = \frac{1}{2} \int_{\rm BZ}\! \frac{{\rm d}^3 k}{8\pi^3}
  \,u_\alpha({\bf k})\, \Phi_{\alpha\beta}({\bf k})\, u_\beta^*({\bf k})
       \end{eqnarray}
with $(\alpha,\beta) = (x,y)$. The $k$-integrals in (4.1) and
(4.2) are over the first Brillouin zone (BZ) of the FLL and
over $-\xi^{-1} \le k_z \le \xi^{-1}$. The coefficient
$\Phi_{\alpha\beta}({\bf k})$ in (4.2), called elastic matrix,
is real, symmetric, and periodic in $k$-space. Within continuum
theory of this uniaxial elastic medium, $\Phi_{\alpha\beta}$
is related to the elastic moduli
$c_{11}$ for uniaxial compression, $c_{66}$ for shear, and $c_{44}$
for tilt, by ({\bf figure} 8)
       \begin{eqnarray} 
  \Phi_{xx} &=& c_{11} k_x^2 + c_{66}k_y^2 + c_{44}k_z^2
                             + \alpha_L({\bf k})\nonumber\\
  \Phi_{yy} &=& c_{66} k_x^2 + c_{11}k_y^2 + c_{44}k_z^2
                             + \alpha_L({\bf k})\nonumber\\
  \Phi_{xy} &=& \Phi_{yx} = (c_{11}-c_{66})k_x k_y \,
       \end{eqnarray}
or in compact form,
 $\Phi_{\alpha\beta} = (c_{11} - c_{66})k_\alpha k_\beta +
  \delta_{\alpha\beta} [\,(k_x^2 +k_y^2)c_{66} + k_z^2 c_{44}
  + \alpha_L({\bf k}) \,]$.  For completeness in (4.3) the Labusch
parameter $\alpha_L$ is added, which describes the elastic
interaction of the FLL with the pinning potential caused by
material inhomogeneities (section 8.2.). A $k$-independent
$\alpha_L$ would mean that all flux lines are pinned
individually (Labusch 1969b). For weak collective pinning
(Larkin and Ovchinnikov 1979) $\alpha_L({\bf k})$ should decrease
when $k_\perp > R_c^{-1}$ or $k_z >L_c^{-1}$  where $R_c$ and
$L_c \approx (c_{44}/c_{66})^{1/2} R_c$ are the radius and length
of the coherent (short-range order) regions of the pinned FLL
(section 8.7).

For {\it uniform} distortions the elastic moduli of
the triangular FLL are (Labusch 1969a, Brandt 1969a, b, 1976c, 1986)
       \begin{eqnarray} 
  c_{11} - c_{66} &=& B^2 \partial^2 F/\partial B^2
            =~ (B^2 /\mu_0) \partial B_a/\partial B \nonumber \\
  c_{44} &=& B \partial F/\partial B ~~~~=~ B B_a/\mu_0 \nonumber \\
  c_{66} &\approx&  (B\phi_0/16\pi \lambda^2\mu_0) (1-\frac{1}{2\kappa^2})
           (1-b)^2 (1- 0.58b +0.29b^2)
       \end{eqnarray}
with $b=B/B_{c2}$. Here $c_{11} -c_{66}$ is the modulus for
isotropic compression and $B_a/\mu_0 =  \partial F/\partial B$
is the applied field which is in equilibrium with the FLL at
induction $B$. The last two factors in $c_{66}$ show that the
shear modulus vanishes when $\kappa = 1/\sqrt 2$ (in this special
case all flux-line arrangements have the same energy), when
$\lambda \to \infty$ (strongly overlapping vortex fields),
and when $B\to B_{c2}$ (strongly overlapping vortex cores).
In addition $c_{66}$ vanishes at the melting transition of the
FLL (\v{S}\'a\v{s}ik and Stroud 1994b, Franz and Teitel 1994, 1995).
 For $B\ll \phi_0/4\pi\lambda^2 \approx B_{c1}/\ln\kappa$
 one has $c_{11} = 3c_{66} \sim \exp(-a/\lambda)$
(Cauchi relation for nearest neighbour interaction by central forces)
and $c_{44} = BH_{c1} \gg c_{11}$.
For $B\gg \phi_0/4\pi\lambda^2$ one has
$c_{66} \ll c_{11} \approx c_{44} \approx B^2/\mu_0$
(incompressible solid). The shear modulus $c_{66}$ of the 2D FLL in
thin isotropic films was calculated by Conen and Schmid (1974),
the bulk $c_{66}(B,T)$ is given by Brandt (1976c), and the elastic
properties of anisotropic films were calculated by
 Martynovich (1993).

  In uniaxially {\it anisotropic} superconductors the free energy
$F(B,\Theta)$ depends explicitly on the angle $\Theta$ between $B$
and the $c$-axis, and thus on the tilt angle $\alpha$ if the tilt
is not in the $ab$-plane. One then has for uniform tilt
       \begin{eqnarray} 
c_{44} = {\rm d}^2F/{\rm d}\alpha^2 = B B_a/\mu_0 +
         \partial^2 F /\partial\alpha^2 \,.
       \end{eqnarray}
The first term in (4.5) originates from the compression of the
field lines when $B_z$ is held constant during the tilt,
$B(\alpha) = B_z /\cos\alpha \approx B_z (1+\alpha^2/2)$.
In the special case where the equilibrium direction $z$ is
along the crystalline $c$-axis one has $\alpha = \Theta$.
In HTSC's $F(\Theta)$ decreases with increasing angle $\Theta$
since the flux lines prefer to
run along the $ab$-plane. In this case the second term in
(4.5) is negative. For large anisotropy and low flux density
$B\ll B_{c1}$ with $B\| c$, the second term nearly compensates the
first (isotropic) term such that $c_{44}$ becomes very small
(section 4.3). However, for larger $B\gg B_{c1}$ the modulus for
uniform tilt is almost isotropic since the free energy of the FLL is
$F \approx B^2/\mu_0 \approx B_a^2/\mu_0$ [\,cf.\ equations (1.7)
and (3.1)\,], which does not depend on any material parameter and
is thus isotropic.  The anisotropic part of $F$ is proportional to
the magnetization $M = B -B_a$, which is a small correction when
$B\gg B_{c1}$. The magnetization in HTSC's strongly
depends on $\Theta$ such that the mechanical torque exerted by
an applied field is sharply peaked when $B_a$ is close to the
$ab$-plane (Kogan 1988, Bulaevskii 1991, Farrell et al.\ 1988,
Okuda 1991, Martinez et al.\ 1992, Iye et al.\ 1992,
Doria and de Oliveira 1994a).
         \\[0.4 cm]
{\it 4.2.~ Nonlocal elasticity}
         \\*[0.2 cm]
The continuum approximation (4.3) for the elastic matrix in general
applies if $k_\perp^2 = k_x^2 +k_y^2 \ll k_{\rm BZ}^2 = 4\pi B/\phi_0$;
as $k_\perp$ approaches $k_{\rm BZ}$, the quadratic $k$-dependence ceases
to hold since  $\Phi_{\alpha\beta}$ has to be
periodic in $k$-space, $\Phi_{\alpha\beta}({\bf k})
=\Phi_{\alpha\beta}({\bf k +K})$  where ${\bf K}$ are the vectors of
the reciprocal FLL, cf.\ (1.4) and {\bf figure} 8. However, for the FLL
the direct calculation of the elastic matrix from GL and London
theories reveals that the quadratic $k$-dependence holds only in the
smaller region $k \ll \lambda^{-1}$, which typically means
a very small central part of the BZ and a small fraction of the
range $0 \le |k_z| \le \xi^{-1}$ (Brandt 1977a).  Within continuum
approximation one may account for this ``elastic nonlocality''  by
introducing $k$-dependent (dispersive) elastic moduli in (4.3).
For isotropic superconductors at $B>B_{c1}$ one finds
for $\kappa \gg 1$, $b=B/B_{c2} \ll 1$ (London limit),
 and $k=(k_x^2 +k_y^2 +k_z^2)^{1/2} \ll k_{\rm BZ}$ (continuum limit),
 \begin{eqnarray} 
 c_{11}(k) \approx c_{44}(k) \approx(B^2/\mu_0)(1+k^2\lambda^2)^{-1}
 \end{eqnarray}
and a $k$-independent $c_{66}$ (4.4).
The physical reason for the strong dispersion of $c_{11}$ and
$c_{44}$ is that the range $\lambda$ of the flux-line interaction
is typically much larger than the flux-line spacing $a$. This
means that the FLL is {\it much softer} for short wavelengths
$2\pi/k < 2\pi\lambda$ of compressional and tilt distortions
than it is for uniform compression or tilt.

The elastic matrix of the FLL was first calculated from GL theory
by solving for $\psi({\bf r})$ and $B({\bf r})$ up to terms linear
in arbitrarily chosen displacements
${\bf u}_i(z)$
(Brandt 1977a, b). Remarkably, it turns out that the correct
expansion parameter near $B_{c2}$ for periodic ${\bf u}_i(z)$
is not the spatial average $\langle|\psi|^2\rangle \approx (1-b)$
but the parameter $\langle|\psi|^2\rangle /(k^2\lambda^2)$,
which diverges for $k\to 0$, i.e.\ for uniform compression or tilt.
This may be seen from the approximate GL result for the tilt and
compressional moduli valid for
$k^2 \lambda^2 \ll \lambda^2/\xi'^2 = 2\kappa^2 (1-b)$,
     \setcounter{equation}{5} { \renewcommand{\theequation}{4.6a}
     \begin{eqnarray}   
 c_{11}(k) \approx c_{44}(k) \approx \frac{B^2 /\mu_0}
         {1 + k^2 \lambda^2 / \langle|\psi|^2\rangle } =
  \frac{B^2 }{\mu_0}  \sum_{n=1}^\infty (-1)^{n-1} \Big(
  \frac{\langle|\psi|^2\rangle } {k^2 \lambda^2} \Big)^n .
    \end{eqnarray}  }  
  From this argument follows  that truncating the infinite series
(4.6a) (or similar expressions) can give qualitatively wrong results
and that it is not always allowed to put $B({\bf r}) = $ const in
such calculations even when $\kappa$ is very large and $B$ is close
to $B_{c2}$. For example, the elastic moduli (4.6a) would then diverge
as $k^{-2}$ in the limit $k\to 0$ as is evident from the term $n=1$.
Only when the infinite summation is performed does one obtain the
unity in the denominator in (4.6a), which has the meaning of a
``mass'' or  ``self energy'' that renormalizes the spectrum of
excitations. This unity originates from the spatial variation of
$B({\bf r})$ as it becomes obvious from London results like
equation (4.6).

This somewhat unexpected finding partly explains erroneous results
which were obtained for the fluctuating FLL, namely, that there exists
a ``mass-less mode'' of the FLL, that the long-wavelength thermal
fluctuations diverge, and that long-range order is destroyed in an
infinite FLL (Eilenberger 1967, Maki and Takayama 1971,
Houghton and Moore 1988, Moore 1989, Ikeda et al.\ 1990, 1991a, 1992),
see also later discussion by Houghton et al.\ (1990), Moore (1992),
and Ikeda (1992b).
In such analytical calculations $B({\bf r}) = {\rm const}$ was
used explicitely, corresponding to the limit $\lambda\to \infty$
and to the neglection of effects of magnetic screening.
Interestingly, even when the diverging wrong tilt modulus
$c_{44} = B^2 \langle|\psi|^2\rangle /(\mu_0 k^2 \lambda^2)$ is
used, the correct average fluctuation $\langle u^2 \rangle$
(section 5.1.) is obtained (Moore 1989) since the main contribution
to $\langle u^2 \rangle$ (5.3) comes from large $k\approx k_{\rm BZ}$.

  It should be remarked here that the limit $\lambda \to\infty$
[or $\kappa \to \infty$, or $B({\bf r}) = {\rm const}$] does not only
mean vanishing tilt stiffness $c_{44} (k \ne 0) \to 0$ and
compressibility $c_{11} (k \ne 0) \to 0$, but also
vanishing shear modulus $c_{66} \to 0$ since all three moduli are
$\sim \lambda^{-2}$. This exact result might appear counter-intuitive
since large $\lambda$ means a long-range interaction between the
flux lines, which sometimes erroneously is believed to lead to a
stiff FLL. A further little known consequence of the academic limit
$\lambda \to\infty$ is that the magnetic flux $\phi$ of each flux line
vanishes since it is compensated by the flux of its image flux line if
$\lambda$ exceeds the distance $x$ of the flux line to the surface; for
a planar surface one has $\phi(x) =  \phi_0 [\,1 -\exp(-x/\lambda)\,]$,
thus the flux of a vortex vanishes when the vortex approaches the surface.
Therefore, the limit $\lambda \to \infty$ has to be taken such that
$\lambda$ still stays smaller than the size of the superconductor.

  But what about vortices in superfluid $^4$He? The interaction between
vortex lines in a neutral superfluid is not screened, i.e., formally
one has $\lambda = \infty$ for $^4$He. In spite of this, the triangular
vortex lattice in $^4$He has a {\it finite} shear modulus as shown by
Tkachenko (1969), see also the stability considerations by Baym (1995)
and the intricate observation of vortex arrangements by trapped ions
(Yarmchuck et al.\ 1979 and Yarmchuck and Packard 1982).
This seeming paradox is solved by
noting that the neutral superfluid $^4$He formally corresponds to the
limit of zero electron charge $e$, i.e., the absence of magnetic
screening.  In this limit both the flux quantum
$\phi_0 =h/2e$ and the London penetration depth
$\lambda_L = (m/2e^2 \mu_0 n_C)^{1/2}$ (in SI units with $2m$, $2e$,
and $n_C$ the mass, charge, and density of the Cooper pairs,
see Tilley and Tilley 1974) {\it diverge}. Therefore, in the
shear modulus per flux line,
$c_{66}/n = \phi_0^2 /(16 \pi \lambda_L^2 \mu_0)$ (4.4) in the
London limit, the factors $ e^{-2}$ in the numerator and
denominator cancel and $c_{66}$ is independent of $e$.

  While the GL calculations of FLL elasticity are rather intricate,
and the subsequent
derivations from the microscopic BCS-Gor'kov theory by Larkin and
Ovchinnikov (1979) and  from the anisotropic GL theory by
Houghton et al.\ (1989) are even more complicated, it is much
simpler to obtain the elastic matrix from London theory. At not too
large inductions $B <0.25 B_{c2}$ and for not too small $\kappa > 2$,
the London and GL results practically coincide;
the main correction from GL theory is that at larger $B$ the
London penetration depth $\lambda$ (or the anisotropic
$\lambda_{ab}$ and $\lambda_c$) have to be multiplied by a factor
$\langle |\psi({\bf r})|^2 \rangle^{-1/2} \approx (1-b)^{-1/2}$,
cf.\ equation (4.6a); the resulting increase of $\lambda$
accounts for the reduced screening near the upper critical
field $B_{c2}$. Another correction from the GL theory is a cutoff
in the interaction (3.6), which originates from the finite vortex core and
enters the vortex self-energy.
  The approximation (4.6) considers only the magnetic interaction
between the flux lines. When $k$ increases above $\approx k_{\rm BZ}/3$
then $c_{11}$ becomes smaller than (4.6) and $c_{44}$ becomes larger
than (4.6), cf.\ {\bf figure} 8: The compression modulus $c_{11}$ has
an additional factor $(1+k^2 \xi'^2)^{-2}$ with $\xi'^2 =\xi^2/(2-2b)$,
which comes from the overlapping vortex cores and is important
only at large $b=B/B_{c2} > 0.3$; $c_{11}$ is reduced further by
geometric dispersion such that at $k_\perp\approx k_{\rm BZ}$, $k_z=0$,
 one has $c_{11}(k)\approx c_{66}$; this is so since in a hexagonal
lattice compressional waves with ${\bf k=K}_{10}$ are identical to
shear waves with ${\bf k =K}_{11}$ and thus must have the same energy.
The tilt modulus $c_{44}$ has an additional term which originates
from the self energy of the flux lines, see equation (4.10) and
section (4.3); this line-tension term has only weak logarithmic
dispersion and exceeds the interaction term (4.6) when
$k^2 > k_{\rm BZ}^2 /\ln\kappa$.

The elastic nonlocality (4.6) has been questioned
by Matsushita (1989, 1994), who also
claims that the general expressions for ${\bf B(r)}$ (2.7) and (2.10)
violate div${\bf B}=0$. While it is easy to show that the
${\bf B(r)}$ (2.7), (2.10), (3.4), (3.8), (3.9), (3.13) do satisfy
div${\bf B} =0$, one can trace back the claim of a local tilt modulus
to the confusion of Abrikosov {\it flux lines} and magnetic
{\it field lines}. When the {\it field lines} are shifted by small
displacements ${\bf u(r)}$ the new induction is
${\bf B + \nabla \times (u \times B)}$, which is a {\it local}
relationship. However, when the {\it flux lines} are shifted then
the resulting {\bf B} follows from the {\it nonlocal} expression
(2.10).  In particular, for small periodic ${\bf u(r)}$ one obtains
thus the new induction ${\bf B + \nabla \times (u \times B)}/
(1+k^2 \lambda^2)$  (Brandt 1977a, b). This fact leads also to the
dispersion of the compressional and tilt moduli.
         \\[0.4 cm]
{\it 4.3.~ FLL elasticity in anisotropic superconductors}
         \\*[0.2 cm]
  Within London theory, the elastic matrix of the FLL in uniaxially
anisotropic superconductors is obtained  straightforwardly by expanding
the free energy $F\{ {\bf R}_i + {\bf u}_i(z) \}$ (3.3) up to
quadratic terms in the flux-line displacements ${\bf u}_i(z)$.
The general result may be written as an infinite sum over all
reciprocal lattice vectors {\bf K}
which guarantees periodicity in $k$ space. Choosing $B\| z$ one gets
for arbitrary orientation of the $c$-axis (Brandt 1977a, Sudb{\o}
and Brandt 1991a, b, Nieber and Kronm\"uller 1993b)
  \begin{eqnarray} 
  \Phi_{\alpha \beta}({\bf k}) &=& \frac{B^2}{2\mu_0}\,\sum_{\bf K}\,
    [\, g_{\alpha\beta} ({\bf K + k}) -
        g_{\alpha\beta} ({\bf K}) \,] \\
  g_{\alpha\beta}({\bf k}) &=& k_z^2 f_{\alpha\beta}({\bf k}) +
k_\alpha k_\beta f_{zz}({\bf k}) -2 k_z k_\beta f_{z \alpha}({\bf k})
  \end{eqnarray}
with $f_{\alpha\beta} ({\bf k})$ given by (3.6).
 The last term  $- 2 k_z k_\beta f_{z \alpha}({\bf k})$ in (4.8)
was first obtained by Sardella (1991, 1992).
This term influences the compression and shear moduli of the
FLL when ${\bf B}$ is not along the $c$-axis, but it does not enter
the tilt moduli.
Taking the limit $k_\perp \ll K_{10}$ in (4.7) one obtains the
continuum approximation as follows:  The term ${\bf K} =0$ yields the
dispersive  compression and tilt moduli while the nondispersive shear
moduli are obtained by integrating over all terms ${\bf K}\ne 0$.
For general orientation of $B\| z$ with respect to the $c$-axis,
there are several moduli for compression and tilt or for
combined compression and tilt (Sudb{\o} and Brandt 1991b, Sardella
1992, Sch\"onenberger et al.\ 1993) and several shear moduli
(Kogan and Campbell 1989). Since the FLL orientation couples to the
crystal lattice anisotropy, there appears also a ``rotation modulus'',
see Kogan and Campbell (1989) and the extension to films of
a biaxial superconductor by Martynovich (1993).
The situation simplifies  when $B$ is along
the $c$-axis. The FLL is then a uniaxial elastic medium like in
isotropic superconductors, and the elastic matrix resumes the
form (4.3) but now with the ${\bf k}$-dependent moduli
  \begin{eqnarray} 
 c_{11}({\bf k}) &=& \frac{B^2}{\mu_0} ~
   \frac{1 +\lambda_c^2 k^2}{(1+\lambda_{ab}^2 k^2)
   (1 +\lambda_c^2 k_\perp^2 +\lambda_{ab}^2 k_z^2)}
         \\
 c_{44}({\bf k}) &=& \frac{B^2}{\mu_0} ~     \Big[
   \frac{1}{1 +\lambda_c^2 k_\perp^2 +\lambda_{ab}^2 k_z^2}
   \,+\, \frac{f(k_z)}
        { \lambda_{ab}^{2} k_{\rm BZ}^{2} } \,\Big]
          \\
   f(k_z) &=& \frac{1}{2\Gamma^2} \ln \frac{\xi_c^{-2}}{\lambda_{ab}^{-2}
            + k_z^2 +\Gamma^2 k_0^2 } + \frac{\ln[1\, + k_z^2/
        (\lambda_{ab}^{-2} + k_0^2)\,] }{2k_z^2 \lambda_{ab}^2}
           \\
  c_{66} & \approx & B\phi_0 /(16\pi \lambda_{ab}^2 \mu_0) \,.
  \end{eqnarray}
These London results apply to $b=B/B_{c2} < 0.25$ and $\kappa =
\lambda_{ab}/\xi_{ab} >2$; in (4.11) $k_0 \approx k_{\rm BZ}$ is a cutoff.
 For the extension to $b>0.5$ the shear modulus $c_{66}$
(4.4) should be used with $\lambda$ replaced by $\lambda_{ab}$,  the
$\lambda_{ab}^2$ and $\lambda_c^2$ in (4.9) and (4.10) should be
divided by $1-b$, and  the second term in the brackets in (4.10)
should be replaced by $(1-b)/(\lambda_c^2 k_{\rm BZ}^2) =
 (1-b)\phi_0 /(4\pi B\lambda_c^2)$. Note that the factors 1-b
provide vanishing of all moduli as $B\to B_{c2}$. An interesting
and useful scaling argument which for $k\gg \lambda^{-1}$ allows to
obtain the anisotropic elastic energy (and other properties) of the FLL
from the isotropic results is given by Blatter, Geshkenbein
and Larkin (1992, 1993).

    The compression modulus $c_{11}$ (4.9) and the first term in
the tilt modulus (4.10) originate from  the vortex--vortex
interaction [\,the term ${\bf K}=0$ in (4.7)\,].
The second term in (4.10) originates mainly from the interaction of
each vortex line with itself and may thus be called isolated-vortex
contribution (Glazman and Koshelev 1991, Brandt and Sudb{\o} 1991,
Fisher 1991).  Due to the $k_z$-dependence
of this term the exact tilt modulus even of isotropic superconductors
depends on the wavevector {\bf k} rather than on $k=|{\bf k}|$.
This correction term dominates for large $k$, namely, when
$k_\perp^2 >k_{\rm BZ}^2/\ln(\Gamma\kappa)$ or when $k_z^2 >
 k_{\rm BZ}^2 \Gamma^2 /\ln(\Gamma/\xi_{ab} k_z)$, and for small $B$.
         \\[0.4 cm]
{\it 4.4.~ Line tension of an isolated flux line}
         \\*[0.2 cm]
  In the limit $B\to 0$ the second term in $c_{44}$ (4.10) yields the
line tension $P=c_{44} \phi_0/B$ of an isolated flux line
oriented along the $c$-axis,
  \begin{eqnarray} 
  P(k_z) = \frac{\phi_0^2}{4\pi \mu_0 \lambda^2} \Big[ \,
 \frac{1}{2\Gamma^2} \ln\frac{\Gamma^2\kappa^2}{1 +k_z^2
 \lambda_{ab}^2} ~+~ \frac{\ln(1 +k_z^2\lambda_{ab}^2) }
  { 2k_z^2 \lambda_{ab}^2} \, \Big] \,.
  \end{eqnarray}
  Due to the factor $1/\Gamma^2 =\lambda_c^2/\lambda_{ab}^2 \gg 1$
in (4.13) the line tension is very small compared to the
self-energy $J = \phi_0 H_{c1}$ of this flux line,
  \begin{eqnarray} 
 J \approx (\phi_0^2/4\pi \mu_0 \lambda_{ab}^2) \ln\kappa\,, ~~~
 P(k_z) \approx (\phi_0^2/4\pi \mu_0 \lambda_c^2)
                \ln(\Gamma/k_z\xi_{ab})\,.
  \end{eqnarray}
  The $P(k_z)$ (4.14) applies to large $\kappa =\lambda_{ab}/\xi_{ab}
\gg 1$ and not too long tilt wavelength, $k_z \lambda_{ab} \gg 1$.
If $k_z \lambda_{ab} \ll 1$, the bracket in (4.13) becomes
$\Gamma^{-2} \ln(\Gamma\kappa) + \frac{1}{2}$. In particular,
in the limit $\Gamma \to \infty$ the line tension of a straight
flux line ($k_z = 0$) is $P = \phi_0^2 /(8\pi\mu_0 \lambda_{ab}^2)$.
This result may also be obtained from Clem's (1991) stack of
pancake vortices, which has a line energy
  \begin{eqnarray} 
 J_{\rm stack}(\Theta) = (\phi_0^2 \cos\Theta /4\pi \mu_0 \lambda_{ab}^2)
  \,\ln\Big( \kappa\,\frac{1 +\cos\Theta}{2\cos\Theta}\Big)
  \end{eqnarray}
if it runs at an angle $\Theta$ with respect to the $c$-axis.
 In general, the line tension of a straight flux line is related
to its self-energy  by (Brandt and Sudb{\o} 1991)
  \begin{eqnarray} 
 P = J + \partial^2 J /\partial \alpha^2 ,
  \end{eqnarray}
 cf.\ equation (4.5). A similar relation holds for the line energy
and line tension of dislocations in anisotropic crystals
(Seeger 1955, 1958). Using the approximate expression for the
self-energy of a flux line (Balatski\u{\i} et al.\ 1986)
$J(\Theta) \approx J(0)(\cos^2\Theta + \Gamma^{-2}\sin^2\Theta)^{1/2}$
one obtains for $\Theta = 0$ (flux line along $c$)
$P(0) = J(0)/\Gamma^2$ since the two terms in  (4.15) almost
compensate each other; for $\Theta=\pi/2$ (flux line in the $ab$-plane)
one has the self-energy $J(\pi/2) = J(0)/\Gamma$ and the line tension
for tilt out of the $ab$-plane $P_{\rm out}(\pi/2) = J(\pi/2) \Gamma^2
 = P(0) \Gamma^3$. For tilt within the $ab$-plane the flux lines do
not feel the anisotropy and thus $P_{\rm in}(\pi/2) = J(\pi/2)$.

   These arguments show that the line tension of flux lines, and the
tilt modulus of the FLL, are {\it strongly reduced by the material
anisotropy} if $B$ is along the crystalline $c$-axis and if the tilt
is not uniform, e.g.\ if tilt or curvature occurs only over a finite length
$L < 2\pi\lambda_{ab}$, which means that values $|k_z| \le L^{-1}$
dominate (Brandt 1992c). Note that the elastic nonlocality is
crucial for this reduction.

  For layered superconductors this means that {\it very little energy
is required to displace one pancake vortex from the stack\/},
since for this distortion the characteristic wavevector
$k_z \approx \pi/s$ is very small ($s \ll \lambda_{ab}$ is the
layer spacing, {\bf figure} 7).  One can show that the above elasticity
theory derived from anisotropic GL and London theory approximately
also applies to layered superconductors. In a more accurate
formulation (Glazman and Koshelev 1991), $k_z^2$ in (4.3) and
in (4.8) to (4.14) should be replaced by $Q^2 = 2(1-\cos q)/s^2$
and all integrals $\int_{-\infty}^{\infty} \! {\rm d}z\, f(k_z^2)$
by $\int_0^{2\pi} \!{\rm d}q\, f(Q^2)$. In general one may say that
the anisotropic London theory gives good results even for layered
superconductors if one excludes resulting lengths which are less than
the layer spacing $s$ and replaces these by $s$.
          \\[0.4 cm]
{\it 4.5.~ Examples and elastic pinning}
         \\*[0.2 cm]
   As discussed in the previous section, the flux line displacements
obtained by the correct nonlocal elasticity theory typically are
much larger than estimates using the usual local theory of
elasticity. This enhancement is mainly due to the dispersion of the
tilt modulus $c_{44}({\bf k})$ or line tension $P=c_{44} \phi_0/B$.
Three examples shall illustrate this.

{\it 4.5.1. Elastic response to a point force.~}
  The maximum elastic displacement $u(0)$ caused in the FLL by a
point-pinning force $f$ acting on the flux-line segment at
${\bf r}=0$ ({\bf figure} 6, middle) is easily obtained in $k$ space
(Schmucker and Brandt 1977). The force balance equation
 $f_\alpha({\bf k}) = \delta F/\delta u_\alpha({\bf k}) =
\Phi_{\alpha\beta}({\bf k}) u_\beta({\bf k})$ is solved by inverting
the elastic matrix, yielding $u_\alpha({\bf k}) =
\Phi_{\alpha\beta}^{-1} ({\bf k}) f_\beta({\bf k})$. Noting that
for a point force ${\bf f( k )} =$ const,  one gets in continuum
approximation (4.3)
  \begin{eqnarray} 
 \frac{u(0)}{f} &=& \frac{1}{2} \int_{\rm BZ}\! \frac{{\rm d}^3 k}{8\pi^3}\,
  \Big[ \frac{1}{c_{66} k_\perp^2 + c_{44}({\bf k}) k_z^2 }
      + \frac{1}{c_{11}({\bf k}) k_\perp^2 + c_{44}({\bf k}) k_z^2 }
  \Big] \nonumber \\
  &\approx& \frac{k_{\rm BZ}}{ 8\pi [c_{66} c_{44}(0)]^{1/2} }
    \times \Big( \frac{4 \pi\lambda_{ab}^2}{\phi_0}
                 \frac{B}{1-b}   \Big)^{1/2}
    \times \frac{\lambda_c}{\lambda_{ab}} \nonumber \\
  &\approx& (4\pi/B\phi_0^3)^{1/2} \mu_0 \lambda_{ab}\lambda_c
            /(1-b)^{3/2} \,.
  \end{eqnarray}
The second term in the first line of (4.17) may be disregarded
since $c_{11} \gg c_{66}$.
The three factors in the middle line of (4.17) have the following
interpretation: The first factor is the result of local
elasticity, which assumes a dispersion-free $c_{44} = c_{44}(0)
\approx B^2/\mu_0$. The second factor
 $\approx [c_{44}(0)/c_{44}(k_\perp \!=\! k_{BZ})]^{1/2}
 \approx k_{BZ} \lambda_{ab}/(1-b)^{1/2} \approx
 (B\ln\kappa/B_{c1})^{1/2}$ is the
correction caused by elastic nonlocality.
The third factor $\lambda_c/\lambda_{ab} = \Gamma$ is the correction
due to anisotropy. Both correction factors are typically much
larger than unity. Therefore, {\it the elastic nonlocality and the
anisotropy enhance the pinning-caused elastic displacements of the FLL}.

{\it 4.5.2. Thermal fluctuations of the FLL.~}
   The thermal fluctuations of the flux-line positions
({\bf figure} 6, right) are given by the same integral (4.17),
$\langle {\bf u}^2 \rangle \approx 2k_B T u(0)/f$,
 and are thus enhanced by nonlocality and anisotropy in the
same way as is the elastic response to a point force.
These thermal fluctuations will be discussed in section 5.1.

{\it 4.5.3. Elastic energy for general non-uniform distortions.~}
  If the tilt wavelength of the flux lines is less than
$2\pi \lambda_{ab}$, i.e.\ for $k_z \lambda_{ab} > 1$, the
elastic energy (4.2) with (4.9)--(4.12) inserted takes an instructive
form. For each Fourier component $u\exp(i{\bf kR}_\nu)$ of the
general displacement field one obtains an energy density
  \begin{eqnarray} 
  \frac{u^2}{2}\, \frac{B^2}{\mu_0 \lambda_{ab}^2} \Big[
  1 + \frac{k_\perp^2}{4k_{\rm BZ}^2} + \frac{k_z^2 \ln(\xi_c^2 k_z^2 +
    \xi_{ab}^2 k_{\rm BZ}^2)^{-1} } {2k_{\rm BZ}^2 \Gamma^2} +
    \frac{\ln(1 \!+\! \frac{k_z^2}{\lambda_{ab}^2 +k_{\rm BZ}^2}) }
    {2k_{\rm BZ}^2 +\lambda_{ab}^{-2} } +
    \frac{\epsilon}{k_{\rm BZ}^2 \xi_{ab}^2 } \,\Big] .
  \end{eqnarray}
 where $k_{\rm BZ}^2 \lambda_{ab}^2 = 2b\kappa^2 \approx
B\ln\kappa /B_{c1} \approx c_{44}/4c_{66}$ and
$k_{\rm BZ}^2 \xi_{ab}^2 = 2b = 2B/ B_{c2}$.
 In (4.18) the first term ($=1$) originates from compression
($c_{11}$) and tilt ($c_{44}$) and presents a global parabolic
potential caused at the position of a flux-line segment by all
other flux lines.
The second term ($\le 1/4$) comes from the shear distortion ($c_{66}$)
and may be disregarded for estimates.
The third and fourth terms originate from the interaction of a
flux-line segment with the other segments of the same flux line:
The third term describes the coherence of a flux line
(``Josephson coupling of the segments''); this term ($\sim k_z^2$)
is a line tension with weak logarithmic dispersion,  a four-point
interaction in the language of pancake vortices (Bulaevski et al.\ 1992b).
The fourth term stems from the magnetic forces between the flux-line
segments of one flux line; for small $k_z < \lambda_{ab}^{-1}$
this term is $\sim k_z^2$ and thus is a line tension, but
for larger $k_z > \lambda_{ab}^{-1}$ the $k_z$-dependence becomes
logarithmic and thus this term is more like a local parabolic
potential, which has no dispersion.
These two terms can be smaller or larger than unity, depending on
$B$ and on the characteristic value of $k_z$.

  The last term ($=\epsilon/2b$) in the energy density (4.18) originates
from elastic pinning of the FLL. Here the pinning strength $\epsilon$
and the Labusch parameter $\alpha_L$ in (4.3) were estimated as follows.
The maximum possible pinning energy $U_{\rm p,max}$ per cm is
achieved when all flux-line cores are centred in long holes
of radius $r_p >\sqrt2 \xi_{ab}$. The gain in condensation
energy is then $U_{\rm p,max} \approx \mu_0 H_c^2 \cdot
2\pi \xi_{ab}^2 = \phi_0^2/(4\pi \mu_0 \lambda_{ab}^2)$.
The same estimate follows if one cuts off the magnetic self-energy
(4.14) of a flux line at the radius $r_p$ of the long pinning cylinder;
this inner cutoff replaces the factor
$\ln \kappa =\ln(\lambda_{ab}/\xi_{ab})$
by $\ln(\lambda_{ab}/r_p)$ and yields the energy difference
(pinning energy)  $U_{\rm p,max} \approx (\phi_0^2/4\pi
\mu_0\lambda_{ab}^2) [\ln(r_p/\xi_{ab}) +0.5]$ [c.f.\ equation (1.1)
and Brandt (1992c)]. Realistic pinning by spatial variation of
material parameters is weaker; it has a pinning energy per cm
$U_p = (\phi_0^2/4\pi \mu_0 \lambda^2)\cdot \epsilon$ with an
unknown parameter $\epsilon \le 1$, typically $\epsilon\ll 1$.
The Labusch parameter $\alpha_L$ in the model of direct summation
(section 8.7.) is estimated as the curvature
$\approx U_p /\xi_{ab}^2$ of this potential times the density
$B/\phi_0$ of the flux lines, $\alpha_L \approx (\phi_0^2/4\pi \mu_0
\lambda_{ab}^2) \epsilon \xi^{-2} B/\phi_0$. From the elastic
pinning energy density $u^2 \alpha_L /2$ one obtains the last term
$\epsilon/2b$  in the brackets in (4.18). This term can be smaller
or larger than unity, depending on the induction $B=bB_{c2}$
and on the pinning strength $\epsilon$.

  Summarizing the discussion of (4.18) one may say that for
non-uniform  distortion each flux-line segment behaves as if it were
sitting in a parabolic potential and experiencing a slightly dispersive
line tension, since the elastic energy has the form
 $(c_1 + c_2 k_z^2) s^2$.
The parabolic potential is caused by the magnetic interaction with
other flux lines (first term) and with the same flux line (fourth
term), and by pinning (fifth term). The line tension comes from the
coherence of the flux line (third term), which is absent in the
limit of very large anisotropy or for decoupled layers.
Depending on $B$, $k_z$, and on the pinning strength $\epsilon$,
each of these terms may dominate.

   From this picture  one can easily reproduce the displacement
(4.17) caused by a point pinning force. Including elastic pinning,
 one finds that the response (4.17) is multiplied by a factor
$(1 + \epsilon/2b)^{-1/2}$, which reduces the displacement
(and the thermal fluctuations, section 5.1.) when the pinning
strength $\epsilon$ exceeds the reduced induction $b =B/B_{c2}$.
%
         \\[0.7 cm]
{\bf  5.~ Thermal fluctuations and melting of the flux-line lattice}
         \setcounter{subsection}{5}  \setcounter{equation}{0}
         \\*[0.5 cm]
{\it 5.1.~ Thermal fluctuations of the flux-line positions}
         \\*[0.2 cm]
  The thermal fluctuations of the FLL in HTSC's can become quite
large due to the higher temperature and the elastic softness which
is caused by the large magnetic penetration depth (elastic nonlocality)
and the pronounced anisotropy or layered structure.
  The mean square thermal displacement of the vortex positions
  can be calculated within linear elasticity
  theory (section 4) by ascribing to each elastic mode of the FLL
  (with discrete {\bf k}-vectors in a finite volume V) the
  average energy $k_BT/2$. With (4.2)  one has
        \begin{equation}   
  \langle u_\alpha ({\bf k})\, \Phi_{\alpha\beta} ({\bf k})\,
  u_\beta({\bf k}) \rangle /2 = k_BT/2 \,, ~~~{\rm thus}~~
  \langle u_\alpha ({\bf k}) \, u_\beta({\bf k})\rangle = k_BT \,
       \Phi_{\alpha\beta}^{-1} ({\bf k}) ,
            \end{equation}
  where $\Phi_{\alpha\beta}^{-1} ({\bf k})$ is the reciprocal elastic
  matrix. From the Fourier coefficients (5.1) the thermal average
  $\langle  u^2 \rangle = \langle |{\bf u}_\nu(z)|^2 \rangle$  is
  obtained as a sum over all $\langle |{\bf u(k)}|^2 \rangle$
  (Parseval's theorem).  Letting $V\to \infty$  this sum becomes an
  integral over the Brillouin zone and over all $k_z$,
      \begin{eqnarray}   
    \langle u^2\rangle = \langle u^2_x + u^2_y \rangle =
    k_BT \int _{\rm BZ} \frac{{\rm d}^2 k_\perp}{4\pi^2}
    \int_{-\infty}^\infty \frac{{\rm d}k_z}{2\pi} \,
     [\, \Phi_{xx}^{-1}({\bf k}) + \Phi_{yy}^{-1}({\bf k}) \,].
            \end{eqnarray}
   With the continuum approximation (4.3) for  $\Phi_{\alpha
  \beta}$, which for consistency requires to replace the hexagonal
 first BZ by a circle of radius $k_{\rm BZ} = (4\pi B/\phi_0)^{1/2} =
  (2b)^{1/2}/\xi$, and omitting the compression modes since
  $c_{11} \gg c_{66}$, one gets explicitly for $B\| c$,
           \begin{equation}   
    \langle u^2\rangle = \frac{k_BT}{2\pi^2} \int_0 ^{k_{\rm BZ}}  \!
   {\rm d}k_\perp \,k_\perp \int_0^\infty  \frac{{\rm d}k_z }
   {c_{66}k_\perp^2 + c_{44}(k)k_z^2 + \alpha_L} \,.
            \end{equation}
  This is the same integral (4.17) which determines the flux-line
  displacement at the position of a point force.
    The main contribution to this integral comes from large  values
  $k_\perp \approx k_{\rm BZ}$ due to the factor
  $k_\perp$ in the numerator. The tilt modulus (4.10) is thus
  strongly reduced with respect to its values for uniform strain
  ($k=0$).  The dispersion (nonlocality) of $c_{44}({\bf k})$
  enhances the fluctuation $\langle u^2 \rangle$ near $B_{c2}$
  approximately by a factor $ k_{\rm BZ} \lambda_c/2(1-b)^{1/2} =
  [b\kappa^2 /2(1-b)]^{1/2} \Gamma$ = $[B\ln\kappa /4B_{c1}
  (1-B/B_{c2})]^{1/2 } \Gamma\gg 1$ (Houghton et al.\ 1989,
  Feigel'man and Vinokur 1990, Brandt 1989a, Moore 1989, 1992).
  The result may be written as the $\langle u^2 \rangle$
  of local elasticity theory,
  obtained first by Nelson and Seung (1989), times typically large
correction factors originating from elastic nonlocality and from
anisotropy,   times a reduction  factor from local pinning,
cf.\ (4.17) and the discussion below equation (4.18).

   A good approximation which slightly overestimates
 $\langle u^2 \rangle$ (5.3) is the replacement of $k_\perp$
by $k_{\rm BZ}$  in the tilt modulus $c_{44}({\bf k})$ (4.10);
the dominating contribution to $c_{44}$ then comes from the
second term in (4.10). Noting that
$k_{\rm BZ}^2 = 4\pi B/\phi_0 = 2b/\xi_{ab} $ and
$\ln(\xi_c ^{-2}/\Gamma^2 k_{\rm BZ}^2) \approx \ln (1/2b)$
(for $\kappa \gg 1$), one has the interpolation formulae
$c_{44}(k_\perp \!=\! k_{\rm BZ})/c_{44}(0) \approx (1-b)
 \ln[2 +(1/2b)^{1/2}] /(2b\kappa^2 \Gamma^2)$ and
$c_{66} \sim b(1-b)^2$ (4.4); note that the replacement of
$\lambda^2$ by $\lambda'^2 = \lambda^2/(1-b)$ in the
low-field $c_{66} \sim b/\lambda^2$ would yield the wrong
high-field limit  $c_{66} \sim b(1-b)$.  One thus obtains for
$0 < b < 1$,  $\kappa \gg 1$, and $B\| c$ the thermal fluctuations
 $ \langle u^2 \rangle \approx  (k_B T /4\pi)[ k_{\rm BZ} -
 (\alpha_L/c_{66})^{1/2} ] [c_{66} c_{44}(k_\perp \!=\! k_{\rm BZ})
  ]^{-1/2}$, or explicitly,
   \begin{equation}   
   \langle u^2 \rangle \approx  k_B T \Big( \frac{16\pi \mu_0^2
   \lambda_{ab}^4 \Gamma^2} {B \phi_0^3 } \Big)^{1/2} \!\! \times
   \Big[ (1-b)^3\, \ln\Big(2 +\frac{1}{\sqrt{2b}} \Big) \Big] ^{-1/2}
   \!\! \times \Big[1 -\Big( \frac{2\epsilon }{b} \Big)^{1/2}\Big].
          \end{equation}
In the following the pinning correction
 $2\epsilon/b =\alpha_L /c_{66}k_{\rm BZ}^2 =
4\alpha_L \lambda_{ab}^2 \mu_0 /B^2 \ll 1$ will be disregarded.
The general result (5.4) is conveniently expressed in terms of the
Ginzburg number $Gi = \frac{1}{2} [\, k_B T_c \Gamma \mu_0 /
 4\pi \xi_{ab}^3(0) B_c^2(0)\, ]^2 $ (9.1),
which determines the relative width of the fluctuation regime
of a superconductor near $B_{c2}(T)$,  $Gi = 1 -T_f/T_c$
(cf.\ section 9.1.),
   \begin{equation}   
  \frac{ \langle u^2 \rangle }{a^2}
   \approx  \Big( \frac{3 ~Gi~ B}
     {\pi^2 B_{c2}(0)} \Big)^{1/2}~ \frac{T}{T_c}~
    \frac{\lambda_{ab}^2(T)} {\lambda_{ab}^2(0)} ~ \Big[ (1-b)^3\,
     \ln\Big(2 +\frac{1}{\sqrt{2b}} \Big) \Big] ^{-1/2} ,
          \end{equation}
where $a^2 = 2B/\sqrt3 \phi_0$ was used.
  The fluctuation $\langle u^2 \rangle$ is thus strongly enhanced
by nonlocality  and anisotropy. It is further enhanced when one
  goes beyond the continuum approximation, e.g.,  by inserting
  in (5.2) the correct (periodic) elastic matrix  from the
  London theory (4.8) or from the GL theory (Brandt 1977a, 1989a).
   $\langle u^2 \rangle$  is  enhanced even further by nonlinear
  elasticity and plasticity, i.e.\ by the creation and motion of
  dislocations and other structural defects in the FLL (Brandt 1986,
  Marchetti and Nelson 1990, Koshelev 1993).
          \\[0.4 cm]
{\it 5.2.~ Time scale of the fluctuations}
         \\*[0.2 cm]
   Time scale  and frequency spectrum of thermal fluctuations of
  the FLL are determined by the elastic restoring force and the viscous
  motion of the flux lines. A FLL moving with velocity ${\bf v}$
  exerts a drag force $\eta{\bf v}$  per unit volume on the atomic
  lattice.   The viscosity  $\eta$ (per unit volume) is
  related to the flux-flow  resistivity
  $\rho_{\rm FF}\approx (B/B_{c2})\rho_n $ and to the  normal-state
  resistivity $\rho_n$ [equal to $\rho$ of the superconductor at
  $B=B_{c2}(T)$] by   (cf.\ section 7.1.)
   \begin{equation}   
         \eta = B^2/\rho_{\rm FF} \approx BB_{c2}/\rho_n.
     \end{equation}

   The FLL has two types of elastic eigenmodes, which are strongly
overdamped (i.e.\ not oscillating) since the viscous drag force by far
exceeds a possible inertial force (up to very high frequencies
  $\gg 10^{11}\,$sec$^{-1}$); the small flux-line mass per cm,
$m_{FL} \approx (2/\pi^3) m_e k_F$ ($m_e$ is the electrom mass
and $k_F$ the Fermi wavevector),  was estimated by Suhl (1965) and
Kuprianov and Likharev (1975), see also Hao and Clem (1991b)
and Blatter and Geshkenbein (1993) (vortex mass in anisotropic
superconductors), Coffey and Clem (1991b) (in layered superconductors),
and \v{S}im\'anek (1992) and Coffey (1994a) (contribution of
atomic-lattice distortions to the vortex mass).
 The elastic eigenmodes of the FLL are obtained
  by diagonalization of the elastic matrix $\Phi_{\alpha\beta}$,
yielding a compressional mode ${\bf u}_1{\bf (k)}$ and a shear mode
${\bf u}_2{\bf (k)}$ (Brandt 1989b, 1991b). After excitation these
eigenmodes relax with exponential time laws
$\sim \exp(-\Gamma_1 t)$ and $\sim \exp(-\Gamma_2 t)$,
respectively. In the continuum  approximation (4.3) one has
       \begin{eqnarray}  
  \Gamma_1 ({\bf k})~ =& [c_{11}({\bf k})k_\perp^2
          +c_{44}({\bf k})k_z^2\,]/\eta & \approx ~\Gamma_1 \\
  \Gamma_2 ({\bf k})~ =& [c_{66}k_\perp^2
     +c_{44}({\bf k})k_z^2\,]/\eta & \approx ~\Gamma_1 k_z^2/k^2
       \end{eqnarray}
with $\Gamma_1 = B^2k_h^2/\mu_0\eta \approx b\,(1-b)\, \rho_n/\mu_0\,
  \lambda^2    = 2\pi\,B (1 - b)\,\rho_n /\phi_0\,\mu_0\,\kappa^2$.
 The power spectra $P(\omega)$ of these viscous thermal fluctuations
 are Lorentzian,
       \begin{eqnarray}  
   P_{1,2}(\omega, {\bf k}) = \pi^{-1} \Gamma_{1,2}({\bf k}) /
       [\omega^2 + \Gamma_{1,2}({\bf k})^2 ].
       \end{eqnarray}
This Lorentzian spectrum should describe the high-frequency
depinning noise in the electric conductivity of type-II superconductors
since each ``plucking'' of a vortex triggers a visco-elastic relaxation
process, see section 7.7. and the review on flux noise by Clem (1981).
These two overdamped elastic modes (Brandt 1989b) recently were
termed ``vortons'' by Bulaevskii and Maley (1993) and
Bulaevskii et al.\ (1994a, b).

Due to the dispersion of the moduli $c_{11}$ (4.9) and $c_{44}$ (4.10),
the compressional mode  $u_1({\bf k})$ for {\it short} wavelengths
$k>k_h = (1-b)^{1/2}\lambda_{ab}^{-1}$ is nearly dispersion free,
$\Gamma_1({\bf k}) \approx \Gamma_1$. The relaxation rate or
fluctuation  frequency $\Gamma_1$  has thus a very high density of
states. At {\it long} wavelengths $k<k_h$  the compressional relaxation
is a {\it diffusion} process since  $\Gamma_1({\bf k}) \approx
 (B^2/\mu_0\eta) k^2$; the diffusivity is
 $B^2/\mu_0\eta = \rho_{\rm FF} /\mu_0$ (Brandt 1990b, see also
  section 7.1.). The relaxation of mere shear strains is always diffusive,
  $\Gamma_2({\bf k}) \propto k_\perp^2$ if $k_z=0$, and much slower
  than the compressional mode, $\Gamma_2({\bf k}) \ll \Gamma_1$
since $c_{66} \ll c_{11}({\bf k})$.
          \\[0.4 cm]
{\it 5.3.~ Simple melting criteria}
         \\*[0.2 cm]
  In analogy to crystal lattices a ``melting temperature'' $T_m$
of the three-dimensional (3D) FLL has been estimated from the
Lindemann criterion for the thermal fluctuations,
$\langle u^2 \rangle^{1/2} = c_La$, where
$c_L \approx 0.1\,...\,0.2$  and  $a$  is the flux-line spacing.
With (5.4), $T_m =t_m T_c$ is implicitly given by
(see also Houghton et al.\ 1989, Brandt 1989a, 1990c, 1991a,
Blatter et al.\ 1994a)
 \begin{equation}    
  c_L^4 = \frac{3 ~Gi}{\pi^2}~ \frac{B}{B_{c2}(0) } ~
  \frac{\lambda_{ab}^4(T_m) }{\lambda_{ab}^4(0) } ~
  \frac{t_m^2}{(1-b)^3 \, \ln(2 +1/\sqrt{2b}) } .
 \end{equation}
This is still a general (for $\kappa \gg 1$, $B\| c$) though
 slightly large estimate of the
fluctuations since $k_\perp \approx k_{\rm BZ}$ was assumed in
$c_{44}({\bf k})$.  To obtain explicit expressions one may
assume $B_{c2}(T) \approx B_{c2}(0) (1-t)$ and
 $\lambda_{ab}(T) \approx  \lambda_{ab}(0)/(1-t)^{1/2}$,
which means a temperature independent
$\kappa = \lambda_{ab}/\xi_{ab} $. Writing
$B=b B_{c2}(T) = \tilde b  B_{c2}(0)$, $ 1-b = 1 -\tilde b /(1-t)
  = (1-t-\tilde b)/(1-t)$, and $t_m = T_m/T_c$ as above, one obtains
 \begin{equation}    
  c_L^4 \approx \frac{3~ Gi}{\pi^2}~ \frac{b}{(1-b)^3}~
  \frac{t_m^2}{1-t_m} = \frac{3~ Gi}{\pi^2}~
  \frac{\tilde b~ t_m^2 \,(1-t_m)} {(1-t_m -\tilde b)^3} .
 \end{equation}
For low inductions $b\ll 1$ the explicit expression for the melting
field $B_m(T)$ reads
 \begin{equation}    
 B_m(T) = B_0 ~\Big(\frac{1-t}{t} \Big)^2 ~\ln\frac{1}{\sqrt{2b}},~~~
 B_0 = \frac{ c_L^4 ~ \phi_0^5 }{12\pi~ k_B^2~ T_c^2 ~\Gamma^2~
       \mu_0^2~ \lambda_{ab}^4(0) } .
 \end{equation}
 This means  $B_m(T) \propto (T_c-T)^2$  near $T_c$ and $B_m(T)
    \propto 1/T^2$  far below $T_c$.
 The thus defined melting temperature $T_m^{3D}(B)$ is rather low.
 As stated above, softening  of the FLL by structural defects or by
 anharmonicity decreases $T_m^{3D}$  even more.
 For $T_m^{3D} \approx T_c$ one has
     \begin{eqnarray} 
 T_m^{3D} = T_c \Big[\,1 -\Big(24\pi B /\ln\frac{1}{2b}\Big)^{1/2}
 k_B T_c \mu_0 \Gamma\lambda_{ab}^2(0) /c_L^2 \phi_0^{5/2} \Big] \,.
     \end{eqnarray}
 Comparing this with the 2D melting temperature of the FLL in
 thin films of thickness $d$, $T_m^{2D} = a^2d c_{66}/4\pi k_B
 = {\rm const}(B)$  (6.2)  one finds that the 3D and 2D
 melting temperatures formally coincide for a film of thickness
 $d= a\, (c_L/0.10)^2 \, \lambda_{ab}/\lambda_c$, thus $d \approx
 a /\Gamma$ for $c_L=0.1$, see also Markiewicz (1990a).
 This length is of the same order as the length $a/\pi\Gamma$
over which the flux lines curve to hit the surface at a right angle
(Hocquet et al.\ 1992, Pla\c{c}ais et al.\ 1993, 1994, Brandt 1993a).
Note, however, that the melting thickness depends also on the
structure-dependent shear modulus, $d^2/a^2 \approx c_L^4 (4\pi)^3
c_{44}(k_\perp \!=\! k_{\rm BZ}) /c_{66}$.

As a matter of principle, it is not clear whether the Lindemann
criterion applies to line liquids; it was shown to fail if the lines are
confined to planes, e.g.\ to the space in between the Cu-O planes in
HTSC's (Mikheev and Kolomeisky 1991).
    Interestingly, the Lindemann criterion has been applied also
 to the melting of a lattice of ``cosmic strings''
 in cosmology  (Nielsen and Olesen 1979) or ``colour magnetic
 flux tubes'' in field theory (Ambj\o rn and Olesen 1989).
 Cosmic strings have a microscopic radius of $\approx 10^{-24}$ cm
 and are thousands of light years long, see the review by
 Hindmarsh and Kibble (1995). They are believed to form a
 line lattice which should melt due to quantum fluctuations
 (section 6.6.).

   Other possible melting criteria consider the thermal
fluctuation of the vortex distances ${\bf |r}_i -{\bf r}_{i+1}|$,
of the vortex separation $|x_i - x_{i+1}|$  (nearest neighbour
direction along $x$), or of the shear strain $\gamma$  (Brandt 1989b).
These fluctuations are related by
         \begin{eqnarray} 
     \langle u^2\rangle \approx \frac{1}{2}\langle|{\bf r}_i -
     {\bf r}_{i  +1}|^2 -a^2\rangle \approx \langle|x_i   -x_
     {i  +1}|^2 -a^2\rangle \approx \frac{\sqrt{3}a^2}{4\pi}
     \langle \gamma^2 \rangle~.
         \end{eqnarray}
  The melting temperatures resulting from these criteria differ
  only by numerical factors of order unity from the
  Lindemann result $T_m(B)$ (5.13).
 At the Lindemann $T_m$ one has large shear-strain fluctuations
 $\langle \gamma^2\rangle^{1/2}\approx 2.7\,c_L$. In metals the
 critical flow strain at which plastic deformation sets in is
 $\gamma_c \approx 1/30$  (Seeger 1958), and a similar flow strain
 (or flow stress $c_{66}\gamma_c$) was assumed for the 2D FLL
 (Schmucker 1977). Equating $\langle \gamma^2 \rangle ^{1/2}$ to
 this small critical shear strain   would yield a much smaller $T_m$;
 but probably local thermal fluctuations of the shear strain will
 not usually produce  permanent structural defects in the FLL.
 After a short time of local entanglement the flux lines will
 straighten again and global melting should not occur at such
 low temperatures.
 Similar arguments apply to the distance criterion. Even when
 the vortices touch locally, they might merge for a short time and
 then separate and stretch again  since the barrier for vortex
 cutting is small (Brandt et al.\ 1979, Wagenleithner 1982,
Sudb{\o} and Brandt 1991, Moore and Wilkin 1994,
Carraro and Fisher 1995, Sch\"onenberger et al.\ 1995).
A periodic FLL is indeed observed up to
rather high $T$ by Lorentz electron microscopy (Harada et al.\ 1993).

    As for the vanishing of the shear modulus $c_{66}$  of the FLL,
 this melting criterion is obvious but not trivial since the
 shear stiffness in principle depends on the time and length scales
 and on pinning. If one shears a {\it large}  FLL uniformly it will
 creep by thermal activation of dislocations, which are always present
 and can move. In particular, screw dislocations  do not
 see a Peierls potential and thus could  move easily along the
 flux lines if there where no pinning. The shear and tilt creep
 rates thus depend crucially on pinning.
    Marchetti and Nelson (1990) show that a {\it hexatic}
(i.e. with short-range order) flux-line liquid, believed to lie in
between the line lattice and normal line liquid states, can be
described by a continuum elasticity theory in which one includes
free dislocation loops; this yields an effective shear modulus
$c_{66}(k_\perp, k_z=0) \sim k_\perp^2$. Thus, $c_{66}$ remains
finite for $k_\perp >0$ but vanishes for $k_\perp =0$ as
expected for a liquid.
Interestingly, when a weak random pinning potential is switched on,
the shear modulus of the FLL is {\it reduced} (Brechet et al.\ 1990).
The influence of pinning and thermal depinning on FLL melting is
discussed by Markiewicz (1990b). A universality of $B_{c2}(T)$
and of the melting line of the FLL in clean HTSC's in terms of the
Ginzburg number $Gi$ is discussed by Mikitk (1995).
       \\[0.4 cm]
{\it 5.4.~ Monte-Carlo simulations and properties of line liquids}
         \\*[0.2 cm]
Numerous theories on a possible melting transition of the 3D FLL
were published since 1990. The melting transition of a 3D
lattice of point vortices was investigated by S. Sengupta et al.\
(1991) and Menon and Dasgupta (1994) using a density functional
method based on the structure
function of liquids and solids. Ryu et al.\ (1992) performed
Monte Carlo (MC) simulations on this point-vortex system and
found {\it two} melting lines where first the translational and then the
bound-angle order parameters drop to zero. Xing and Te{\v s}anovi{\'c}
(1990) in their MC simulation found a melting transition
very close to the $H_{c1}(T)$-line; however, due to the large
distance and week interaction between flux lines near $H_{c1}$,
this transition should be suppressed by even extremely weak pinning.
Ma and Chui (1991, 1992) discussed dislocation-mediated melting of the
3D FLL and performed MC simulations of a small system of flux lines
interacting by the correct 3D potential (3.3) with periodic boundary
conditions.

The probably most realistic MC study of the 3D FLL in anisotropic
London superconductors uses the correct interaction (3.3)
and discretizes the space by cubic cells into which vorticity can
enter from all six sides (Carneiro et al.\ 1993, Carneiro 1992, 1994,
Chen and Teitel 1995);
in this way a melting line $B(T)$ is obtained; for weak anisotropy,
entangled and disentangled vortex-liquid phases are observed,
but for strong anisotropy, a nearly 2D point-vortex liquid results.
Other MC simulations of the 3D FLL consider a 3D $xy$-model with the
Hamiltonian (Ebner and Stroud 1985)
\begin{eqnarray} 
  H = -\sum J_{ij} \cos(\phi_i -\phi_j -A_{ij}), ~~~~~ A_{ij} =
  \frac{2\pi}{\phi_0} \int_i^j \! {\bf A}\,{\rm d}{\bf l}
\end{eqnarray}
where $J_{ij}$ are nearest neighbour exchange interactions on a
cubic or triangular lattice,  $\phi_i$ are phase angles, and
 ${\bf \nabla \times A = H} =$ const  is assumed. With this model
Hetzel et al.\ (1992) found a first order melting transition with
a hysteresis.
Recent analytical solutions of the anisotropic 3D $xy$-model were
obtained by Choi and Lee (1995) and Shenoy and Chattopadhyay (1995).
Li and Teitel (1991, 1992, 1993, 1994) in their simulations find two
sharp phase transitions: from an ordered FLL to a disentangled
flux-line liquid and then into a normal-conducting phase with much
entangling, cutting, and loop excitations.
Hikami et al.\ (1991) investigated the perturbation series for the
3D GL free energy in a random pinning potential at high fields
near $H_{c2}$, see also Fujita and Hikami (1995).
Blatter and Ivlev (1993, 1994) showed that quantum
fluctuations modify the theoretical melting line $B(T)$.
Quantum fluctuations of the FLL were also calculated by
Bulaevskii and Maley (1993) and Bulaevskii et al.\ (1994a, b),
cf.\ section 6.6.

    An excellent  realistic 3D MC study of YBCO was performed
by \v{S}\'a\v{s}ik and Stroud (1993, 1994a, b). These authors
start from the GL free energy and insert a spatially varying
critical temperature $T_c({\bf r})$, which describes the layered
structure (CuO$_2$ double layers) and pinning in the layers,
and then use an expansion in terms of lowest-Landau-level
basic functions in a system of $8\times 8$ flux lines and
28 double layers. To check for a melting transition
\v{S}\'a\v{s}ik and Stroud (1993, 1994a) consider
the ``helicity modulus tensor'', defined as the second derivative
of the fluctuating free energy $F$ with respect to the
supervelocity ${\bf Q}= (\phi_0/2\pi) {\bf \nabla} \phi -{\bf A}$
(section 2.1), a kind of superfluid density tensor;
note that $\delta F/\delta {\bf Q} = {\bf J}$ is the current density
and, with (2.1a), $\delta^2 F /\delta |{\bf Q}|^2 \sim |\psi|^2$.
 In the pin-free case, the tensor
$\delta^2 F/\delta Q_\alpha \delta Q_\beta$ is found to have zero
components parallel to the layers
but a finite component along the $c$-axis; with pinning, all
diagonal elements are finite. Without pinning and with weak pinning,
this tensor vanishes at the same melting temperature where the
structure factor of the FLL looses its Bragg peak and also the
shear modulus of the FLL is found to vanish
(\v{S}\'a\v{s}ik and Stroud 1994b). The nearly
approximation-free (apart from the periodic boundary conditions)
3D computations of \v{S}\'a\v{s}ik and Stroud
(1993, 1994a, b) thus indicate a sharp melting transition in YBCO.
        \\[0.4 cm]
{\it 5.5.~ Flux-line liquid.~}
         \\*[0.2 cm]
  The entanglement of flux lines in a liquid state has been
discussed by Nelson (1988, 1989), Marchetti and Nelson (1990, 1991)
(hexatic line liquid and hydrodynamic approach), Marchetti (1991),
and with pinning included
by Nelson and Le Doussal (1990) and Marchetti and Nelson (1993)
(entanglement near a surface, slab geometry). A kinetic theory of
flux-line hydrodynamics was
developed by Radzihovsky and Frey (1993). Using a similar
approach, Glyde et al.\ (1992a, b) studied the stability and dynamics
of the FLL and its dependence on film thickness.

 In order to take advantage of Nelson's (1989) analogy of his model
FLL with an interacting 2D Bose superfluid, these
theories use a local flux-line tension and a simplified 2D
interaction between the flux lines suggested by Nelson (1989),
$\sim K_0[|{\bf r}_i(z) - {\bf r}_j(z)| /\lambda]$, cf.\ (2.5),
rather than the full 3D interaction (2.8) or (3.3). This quasi-2D
interaction is good for {\it long} tilt wavelengths
$2\pi/k_z > 2\pi\lambda$, whereas the non-dispersive weak line
tension  (reduced due to anisotropy, cf.\ section 4.3) is valid for
 {\it short} tilt wavelengths $< 2\pi\lambda$  (Brandt 1992c).
Nelson's model interaction yields elastic moduli of the FLL which have
the form of (4.9) to (4.12) but with $k_z = 0 $ inserted. Fortunately,
this simplification does {\it not} change the resulting thermal
 fluctuations of the FLL too much, since in $\langle u^2 \rangle$
(5.3) the main contribution comes from large
$k_\perp \approx k_{\rm BZ}$  and from small $k_z$, because the
isolated-vortex-term  (4.11) dominates in $c_{44}$ (4.10) when
$k_\perp \approx k_{\rm BZ}$; this dominance was stated first by
Glazman and Koshelev (1991a). Note, however, that the quasi-2D
model interaction between flux lines is always {\it repulsive},
while the correct 3D London interaction [the first term in (2.8),
or (2.8) with $\xi \to 0$]  becomes {\it attractive} when the angle
between the flux lines exceeds $90^\circ$.
 Recently Feigel'man et al.\ (1993a) argued that
the FLL with the correct 3D London interaction
indeed can be mapped onto a boson model, from which they conclude
that the flux-line liquid state is a genuine thermodynamic state
occurring between the FLL state and the normal conducting state.
Fleshler et al.\ (1995) measure the resistivity of the vortex liquid
in YBCO crystals with and without electron irradiation.

Comparing entangled flux lines with polymeres, Obukhov and Rubinstein
(1990) obtained astronomically large relaxation times for the
disentanglement of a flux-line liquid. These times probably do not
apply since the barrier for vortex cutting is rather low
(Sudb{\o} and Brandt 1991c, Moore and Wilkin 1994,
Carraro and Fisher 1995). Sch\"onenberger et al.\ (1995) recently
have investigated vortex entanglement and cutting in detail for a
lattice and liquid of London flux lines; they find that the twisted
(by $180^\circ$) vortex pair is unstable when embedded in a FLL or
liquid, but the twisted (by $120^\circ$) vortex triplet presents a
metastable, confined configuration of high energy.
Chen and Teitel (1994, 1995) consider the
``helicity modulus'' of a fluctuating elastic FLL and find that at
the FLL melting transition there remains superconducting
coherence (zero resistivity) parallel to the applied magnetic field,
provided the finite-wavelength shear modulus $c_{66}(k_\perp)$
stays finite for $k_\perp >0$.
  Statistical pinning of flux-line liquids and of defective FLL
is considered by Vinokur et al.\ (1990a, 1991b) and Koshelev (1992a),
note also the analogies of strongly pinned flux lines with a
``vortex glass'' (section 9.5.) and a ``bose glass'' (section 9.2.).
         \\[0.4 cm]
{\it 5.6.~ First-order melting transition.~}
         \\*[0.2 cm]
   So far the experimental evidence for a sharp melting transition
is not compelling but quite likely. Probably the strongest evidence
for a first-order melting (or other) transition in BSCCO crystals is
an abrupt jump in the local induction $B$ of small amplitude
$\approx 0.3$ Gauss, measured by Hall probes when $H_a$ or $T$ is
swept (Zeldov et al.\ 1995a).
First indication of FLL melting in vibrating YBCO
crystals by Gammel et al.\ (1988) can also be explained by
thermally activated depinning (Brandt et al.\ 1989), cf.\ section 10.
Recently, first order transitions were indicated by a sharp jump with
hysteresis in the resistivity of very weak-pinning untwinned YBCO
single crystals by  Safar et al.\ (1992b) and
Kwok et al.\ (1992, 1994a). Charalambous et al.\ (1993) find that the FLL
superheats at low current densities, but supercooling is
negligible. Stronger pinning suppressed the observed hysteresis.
Geshkenbein et al.\ (1993) explain the observed
hysteresis by vortex glass transitions of the first order.
In general, a hysteresis may also be caused by dynamic processes
or disorder.

It might be argued that even weak pinning may cause a similar
hysteretic jump in the current-voltage curves. For example, a sharp
transition from weak 2D to strong 3D pinning in films of amorphous
Nb$_x$Ge and Mo$_x$Si, with an abrupt jump of the critical
current density $J_c(B)$ by a factor $> 10$ has been observed by
W\"ordenweber and Kes (1985, 1986). However, hysteresis was not
observed in that experiment. Farrell et al.\ (1991, 1995) in a
sensitive torsional oscillator experiment on a very clean untwinned
YBCO crystal see no hysteresis and no jump in the magnetization
versus temperature. Jiang et al.\ (1995) in untwinned YBCO crystals
observe a decreasing hysteresis with increasing angle between the
applied field and the $c$-axis and propose a current-induced
nonequilibrium effect as the possible origin of this hysteresis.
 Worthington et al.\ (1992) measured resistivity curves
over large ranges of current density, magnetic field, and
temperature in various YBCO samples, finding evidence for {\it two}
phase transitions at the same field, separated by a phase of
``vortex slush'' and interpreted as disorder-dependent
melting and a vortex-glass transition (section 9.4.).
The influence of twin-boundaries
on FLL melting in YBCO is measured by Fleshler et al.\ (1993).
Kwok et al.\ (1994b) in a YBCO single crystal containing only two
twin planes observe a pronounced peak in the critical current density
$J_c(T)$ in fields between 0.3 and 1.5 T for $H_a \| {\bf \hat c}$,
which is indicative of FLL softening prior to melting.

 Melting transitions were even observed in conventional
superconductors. Berghuis et al.\ (1990) in thin films of
amorphous Nb$_3$Ge  observed the dislocation-mediated 2D melting
predicted by Feigel'man et al.\ (1990) and Vinokur et al.\
(1990b). Similar resistivity data by Schmidt et al.\ (1993)
suggest 2D melting of the FLL in thin Nb films.

  All these experiments actually measure resistivities and
dissipation and thus require at least some pinning of the flux
lines by material inhomogeneities (section 8.2.). In a completely
pin-free superconductor, the dissipation comes from uniform
free flux flow, which is not expected to exhibit any
discontinuity or extremum at the proposed melting transition.
Besides this, dissipation by uniform flux flow is very small
at low frequencies $\omega/2\pi < 10^4\,$s$^{-1}$, much smaller
than the hysteresis losses caused by even extremely weak pinning. These
hysteresis losses in principle also originate from flux-line motion, but
here from the highly non-uniform jerky motion triggered by depinning
processes. A different dissipation mechanism, which seemingly is
independent of pinning, was suggested by Marchetti and Nelson
(1990, 1991), the plastic deformation or viscous shearing of
an entangled FLL. However, shear deformation of the FLL can only
be excited by very local forces acting on {\it one} flux line or on
{\it one} flux-line row (Brandt 1978). Such local forces are exerted by
linear or point pins, by planar pins, or by the specimen surface,
but not by applied
magnetic fields or currents since these act over the penetration
depth $\lambda$, which typically is larger than the flux-line
spacing. Therefore, all these dissipation mechanisms are caused
by pinning. Since pinning in HTSC's strongly depends on temperature
(due to thermally activated depinning, section 9), resistivity
experiments in principle will not give unique evidence for a
melting transition of the FLL.
         \\[0.4 cm]
{\it 5.7.~ Neutrons, muons, and FLL melting.~}
         \\*[0.2 cm]
  In contrast to resistivity measurements,
 small-angle neutron-scattering (SANS) and
muon spin rotation ($\mu$SR) experiments in principle can provide
direct information on the structure of the FLL.
Transverse rotation experiments with positive muons measure the
probability $P(B)$ to find a magnetic induction value $B$ inside
the specimen. The field distribution
$P(B') = \langle \delta[\,B({\bf r})-B'\,] \rangle$
($\langle \dots \rangle$ denotes spatial averaging)
of a perfect FLL exhibits van Hove singularities (Seeger 1979,
Brandt and Seeger 1986), which will be smeared by
the random distortions of the FLL caused by pinning (P\"umpin et al.\
1988, 1990, Brandt 1988a,b, 1991c, Herlach et al.\ 1990,
Harshman et al.\ 1991, 1993, Weber et al.\ 1993, Riseman et al.\ 1995)
or by thermal fluctuation (Song et al.\ 1993, Song 1995) or
flux diffusion (Inui and Harshman 1993).
By $\mu$SR Cubitt et al.\ (1993a) measure the angular dependence
of the field distribution in BSCCO and find evidence for the
presence of weakly coupled point vortices. The same conclusion is
drawn by Harshman et al.\ (1993) from the very narrow $\mu$SR line
observed in BSCCO at low $T$ as a consequence of the pinning-caused
reduction of the correlation between the 2D vortex lattices in
adjacent layers. Complete longitudinal disorder {\it reduces} the
field variance
$\sigma = \langle [{\bf B(r)} - \langle {\bf B}\rangle ]^2 \rangle$
(the line width squared)
from the value $\sigma_{\rm tri} = 0.0609 \phi_0 /\lambda_{ab}^2$ for
the ideal triangular FLL along the $c$-axis (Brandt 1988a) to the much
smaller value $ \sigma_{2D} = 1.4 (s/a)^{1/2} \sigma_{\rm tri}$
where $s$ is the layer spacing and $a$ the flux-line spacing.
[There is a misprint in $\sigma_{2D}$ of Harshman et al.\ (1993)].
In contrast,  pinning-caused shear or compressional disorder
{\it within} the 2D vortex lattices, or in the 3D FLL, {\it enhances}
the variance, and a completely random disorder also in the layers
has the variance $\sigma_{\rm rand} = (B \phi_0 s /8 \pi
 \lambda_{ab}^3)^{1/2}$, which increases with $B$
 (Brandt 1988a,b, 1991c, Harshman et al.\ 1993).

Recent SANS experiments by Cubitt et al.\  (1993b) and $\mu$SR studies
by Lee et al.\ (1993, 1995) gave evidence for FLL melting in BSCCO
when $T$ was increased, and for a crossover from 2D to 3D behaviour
(decoupling of layers, section 6.3.) when $H_a$ was increased.
With increasing temperature the Bragg peaks vanished at about the same
 $T(B)$ value where the $\mu$SR line abruptly became narrower and
changed its shape (lost its high-field tail).
By Lorentz microscopy Harada et al.\ (1993) saw that in
BSCCO films the FLL averaged over the exposure time of 30 sec stays
{\it regular} even slightly above the irreversibility line obtained from
dc magnetization measurements; this finding shows that the FLL did not
melt at the irreversibility temperature.
In their SANS experiments on Nb, Lynn et al.\ (1994)
interpret the increase in transverse width of the Bragg peaks
as melting of the FLL in Nb. However, as argued by
Forgan et al.\ (1995), this observation more likely reflects a slight
loss of orientational perfection of the weakly pinned FLL, which is
caused by the softening of the FLL (peak effect, Pippard 1969,
see section 9.8.).
          \\[0.7 cm]
{\bf 6~ Fluctuations and phase transitions in layered superconductors}
          \setcounter{subsection}{6}  \setcounter{equation}{0}
          \\*[0.5 cm]
  Due to their layered structure, HTSC's may exhibit two-dimensional
character. When the temperature is lowered from $T_c$, a transition
from 3D to 2D superconductivity is expected when the coherence
length along the $c$-axis, $\xi_c(T) \sim (T_c -T)^{-1/2}$, becomes
smaller than the layer spacing $s$ (Lawrence and Doniach 1971).
Various phase transitions have been predicted for this quasi 2D
system.
          \\[0.4 cm]
{\it 6.1.~ 2D melting and the Kosterlitz-Thouless transition}
         \\*[0.2 cm]
  An infinite pin-free 2D vortex lattice is {\it always unstable}
because its free
energy can be reduced by thermal nucleation of edge dislocations even
when $T\to 0$. But even weak pinning stabilizes a 2D lattice.
This can be seen from the estimate of 2D fluctuations
\begin{eqnarray}  
    \langle u^2\rangle_{\rm film} =  k_BT \int _{\rm BZ}
    \frac{{\rm d}^2 k_\perp /4\pi^2 d}
    { c_{66} k^2_\perp +\alpha_L} = \frac{k_B T}{4\pi d c_{66}}
    \ln\Big( 1 +\frac{k_{\rm BZ}^2 c_{66}}{\alpha_L} \Big) ,
\end{eqnarray}
cf.\ (5.3) with $\int_{-\pi/d}^{\pi/d} \! {\rm d}k_z = 1/d$ inserted.
Thus, for large elastic pinning restoring force
$\alpha_L \gg k_{\rm BZ}^2 c_{66} = B^2 /(4 \lambda^2 \mu_0)$ one
has $\langle u^2\rangle_{\rm film} = k_BT k_{\rm BZ}^2 /(4\pi\alpha_L d)
 = k_B TB/(\phi_0 \alpha_L d) \approx k_B T/(\alpha_L d a^2)$, but for
small $\alpha_L$ the fluctuations diverge logarithmically with
decreasing $\alpha_L$ or $d$ for films of infinite size.

   As argued by Hubermann and Doniach (1979) and Fisher (1980),
in the 2D FLL of a thin film of thickness $d$ and radius $r$,
spontaneous nucleation of edge dislocations in the FLL of
spacing $a$ costs the energy $U= (da^2 c_{66}/4\pi)\ln(R/a)$
and enhances the entropy by $S= k_B \ln(R/a)$; here $R^2/a^2$
is the number of possible positions of the dislocation cores.
The free energy $F=U-TS$ can, therefore, be lowered by spontaneous
creation of dislocations, leading to dislocation-mediated
melting at $T=T_m^{2D} = a^2 d c_{66} /(4\pi k_B)$ (5.13).
To obtain melting in a stack of films or layers, as in HTSC's,
one has to replace $d$ by the layer spacing $s$. Inserting
the shear modulus $c_{66}$ (4.12), one explicitly gets
\begin{eqnarray} 
  T_m^{2D} = \frac{a^2 s c_{66}}{4\pi k_B} =
  \frac{\phi_0^2 s}{64 \pi^2 k_B \mu_0 \lambda^2_{ab}}.
\end{eqnarray}
  Note that $T_m^{2D}$ does not depend on $B$.

  Earlier, a topological phase transition in a 2D superfluid was
predicted by Berezinskii (1970) and Kosterlitz and Thouless (1973)
(BKT) and elaborated by Halperin and Nelson (1979). At $T>0$,
vortex-antivortex pairs nucleate spontaneously in a single
superconducting layer or film even in zero applied field. These
pairs dissociate at a temperature
\begin{eqnarray} 
  T_{\rm BKT} =  \frac{\phi_0^2 s}{8 \pi k_B \mu_0 \lambda^2_{ab}}.
\end{eqnarray}
  Formula (6.3) exceeds (6.2) by a factor of $8\pi$. Accounting
for the temperature dependence of the penetration depth
$\lambda_{ab}(T)$ one obtains (Feigel'man et al.\ 1990)
\begin{eqnarray} 
  \frac{T_m^{2D}}{T_{\rm BKT}} = \frac{\lambda^2_{ab}(T_m^{2D})}
        { 8\pi\,\lambda^2_{ab}(T_{\rm BKT}) } \approx
   \frac{1}{8\pi} \,\frac{T_c -T_{\rm BKT}}{T_c -T_m^{2D}}.
\end{eqnarray}
 One may say that the 2D melting (6.2) is also a Kosterlitz-Thouless
transition, with the dissociating objects being not
vortex--antivortex pairs but dislocation pairs with opposite
Burgers vectors. The 2D melting temperature is lower than $T_{\rm BKT}$
since the binding energy of dislocation pairs is lower than that of
vortex pairs by a factor of $1/(8\pi)$.

A Kosterlitz-Thouless transition in layered HTSC's was predicted
by Stamp et al.\ (1988).
In a nice work Blatter, Ivlev and Rhyner (1991b) discuss that in the
presence of a large field parallel to the layers the FLL may undergo
a transition to a smectic state and this would lead to a vanishing
inter-layer shear stiffness. The changed elastic properties of
this FLL then would modify the Kosterlitz-Thouless type behaviour and
would lead to an algebraic current--voltage law down to zero
current density.

  Clear evidence for a BKT transition is observed in current--voltage
curves of thin YBCO films by Ying and Kwok (1990).
Current--voltage curves by Gorlova and Latyshev (1990), originally
interpreted as evidence for a BKT transition, are shown by Minnhagen
(1991) to obey Coulomb gas scaling associated with 2D vortex-pair
fluctuations, see also Jensen and Minnhagen (1991), Minnhagen
and Olsson (1992), and the review by Minnhagen (1987).
A renormalization group theory of vortex-string pairing in thin BSCCO
films yields a BKT transition (Ota et al.\ 1994).
The effect of a magnetic field on the BKT transition is investigated
by Martynovich (1994), who finds that the transition can become of
first order at high fields.
Fischer (1995b) shows that point pinning practically
does not change the BKT transition, see also Fischer (1991, 1992a).
Yazdani et al.\ (1993) observe that in an amorphous MoGe film the
BKT melting is not strongly influenced by pinning when the order
in the vortex lattice is sufficiently large, but
 Hsu and Kapitulnik (1992) in detailed studies of ultrathin
clean monocrystalline Nb films see a strong influence of pinning
on the BKT transition. Chen et al.\ (1993) observe 2D and 3D
flux unbinding in YBCO thin films at small fields.

Melting of the 2D lattice of flux lines in a film with the
interaction (2.5) was simulated by Brass and Jensen (1989).
Using the solution for the complex GL-function $\psi(x,y)$ (2.2)
which describes a 2D FLL with arbitrary flux-line positions near
$H_{c2}$ (Brandt 1969b), Te{\v s}anovi{\'c} (1991),
Te{\v s}anovi{\'c} and Xing (1991)
and Te{\v s}anovi{\'c} et al.\ (1992)
obtained analytical expressions for the partition function and
melting line of 2D superconductors near the upper critical field
$H_{c2}$. For new ideas on the critical behaviour at low fields
see Te{\v s}anovi{\'c} (1995).
In 2D MC simulations using periodic boundary conditions
O'Neill and Moore (1992, 1993) and Lee and Moore (1994) found
no evidence for a liquid-to-solid transition. Hikami et al.\
(1991) and Wilkin and Moore (1993) developed a perturbation
series for the free energy from GL theory, but their
Pad{\'e}-Borel resummation still does not converge to the
exact low-temperature limit. In other 2D MC simulations
Kato and Nagaosa (1993) and Hu and McDonald (1993) obtained
evidence for a first-order transition in the positional
correlation function of a 2D superconductor.
In the London limit and in extensive 2D MC simulations,
Franz and Teitel (1994, 1995) find a weak first order melting
transition with a discontinuous vanishing of the shear modulus.
Yates et al.\ (1995) in their 2D MC simulations using periodic
hexagonal boundary conditions find a melting transition of the FLL.
         \\[0.4 cm]
{\it 6.2.~ Evaporation of flux lines into point vortices}
         \\*[0.2 cm]
   Clem (1991) and Bulaevskii et al.\ (1991) predicted that a
thermally fluctuating vortex line ({\bf figure} 6, right)
``evaporates'' into independent point vortices ({\bf figure} 7)
at a temperature $T_{\rm evap}$ which coincides with the
BKT transition temperature (6.3). At this temperature the mean
square radius $\langle r^2 \rangle$ of the vortex line diverges.
Hackenbroich and Scheidl (1991) and Scheidl and Hackenbroich (1992)
considered this problem in  detail and confirmed the identity
$T_{\rm evap} = T_{\rm BKT}$ even when the virtual excitation of
point-vortex pairs is accounted for and even in the case of
nonequivalent layers.
         \\[0.4 cm]
{\it 6.3.~ Decoupling of the layers}
         \\*[0.2 cm]
  One expects that the 2D pancake vortices in neighbouring layers
of distance $s$ decouple when the typical shear energy of the FLL
$\frac{1}{2} k_{\rm BZ}^2 c_{66}$ starts to exceed the tilt energy
$\frac{1}{2} k_z^2  c_{44}(k_z) $ at $k_z =\pi/s$ (at the shortest
possible shear and tilt wavelengths). With the elastic moduli
from (4.10) to (4.12) this estimate gives a decoupling field
$B_{2D}$ above which the layers are decoupled and the 2D lattices
of pancake vortices are uncorrelated,
\begin{equation}   
 B_{2D} =(\pi\phi_0 /\Gamma^2 s^2) \ln(\Gamma s /\pi\xi_{ab})\,.
\end{equation}

The thermal fluctuations of the point-vortex lattice in the
superconducting layers lead to a misalignment of the point vortices
in adjacent layers ({\bf figure} 9, middle). With perfect alignment,
i.e.\ if the vortex lines are straight and perpendicular to the layers,
the phase difference between neighbouring layers [Eq.\ (3.12)] is
$\delta_n ({\bf r}_\perp) = 0$. With increasing temperature, the
space- and time average $\langle \delta_n^2 \rangle$ increases
until at $\langle \delta_n^2 \rangle \approx 1$ the phase coherence
between the layers is destroyed and the layers thus {\it decouple}.
Glazman and Koshelev (1991) find for this decoupling transition
$T=T_{\rm dec}^{\rm GK} \approx a\phi_0^2 (4\pi)^{-3/2}/ ( \mu_0 k_B
\lambda_{ab}\lambda_c) $. In terms of the 2D melting
transition temperature (6.2) one may write this as
\begin{equation}   
  T_{\rm dec}^{\rm GK} \approx T_m^{2D} (B_{cr}/B)^{1/2}
   ~~~{\rm with}~~~ B_{cr}\approx 64\pi \phi_0/\Gamma^2 s^2.
\end{equation}
Thus, at sufficiently large fields $B>B_{cr}$, decoupling of the
layers should occur at $T$ below $T_m^{2D}$. However, at present
it is not clear if the decoupling is a true phase transition, see
the review by Blatter et al.\ (1994a) for a more detailed discussion
and the recent careful renormalization group analysis of coupled
2D layers by Pierson (1994, 1995). By Monte-Carlo simulations
\v{S}\'a\v{s}ik and Stroud (1995) find a smooth crossover from
3D to 2D behaviour characterized by a continuous change of the
helicity modulus ${\bf Q}$ (section 5.4.) in the $c$-direction,
and of the shear modulus $c_{66}$ in the $ab$-plane, when the
effective inter-layer coupling was decreased.

  A different decoupling transition, driven by topological excitations,
has been suggested by Feigel'man, Geshkenbein and Larkin (1990) (FGL).
The excited objects, called ``quarters'', consist of four 2D edge
dislocations and correspond to a pair of dislocation loops in a 3D
vortex lattice. With increasing $T$, the density of these quartets
increases, and a decoupling transition is expected to occur at
     \setcounter{equation}{5} { \renewcommand{\theequation}{6.6a}
 \begin{equation}   
    T_{\rm dec}^{\rm FGL} \approx T_m^{2D}/\ln(B/B_{2D}).
 \end{equation} }
Daemen, Bulaevskii et al.\ (1993a) (DB) have followed the decoupling
idea further and suggest that  due to these phase fluctuations the
Josephson coupling between
neighbouring layers is renormalized to a  smaller value as $T$ increases.
As consequences, the effective penetration depth
$\lambda_{c {\rm, eff}}$ and
anisotropy ratio  $\Gamma_{\rm eff}= \lambda_{c {\rm, eff}}/\lambda_{ab}$
increase and the maximum possible current density  along the $c$-axis
$J_{z,{\rm max}} = J_0 \langle \cos \delta_n(x,y)  \rangle =
(\Phi_0 /2\pi\mu_0 s\lambda^2_{\it c, {\rm eff}}) \propto
 J_0 \exp[-B/B_{\rm dec}^{\rm DB}(T)]$  decreases and vanishes at a
decoupling field
\begin{eqnarray}  \nonumber 
B_{\rm dec}^{\rm DB}(T) = [\,8\Phi_0^3 / \mu_0 k_B T s\,e\lambda^2_{c}(T)\,]
      ~~~{\rm for}~~~ \Gamma \ll \lambda_{ab}/s,  \\
B_{\rm dec}^{\rm DB}(T) =
              [\, s \Phi_0^3 /8\pi^2 \mu_0 k_B T \lambda^4_{ab}(T)\,]
              \ln(2 \pi \mu_0 k_B T\lambda^4_{ab} /\Phi_0^2 s^3)
      ~~~{\rm for}~~~ \Gamma \gg \lambda_{ab}/s.
\end{eqnarray}
However, this self-consistent treatment of the decoupling is
probably not allowed. Formula (6.7) at low fields $B \ll B_{2D}$
 coincides with the ``second melting line'' of
 Glazman and Koshelev (1991), but at higher $B$ it does not apply since
 $c_{66}$ was neglected in the calculation of the phase-difference.

Several authors find decoupling (3D--2D crossover) in zero
applied field from $xy$-models using Monte-Carlo simulations
(Minnhagen and Olsson 1991, Weber and Jensen 1991) or a
renormalization group treatment (Chattopadhyay and Shenoy 1994,
Shenoy and Chattopadhyay 1995, Pierson 1994, 1995, Friesen 1995).
The exact vortex-antivortex interaction energy within the
anisotropic $xy$-model is calculated by Choi and Lee (1995).
Genenko and Medvedev (1992) predict
a suppression of the critical current by 2D vortex fluctuations
and an avalanche-like generation of free vortices.

  Apart from thermal fluctuations, decoupling of the superconducting
layers may also be caused by random pinning of the point-vortices.
This problem has been considered in detail by Koshelev,
Glazman and Larkin (1995), who find decreasing effective coupling
as the pinning strength or the magnetic field increase. However,
above a ``decoupling field'' this decrease comes to a halt and the
coupling strength saturates to a value which is proportional to the
{\it square} of the original Josephson coupling. These authors also
give an explicit expression for the point-vortex wandering on the
tail of the field distribution, which can be measured by muon-spin
rotation. As shown by Brandt (1991c),  while in conventional
superconductors random pinning distorts the FLL such that the field
distribution becomes {\it broader} (Brandt 1988a,b), in layered
HTSC's pinning can {\it narrow} this distribution since the
flux lines (as stacks of point vortices) become broader and
overlap more strongly such that the magnetic field variance is
reduced, cf.\ section 5.7. The resulting strong narrowing of the $\mu$SR line
was
observed in BSCCO by Harshman et al.\ (1993).

  A further decoupling problem occurs when the field $H_a$ is applied
in the $ab$-plane. Decoupling of the superconducting Cu-O layers then
is equivalent to the melting of the lattice of Josephson vortices
between the layers (Efetov 1979). It was shown both for low fields
(Mikheev and Kolomeisky 1991) and high fields
(Korshunov and Larkin 1992) that such a decoupling does not occur
before total destruction of superconductivity.
  Thermal nucleation of vortex rings between the
Cu-O planes, or crossing these planes, is calculated with a
renormalization group technique by Korshunov (1990, 1993),
Horovitz (1991, 1992, 1993, 1994), and Carneiro (1992).
The importance of vortex rings in layered superconductors was also
stressed by Friedel (1989) and Doniach (1990).
         \\[0.4 cm]
{\it 6.4.~ Measurements of decoupling}
         \\*[0.2 cm]
Decoupling and depinning transitions have been studied in
artificially grown superlattices of                 
YBa$_2$Cu$_3$O$_7$/PrBa$_2$Cu$_3$O$_7$              
(Triscone et al.\ 1990, 1994, Brunner et al.\ 1991,
Fischer et al.\ 1992, Qi Li et al.\ 1992, Fu et al.\ 1993b,
Obara et al.\ 1995)
where the coupling between the superconducting layers can be
gradually decreased to zero by increasing the thickness of the
Praseodymium layers.
Neerinck et al.\ (1991) investigate Ge/Pb superlattices. Decoupling
in BSCCO crystals is observed by Hellerquist et al.\ (1994) and
in a nice work by Pastoriza et al.\ (1994) who find a drop of the
resistivity in $c$-direction by five decades in a temperature
interval of 0.1 K at $B_a = 360$ G. White et al.\ (1991, 1994)
and Steel et al.\ (1993) observe decoupling in synthetic MoGe/Ge
multilayers, see also Urbach et al.\ (1994). In micrometer-thick
such layers with various anisotropies Klein et al.\ (1995)
observe a strong decrease of the critical current with increasing
anisotropy.
Korevaar et al.\ (1994) in NbGe/Ge multilayers with varying
Ge thickness observe a crossover from  2D melting in the full
sample to 2D melting in single layers at high fields, but a 3D
melting at intermediate fields. A decoupling transition is found in
ultrathin and multilayered BSCCO by Raffy et al.\ (1991, 1994)
and in fully oxidized (and thus extremely anisotropic)
YBa$_2$Cu$_3$O$_7$ by Gao et al.\ (1993).
Mikheenko et al.\ (1993) show  that decoupling of the Cu-O layers
can be induced by an in-plane current density whose critical value
$J_c(T)$  is found to vanish at $T_{\rm BKT}$. A similar  current-induced
decoupling in BSCCO films is observed by Balestrino et al.\ (1995).
A review on thermal softening of the interplane Josephson coupling
in HTSC's is given by Kopelevich et al.\ (1994).

  A nice six-terminal method applies a current between two contacts
on the top surface of a HTSC platelet in a perpendicular field
$H\| c$ and measures the ``primary'' and ``secondary'' voltage
drops on the top and bottom by two further pairs of contacts
({\bf figure} 9)
(Busch et al.\ 1992, Safar et al.\ 1992c, 1994, Wan et al.\ 1994,
L\'opez et al.\ 1994, de la Cruz et al.\ 1994a, b,
Eltsev et al.\ 1994, Eltsev and Rapp 1995).
Above some temperature (or at higher
current densities) the bottom voltage drop falls below that of
the top surface since the vortex lattices in the Cu-O layers decouple.
This effect thus has to do with decoupling and with currents along the
$c$-axis (Brice\~no et al.\ 1991, 1993, Charalambous 1992,
Rodr\'{\i}guez et al.\ 1993), but also with flux-line cutting
and with the helical instability of flux lines in a longitudinal
current (section 7.2.). It is thus certainly not an exactly
linear effect. A nice phenomenological explanation is given by
L\'opez et al.\ (1994): The 2D FLL in one layer drags the 2D FLL
in the adjacent layer;  when the coupling force decreases
below the pinning force on the 2D FLL, the adjacent lattices slip
and decouple. This decoupling was calculated for two layers by
Yu and Stroud (1995) and for $N$ layers by
Uprety and Dom\'{\i}nguez (1995).
A similar dragging of magnetically coupled flux-line lattices
in two films separated by an insulating layer was observed by
Giaever (1966) and by Deltour and Tinkham (1968) and was termed
``flux transformer effect'' since a voltage drop can appear at
one film  when the current is applied to the other film.
This effect was investigated in detail by Ekin et al.\ (1974),
Ekin and Clem (1975), and Clem (1974b, 1975b).
Glazman and Fogel' (1984b) consider the coupling between two films
with solid and melted flux-line lattices. An interesting related
``cross-talk'' effect, possibly also caused by moving vortices,
is observed in metal/insulator/metal trilayers near the
superconducting transition of one of the layers (Shimshoni 1995).

In my opinion it seems not appropriate to ascribe the
coincidence between the top and bottom voltage drop to a nonlocal
resistivity as suggested by Safar et al.\ (1994).
Busch et al.\ (1992) and Eltsev and Rapp (1995) successfully
explain their six-terminal measurements on YBCO crystals by
a {\it local} anisotropic conductor.  A {\it nonlocal} resistivity
in principle could follow, e.g., from a  shear viscosity
$\eta_{\rm shear}$ of an entangled flux-line liquid
(Marchetti and Nelson 1990, 1991) in combination with the
viscous drag force exerted by the atomic lattice, usually also
termed  ``viscosity'' $\eta = B^2/\rho_{\rm FF}$ (5.6).
   The characteristic length of this nonlocality
$l_{\rm nloc} = (\eta_{\rm shear} /\eta)^{1/2}$
(Huse and Majumdar 1993) is not known at present; possibly it is
smaller than the flux-line spacing $a$ and thus does not play a
r\^ole. Other possible reasons for nonlocal conductivity in
superconductors are discussed by Mou et al.\ (1995) and
Feigel'man and Ioffe (1995), see also Blum and Moore (1995).
 However, note that any lattice is
 ``nonlocal'' on the length scale of the lattice spacing $a$ if it
is described by a continuum theory: The elastic
moduli become dispersive when the wavevector $k$ of the distortion
approaches the radius of the Brioullin zone $k_{\rm BZ} \approx \pi/a$,
since the elastic energy of a lattice has to be periodic in
${\bf k}$-space, see {\bf figure} 8. For example, in the simple model of a
1D lattice with nearest-neighbour interaction, the factor $k^2$ in
the elastic energy is replaced by the periodic function
$[2\sin(ak/2)/a ]^2$, which formally yields a dispersive
compression modulus $c_{11}(k) =c_{11}(0) [2\sin(ak/2)/ka]^2$.
This natural ``geometric'' nonlocality of a lattice should not
be confused with the nonlocality discussed in section 4.2., which
is due to the large range $\lambda \gg a/\pi$ of the forces
between flux lines. Note also that the top and bottom voltage drops
must exactly concide when the flux lines are coherent and aligned,
since the voltage drop along flux lines is zero (the space between
stationary vortex cores is ideal conducting). The linear ac resistivity
of a coherent FLL is indeed found to be local (section 9.5.), and
Busch et al.\ (1992) could explain their six-terminal
measurements by an anisotropic {\it local} dc conductivity. This
anisotropic conductivity depends on the material anisotropy and also
on the {\it direction of the flux lines}, which varies with the
measuring current due to the self-field of the current. Furthermore,
with pinned flux lines, a weak current typically will not flow through
the bulk but along the surface due to the Meissner effect. These facts
complicate the interpretation of six-terminal measurements.
        \\[0.4 cm]
{\it 6.5.~ Vortex fluctuations modify the magnetization}
         \\*[0.2 cm]
  The thermal fluctuation of the vortex lines gives an entropy
contribution to the free energy and thus to the magnetization
$M=\partial F/\partial H_a$. Bulaevskii et al.\ (1992c, 1994c)
predicted that all magnetization curves $M(H_a,T)$ should cross in
{\it one} point if $M$ is plotted  versus the temperature $T$ with
the applied field $H_a$ as a parameter.
 At this crossing point one has $T=T^* \approx T_c$
and $M(H_a,T^*) = M^* = (c_m  k_B T^* /\phi_0 s) (4\pi /\mu_0)$
with $c_m \approx 1$. This result was also obtained from GL theory
 for $H_a \approx H_{c2}$ by Te\v{s}anovi\'c et al.\ (1992). In a
nice overview on the fluctuation magnetization of 2D superconductors
Koshelev (1994b) derived the correction factor as $c_m =0.346573...~$.
 This relation was experimentally confirmed,
e.g., by Kes et al.\ (1991), Qiang Li et al.\ (1992, 1993),
Zuo et al.\ (1993), Kogan et al.\ (1993b), Thompson et al.\ (1993),
Jin et al.\ (1994), Wahl et al.\ (1995), and Genoud et al.\ (1995).
 It also agrees
with earlier data $M(H_a, T)$ of Welp et al.\ (1991), which
depend only on {\it one scaling variable} $[T-T_c(H)] /(TH)^{2/3}$
as explained by 3D  GL fluctuation theory, see also the Comment by
Salamon and Shi (1992).
Buzdin and Feinberg (1994a) calculate the angle dependence of the
fluctuation magnetization and specific heat for anisotropic
superconductors, and Baraduc et al.\ (1995) extend this to
include the 2D-3D crossover regime in YBCO.
The influence of thermal fluctuations on the vortex-caused
$^{205}$TL NMR frequency shift is discussed by
Song et al.\ (1993) and Bulaevskii et al.\ (1993b),
see also Song (1995).
  Kogan et al.\ (1993b) show that thermal fluctuations of vortices
in layered superconductors influence also the penetration depth
$\lambda_{ab}(T)$, the GL parameter $\kappa(T)$, and the upper
critical field $H_{c2}(T)$ extracted from magnetization data.

Martinez et al.\ (1994) find experimental evidence that both quantum
and thermal fluctuations modify the reversible magnetization of a
BSCCO crystal. Blatter, Ivlev and Nordborg (1993) consider the
fluctuations of vortices in HTSC's  and in thin films accounting also
for the fluctuations of the amplitude of the order parameter $\psi$;
they argue that a spontaneous creation of flux lines cannot occur in a
3D superconductor.
         \\[0.4 cm]
{\it 6.6.~ Quantum fluctuations of vortices}
         \\*[0.2 cm]
The possible melting of the vortex lattice in HTSC's due to quantum
fluctuations was considered by Narahari Achar (1991) and by
Blatter and Ivlev (1993, 1994); a melting line with quantum
correction was observed in YBCO by de Andrade and de Lima (1995).
 A related effect, the macroscopic
tunnelling of vortices out of pins, is discussed in section 9.9.
A particularly elegant formulation of the quantum theory of superfluid
undamped vortices was given by Fetter (1967) and is extended to
undamped point vortices in layered superconductors by Fetter (1994),
see also the calculations of helicon modes in HTSC by Blatter and
Ivlev (1995). Following Blatter and Ivlev (1993, 1994) and  Blatter
et al.\ (1994b), the combined thermal and quantum fluctuations
 $\langle u^2 \rangle$ are given by a sum over Matsubara frequencies
$\omega_n = 2\pi n kT/\hbar$ ($n=$0, 1, 2, ...) with the term $n=0$
corresponding to the pure thermal fluctuations. The cutoff of this
sum crucially enters the resulting $\langle u^2 \rangle$.

  Bulaevskii and Maley (1993) calculate the low temperature specific
heat of the vortex lattice, introducing the term ``vorton'' for the
viscously relaxing elastic modes considered in section 5.2.
In subsequent work Bulaevskii et al.\ (1994a, b, c) consider the effect
of quantum fluctuations on the reversible  magnetization in HTSC's
starting from the classical Langevin equation (Schmid 1982) and adding
the vorton contribution obtained from Caldeira and Leggett's (1981)
action for a damped oscillator.
The Hall effect of moving vortices yields a force term
$\alpha {\bf v\times \hat z}$ which is linear in the velocity
${\bf v}$ like the viscous drag force $\eta {\bf v}$, but which does not
dissipate energy. This Hall term was considered in quantum fluctuations
(Fetter 1967, 1994, Blatter and Ivlev 1994,
Morais et al.\ 1994, 1995,  Horovitz 1995)
and quantum depinning (Feigel'man et al.\ 1993b, Ao and Thouless 1994,
Stephen 1994, Sonin and Horovitz 1995, Chudnovsky 1995).
%
         \\[0.7 cm]
{\bf 7.~ Flux motion}
         \setcounter{subsection}{7}  \setcounter{equation}{0}
         \\*[0.5 cm]
{\it 7.1.~ Usual flux flow}
         \\*[0.2 cm]
An  electric current density ${\bf J}$ flowing through the
superconductor exerts a Lorentz force
density ${\bf B \times J}$ on the flux-line lattice which
causes the vortices to move with mean velocity ${\bf v}$. This
vortex drift generates an electric field ${\bf E = B\times v}$
where {\bf B} is the flux density or magnetic induction in the sample,
see the review by Kim and Stephen (1969).
An interesting discussion of the voltage generated along normal
conductors and superconductors is given by Kaplunenko et al.\ (1985),
who show that the ``Faraday voltage'' $\partial \Phi/\partial t$
induced by the changing magnetic flux $\Phi$ can be compensated by the
``Josephson voltage'' $(\hbar /2e)\partial\phi/\partial t$ caused by the
changing phase difference $\phi$ of the order parameter $\psi$ at the
contacts, see also Clem (1981). Moving flux lines dissipate energy by
two effects which give approximately equal contributions:

(a) By dipolar eddy currents that surround each moving flux line
and have to pass through the vortex core, which in
the model of Bardeen and Stephen (1965) is approximated by a normal
conducting cylinder.

(b) By the retarded recovery of the order parameter $\psi({\bf r})$
at places where the vortex core (a region of suppressed  $|\psi|$)
has passed by (Tinkham 1964).

In the solution of this flux-flow problem by the time-dependent
Ginzburg-Landau theory (TDGL), achieved for $B\approx B_{c2}$ and
for curved flux lines
by Thompson and Hu (1971, 1973) and for $B\ll B_{c2}$ by
Hu and Thompson (1972), both dissipation effects are combined.
Applying the TDGL theory to HTSC's, Ullah and Dorsey (1990, 1991),
Vecris and Pelcovits (1991),
and Troy and Dorsey (1993) consider the influence of fluctuations
on the transport properties of type-II superconductors and
on the moving FLL;  Dorsey (1992) and Kopnin et al.\ (1993)
obtain the  Hall effect during flux flow, and Ivlev
et al.\ (1990), Hao and Clem (1991a),
Golosovsky et al.\ (1991, 1992, 1993, 1994),
Hao and Hu (1993), and Krivoruchko (1993) give
the flux-flow resistivity of anisotropic superconductors.
 The Bardeen-Stephen model is generalized
to layered superconductors by Clem and Coffey (1990).
 For reviews on moving vortices see Larkin and Ovchinnikov (1986),
Gor'kov and Kopnin (1975), Shmidt and Mkrtchyan (1974), and
Tinkham's book (1975).

  As observed by neutron scattering (Simon and Thorel 1971,
Schelten et al.\ 1975, Yaron et al.\ 1994, 1995) and NMR
(Delrieu 1973),  during flux flow the FLL can become {\it virtually
perfect over the entire specimen}, in spite of pinning; see also
the NMR experiments on rf penetration and flux flow by Carretta (1993)
and recent decoration experiments by Duarte et al.\ (1995).
This means the FLL flows past the random pinning potential
averaging out its effect such that the FLL correlation length
increases. This ``shaking'' and ordering of a FLL flowing across pins,
and maximum disorder just above $J_c$ when depinning starts,
is also seen in computer simulations (Brandt unpublished) and is
calculated explicitly by Koshelev and Vinokur (1994), see also
Shi and Berlinsky (1991) and Bhattacharya and Higgins (1993, 1994).

In thin superconducting films in the presence of a driving current and
an rf field, the differential resistivity ${\rm d}V/{\rm d}I$ exhibits
peaks caused by the interaction of the moving periodic FLL with the
pins; this effect allows the determination of the shear modulus of
the FLL (Fiory 1971, 1973). Fiory's nice experiments were explained by
Larkin and Ovchinnikov (1973) and Schmid and Hauger (1973).
Recent new such experiments on the ``washboard frequency'' during
flux flow by Harris et al.\ (1995a) show that the peaks in
${\rm d}V/{\rm d}I$ vanish rather abruptly at a ``melting field''
$B_m(T)$ of the FLL, e.g., $B_m = 3.5$ T at $T=86$ K, cf.\ section 5.3.
 Mathieu et al.\ (1993) investigate the
dissipation by short vortices or ``flux spots'' in a thin surface
sheath above $H_{c2}$. Tilted vortices in thin films in an oblique field
are considered in a brilliant theoretical paper by Thompson (1975),
and are reviewed by Minenko and Kulik (1979). In thin films with
a current the curvature of the {\it flux lines} can be much less than
the curvature of the magnetic {\it field lines} (Brandt 1993a).

Since at low $B$ the dissipation of the vortices is additive and
since at the upper critical field $B_{c2}(T)$ the flux-flow
resistivity $\rho_{\rm FF}$ has to reach the normal conductivity
$\rho_n$, one expects $ \rho_{\rm FF}(B,T) \approx \rho_n B/B_{c2}(T)$.
This linear relationship was observed by Kim, Li and Raffy (1965) in
a NbTa alloy at low $T\ll T_c$, while $\rho_{\rm FF}(B)$ became
strongly concave as $T\to T_c$. Raffy et al.\ (1991) carefully
measured the flux-flow resistivity of thin BSCCO films.
Recently  Kunchur et al.\  (1993, 1994) nicely confirmed the linear
$\rho_{\rm FF}(B)$ in experiments using short pulses of extremely
high current density {\bf J},  which provides spatially constant
${\bf J}$ and ``free flux flow'', i.e.\  absence of pinning effects,
but still does not heat the sample, such that even the depairing
current density $J_0 = (\phi_0 /3\sqrt{3} \pi \lambda^2 \xi\mu_0)$
can be reached.

At high flux-flow velocities in YBCO
Doettinger et al.\ (1994) observe an instability and a sudden
jump to a state with higher resistivity, which was predicted by
Larkin and Ovchinnikov (1975) as caused by the nonequilibrium
distribution of the quasiparticles,
see also Guinea and Pogorelov (1995).
This electronic instability should not be confused with the
pinning-caused thermo-magnetic instability predicted as the
solution of a system of nonlinear differential equations by
Maksimov (1988); this explosive instability occurs during flux jumps
(see, e.g.\ McHenry  et al.\ 1992, Gerber et al.\ 1993,
and M\"uller and Andrikidis 1994) and may lead to
the nucleation and dendritic fractal growth of normal spots, which
was observed by Leiderer et al.\ (1993) in YBCO films and
explained by Maksimov (1994). Magnetic instabilities in type-II
superconductors were also investigated by Wipf (1967) and
Ivanchenko et al.\ (1979) and were reviewed by
Mints and Rakhmanov (1981) and Wipf (1991).
         \\[0.4 cm]
{\it 7.2.~ Longitudinal currents and helical instability}
         \\*[0.2 cm]
When the current flows exactly parallel to the flux lines it does not
exert a Lorentz force and thus should flow without dissipation.
Flux-line arrangements which satisfy ${\bf B \times J}=0$
(or ${\bf B\times} {\rm rot}{\bf B} =0$ in the bulk) are called
``force-free'', see Walmsley (1972) and Campbell and Evetts (1972)
and references therein.  For example, a twisted FLL with
$B_x = B\cos(kz)$ and $B_y=B\sin(kz)$ exhibits
${\rm rot}{\bf B}= -k{\bf B}$ and thus is a force-free configuration.
In real, finite-size superconductors such force-free configurations
develop an instability near the surface since there the condition
${\bf B \| J }$ is violated.
Longitudinal currents are limited by the onset of a
{\it helical instability} of the single flux line (Clem 1977) or of
the FLL in the bulk or near a planar surface (Brandt 1980c, 1981b),
or of the macroscopic current path similarly as a wire carrying a
current between the poles of a magnet will bend into an arc
(Campbell 1980).
  This instability (see also section 3.5.) defines a longitudinal
critical current density $J_{c \|}$, which has been measured, e.g.,
by Irie et al.\ (1988a), see also D'Anna et al.\ (1994a, b).
  The onset of spontaneous helical deformation of the flux lines
is delayed by the finite shear modulus $c_{66}$ of the FLL or
by elastic pinning (Labusch parameter $\alpha_L$): For zero pinning
one has $J_{c\|} \sim (c_{66}/c_{11})^{1/4}$, and for weak pinning
one has $J_{c\|} \sim (\alpha_L /c_{11})^{1/4}$ (Brandt 1981b).
 The low power $1/4$ means that even very weak pinning can stabilize
longitudinal currents. Khalfin et al.\ (1994) considered a layered
superconductor with a vortex line driven by a longitudinal ac current.
 Genenko (1993, 1994, 1995) calculates the relaxation of vortex rings
in a cylindre, the irreversible right-handed vortex-helix entry into a
current carrying cylindrical wire in a longitudinal field, and the
left-handed helical instability near the surface.

  For $J > J_{c \|}$  the FLL deforms spontaneously and
the flux-line helices grow until they cut other flux lines or
hit the specimen surface. The resulting dissipative, oscillating,
and possibly chaotic state, with spiral structure at the surface of a
cylindrical specimen, has been the subject of intricate experimental
(Timms and Walmsley 1975, Ezaki and Irie 1976, Walmsley and
Timms 1977, Cave and Evetts 1978, Blamire and Evetts 1985,
Le Blanc et al.\ 1993) and theoretical
(Clem 1980, 1981a, 1992, Brandt 1980b, 1982,
Perez-Gonzales and Clem 1985, 1991, Matsushita and Irie 1985)
investigation, which recently was resumed by Marsh (1994).
  The basic problem with this resistive state, say in a cylindre
with longitudinal current, is that a continuous migration
of helical flux lines from the surface leads to a pile-up
of longitudinal flux in the centre of the specimen, since there
the azimuthal flux component of a helix vanishes (vortex rings
contract to a point) but the longitudinal component remains.
The diverging longitudinal flux cannot be removed from the centre
simply by outgoing helices since these will cut and reconnect with
the incoming helices such that the transport of azimuthal flux
comes to a halt and thus the observed longitudinal voltage drop
cannot be explained. An intricate double-cutting model which solves
this problem was suggested by Clem (1981a).
{}From these problems it is clear that there are situations where the
magnetic flux, carried by continuous Abrikosov flux lines,
resists cutting and reconnection, in contrast to the imaginary magnetic
fields lines in a normal metal which can easily go through each other.
         \\[0.4 cm]
{\it 7.3.~ Cross-flow of two components of magnetic flux}
         \\*[0.2 cm]
In a long strip the diffusion (or flow) of transverse flux
slows down when longitudinal flux is present (Brandt 1992b).
Similarly, in a strip
with pinning the penetration of transverse flux is delayed
when a longitudinal field is applied. This induced anisotropy is
discussed in a nice paper by Indenbom et al.\ (1994c), see also
Andr{\'e} et al.\ (1994).
The resulting enhancement of the critical current by this ``cross-flow''
is measured by LeBlanc et al.\ (1993) and Park and Kouvel (1993).
This enhancement is closely
related to the stabilization of longitudinal currents and the
delay of helical instability by pinning (Brandt 1981b).
For a long strip containing a parallel flux density
$B_\| \approx \mu_0 H_{a \|}$, the obstruction of the penetration of
a perpendicular flux density $B_\perp$ from the long edges
can be understood in three different ways ({\bf figure} 10):

 (a) The application of a small perpendicular field $H_{a \perp}$
tilts the flux lines near the edges. Penetration of $B_\perp$
means that this tilt has to penetrate, but for long strips even small
tilt causes large flux-line displacements, which are prevented by
even weak pinning. The flux lines thus stay parallel to the strip,
and $B_\perp$ cannot penetrate if the flux lines are coherent.

 (b) Before $B_\perp$ has penetrated, the application of $H_{a \perp}$
induces a Meissner screening current along the four edges of the strip
and over its entire surface. At the {\it short} edges, this
screening current is {\it perpendicular} to $B_\|$  and thus tilts
the flux lines, which means that $B_\perp$ penetrates from there.
However, at the {\it long} edges, the screening current flows
{\it parallel} to the flux lines and does not exert a force or
torque on these until helical instability sets in, which can be
stabilized by even weak pinning.

 (c) In a more local picture, the penetrating $B_\perp$ has to
cross the large $B_\|$, which means that flux lines have to cut
each other at an angle of $90^0$. After each cutting process the
new flux-line ends reconnect. The resulting flux lines are thus
tilted and will prevent further penetration of perpendicular flux
if they are pinned and coherent.

  These three pictures show that the presence of weakly pinned
longitudinal coherent flux lines causes a high threshold for the
penetration of perpendicular flux, but if the flux lines consist of
almost uncoupled point vortices and of Josephson strings like
in BSCCO, this threshold will be low. At $T=0$ the flux
penetration (or initial magnetization) will be highly nonlinear.
At higher temperature, when thermally activated pinning occurs, the
flux penetration may be described as a linear anisotropic
diffusion of flux (Brandt 1992b).

 The cross-flow problem may be illustrated also with the simpler
example of a slab in the $y,z$-plane  containing a parallel
flux density $B_z$ (Campbell and Evetts 1972). A penetrating
$B_y$ will tilt the flux lines at the surface, but even weak pinning
can prevent this. In fact, as discussed by Brandt and Indenbom (1993),
in the above example of a long strip with pinning, the penetration of
perpendicular flux will start at the {\it flat surfaces}. From there
a Bean critical state penetrates from both sides like with the
slab. When the fronts of antiparallel flux meet at the midplane
of the strip, the flux lines will close first at the long edges and
$B_\perp$ will then penetrate from there as discussed in section 8.4.
         \\[0.4 cm]
{\it 7.4.~ Dissipation in layered superconductors}
         \\*[0.2 cm]
In HTSC's the flux-flow resistivity is highly anisotropic, see,
e.g.\ the beautiful angle-dependent data of
Iye et al. (1990a,b, 1991a, b, 1992)  ({\bf figure} 11)
and the careful comparison of the resistivity in the two cases
$B\perp J$ and $B\| J$ by Kwok et al.\ (1990).
Most remarkably, if both $B$ and $J$ are in the $ab$-plane, the
resistivity in BSCCO is found to be {\it independent of the angle
between $B$ and $J$}, in contradiction to the prediction of the
flux-flow picture (Iye et al.\ 1990a,b, Raffy et al.\ 1991).
This puzzling observation can be understood in various pictures: By the
only relevance of the induction component perpendicular to the layers
(Kes et al.\ 1990); by vortex segments hopping across the layers,
thereby nucleating pancake-antipancake pairs which are
driven apart by the Lorentz force (Iye et al.\ 1990b); by the
correct angular scaling of anisotropic resistivity
(Blatter, Geshkenbein and Larkin 1992, Iye et al.\ 1992,
Fu et al.\ 1993a); or by the motion of
vortex kinks (Ando et al.\ 1991, Iye et al.\ 1991a).
Hao, Hu and Ting (1995) show that this Lorentz-force independence of
the resistivity in BSCCO (observed to a lesser extend in the less
anisotropic YBCO, but not at all in isotropic MoGe films)
is just a consequence of the large material anisotropy
when one accounts for the fact that the electric field induced
by vortex motion is always perpendicular to ${\bf B}$. Note that
only the perpendicular part ${\bf J_\perp}$ of the current density
${\bf J = J_\perp +J_\| = \hat B (J \hat B)+
          \hat B \times (J \times \hat B)}$
 causes dissipation, while the parallel current density ${\bf J_\|}$
is dissipation-free or may  cause a helical instability
(Clem 1977, Brandt 1980c, 1981b), which distorts the FLL.

  A further explanation of the Lorentz-force independence of the
resistivity near $T_c$ (Ikeda 1992a) is based on the renormalized
fluctuation theory of the order parameter developed by
Ikeda et al.\ (1991b). These fluctuations also modify the
conductivity (Aslamazov and Larkin 1968,
Schmidt 1968, Maki and Thompson 1989, Dorsey 1991,
Ullah and Dorsey 1990, 1991, Dorin et al.\ 1993),
the magnetoresistance (Hikami and Larkin 1988), and the diamagnetism
(Hopfeng\"artner et al.\ 1991, Koshelev 1994b) just above and below
$T_c$, and they broaden the resistive transition in multilayer HTSC's
(Koyama and Tachiki 1992), see also section 6.5.
 For a comprehensive review on the theory
and experiments of fluctuation effects in superconductors see
Tinkham and Skocpol (1975).
Lang et al.\ (1995) recently gave a detailed study of the
normal-state magnetotransport in BSCCO thin films with regard to
contributions of the superconducting fluctuations.

  Current transport along the $c$-axis (see also section 3.5.)
was investigated in YBCO for $B\| c$ and $B\perp c$
by Charalambous et al.\ (1992) and
in BSCCO by Brice\~no et al.\ (1991, 1993) and
Rodriguez et al.\ (1993). The pronounced maximum in the
$c$-axis resistivity at a temperature below $T_c$ was interpreted and
well fitted by Brice\~no et al.\ (1993) using the
Ambegaokar and Halperin (1969) resistivity of weak links at
finite temperature; the same model was found to fit the resistivity of
various HTSC's very well (Soulen et al.\ 1994); see also the
explanations of the magnetic-field-caused broadening of the
resistive transition by Tinkham (1988) and Inui et al.\ (1989), and
the phenomenology of flux motion in HTSC's by Doniach et al.\ (1990).
Balestrino et al.\ (1993) fit the enhanced $c$-axis resistivity
in thin BSCCO films to the fluctuation model of Ioffe et al.\ (1993).
For recent work on the $c$-axis resistivity of BSCCO see
Zhao et al.\ (1995), Luo et al.\ (1995), and Cho et al.\ (1995).

Woo et al.\ (1989) and Kim et al.\ (1990b) in
Tl$_2$Ba$_2$CaCu$_2$O$_x$,
and Kim et al.\ (1990a) in granular NbN films, observe a Lorentz-force
independent resistance tail in magnetic fields near $T_c$ and
explain this quantitatively by Josephson coupling between grains
using the Josephson critical current derived by Ambegaokar and
Baratoff (1963). Recent experiments on fluctuations in Tl-based
HTSC's were performed by Ding et al.\ (1995) and Wahl et al.\ (1995).
Another dissipation mechanism in terms of 2D
vortex--antivortex pair breaking is suggested by Ando et al.\ (1991).
In NbN/AlN multilayer films Gray et al.\  (1991)
find a crossover between flux pinning and Josephson coupling in the
current-voltage curve as $H_a$ is increased.  Iye et al.\ (1990b)
carry out a comparative study of the dissipation processes in HTSC's.
A comprehensive review on Lorentz-force independent dissipation in
HTSC's is given by Kadowaki et al.\ (1994).
         \\[0.4 cm]
{\it 7.5.~ Flux-flow Hall effect}
         \\*[0.2 cm]
 A much debated feature of HTSC's and of some conventional
superconductors [see Hagen et al.\ (1993) for a compilation]
is that near $T_c$ the Hall effect observed during flux flow
changes sign as a function of $B$ or $T$. This {\it sign reversal
of the Hall angle} is not yet fully understood though numerous
explanations were suggested,
e.g., large thermomagnetic effects (Freimuth et al.\ 1991),
flux flow and pinning (Wang and Ting 1992 and Wang et al.\ 1994),
opposing drift of quasiparticles (Ferrell 1992),
complex coefficient in the time-dependent GL equation (Dorsey 1992),
bound vortex-antivortex pairs (Jensen et al.\ 1992),
the energy derivative of the density of states of quasiparticles at
the Fermi energy may have both signs (Kopnin et al.\ 1993),
segments of Josephson strings between the Cu-O layers
(Harris et al.\ 1993, 1994; see the criticising Comments by
Geshkenbein and Larkin 1994 and Wang and Zhang 1995),
renormalized tight-binding model (Hopfeng\"artner et al.\ 1993,
and unpublished 1992; Hopfeng\"artner notes that the sign reversal
occurs at a temperature where the longitudinal resistivity is
dominated by fluctuations of the order parameter $\psi$),
fluctuations (Aronov and Rapoport 1992),
vortex friction caused by Andreev reflection of electrons at the
vortex core boundary (Meilikhov and Farzetdinova 1993),
unusual Seebeck effect (Chen and Yang 1994),
motion of vacancies in the FLL (Ao 1995), and the
difference in the electron density inside and outside the vortex core
(Feigel'man et al.\ 1994, 1995). See also the recent calculations of
the  Hall effect in dirty type-II superconductors by
Larkin and Ovchinnikov (1995) and Kopnin and Lopatin (1995)
and the microscopic analysis by van Otterlo et al.\ (1995).

Samoilov (1993) points out a universal scaling behaviour of the
thermally activated Hall resistivity. Vinokur et al.\ (1993) prove the
universal scaling law $\rho_{xy} \sim\rho_{xx}^2$ for the resistivity
tensor in HTSC's, which is confirmed, e.g.,  by the experiments of
W\"oltgens et al.\ (1993b) and Samoilov et al.\ (1994).
  Pinning of flux lines influences the Hall signal
(Chien et al.\ 1991):  the thermally activated nature
of flux flow renormalizes the effective viscosity $\eta$ but leaves
the Hall parameter $\alpha$ unchanged (Vinokur et al.\ 1993).
Direct evidence of the independence of the Hall resistivity on
the disorder caused by heavy-ion irradiation of both YBCO
and Tl$_2$Ba$_2$CaCu$_2$O$_8$ (with negative and positive sign of
the Hall effect in the pinned region, respectively) is observed
by Samoilov et al.\ (1995).
  Including the Hall effect into collective pinning theory, Liu,
Clinton and Lobb (1995) also find the scaling
$\rho_{xy} \sim\rho_{xx}^2$ but no sign reversal, suggesting that
the sign reversal is not a pinning effect. Indeed, the sign reversal
of $\rho_{xy}$ was found to be even more pronounced at very high current
densities where pinning is unimportant (Kunchur et al.\ 1994).
Budhani et al.\ (1993) find suppression of the sign anomaly by strong
pinning at columnar defects.

A double sign change [explained by Feigel'man et al.\ (1994, 1995)]
is observed in YBa$_2$Cu$_4$O$_8$ by Schoenes et al.\ (1993),
and a sign change in MoGe/Ge multilayers by Graybeal et al.\ (1994)
and in amorphous isotropic Mo$_3$Si by Smith et al.\ (1994).
 Bhattacharya et al.\ (1994) observe a
sharp minimum in $\rho_{xy}$ just below $H_{c2}$.
Ginsberg and Mason (1995) in twin-free YBCO crystals observe a
Hall conductivity $\sigma_{xy} = -c_1\tau^2/H_a + c_2 \tau H_a^2$
which has the dependence on field $H_a$ and reduced temperature
$\tau= 1-T/T_c$ predicted by Dorsey (1992), Troy and Dorsey (1993),
and Kopnin et al.\ (1993).
Harris et al.\ (1995b) corroborate the additivity of the Hall
conductivities of quasiparticles ($\sigma_{xy}^n \sim H$) and of
vortices ($\sigma_{xy}^f \sim 1/H$) with $\sigma_{xy}^n >0$ and
with $\sigma_{xy}^f >0 $ in  LaSrCuO,
but  $\sigma_{xy}^f <0 $ in untwinned YBCO, resulting in a
sign reversal of the total $\sigma_{xy}$ in YBCO but not in LaSrCuO.
So far in all these experiments and theories it remains open why
one of the contributions to the Hall conductivity is negative.
A mechanism for an even Hall effect in ceramic or inhomogeneous
superconductors is proposed by Meilikhov (1993). Horovitz (1995)
predicts a quantized Hall conductance  when the vortex dynamics
is dominated by the Magnus force.
         \\[0.4 cm]
{\it 7.6.~ Thermo-electric and thermo-magnetic effects}
         \\*[0.2 cm]
  Another interesting group of phenomena which for brevity
cannot be discussed here in detail, are the thermo-electric
(Seebeck and Nernst) effects and the thermo-magnetic (Peltier and
Ettingshausen) effects, i.e., the generation of a temperature gradient
by an electric current, or vice versa, and the dependences on the
applied magnetic field, see, e.g.,  Huebener et al.\ (1991, 1993),
Hohn et al.\ (1991),  Fischer (1992b), Oussena et al.\ (1992),
Ri et al.\ (1993), Zeuner et al.\ (1994), and Kuklov et al.\ (1994).
The Peltier effect is discussed by Logvenov et al.\ (1991).
Meilikhov and Farzetdinova (1994) present a unified theory of
the galvanomagnetic and thermomagnetic phenomena based on a
phenomenological three-fluid model (superconducting and normal
quasiparticles plus flux lines) in analogy to the two-fluid
model of superconductors by Gorter and Casimir (1934).
The thermo effects originate from the {\it transport of entropy by
moving flux lines} (Maki 1969b, 1982, Palstra et al.\ 1990a,
Huebener et al.\ 1990, Kober et al.\ 1991a,
Golubov and Logvenov 1995).
A comprehensive comparative study of Nernst,
Seebeck, and Hall effects was recently given by Ri et al.\ (1994),
see also the review paper on superconductors in a temperature
gradient by Huebener (1995).
         \\[0.4 cm]
{\it 7.7.~ Flux-flow noise}
         \\*[0.2 cm]
The motion of flux lines causes both magnetic and voltage noise
originating from pinning in the bulk and at surfaces. Measurements
and theories of the flux-flow noise were reviewed by Clem (1981b).
 Ferrari et al.\ (1991, 1994) and Wellstood et al.\ (1993)
find that noise in thin YBCO films can be explained by individual
vortices hopping between two potential wells in a confined region
and that flux noise is suppressed by a very weak current.
Lee at al.\ (1995) observe correlated motion of vortices in
YBCO and BSCCO films up to 30 $\mu$m thick by measuring the noise
at opposing surfaces; depending on temperature, the flux lines
move as rigid rods, or they are pinned at more than one pin along
their length, thus destroying the correlation.
 Marley et al.\ (1995) and D'Anna et al.\ (1995)
measure an increase of flux-flow noise near the peak effect
(section 9.8.) as an indication of a dynamic transition in the FLL.
Marx et al.\ (1995) observe and discuss $1/f$ noise in a BSCCO
bicrystal grain-boundary Josephson junction,
and Kim, Kang et al.\ (1995) observe voltage noise in YBCO films.
  For interesting
ideas related to $1/f$ noise and to noise during self-organized
criticality see Jensen (1990) and Wang et al.\ (1990).
Ashkenazy et al.\ (1994, 1995a) give theories of random telegraph
noise and (1995b) of flux noise in layered and anisotropic
superconductors. Tang et al.\ (1994) find the noise of a single
driven flux line by computer simulation, see also the Comment by
Brandt (1995b) and new work by Denniston and Tang (1995).
Houlrik et al.\ (1994) and Jonsson and Minnhagen (1994)
calculate the flux-noise spectrum for 2D superconductors from a
$xy$-like model with time-dependent GL dynamics.
         \\[0.4 cm]
{\it 7.8.~ Microscopic theories of vortex motion}
         \\*[0.2 cm]
The flux-flow resistivity, the Hall effect, and the effects of a
temperature gradient on flux flow have to do with the forces on moving
vortex lines in superconductors (Huebener and Brandt 1992).
This complex problem is still not fully understood.
A nonlocal generalization of the model of Bardeen and Stephen (1965)
to low-$\kappa$ superconductors was given by Bardeen and Sherman (1975)
and Bardeen (1978). The equation of motion of a vortex in the
presence of scattering was derived by Kopnin and Kravtsov (1976),
Kopnin and Salomaa (1991), and Kopnin and Lopatin (1995).
The modern concept of the Berry phase was applied to reproduce the
Magnus force on a moving vortex (Ao and Thouless 1993, Gaitan 1995,
Ao and Zhu 1995).
The classical theory of vortex motion by Nozi\`eres and Vinen (1966)
and Vinen and Warren (1967) recently was confirmed quantitatively
by Hofmann and K\"ummel (1993) who show that half of the Magnus force
on the vortex originates from the electrostatic field derived from
Bernoulli's theorem, and the other half from
Andreev scattering of the quasiparticles inside the vortex core.
During Andreev scattering at an interface or at inhomogeneites
   of the order parameter, a Cooper pair dissociates into
an electron and a hole, and one of these two particles is reflected
and the other one passes; see K\"ummel et al.\ (1991) for a review on
Andreev scattering in weak links.
Andreev scattering at the vortex core also changes the thermal
conductivity and ultrasonic absorption in the presence of a FLL
(Vinen et al.\ 1971, Forgan 1973).
Accounting for the excitations in the vortex core (Caroli et al.\ 1964,
Klein 1990, Ullah et al.\ 1990),
Hsu (1993) and Choi et al.\ (1994) calculate the frequency-dependent
flux-flow conductivity up to infrared frequencies, cf. also section 9.6.
For further ideas on moving vortices see the reviews by
Gorkov and Kopnin (1975) and Larkin and Ovchinnikov 1986, and the work
by Gal'perin and Sonin (1976), Ao and Thouless (1993, 1994),
 Niu et al.\ (1994), and Gaitan (1995).
The analogy between the vortex dynamics in superfluids and
superconductors is only partial, see Vinen (1969) and
Sonin and Krusius (1994) for overviews and the book by
Tilley and Tilley (1974). Recently Dom\'{\i}nguez et al.\ (1995)
showed that the Magnus force on the flux lines results in the
{\it acoustic Faraday effect}, i.e., the velocity of ultrasound
propagating along the magnetic field depends on the polarization.

 Another interesting problem is the low-temperature viscosity
$\eta$ of flux lines. Matsuda et al.\ (1994) in YBCO measure extremely
large $\eta$ (or low $\rho_{FF} =B^2/\eta$), which hints to a large
Hall angle and to extremely clean materials ($l \gg \xi$),
see also Harris et al.\ (1994). This
vortex viscosity enters also the theories of the low-temperature
specific heat of the FLL (Bulaevskii and Maley 1993, Fetter 1994),
which thus may be used to measure $\eta$. Guinea and Pogorelov (1995)
find that HTSC's with their short coherence length are in the
``ultra\-clean'' limit, with electronic levels in the vortex core
separated by energies comparable to the energy gap, resulting in
a large vortex viscosity.
%
         \\[0.7 cm]
{\bf 8.~ Pinning of flux lines}
         \setcounter{subsection}{8}  \setcounter{equation}{0}
         \\*[0.5 cm]
{\it 8.1.~ Pinning of flux lines by material inhomogeneities}
         \\*[0.2 cm]
  In real superconductors at small current densities  $J<J_c$ the
flux lines are pinned by inhomogeneities in the material, for example
by dislocations, vacancies, interstitials, grain boundaries, twin planes,
precipitates, irradiation-caused  defects (e.g.\ by fast neutrons,
Weber 1991, Sauerzopf et al.\ 1995), by planar dislocation networks
(e.g.\ competing with oxygen vacancies in BSCCO, Yang et al.\ 1993),
or by a rough surface, which causes spatial variation
of the  flux-line length and energy.  HTSC's and some organic
superconductors exhibit intrinsic pinning by their layered structure
(Tachiki and Takahashi 1989). Only with pinned flux lines does
a superconductor exhibit zero resistivity.
When $J$ exceeds a  critical value $J_c$, the vortices
move and dissipate energy. Pinning of flux lines was predicted by
Anderson (1962) and is described in a comprehensive review by
Campbell and Evetts (1972), see also the books by Ullmaier (1975) and
Huebener (1979), the reviews by Campbell (1991, 1992a, b),
Zhukov and Moshchalkov (1991), Malozemoff (1991), and Pan (1993),
and the general articles on high-field high-current superconductors by
Hulm and Matthias (1980), on critical currents and magnet applications
of HTSC's by Larbalestier (1991), and on pinning centres in HTSC by
Raveau (1992). Pinning has several consequences:

  (i) The current-voltage curve of a superconductor in a magnetic
field is highly nonlinear. At $T=0$ or for conventional
superconductors one has the electric field $E=0$ for $J<J_c$
and $E =\rho_{\rm FF}J$ for $J\gg J_c$.
 For $J$ slightly above $J_c$ various concave or convex shapes of
$E(J)$  are observed, depending on the type of pinning and on the
geometry of the sample. Often a good approximation is (with, e.g.\,
$p=1$ or $p=2$)
   \begin{equation}   
E(J) = 2\rho_{\rm FF} [1 - (J_c/J)^p]^{1/p}
       ~~~~(J \ge J_c).
   \end{equation}

   (ii) The magnetization curve $M(H_a)$ is irreversible
and performs a hysteresis loop when the applied magnetic field
$H_a= B_a/\mu_0$ is cycled.
When $H_a$ is increased or decreased the magnetic flux enters or
exits until a {\it critical slope} is reached like with a
pile of sand, for example, a maximum and nearly constant gradient
of $B=|{\bf B}|$. More generally spoken, in this {\it critical
state} the {\it current density} $J ={\bf |J|}$ attains its
maximum value $J_c$. Averaged over a few flux-line spacings, the
current density of the FLL is
${\bf J} = (\partial H/\partial B){\bf \nabla\times
B} \approx \mu_0^{-1} {\bf \nabla \times B}$, where
$H(B) \approx B/\mu_0$  is the (reversible) magnetic field $H_a$
which would be in equilibrium with the induction $B$.

  (iii) In general, the current density in  type-II
superconductors can have three different origins:
(a) {\it surface currents} (Meissner curents) within the penetration
    depth $\lambda$,
(b) a {\it gradient} of the flux-line density, or
(c) a {\it curvature} of the flux lines (or field lines).
The latter two contributions may be seen by writing
   $\nabla \times {\bf B} =  \nabla B\times {\bf \hat{B}} +
B\nabla \times {\bf \hat{B}}$ where ${\bf \hat{B} = B}/B$.
In bulk samples typically the gradient term dominates,
$J\approx \mu_0^{-1} \nabla B$, but in thin
films the current is carried almost entirely by the
{\it curvature} of the flux lines. Since in simple geometries a
deformation of the FLL is caused mainly by pinning, it is sometimes
said that ``the current flows in regions where the FLL is pinned''.
However, due to a finite lower critical field $H_{c1}$, the
current density in type-II superconductors may considerably
exceed the critical current density $J_c$ required to move the
vortices. Such ``overcritical current'' flows not only near the
specimen surface (as a magnetization current or as a Meissner
current, which remains also above $H_{c1}$), but also {\it inside}
the specimen at the flux front when $H_{c1} >0$
(Campbell and Evetts 1972, Ullmaier 1975). Recently an excess volume
current was observed in the form of a ``current string'' in YBCO and
BSCCO platelets by Indenbom et al.\ (1993, 1995).

  (iv) A kind of opposite situation, with a finite driving force on
the flux lines in spite of zero average current, occurs in the
``flux transformer'' (section 6.4.\ and figure 9) when the flux lines
in an isolated film are dragged by the moving  periodic field
caused by the moving flux lines in a very close-by parallel film
to which a current is applied. In general, the driving force is given
by the average of the current density {\it at the vortex cores}, which
may deviate from the global average (transport) current density. This
deviation is largest at low inductions $B<B_{c1}$ where the
relative spatial variation of the periodic $B({\bf r})$ is not small.
A further ``current-free'' driving force on flux lines is exerted
by a temperature gradient (section 7.6.), see the review paper by
Huebener (1995). In all these cases,
flux flow induces a voltage drop and can be suppressed by pinning.

  (v) Pinned flux lines exert a force on the atomic lattice
which leads to magneto-mechanical effects, e.g., to
``suprastriction'' (Kronm\"uller 1970) or  ``giant
magnetostriction'' (Ikuta et al.\ 1993, 1994) and to a change of the
velocity and damping of ultrasound (Shapira and Neuringer 1967,
Forgan and Gough 1968, Liu and Narahari Achar 1990,
Pankert et al.\ 1990, Lemmens et al.\ 1991, Gutlyanskii 1994,
Zavaritski{\u{\i}} 1993, Dom\'{\i}nguez et al.\ 1995).
The pinning force can enhance the
frequency of flexural or tilt vibrations of superconductors
while depinning will attenuate these vibrations, see section 10
and the reviews by Esquinazi (1991) and Ziese et al.\ (1994c).
Another mechanical effect
of pinning is a strong ``internal friction'' which enables
superconductors to levitate above (or below) an asymmetric
permanent magnet motionless, without oscillation or rotation,
 {\bf figure} 12,  since
part of the magnetic field lines is pinned inside the specimen
in form of flux lines (Moon et al.\ 1988, Peters et al.\ 1988,
Huang et al.\ 1988, Brandt 1988c, 1989c, 1990a, Kitaguchi et al.\ 1989,
Murakami et al.\ 1990). Levitation forces were calculated by
various authors (Nemoshkalenko et al.\ 1990, 1992,
Yang et al.\ 1992, Huang et al.\ 1993, Barowski et al.\ 1993,
Chan et al.\ 1994, Xu et al.\ 1995, Haley 1995, Coffey 1995).
         \\[0.4 cm]
{\it 8.2.~ Theories of pinning}
         \\*[0.2 cm]
 Theories of pinning deal with several different groups
of problems:

{\it 8.2.1. Models for the magnetization.~}
The distribution of flux and current inside the
superconductor and the (more easily measurable) total magnetic
moment or nonlinear ac response were calculated for various models
of the {\it reversible} magnetization curve $B(H_a)$, e.g.\ the model
 $B=0$ for $H_a <H_{c1}$, $B=\mu_0 (H_a -cH_{c1})$ for $H_a >H_{c1}$
with $c\approx \frac{1}{2}$ (Clem and Hao 1993, Indenbom et al.\ 1995),
or models of the {\it irreversible} magnetization or
field dependence of the critical current, e.g.\ the Kim-Bean model
$J_c(B) = J_c(0)/(1+|B|/B_0)^{1/2}$ (Kim et al.\ 1963,
Cooley and Grishin 1995) or the Anderson-Kim model
$J_c(B) = J_c(0)/(1+|B|/B_0)$ (Anderson and Kim 1964,
Lee and Kao 1995) or other models
 (Yamafuji et al.\ 1988, Chen and Goldfarb 1989,
Bhagwat and Chaddah 1992, Yamamoto et al.\ 1993,
Gugan and Stoppard 1993, G\"om\"ory and Tak\'acs  1993,
Shatz et al.\ 1993).
In addition, these results depend on the shape of the
specimen and on the orientation of the homogeneous applied field,
 cf.\ section 8.4. They are further modified by the assumptions of
inhomogeneous or anisotropic pinning and of a surface barrier
(Clem 1979) (section 8.5.).

{\it 8.2.2. Granular superconductors.~}
Various effects are responsible for the low $J_c$ of polycrystalline
HTSC's (Mannhardt and Tsuei 1989). The granular structure of
ceramic HTSC's  was modelled as a percolative network of inter-grain
Josephson junctions (weak links) (Deutscher et al.\ 1974,
Clem et al.\ 1987, Clem 1988, Peterson and Ekin 1988,
Tinkham and Lobb 1989). The almost constant $J_c(B)$ at very low
fields, and the rapid decrease at fields between 10 and 30 Gauss, have
been well fitted (Kugel et al.\ 1991, Schuster et al.\ 1992b)
by assuming that the current through each junction
is defined by its own Fraunhofer diffraction pattern whose
oscillations are smeared out after averaging the junction lengths
and orientations. Pereyra and Kunold (1995) observe quantum
interference in the critical current of polycrystalline YBCO.
  In a nice work Schindler et al.\ (1992) measured a single grain
boundary in DyBa$_2$Cu$_3$O$_x$ by micro contacts and find that it
acts as a Josephson junction.
Dong and Kwok (1993) observe $J_c(H)$ oscillations in bitextured YBCO
films caused by the weak-link character of 45$^\circ$ grain boundaries.
Amrein et al.\ (1995) measure the orientation dependence of
$J_c$ across the grain boundary in a BSCCO bicrystal and find
a similar exponential decrease (for angles $ >45^\circ$) as
Dimos et al.\ (1988) saw in an YBCO bicrystal.

  L.\ M.\ Fisher et al.\ (1990) and Gilchrist (1990) explain
critical current and magnetic hysteresis of granular YBCO as due to
Josephson junctions with randomly distributed vortices and with a
Bean-Livingston surface barrier, see also Kugel and Rakhmanov (1992).
  Nikolsky (1989) discusses the possibility that the intergrain
junctions are non-tunneling but of metallic kind, at which Andreev
reflection of the quasiparticles occurs. Dobrosavljevi\'{c}-Gruji\'{c}
and Radovi\'{c} (1993) and Proki\'{c} et al.\ (1995) calculate critical
currents of superconductor-normal metal-superconductor junctions;
see Deutscher and DeGennes (1969) and Deutscher (1991) for reviews
on the related proximity effect and recent experiments on the
proximity effect in HTSC edge-junctions (Antognazza et al.\ 1995)
and Nb/insulator/Nb junctions (Camerlingo et al.\ 1995).
  The r\^ole of grain boundaries in the high-field high-current
superconductors NbTi and Nb$_3$Sn  was comprehensively studied by
Dew Hughes (1987).
Evetts and Glowacki (1988) consider the flux trapped in the grains
of ceramic YBCO. Inter- and intra-grain critical currents in HTSC's
were measured by K\"upfer et al.\ (1988) and Zhukov et al.\ (1992).
The competition between the demagnetization factors of the sample
and of the grains was investigated by Senoussi et al.\ (1990) and
Yaron et al.\ (1992).  Schuster et al.\ (1992b) investigate
$H_{c1}(T)$ and $J_c(T)$ of YBCO with various grain sizes.
Isaac et al.\ (1995) measure relaxation of the persistent current
and energy barrier $U(j)$ (section 9.7.) in a ring of
 grain-aligned YBCO.
For further work on ceramic HTSC's see, e.g.,
Dersch and Blatter (1988), Rhyner and Blatter (1989),
Stucki et al.\ (1991) (low-field critical state),
M\"uller (1989, 1990, 1991), Murakami (1990),
Gilchrist and Konczykowski  (1990), Senoussi et al.\ (1991),
Halbritter (1992, 1993), Kugel and Rakhmanov (1992),
Jung et al.\ (1993, 1994), Castro and Rinderer 1994,
Lee and Kao 1995, and the review on magnetic ac susceptibility
by Goldfarb et al.\ (1991).

{\it 8.2.3. Elementary pinning forces.~}
 Most pinning forces act on the vortex core and thus typically
vary spatially over the coherence length $\xi$. Pinning may be
caused by spatial fluctuations of $T_c({\bf r})$ (Larkin 1970)
or of the mean free-path of
the electrons $l({\bf r})$, corresponding to fluctuations of the
prefactors of $|\psi|^2$ or of $|(-i\nabla/\kappa -{\bf A}) \psi|^2$
in the GL free energy (2.1), respectively.
Griessen et al.\ (1994) find evidence that in YBCO films pinning
is caused mainly by fluctuations of the mean free path.
The elementary pinning force between a flux line and a given
material defect was estimated from GL theory by numerous authors
(e.g.\ Seeger and Kronm\"uller 1968, Labusch 1968, Kammerer 1969,
 Kronm\"uller and Riedel 1970,
 Schneider and Kronm\"uller 1976, F\"ahnle and Kronm\"uller 1978,
F\"ahnle 1977, 1979, 1981,
Shehata 1981, Ovchinnikov 1980, 1982, 1983), see also the reviews
by Kronm\"uller (1974) and Kramer (1978).
Trapping of flux lines at cylindrical holes of radius $r_p$ was
calculated by Mkrtchyan and Schmidt (1971), Buzdin (1993),
Buzdin and Feinberg (1994b) (pointing out the analogy with
an electrostatic problem when $\lambda \to \infty$),
and Khalfin and Shapiro (1993); cylindrical holes can pin
multiquanta vortices with the number of flux quanta $n_v$
up to a saturation number $n_s \approx r_p/2\xi(T)$; when more
flux lines are pinned ($n_v > n_p$), the hole acts as a repulsive
centre, see also Cooley and Grishin (1995).
Anisotropic pinning by a network of planar defects is calculated
and discussed in detail by Gurevich and Cooley (1994).
  From microscopic BCS theory pinning forces were calculated by
Thuneberg et al.\ (1984) and Thuneberg (1985, 1986, 1989),
see also Muzikar (1991). The most comprehensive and correct
discussion of the microscopic theory of pinning forces is given
by Thuneberg (1989), who also points out that some previous
estimates of pinning forces  from GL theory were wrong since they
used reduced units, which in inhomogeneous superconductors should
vary spatially. The main result from microscopic theory is that the
pinning energy of a small pin of diameter
$D\ll \xi_0$ (the coherence length at $T=0$) is enhanced by a
factor $\xi_0 /D \gg 1$ (or $l/D$ for impure superconductors
if $D < l \ll \xi_0$) with respect to the GL estimate
 $U_{\rm p}^{\rm GL} \approx \mu_0 H_c^2 D^3$
due to the ``shadow effect'' caused by the scattered Cooper pairs.
  Kuli\'c and Rys (1989),  Blatter, Rhyner and Vinokur (1991c),
and Marchetti and Vinokur (1995)
calculate pinning by parallel twin planes, which occur in YBCO
crystals, and Svensmark and Falikov (1990) consider equilibrium
flux-line configurations in materials with twin boundaries.
Shapoval (1995)
considers the possibility of localized superconductivity at a
twin boundary due to the modification of the electron spectrum.

  Pinning by twin boundaries in YBCO crystals was measured e.g.\ by
Swartzendruber et al.\ (1990), see also Roitburd et al.\ (1990).
Flux lines pinned near a saw-tooth twin were observed in decoration
experiments by Gammel et al.\ (1992b), and the flux pile-up at a
twin plane was seen magneto-optically by Vlasko-Vlasov (1994b)
and Welp (1994). Oussena et al.\ (1995) observe that if bulk pinning
is strong, flux lines can move along the twin planes (channelling),
but only if the angle between pins and flux lines is less than a
lock-in angle of 3.5$^\circ$.
Random arrays of point pins have been studied in detail, e.g.\
by van Dover (1990). The elastic response of the pinned
FLL (the Labusch parameter $\alpha_L$) has been measured by
vibrating reed techniques (Esquinazi et al.\ 1986, Gupta et al.\ 1993,
Kober et al.\ 1991b, Gupta et al.\ 1993, Ziese and Esquinazi 1994,
Ziese et al.\ 1994c) and torque magnetometry (Farrell et al.\ 1990).
  Doyle et al.\ (1993, 1995), Tomlinson et al.\ (1993), and
Seow et al.\ (1995) measure the linear and nonlinear
force--displacement response of pinned flux lines by a directed
transport current, which gives more information than the
circulating currents flowing in the usual magnetic measurements.

{\it 8.2.4. Attractive and repulsive pins.~}
 Typical pins attract flux lines. For example, the maximum possible
(negative) pinning energy which a cylindrical normal inclusion or
hole with radius $r_p >\xi$ can exert on a parallel flux-line core is
per unit length $U_{\rm p,max} \approx (\phi_0^2/4\pi \mu_0\lambda^2)
[\ln(r_p/\xi) +0.5]$  (cf.\ section 4.5.\ and Brandt 1980a, 1992c);
from this energy the maximum pinning force is estimated as
$f_{\rm p,max} \approx U_{\rm p,max}/\xi$.
 A normal cylinder or hole {\it attracts} flux lines since a flux line
saves core energy when it is centred at the defect. However,
at high inductions close to $B_{c2}$ the same cylinder {\it repels}
the flux lines. This is so since near the surface of a normal inclusion
or hole the order parameter $|\psi|^2$ is enhanced (Ovchinnikov 1980)
by the same reason which also leads to the surface sheath
of thickness $\approx \xi$ at a planar surface in the field range
$B_{c2} \le B \le B_{c3}$ (Saint-James and DeGennes 1963,
DeGennes 1966), see also section 1.3.
The (strongly overlapping) cores of the flux lines try to avoid the
region of enhanced $|\psi|^2$ around the pin similarly as they are
repelled from the surface sheath. Another type of repulsive pins are
inclusions of a material with higher $T_c$ than the matrix; these
enhance the order parameter locally (Brandt and Essmann 1995).
Such inclusions of various shape have been considered within GL theory
by Brandt (1975), and within BCS theory by Thuneberg (1986), see also
Zwicknagl and Wilkins (1984).
Remarkably, the region of enhanced $|\psi|^2$ may stretch along
$B$ to far outside the pin. For example, a small spherical inclusion
of radius $\le \xi$ above $B_{c2}$ creates a cigar-shaped
superconducting nucleus of radius $\approx \xi$ and length
$\approx \xi/(B/B_{c2} -1)^{1/2}$ which diverges as $B$ approaches
$B_{c2}$ from above (Brandt 1975). Below $B_{c2}$ this nucleus
qualitatively survives and repels the surrounding flux lines.
A repulsive pinning force is also exerted by a pin that has trapped
a flux line which now repels other flux lines magnetically.
Repulsive pins typically are less effective in pinning the FLL than
are attractive pins, since the flux lines may flow around such pins.
This effect was seen in early simulations of flux-line pinning
(Brandt 1983a, b).
However, if the pins are very dense such that their distance becomes
less than $\xi$, the notions repulsive and attractive lose their
sense and the pinning potential becomes a 2D random function in each
plane perpendicular to the flux lines.

{\it 8.2.5. Statistical summation of pinning forces.~}
  Numerous theories try to sum the random pinning forces
that act on the flux lines, e.g.\ the theory of collective
pinning discussed in section 8.7.
Early summation theories should be understood as just parametrizing
the measured critical current density $J_c(B,T)$, since the
assumptions of a theory remain either unclear or are not
fulfilled in experiments. For example, the scaling law
by Kramer (1973), $J_c(B,T) \sim b^p (1-b)^q$ were $b=B/B_{c2}(T)$
and $p$ and $q$ are constants depending on the pinning mechanism,
typically fits data very well. For $p=1$ or $1/2$ and $q=2$
the Kramer law follows from the assumption that the FLL flows
plastically around large pins, see also W\"ordenweber (1992);
therefore, a softer FLL yields {\it weaker}
pinning and one has  $J_c \sim c_{66}(B,T)$ when
the flow stress of the FLL is proportional to its shear modulus
 $c_{66}$ (4.4). As opposed to this idea, the statistical theories of
weak pinning yield {\it stronger} pinning for softer FLL,
e.g.\ $J_c \sim 1/c_{66}$, section 8.7.

{\it 8.2.6. Thermally activated depinning.~}
  After the discovery of HTSC's the collective pinning theory,
which applies at zero temperature,  has been extended
to include thermally activated depinning, section 9.
Various phase diagrams of the FLL in the $B$-$T$-plane were
suggested by combining this collective creep theory (section 9.4.),
 or the vortex-glass (section 9.5.) and Bose-glass (section 9.2.)
pictures, with the melting criteria and with the phase transitions
of layered superconductors discussed in section 6, see, e.g.,
the general article by Bishop et al.\ (1992) and the comprehensive
review by Blatter et al.\ (1994a). Transitions from 3D pinning of
vortex lines to 2D pinning of point vortices in BSCCO were observed
by many authors, e.g., by Gray et al.\ (1992), Almasan et al.\ (1992),
Marcon et al.\ (1992, 1994) and Sarti et al.\ (1994),
see also section 6.4.

{\it 8.2.7. Anisotropic pinning.~}
  The critical current density $J_c$ in layered HTSC's is highly
anisotropic (Barone et al.\ 1990, Pokrovsky et al.\ 1992,
Blatter and Geshkenbein 1993, Balents and Nelson 1994)
 since flux lines are pinned  much more strongly when
they are parallel to the Cu-O layers (Tachiki and Takahashi 1989,
Tachiki et al.\ 1993, Schimmele et al.\ 1988, Senoussi et al.\ 1993,
Nieber and Kronm\"uller 1993a, Koshelev and Vinokur 1995).
In YBCO planar twin-boundaries cause anisotropic pinning
(Kwok et al.\ 1990, Liu et al.\ 1991, Crabtree et al.\ 1991,
Hergt et al.\ 1991, Fleshler et al.\ 1993, Vlasko-Vlasov et al.\ 1994b).
Anisotropic pinning by linear defects (e.g.\ Crabtree et al.\ 1994)
 is discussed in section 9.2.
An anisotropic depinning length was observed by Doyle et al.\ (1993).
Nice angle-dependent $J_c(\Theta)$ with sharp peaks were measured in
HTSC's by Roas et al.\ (1990b), Schmitt et al.\ (1991),
Kwok et al.\ (1991), Labdi et al.\ (1992), Schalk et al.\ (1992),
Theuss et al.\ (1994), Kraus et al.\ (1994b) ({\bf figure} 13),
and Solovjov et al.\ (1994),
and in Nb/NbZr multilayers by Nojima et al.\ (1993). The anomaly in
the resistivity observed by Kwok et al.\ (1991) for $B$ nearly
parallel to the $ab$ plane, was recently explained by
Koshelev and Vinokur (1995).  Anisotropic $J_c$ was also observed
by magneto-optics by Schuster et al.\ (1993a).
A similar angular dependence of pinning and flux flow in thin type-II
superconducting Sn films was investigated by
Deltour and Tinkham (1966, 1967), see also Tinkham (1963). Thin
films of type-I superconductors behave like type-II superconductors,
e.g.\ a Pb film of thickness $< 0.16\,\mu$m (Dolan 1974).
  Gurevich (1990a, 1992b) shows that uniform
current flow in HTSC's with anisotropic $J_c$ may become unstable and
that current domain walls and macro vortex patterns may occur. A
similar macroturbulence was observed by Vlasko-Vlasov et al.\ (1994a),
which may also be related to the ``current string'' observed and
explained as a consequence of finite $H_{c1}$ by Indenbom et al.\ (1995).

{\it 8.2.8. Periodic pinning structures.~}
Besides layered superconductors such as NbSe$_2$ (Monceau et al.\ 1975)
and HTSC's (Ivlev et al.\ 1991a, Levitov 1991, Watson and Canright 1993),
 artificial periodic pinning structures have been
investigated extensively, e.g., films with spatially modulated
composition (Raffy et al.\ 1972, 1974), films with periodically varying
thickness (Martinoli 1978, Martinoli et al.\ 1978),
NbTi-alloy wires containing a hexagonal array of cylindrical pins
(Cooley et al.\ 1994), and a lattice of holes (Fiory et al.\ 1978,
Baert et al.\ 1995, Metlushko et al.\ 1994b,
Bezryadin and Pannetier 1995).  For recent reviews see Lykov (1993)
and Moshchalkov et al.\ (1994b). Very recently Cooley and Grishin (1995)
showed that in the critical state in superconductors with periodic
pins, the flux density forms a terrace structure, causing
stratification of the transport current into the terrace edges where
the flux-density gradient is large; the magnetization thus changes
abruptly at rational or periodic fields when a new terrace appears
inside the superconductor. This effect might explain the periodic
magnetization observed by  Metlushko et al.\ (1994b) and
Cooley et al.\ (1994), and the rational fractions observed in
YBCO crystals with field parallel to the Cu-O planes
(Raffy et al.\ 1972, 1974) and in multilayer films in parallel
field (Brongersma et al.\ 1993, H\"unneckes et al.\ 1994).
A further interesting behaviour exhibit micro-networks of Josephson
junctions (Th\'eron et al.\ 1993, 1994, Korshunov 1994)
and of superconducting wires (Fink et al.\ 1985, 1988, 1991,
Moshchalkov et al.\ 1993, Itzler et al.\ 1994),
which may contain vortices as seen by decoration
(Runge and Pannetier 1993) and which exhibit a resistive
transition at a critical current density (Giroud et al.\ 1992,
Fink and Haley 1991), see Pannetier (1991) for a review.

{\it 8.2.9. Conductors.~}
 Some theories or models are more application oriented.
For example, Campbell (1982) and Khalmakhelidze and Mints (1991)
calculate the losses in multifilamentary superconducting wires and
Wipf (1967, 1991) and  Mints and Rakhmanov (1981) consider the
stability of the critical state in such wires.
The limits to the critical current in BSCCO tapes were estimated
by Bulaevskii et al.\ (1992d) within the ``brick-wall model'',
which assumes that the current flows in the $ab$-planes until
a structural defect interrupts this path and forces the current
to flow in $c$-direction into the next ``brick''; the current
is thus limited by the conductivity between the layers, i.e.\
by ideal $c$-axis grain boundaries. A better description of BSCCO
tapes is the ``railway switch model'' by Hensel et al.\ (1993, 1995),
which considers realistic small-angle $c$-axis grain boundaries
as current limiters. A further theory of inter- and intra-cell
weak links in BSCCO is given by Iga et al.\ (1995), see also
the magneto-optical imaging of flux penetration and current
flow in silver-sheathed BPbSCCO by Pashitski et al.\ (1995)
and the special impurity phases of Ullrich et al.\ (1993).
          \\[0.4 cm]
{\it 8.3.~ Bean's critical state model in longitudinal geometry}
         \\*[0.2 cm]
The critical state of a pinned FLL is approximately described by the
``Bean model'', which in its original version (Bean 1964, 1970)
assumes a $B$-independent critical current density $J_c$ and disregards
demagnetizing effects that become important in flat superconductors
in perpendicular magnetic field. It further assumes zero
reversible magnetization, namely, $B(H_a) =\mu_0 H_a$ and zero lower
critical field $H_{c1}=0$.
When finite $H_{c1}$ is considered, the penetrating flux front performs
a vertical jump to $B=0$ (Campbell and Evetts 1972, Ullmaier 1975,
Clem and Hao 1993, Indenbom et al.\ 1995).
In flat specimens of finite thickness the consideration of a finite
$H_{c1}$ leads to the appearance of a ``current string'' in the centre
of the specimen as predicted and observed by Indenbom et al.\ (1995).
   The original Bean model further
disregards the nonlocal relationship between the flux density
$B({\bf r})$ and the flux-line density $n({\bf r}) \approx
B({\bf r})/\phi_0$ (Brandt 1991c); this nonlocality originates from
the finite penetration depth $\lambda$ and introduces discontinuities
in the flux-line density in  the Bean critical state as shown by
Voloshin et al.\ (1994) and Fisher et al.\ (1995).

 For long cylinders or slabs in
longitudinal applied field $H_a(t)$ along $z$, the Bean model predicts
that in regions where magnetic flux has penetrated, the flux density
 $B(x,y,t)$ is piecewise linear with slope $|\nabla B| = \mu_0J_c$,
and in the central flux-free zone one has
$B=0$ and $J=0$, {\bf figure} 14.
The detailed pattern of $B$ depends on the magnetic history. Explicitly,
for a slab of thickness $2a$ ($-a \le x \le a$), when $H_a$ is
increased from zero, the Bean model yields
$J(x)=0$,          $H(x)=B(x)/\mu_0=0$  for $|x|\le b$, and
$J(x)=J_c x/|x|$, $H(x)=J_c (|x|-b)$  for $b\le |x|\le a$.
The {\it negative} magnetic moment $M$ per unit area of the slab is
for $H_a \le H_c$
\begin{equation}  
M(H_a)=\int_{-a}^a \!\! x\,J(x)\, {\rm d}x = J_c (a^2 -b^2) =J_c a^2
    \left( \frac{2H_a}{H_c} -\frac{H_a^2}{H_c^2} \right) ,
    \end{equation}
and for $H_a \ge H_c$ it saturates to $M = J_c a^2$.
Here $H_c=J_c a$ is the field at which full penetration is achieved
and $a-b = aH_a/H_c = H_a/J_c$ is the penetration depth of the
flux front. Interestingly, the longitudinal
currents and the transverse currents of the U-turn at the far-away
ends of the slab {\it give identical contributions} to $M$, thus
canceling the factor $1/2$ in the general definition of $M$.
Note also that in this longitudinal geometry the magnetic moment
coincides with the local magnetization $H(x)-H_a$ integrated
over the sample cross section.
The correctness of (8.2) is easily checked by considering the
Meissner state ($H_a \ll H_c$, $b \to a$, ideal shielding)
in which $M =M_{\rm ideal} =2a H_a $ equals the specimen width
(volume per unit area) times the expelled field.
The deviation of $M$ from the ideal Meissner value initially
is $M - 2aH_a \sim H_a^2$. For $|H_a| \ge H_c$ one has complete
penetration ($b=0$) and thus $M =M_{\rm max}= a^2 J_c$.

When $H_a$ is cycled with amplitude $0 < H_0 \le H_c =J_c a$,
the flux profile is as shown in {\bf figure} 11. The magnetic moment
for $H_a$ decreasing from $+H_0$ to $-H_0$ is then
\begin{equation}  
M_{\downarrow}(H_a, H_0, J_c) =
              (H_a^2 +4H_aH_c -2H_aH_0 -H_0^2)/2J_c .
    \end{equation}
The corresponding branch $M_\uparrow$
for $H_a$ increasing from $-H_0$ to $+H_0$ follows from
symmetry, $M_\uparrow (H_a,H_0,J_c) = -M_\downarrow
(-H_a,H_0,J_c)$. The dissipated power $P(H_0)$ per unit area of
the slab is the frequency $\nu$ times the area of the hysteresis loop,
$P = \nu \mu_0\oint M\,{\rm d}H_a$. Introducing a function
$f(x)$ with $f(x \le 1)=x^3/3$, $f(x\ge 1) =x- 2/3$, one gets
for all amplitudes $H_0$ the dissipation
 $P(H_0) = 4 a\nu \mu_0 H_c^2 f(H_0/H_c)$.
Exact solutions of the Bean model to the penetration of a rotating
field are given by  Gilchrist (1972, 1990a, 1994b).

  In general, the flux penetration
obeys a nonlinear diffusion equation, which for slabs or a half space
has been solved numerically (van der Beek and Kes 1991b,
van der Beek et al.\ 1992b, Rhyner 1993, Bryksin and Dorogovtsev 1993,
Calzona 1993, Gilchrist and van der Beek 1994).
 Defining $h(x,y,t) = H(x,y,t) - H_a(t)$ one has for
specimens with general cross section in parallel field the
boundary condition $h=0$ at the surface and the equation of motion
(Brandt 1995c) [see also Eq.\ (8.18) below]
\begin{equation}  
 \dot h(x,y,t) = \nabla\cdot(D \nabla h) - \dot H_a(t) \,,
\end{equation}
where $D=E(J)/\mu_0 J$ is a {\it nonlinear flux diffusivity} with
$E(J)$ the current--voltage law of the superconductor. This diffusivity
in general depends on space and time.

 Remarkably, even in the fully
penetrated critical state where the current density $J = |\nabla H|$
is practically constant, the electric field
${\bf E} = \rho(J) {\bf J}$ in general is {\it not} constant;
${\bf E}$ is a linear function of space satisfying
$\nabla \times {\bf E} = -\mu_0 {\bf \dot H_a} =$ const. For
rectangular specimen cross section (both in longitudinal and
transverse geometries) the magnitude $E$ exhibits a
strange zig-zag profile which is independent of the detailed form
of $E(J)$ as long as this is sufficiently nonlinear (sharply bent
at $J=J_c$) (Brandt 1995c). In particular, for a logarithmic
activation energy $U(J) = U_0 \ln(J_2/J)$ (section 9.7.) one gets
for $E\sim \exp[-U(J)/k_BT]$ the power-law  $E(J) =E_c (J/J_c)^n$
with $n=U_0/k_BT \gg 1$ and $E_c=E(J_c)$. Assuming an integer odd
exponent $n$, equation (8.4) becomes particularly simple and
reads for slabs and cylinders, respectively,
                  \setcounter{equation}{3}
                { \renewcommand{\theequation}{8.4a}
    \begin{eqnarray}   
 \dot h(x,t) = c [\, (h')^n \,]' - \dot H_a(t) \,, ~~~~~
 \dot h(r,t) = (c/r) [\,r (h')^n \,]' - \dot H_a(t)\,,
    \end{eqnarray}  }
where $c= E_c/J_c^n \mu_0$.
The equations for the current density $J=h'$ are even simpler but
do not contain the driving force $H_a(t)$,
 $\dot J(x,t) = c (J^n  )''$,~
 $\dot J(r,t) = c [\,(r J^n)'/r \,]'$.
The nonlinear diffusion equations (8.4) or (8.4a)
are easily integrated on a Personal Computer, e.g., starting
with $H_a(t) =h(x,y,t) = 0$ and then increasing $H_a(t)$.
The numerical integration of (8.4a) is facilitated noting that
$h(x,t)$ and $h(r,t)$ are even functions of $x$ or $r$ which
may be represented by Fourier series, and that one of the
derivatives acts on an even function and one on an odd function.
In the limit $n\to \infty$ one obtains the Bean model;
finite $n$ allows for flux creep. Ryhner (1993) studied such
power-law superconductors in detail and obtained scaling laws for
the dependence of their ac losses on frequency and amplitude.
Corrections to the critical state theory of the surface impedance
were derived by Fisher et al.\ (1993) due to the finite resistivity
and by Voloshin et al.\ (1994) and Fisher et al.\ (1995) due to the
nonlocal relation between
the vortex density $n_v$ and the flux density $B$.

The nonlinear diffusion equations (8.4) and (8.4a) are closely related
to the concept of ``self-organized criticality'' (Bak, Tang and
Wiesenfeld 1988, Jensen 1990, Pla and Nori 1991, Ling et al.\ 1991,
Richardson et al.\ 1994, DeGennes 1966) and to Gurevich's idea of
the universality of flux creep (section 9.8.), or to problems
with logarithmic time scale (Geshkenbein et al.\ 1991,
Vinokur et al.\ 1991a, van der Beek et al.\ 1993).
{}From the universal electric field
during creep ${\bf E(r},t) ={\bf F(r})/t$ (Gurevich and K\"upfer 1993,
Gurevich and Brandt 1994) follows that Eq.\ (8.4) is a {\it linear
diffusion equation in the logarithmic time} $\tau = \ln t$
since $D \sim E \sim 1/t$. This universal behaviour is
{\it independent} of the explicity forms of $E(J, B)$ or $J_c(B)$
and of the specimen shape, and it applies even to transverse
geometry where the diffusion equation becomes nonlocal (section 9.8.).
         \\[0.4 cm]
{\it 8.4.~ Extension to transverse geometry}
         \\*[0.2 cm]
In typical magnetization experiments the above longitudinal geometry
is not realized. Measurements are often performed on films or
flat monocrystalline platelets in {\it perpendicular} magnetic field
in order to get a larger magnetization signal. In this perpendicular
or transverse geometry the original Bean model does not apply and
has to be modified. The critical state of circular disks was computed
by Frankel (1979), D\"aumling and Larbalestier (1989), Conner and
Malozemoff (1991), Theuss et al.\ (1992), Hochmut and Lorenz (1994),
and Knorpp et al.\ (1994), see also the reviews on flux pinning by
Huebener (1979) and Senoussi (1992).

 The problem of flux penetration in
transverse geometry recently was solved analytically for
thin circular disks (Mikheenko and Kuzovlev 1993, J. Zhu et al.\ 1993)
and for thin long strips (Brandt et al.\ 1993).
For strips carrying a transport current in a magnetic field see
Brandt and Indenbom (1993) and Zeldov et al.\ (1994a).
These theories, like Bean, still assume constant $J_c$ and $B_{c1}=0$.
They consider the
sheet current ${\bf J}_s(x,y) =\int {\bf J}(x,y,z)\,{\rm d}z$
(the current density integrated over the specimen thickness $d$)
and the perpendicular component $H_z$ of the magnetic field at the flat
surface, related to the flux density $B_z$ by $B_z=\mu_0 H_z$. For the
strip with width $2a$ ($-a\le x \le a$) in the ideal
Meissner state the sheet current along $y$,
    \begin{equation} 
J_s^{\rm ideal}(x) = 2xH_a/(a^2 -x^2)^{1/2},
    \end{equation}
{\it diverges} at the strip edges in the limit $d\to 0$. Inside the
strip this sheet current exactly compensates the constant applied
field $H_a$. Thus one has $H_z =0$ for $|x|\le a$, and outside the
strip one finds in the plane $z=0$
    \setcounter{equation}{4} { \renewcommand{\theequation}{8.5a}
    \begin{eqnarray}   
 H_z^{\rm ideal}(|x|>a) = |x| H_a/(x^2 -a^2)^{1/2}.
    \end{eqnarray}  }
The negative  magnetic moment per unit length of the strip
$M =  \int_{-a}^a x J_s (x)\,{\rm d}x$ in this ideal shielding state
is $M_{\rm ideal} =\pi a^2 H_a$.
In general, the perpendicular field $H_z(x)$ generated in the film
plane $z=0$ by the sheet current $J_s(x)$ follows from the London
equation (2.4) by integration over the film thickness $d$ using
the flux-line density $n(x) =\mu_0 H_z/\phi_0$
(Larkin and Ovchinnikov 1971),
    \begin{equation} 
 H_z(x) = H_a +\frac{1}{2\pi} \int_{-a}^{a} \frac{J_s(x')}{x-x'} ~
    {\rm d}x'  -\frac{\lambda^2}{d} \,\frac{{\rm d} J_s(x)}{{\rm d}x} .
    \end{equation}
If $a \gg \lambda_{\rm film} = 2\lambda^2 /d$ applies, the gradient
term in (8.6) may be disregarded and (8.6) reduces to Amp\`ere's
law. Note that (8.6) is a {\it nonlocal} relationship, whereas for
longitudinal geometry (for $d\to \infty$) the {\it local} relation
$J(x) = H'(x)$ holds.

Remarkably, both unknown functions $J_s(x)$
and $H_z(x)$ can be determined from {\it one} equation (8.6)
if one adds the Bean-condition $|J_s(x)| \le J_c d$ (constant bulk
pinning). A different such condition, which also allows for a
unique solution, is the assumption of zero bulk pinning but a finite
edge barrier (section 8.5.), which means
$|J_s(|x| = a)| = J_{\rm edge}$; for thin films with
$d\ll \lambda$ one may assume $J_{\rm edge} = J_0 d$ with $J_0$ the
depairing current (Maximov and Elistratov 1995). For films which
exhibit both bulk pinning and an edge barrier, solutions of (8.6)
were obtained by Zeldov et al.\ (1994b, c) and Khaykovich et al.\ (1994).
The main effect of the disregarded gradient term
 $\frac{1}{2} \lambda_{\rm film} J'(x)$ in (8.6) (or of the finite
thickness $d$ if $d>\lambda$) is to remove the infinity of
$H_z(x)$ at the edges $x=\pm a$ (Ivanchenko et al.\ 1979).
Flux penetration into thin circular disks with $d\ll\lambda$ is
calcuated by Fetter (1980) and Borovitskaya and Genkin (1995).

  When the sheet current is limited to the constant critical value
$J_c d$ as in the Bean model, then, when $H_a$ is increased from
zero the shielding current near the specimen edges saturates to
$J_s(x> b) = J_cd$ and $J_s(x< -b) = -J_cd$ since flux starts
to penetrate in the form of flux lines. One thus has
$H_z(x) \neq 0$ for $|x| > b$ and $H_z(x) \equiv 0$ for $-b <x< b <a$.
The current distribution $J(x)$ of this two-dimensional
problem with field-independent $J_c$ can be obtained by
conformal mapping as shown by Norris (1970) for the similar problem of
a strip with transport current.
 From the condition that $J_s(x)$ is
continuous, or $H(x)$ finite inside the specimen, one finds the
flux-front position $b = a/ \cosh ( H_a/H_c)$
where $H_c = J_c d/\pi$ is a critical field. With the
abbreviation $c =(1 -b^2/a^2)^{1/2} = \tanh\, ( H_a/H_c)$
the sheet current $J_s$, the magnetic field $H_z$ in the plane $z=0$,
the negative magnetic moment $M$, and the penetrated flux
 $\Phi = 2\mu_0 \int_0^a H(x){\rm d}x$ for $H_a$ increased from zero
take the form (cf.\ {\bf figure} 15)
    \begin{equation} 
 J_s(x) = \left\{ \begin{array}{lr}
 (2 J_c d/\pi) \arctan\,[cy/(b^2 -x^2)^{1/2}], &\mbox{$|x|<b$}\\[8 pt]
    J_cd\, x/|x|,                              &\mbox{$b< |x| < a$}
    \end{array} \right.  \end{equation}
    \begin{equation} 
 H_z(x) = \left\{ \begin{array}{lr}
    0,                             &\mbox{ $  |x| <b$} \\[8 pt]
    H_c \,{\rm artanh}\, [(x^2 -b^2)^{1/2} /c\,|x|],
                                   &\mbox{$b< |x| <a$} \\[8 pt]
    H_c \,{\rm artanh}\, [c\,|x|/ (x^2 -b^2)^{1/2}],
                                   & \mbox{ $ |x| >a$}
    \end{array} \right. \end{equation}
    \begin{equation} 
 M(H_a) =   J_c da^2 c =
       J_c da^2 \tanh\, (H_a /H_c) ,
    \end{equation}
    \begin{equation} 
 \Phi(H_a) = 2\mu_0 H_c a \ln(a/b) = 2\mu_0 H_c a \ln\cosh \,(H_a/H_c).
    \end{equation}
In the limit $J_c \to \infty$ these equations reproduce the
ideal-shielding results (8.5), (8.6),
$M_{\rm ideal} =\pi a^2 H_a$, and $\Phi_{\rm ideal} =0$.
For {\it weak penetration} ($H_a \ll H_c$) one gets
$b = a - a H_a^2 /2H_c^2$,
$M =\pi a^2 H_a(1 - H_a^2/3 H_c^2)$, and the flux
$\Phi =  \mu_0 a H_a^2 /H_c$.
For {\it almost complete penetration} ($H_a \gg H_c$) one gets
$b = 2a \exp(- H_a/H_c) \ll a$,
$M = J_c da^2 [1 - 2\exp(-2 H_a/H_c)]$,
$J_s(   |x| < b )   = (2J_c d/\pi) {\rm arcsin}(x/b) $,
$H_z( b<|x|< a/2)   = H_c {\rm arcosh}|x/b| $,
$H_z(2b<|x|<\infty) = H_a + H_c \ln(|a^2/x^2 - 1|^{-1/2})$, and
$\Phi = 2\mu_0 a(H_a - 0.69 H_c)$.

Thus in transverse geometry full penetration is realized only at
large fields $H_a \approx H_c\ln (4a/d)$ obtained from the
condition $b\approx d/2$. This estimate is confirmed by
computations for finite thickness $d$ (Forkl and Kronm\"uller 1994).
Note also the vertical slopes of $H_z(x)$ and $J(x)$ at $|x| = b$ and
the logarithmic infinity of $H_z(x)$ at $|x| \to a$. The analytical
solution $H_z(x)$ (8.8) thus proves a {\it vertical slope}
of the penetrating flux front as opposed to the finite and
constant slope in the original longitudinal Bean model.

  In a disk of radius $a$, the sheet current $J(r)$ is also given
by the equation (8.7) for the strip, but with $x$ replaced by $r$ and
 $H_c=J_c/\pi$  by $H_c= J_c/2$ everywhere.
Analytical solutions for the perpendicular field $H_z(r)$ at the
disk surface are not available, but the negative magnetic moment
$M=\pi \int_0^a \! r^2\, J(r)\, {\rm d}r$  of the disk is
    \begin{eqnarray} 
  M = \frac{8}{3} H_a a^3 S(\frac{H_a}{H_c})  ~~~~{\rm with}~~~~
  S(x) = \frac{1}{2x} \Big[ {\rm arccos}\frac{1}{\cosh x}
   + \frac{\sinh |x|}{\cosh^2 x} \Big]
    \end{eqnarray}
and $H_c=J_c/2$. For $H_a \ll H_c$ (ideal shielding) one has
$M=8 H_a a^3/3$  and for $H_a \gg H_c$ (full penetration)
$M=\pi J_c a^3 /3$ since $S(0)=1$ and $S(x\gg 1)=\pi/(4x)$.
The magnetization curves of strips, disk, and squares in the Bean
model are very similar when normalized to the same initial slope
and same saturation value; the magnetization curve $M(H_a)$ of the
strip (8.9) then exceeds that of the disk (8.11) only by
$0.011 M(\infty)$, and the magnetization of the square
(Brandt, unpublished) differs from the disk magnetization only
by less than $0.002 M(\infty)$ along the entire curve.

{}From these virgin solutions one may obtain the solutions
for arbitrary magnetic history $H_a(t)$. For example, if $H_a(t)$
is cycled between the values $+H_0$ and $-H_0$, one has in
the half period with decreasing $H_a$
  \begin{eqnarray}  
M_\downarrow ( H_a)   &=& M( H_0) -2 M( H_0/2 -H_a/2)\,.
    \end{eqnarray}
At $H_a = -H_0$ the original virgin state is reached again but with
$J$, $H$, $M$, and $\Phi$ having changed sign.
In the half period with increasing $H_a$ one  has
$J_\uparrow (y, H_a, J_c) = -J_\downarrow (y,-H_a, J_c)$,
$H_\uparrow (y, H_a, J_c) = -H_\downarrow (y,-H_a, J_c)$,
$M_\uparrow ( H_a, J_c) = -M_\downarrow (-H_a, J_c)$,
and $\Phi_\uparrow ( H_a, J_c)$                       \linebreak
$=-\Phi_\downarrow (-H_a, J_c)$.
When $H_a$ oscillates with frequency $\nu$ and amplitude $H_0$
the dissipated power per unit length of the strip is
 $P =\nu \mu_0 \oint M(H_a)\,dH_a$
(frequency times area of the  hysteresis loop) yielding
    \begin{equation} 
 P(H_0) =4\nu \mu_0 a^2 J_c H_0\,[\,(2/h)\, \ln\cosh h -\tanh h\,]
    \end{equation}
with $h =  H_0 /H_c$. At low amplitudes  $H_0 \ll H_c$ this gives
$P =( 2 \pi \nu\mu_0 a^2 /3 H_c^2 )\,H_0^4$, and for large
amplitudes $H_0 \gg H_c$ (full penetration) one has
$P =4 \nu\mu_0 a^2 J_c \,(H_0 -1.386 H_c )$. The energy loss is thus
initially very small, $P \sim   H_0^4$.

The nonlinear
ac susceptibility for disks is obtained from (8.11) and (8.12)
by Clem and Sanchez (1994). Magnetization curves of disks of
varying thickness are measured by Oussena et al.\ (1994a).
Irreversible magnetization curves and nonlinear ac susceptibilities
of the pinned FLL for various geometries are compared by
Bhagwat and Chaddah (1992), Stoppard and Guggan (1995), and
in a particularly nice short overview by Gilchrist (1994a).
Gilchrist and Dombre (1994) calculate the nonlinear ac response
and harmonics of a slab with power-law resistivity, see also
Rhyner (1993).
Gugan (1994) presents a model which extends the superposition (8.12)
to general induction-dependent $J_c(B)$, see also the ideas on
critical currents in disks and films by Caplin et al.\ (1992),
Atsarkin et al.\ (1994), and Porter et al.\ (1994).
Rogacki et al.\ (1995) confirm the magnetization curve (8.9)
by the vibrating reed technique. The field profile (8.8) is
confirmed, e.g., by magneto-optics (Schuster et al.\ 1994d),
by Hall probes (Zeldov et al.\ 1994b), and by an
electron-spin resonance microprobe (Khasanov et al.\ 1995),
see also section 1.2.
          \\[0.4 cm]
{\it 8.5.~ Surface and edge barriers for flux penetration}
         \\*[0.2 cm]
Interestingly, all
low-field results contain one power of $H_0$ or $H_a$ more in
perpendicular geometry than in longitudinal geometry, namely, the
hysteresis losses are $P  \sim H_0^4$ ($\sim H_0^3$), the initial
magnetization change is $M-M_{\rm ideal} \sim H_a^3$ ($\sim H_a^2$),
and the penetration depth is $a-b\sim H_a^2$ ($\sim H_a$).
This means that perpendicular flux penetrates into flat superconductors
with bulk pinning {\it delayed} as if there were a surface barrier,
with penetration  depth $a-b \sim   H_a^2$.
 This {\it pinning-caused} surface barrier in transverse field
(or better, edge barrier) was investigated magneto-optically by
Indenbom et al.\ (1993). The flux penetration over an artificial surface
barrier in single crystal BSCCO platelets which were irradiated
near the edges to enhance pinning, was observed magneto-optically
by Schuster et al.\ (1994b) and with an array of Hall sensors
by Khaykovich et al.\ (1994), and in both cases it was successfully
compared with theory.

A similar surface or edge barrier, which does not require pinning, is
related to the rectangular (or in general, non-elliptic) cross section of
flat superconductors in transverse fields (Grover et al.\ 1990, 1992,
Indenbom et al.\ 1994a,b, d,
Indenbom and Brandt 1994, Zeldov et al.\ 1994b, c,
Doyle and Labusch 1995), {\bf figure} 16.
 Solutions of the London theory for a vortex penetrating into thin films
of constant thickness $d$ were given
for strips by Larkin and Ovchinnikov (1971), Likharev (1971, 1979), and
Kuprianov and Likharev (1974), for disks by Fetter (1980),
and for an infinite film by Ivanchenko et al.\ (1979),
see also Kogan (1994).
A similar edge-shape barrier delays flux penetration also with type-I
superconductors, which do not contain flux lines but domains
of normal and superconducting phase; the magnetic irreversibility
caused by this edge-barrier is investigated in detail by
Provost et al.\ (1974),  Fortini and Paumier (1976),
Fortini et al.\ (1980), and was rediscovered by
 Grover et al.\ (1990, 1992).
Superconductors of ellipsoidal shape do not exhibit this type
of surface barrier and have thus been used to determine the lower
critical field $H_{c1}$ (Liang et al.\ 1994).
Exact analytical solution for the current and field distribution in
the limit of large width $a \gg \lambda_{\rm film} = 2\lambda^2/d$
were obtained for thin long strips with a surface barrier for
homogeneous (or zero) pinning by Zeldov et al.\ (1994b, c) and
Maksimov and Elistratov (1995), and for
inhomogeneous pinning by Khaykovich et al.\ (1994).

These important and often overlooked
{\it geometric} surface barriers  should be compared with the
microscopic surface barrier of Bean and Livingston (1964) for the
penetration of straight flux lines through a perfectly flat parallel
surface. The Bean-Livingston barrier results from the
superposition of the attractive image force $\sim K_1(2x/\lambda)$
and the repulsive force $\sim H_a\exp(-x/\lambda)$ exerted by the
Meissner current on a flux line at a distance $x$ from the surface.
The resulting potential has a maximum at a distance $\approx \xi$
($\approx$ vortex-core radius) from the planar surface. In HTSC's,
flux lines or point vortices can overcome this microscopic barrier
by thermal activation, section 9, which reduces the field of first
penetration (Buzdin and Feinberg 1992, Burlachkov 1993,
Koshelev 1992b, 1994a, Mints and Snapiro 1993, Kogan 1994,
Burlachkov et al.\ 1994).
In contrast, the {\it geometric} barrier cannot be overcome by
thermal activation and is not sensitive to surface roughness.
Extending previous work by Ternovski\u{\i} and Shehata (1972) and
Clem (1974a), Burlachkov (1993) very carefully analyses the
thermally activated nucleation of half circular flux-line loops
at a planar surface and the penetration of flux lines into a
superconductor which already contains flux lines.
Burlachkov and Vinokur (1994) apply these ideas to thin strips with
transport current and find the self-consistent current distribution
between flux entry and flux exit over a barrier at the two edges.
An observed asymmetric current--voltage curve in films in parallel
field is explained by a Bean-Livingston surface barrier and
inhomogeneous bulk pinning (Jiang et al.\ 1994).
The onset of vortices at the surface of strips and wires with
cross section $\ll \lambda^2$ is calculated from time-dependent
GL theory by Aranson et al.\ (1995), and the nucleation and
instability of vortex helices at the surface of a cylindrical wire
by Genenko (1994, 1995).

  At least seven different surface or edge barriers for the penetration
of transverse flux into thin superconductors may be distinguished:
    \\[0.2 cm]
(i) The Bean-Livingston barrier.
    \\[0.2 cm]
(ii) Clem's Gibbs free energy barrier for penetration of flux tubes
or flux lines into superconducting strips of elliptical cross section
(Clem et al.\ 1973, Buck et al.\ 1981).
    \\[0.2 cm]
(iii) Likharev's (1971) barrier for the penetration of a perpendicular
vortex into a thin film of width $2a$ and thickness $d \ll \lambda$.
The barrier width is $\approx \lambda_{\rm film} =2\lambda^2 /d$
and the barrier height is of the order of $H_{c1} (\lambda/d)^2$
(Indenbom unpublished). Depending on the
parameter $\gamma = da/(2\pi\lambda^2) = a/(\pi \lambda_{\rm film})$,
Likharev (1971) finds that vortex penetration over the barrier
becomes energetically favorable at $H_a \ge H_{c1}^\perp$, where
 $\mu_0 H_{c1}^\perp = (\phi_0\gamma /4a^2)\ln (\lambda^2 /\xi d) =
  (\phi_0 /4\pi a \lambda_{\rm film}) \ln(\lambda_{\rm film}/2 \xi)$
for $\gamma \gg 1$ and
 $\mu_0 H_{c1}^\perp = (\phi_0 /2\pi a^2)\ln (a/\xi)$ for $\gamma \ll 1$.
   \\[0.2 cm]
(iv) The edge shape barrier caused by the rectangular edges and by the
constant thickness of the sample as compared to the ellipsoid
(Provost et al.\ 1974, Indenbom et al.\ 1994a,b, d,
Doyle and Labusch 1995,  Labusch and Doyle 1995). This barrier can
be reduced by rounding the sharp rectangular edges.
    \\[0.2 cm]
(v) A barrier caused in the Bean model by the consideration of the
nonlocal relationship between the flux density $B({\bf r})$ and the
flux-line density $n({\bf r})$ (Fisher et al.\ 1995).
    \\[0.2 cm]
(vi) Surface pinning by the different structure or composition of the
material near the edge (see, e.g., Flippen et al.\ 1995).
    \\[0.2 cm]
(vii) The apparent barrier caused in the presence of bulk pinning
with $J_c d =$const by the slow initial increase of the
flux penetration depth $a-b \sim H_a^2$ (Brandt, Indenbom and
Forkl 1993), see the discussion following equation (8.10).

  Notice that the barriers (i) to (iv) delay only the {\it penetration}
of flux but not the exit of flux; the hysteresis loop of the
irreversible magnetization thus becomes {\it asymmetric}.
This is so since these four pin-free surface barriers are essentially
caused by the shielding current, which flows along the edge in
{\it increasing} field $H_a$,  but which vanishes  when in
{\it decreasing} field the magnetization goes through zero.
Notice also that only (i) and (v) are real surface barriers
while the other barriers are edge-shape barriers and depend
on the profile of the edge.
  Interestingly, in type-I superconductors in the intermediate state
which occurs for demagnetization factors $N>0$, the positive
wall energy of the domains (e.g., lamellae) also enhances the
field of flux penetration into an ellipsoid above the value
$(1-N)B_c$ which would hold
for zero wall energy, see {\bf figure} 3 and the discussion preceding
equation (1.1). This effect is largest in perpendicular geometry
($1-N \ll 1$), but in contrast to edge or surface barriers, it occurs
also in ellipsoidal specimens where it yields a {\it reversible}
magnetization curve if domain-wall pinning is absent.

  Steji\'c et al.\ (1994) give detailed measurements and calculations
of critical currents in thin films in parallel and oblique field,
see also Brongersma et al.\ (1993) and Mawatari and Yamafuji (1994).
Sharp matching peaks in the attenuation of thin vibrating YBCO films
in parallel field were observed by H\"unneckes et al.\ (1994).
In SQUID magnetometers Sun et al.\ (1994) observe the threshold field
caused by bulk pinning (section 8.4.) and by the edge barrier.
Surface barriers in weak-pinning HTSC single crystals in perpendicular
field were seen in magnetization measurements, e.g., by
Kopylov et al.\ (1990), Chikumoto et al.\ (1992),
Konczykowski et al.\ (1991), Chen et al.\ 1993,
and Xu et al.\ (1993), and by magneto-optics as a surface step of
$B$ by Dorosinski et al.\ (1994).  It is now believed that the observed
surface barrier is not the microscopic Bean-Livingston barrier
but a geometric barrier. Kugel and Rakhmanov (1992) and
Sun et al.\ (1995) study the flux penetration into granular HTSC's
accounting for the Bean-Livingston barrier of the grains.

F\`abrega et al.\ (1994b, 1995)  measured the magnetic
relaxation of Ce-based HTSC's near the first flux penetration,
finding a strong dependence on the applied field, which indicates
thermal overcoming of surface or geometric barriers.
Careful measurements in weak fields {\it parallel}
to the {\it ab}-plane by Nakamura et al.\ (1993) in BSCCO,
Zuo et al.\ (1994a, b) in Nd- and Tl-based HTSC's, and
Hussey et al.\ in Tl-based HTSC, yield abrupt jumps of the
magnetization and a large barrier for flux penetration, but nearly
no barrier for flux exit; D\"aumling et al.\ (1994) measure
different magnetization curves of YBCO platelets in parallel and
perpendicular field. These findings were explained, respectively, by a
Bean-Livingston barrier, by decoupling of the layers (section 6.3.),
by a lock-in transition (section 3.3.3.), or by material anisotropy,
but they may also
be caused by the edge-shape barrier (Indenbom and Brandt 1994).
In YBCO films of thickness $d$, Liu, Schlenker et al.\ (1993)
 measured an irreversibility field $B_{irr} \sim d^{1/2}$, see also
the newer measurements and discussion by Neminsky et al.\ (1994).
As argued by Grover et al.\ (1990, 1992), Indenbom et al.\ (1994a,b,d),
 and Zeldov et al.\ (1994b, c, 1995b), the measured
``irreversibility  line'' in HTSC in transverse geometry may well be
due to the edge-shape barrier, at least a section of this line
in the $BT$-plane.
         \\[0.4 cm]
{\it 8.6.~ Flux dynamics in thin strips, disks, and rectangles}
         \\*[0.2 cm]
The flux profiles observed magneto-optically during penetration and
exit of flux  by Schuster et al.\ (1994b, d, 1995a) (see also
Koblischka and Wijngaarden 1995) were quantitatively
explained by computer simulations based on a one-dimensional integral
equation for the sheet current $J_s({\bf r},t)$ in thin strips
(Brandt 1993b, 1994a) and disks (Brandt 1994b).
  The only material properties entering these computations are the
reversible magnetization (usually $B=\mu_0 H$ and $H_{c1}=0$ is assumed)
and a nonlinear resistivity $\rho(J) =E/J$ or sheet resistivity
$\rho_s = \rho/d = E/J_s$, which may depend on the position explicitly
(e.g.\ because of inhomogeneous pinning or varying thickness $d$)
or implicitly via $B({\bf r})$ and $J_s({\bf r})$.
For homogeneous pinning, the above quasistatic results (8.5) to (8.13)
for strips and disks in transverse field $H_a(t)$ are reproduced
by inserting a sharply bent model resistivity, e.g.\ a power law
$\rho \sim (J/J_c)^{n-1}$ with $n\gg 1$ (Evetts and Glowacki 1988,
Ries et al.\ 1993, Rhyner 1993, Gilchrist and van der Beek 1994),
and the desired history $H_a(t)$ with $H_a(0)=0$, and then
integrating over time (Brandt 1994d), {\bf figure} 17.
The same equation for the sheet current can be used to compute
{\it flux creep} in transverse geometry (Gurevich and Brandt 1994)
(section 9.8.) and the linear response (section 9.6.2.) of HTSC's in the
TAFF regime (where $\rho$ is Ohmic) or in the vortex-glass regime with
arbitrary complex and dispersive ac resistivity $\rho(\omega)$
(K\"otzler et al.\ 1994b, Brandt 1994c).

This method recently has been extended to describe the time-  and
space-dependence of the sheet current ${\bf J}_s(x,y)$ in thin squares
and rectangles with linear or nonlinear resistivity in transverse
magnetic field (Brandt 1995a, b).
Superconducting (or nonlinearly conducting) squares or rectangles
during penetration or exit of flux exhibit sharp diagonal lines
which mark the positions where the sheet current bends sharply since
it has to flow parallel to the edges and has to keep constant modulus
$J_s =|{\bf J}_s| = J_c d$. These discontinuity lines (Campbell and
Evetts 1972, Gyorgy et al.\ 1989, Zhukov and Moshchalkov 1991,
Br\"ull et al.\ 1991,
Vlasko-Vlasov et al.\ 1992,  Schuster et al.\ 1994a, b, c,
Grant et al.\ 1994, and Xing et al.\ 1994) are clearly seen as
logarithmic infinities in the perpendicular field $H_z(x,y)$
({\bf figure} 18, 19, 20).
In the fully penetrated critical state with $J=J_c$ everywhere
in a rectangle of extension $-a\le x \le a$, $-b \le y \le b$,
$b\ge a$, the current stream lines are concentric rectangles, and
the Biot-Savart law yields
\begin{eqnarray}    
   H_z(x,y) = H_a +\frac{J_c d}{4\pi}
       [\, f(x,y) +f(-x,y) +f(x,-y) +f(-x,-y)\,] \nonumber \\
   f(x,y)=\sqrt2 \ln\frac{\sqrt2 P +a+b-x-y}{\sqrt2 Q -a+b-x-y}
 +\ln\Big| \frac{(P+y-b)(y-b+a) (P+x-a) \,x}{(y-b)(Q+y-b+a)(x-a)(Q+x)}
      \Big|
\end{eqnarray}
  with $P=[(a-x)^2 +(b-y)^2]^{1/2}$ and $Q=[x^2+(b-a-y)^2]^{1/2}$,
{\bf figure} 18. Note the star-like penetration pattern, the flux
concentration in the middle of the sample edges, and the
logarithmic infinities near the diagonal discontinuity lines
($|y|-|x| = b$) and near the edges ($|x|=a$ or $|y|=b$), cf.\
the equation (8.8) for $b\to 0$.

    The determination of the time-dependent sheet current
${\bf J}_s(x,y,t)$ and perpendicular field component $H_z(x,y,t)$
for thin planar conductors or superconductors of {\it arbitrary shape}
in a time dependent applied field $\hat{\bf z}\, H_a(t)$ is an
intricate problem which can be solved as follows. First, one has
to express the sheet current by a scalar function $g(x,y)$ as
${\bf J}_s(x,y) = -{\bf \hat z}\times \nabla g(x,y) = \nabla \times
{\bf \hat z}\, g(x,y)$. This substitution guarantees that
div${\bf J}_s = 0$ and that the current flows along the specimen
boundary if one puts $g(x,y) = {\rm const} = 0$ there.
For rectangular specimens $g(x,y)$ is conveniently expressed as a 2D
Fourier series in which each term vanishes at the edges. The contour
lines $g(x,y)=$ const coincide with the current stream lines.
 The physical meaning of $g(x,y)$ is the {\it local magnetization}
or density of tiny current loops. The integral of $g(x,y)$ over the
specimen area (or the appropriately weighted sum over all Fourier
coefficients) yields the total magnetic moment,
\begin{equation} 
 {\bf m}= \frac{1}{2} \int\! {\bf r \times J} {\bf(r)}\,{\rm d}^3 r
        = \frac{1}{2} \int\! {\bf r \times J}_s{\bf(r)}\,{\rm d}^2 r
  = {\bf \hat z} \int\! g({\bf r})\, {\rm d}^2 r \,.
\end{equation}
Next one determines the integral kernel $K({\bf r, r'})$
(${\bf r} = x,y$) which relates the perpendicular field $H_z(x,y)$
in the specimen plane $z=0$ to $g(x',y')$ by (for $H_a = 0$)
\begin{equation} 
 H_z({\bf r}) = \int\! K({\bf r,r'})\, g({\bf r'})\, {\rm d}^2 r',~~~
 g({\bf r}) = \int\! K^{-1}({\bf r,r'})\, H_z({\bf r'})\,{\rm d}^2 r'
\end{equation}
 where the integrals are over the specimen area. This
is a nontrivial problem, since when one performs the limit of
zero thickness in the Biot-Savart law, the kernel becomes highly
singular, $K =-1/4\pi |{\bf r-r'}|^{3}$; this form of the kernel only
can be used if ${\bf r}$ lies {\it outside} the specimen; but
{\it inside} the specimen area (where ${\bf r =r}'$ can occur) one
has to perform part of the integration analytically to obtain a
well behaved kernel (Brandt 1992d), or numerically for a small but
finite height $z$ above the specimen (Roth et al.\ 1989,
Xing et al.\ 1994). Explicitly one has
\begin{equation}  
 K({\bf r}, {\bf r}') = - \frac{1}{4\pi}\, \lim_{z\to 0} \frac{2z^2
 -\rho^2}{ (z^2 +\rho^2)^{5/2}}  ~~~{\rm with}~~~
                      \rho^2 = (x-x')^2 +(y-y')^2 \,.
\end{equation}
   The inverse kernel $K^{-1}$ in (8.16) may be obtained by
Fourier transform or by performing the integration on a grid with
positions ${\bf r}_i =(x_i, y_i)$, weights $w_i$, the tables
$H_i =H_z({\bf r}_i)$ and $g_i = g({\bf r}_i)$, and the matrix
$K_{ij} =K({\bf r}_i, {\bf r}_j)\, w_j$. The integrals (8.16) then
are approximated by the sums $H_i=\sum_j K_{ij} g_j$ and
$g_i=\sum_ jK_{ij}^{-1}H_j$ where $K_{ij}^{-1}$ is the inverse matrix
of $K_{ij}$ (Brandt 1994a, b, Xing et al.\ 1994).

As the last step the equation of motion for $g(x,y,t)$ is obtained
from the (3D) induction law $\nabla\times {\bf E} = -{\bf \dot B}$
(the dot denotes $\partial /\partial t$) and from the material
laws ${\bf B} =\mu_0 {\bf H}$ and ${\bf E} = \rho{\bf J}$ valid
inside the sample where
${\bf J} = {\bf J}_s/d = -\hat{\bf z}\times \nabla g(x,y)/d$. In
order to obtain an equation of motion for the 2D function $g(x,y,t)$,
it is important to note that the required $z$-component
$\dot B_z = \hat{\bf z}\, \dot{\bf B} =  -\hat{\bf z}\,
 (\nabla\times{\bf E}) = -(\hat{\bf z}\times\nabla)\, {\bf E}=
 -\hat{\bf x\, \partial E}/\partial y
 +\hat{\bf y\, \partial E}/\partial x$ does not depend on the
 (unknown) derivative $\partial{\bf E}/\partial z$.
With the sheet resistivity $\rho_s = \rho/d$  one may write inside
the sample ${\bf E}=\rho {\bf J} =\rho_s {\bf J}_s = -\rho_s
 (\hat{\bf z} \times \nabla)g$ and thus
$\dot B_z = (\hat{\bf z}\times \nabla)(\rho_s \hat{\bf z}\times
 \nabla g)= \nabla(\rho_s \, \nabla g)$.
 Using (8.16) with an applied field $\hat{\bf z}\, H_a(t)$
added,  $\dot B_z =\mu_0 \dot H_a +
 \mu_0 \int \!K({\bf r,r'})\, \dot g({\bf r'})\, {\rm d}^2 r'$,
one obtains the equation of motion for $g(x,y,t)$ in the form
(Brandt 1995a, b)
\begin{equation} 
  \dot{g}({\bf r},t)
   = \int \! K^{-1}({\bf r,r'})\, [\, f({\bf r'},t)
  -\dot H_a(t) \,]\, {\rm d}^2r' ~~~{\rm with}~~~
   f({\bf r}, t) = \nabla (\rho_s \,\nabla g)/\mu_0.
\end{equation}
This general equation, which applies also when $\rho_s =\rho/d$ depends
on ${\bf r}$, ${\bf J}$, and $H_z$, is easily integrated over time on
a personal computer. Some results obtained from (8.18) for flux
penetration into a superconducting rectangle are depicted in
{\bf figure} 19. Since $\rho/\mu_0$ has the meaning of a diffusivity,
equation (8.18) describes nonlocal (and in general nonlinear)
diffusion of the  magnetization $g(x,y,t)$.
In the limit of a long strip of a London superconductor with (formally)
$\rho = i\omega \mu_0 \lambda^2$ [cf.\ equation (9.10) below],
(8.18) reduces to the static (time-independent) equation (8.6).
 If the resistivity is anisotropic ($E_x =\rho_{xx} J_x$,
     $E_y =\rho_{yy} J_y$) then equation (8.18) still applies but with
modified function $f({\bf r},t)  = \nabla_x [(\rho_{yy}/d) \nabla_x g]
                         +\nabla_y [(\rho_{xx}/d) \nabla_y g]$.
Interestingly, similar distributions of magnetic flux as in figures
18 and 19 are obtained for a square network of Josephson junctions
in an  external magnetic field (Reinel et al.\ 1995).
        \\[0.4 cm]
{\it 8.7.~ Statistical summation of pinning forces}
         \\*[0.2 cm]
  The statistical summation amounts to the calculation of the average
pinning force density $J_cB$ exerted by random  pinning forces,
or more precisely, by the forces exerted by a random pinning
potential. These forces thus depend on the flux-line positions,
and the  problem has to be solved self-consistently.
If the FLL were ideally soft, each pin could exert its maximum force
and the direct summation would apply, yielding a pinning force
per unit volume $J_c B = n_p f_p$ were $n_p$ is the number density
of pins and $f_p$ is the elementary pinning force.
In the opposite limit, if the FLL were
ideally stiff, there would be no correlation between the
positions of the flux lines and pins, and thus the pinning forces
would be truly random and average to zero. Obviously, the {\it elastic
or plastic distortion of the FLL by the pins is crucial for the
pinning summation}. A softening of the shear stiffness of the FLL
may lead to a steep increase of $J_c(B)$ near the upper critical
field $B_{c2}$ were $c_{66}\sim (B_{c2}-B)^2$ (4.4) decreases
(Pippard 1969).

It turns out that the ``dilute limit'' considered in early summation
theories  by Yamafuji and Irie (1967) and Labusch (1969b) and
discussed by Campbell and Evetts (1972) does {\it not} exist
(Brandt 1980a). For random pins with force $f_p$ and number density
$n_p$ this limit yields
$J_cB \sim n_p f_p^2/c_{\rm eff}$ for $f_p \gg f_{\rm thr}$
and $J_cB = 0$ for $f_p \le f_{\rm thr}$. Here $c_{\rm eff}$ is
an effective elastic modulus of the FLL [$c_{66}$ or
$(c_{66}c_{44})^{1/2}$ for linear or point pins, respectively]
 and $f_{\rm thr} $ is
a threshold force which is an artifact of the non-existing dilute
limit: In an infinite system, even for weak dilute pins the correct
$J_cB$ is never proportional to the number of pins since each pin
``feels'' the FLL distortions caused at its position by the other
pins, even when these are far away. A single pin or a few pins in fact
do exhibit this threshold effect, but with increasing number of pins
the threshold decreases. One may see this by formally combining
$N$ random pins of strength $f_p$ into one ``superpin'' which has
the larger force $N^{1/2} f_p$ that will exceed the threshold if
$N$ is large. A summation over point pins by considering ``bundelling''
of flux lines was given by Schmucker and Kronm\"uller (1974),
see also Lowell (1972), Campbell (1978), Matsushita and Yamafuji (1979),
and Tak\'acs (1982).

   The existence of a
 finite average pinning force requires the occurrence of elastic
instabilities in the FLL when the FLL moves past the pins and the
flux lines are plucked. Without such instabilities the total pinning
force would be a unique smooth periodic function $F(X)$ of the
(uniformly driven) center of gravity coordinate $X=vt$, with
zero average $\langle F(X) \rangle = 0$ (Brandt 1983a, b). The
time-averaged pinning force (friction) would thus be zero.
  Instabilities of the FLL in the presence of a few pins explicitly were
calculated by Ovchinnikov (1982, 1983) and for planar pins by
Brandt (1977c), see also Gurevich and Cooley (1994) for a detailed
consideration of planar pinning and Martinoli (1978) for
periodic pins.

   The first correct statistical summation
was presented by Larkin and Ovchinnikov (LO) (1979), after they had
arrived at similar results earlier (Larkin and Ovchinnikov 1973) by a
more complicated calculation using Feynman graphs and disregarding the
important elastic nonlocality of the FLL. For similar summation
theories see Kerchner (1981, 1983), and for pinning of
charge-density waves see
 Lee and Rice (1979) and the detailed theory by Fisher (1985).
 A general scaling law for pinning summation in arbitrary dimension
was suggested by Hilzinger (1977) and applied to collective pinning
of flux lines and of crystal-lattice dislocations
(``solution hardening'') by Brandt (1986c). A renormalization group
argument for this scaling law is given by Fisher (1983).

  The LO ``theory of collective pinning'' applies to weak
random pins in an elastic FLL {\it without dislocations} where it
predicts for 3D and 2D pinning in a film of thickness $d$
\begin{eqnarray} 
 J_cB = (1.5^{1/2}/32\pi^2) W^2 /(r_p^3 c_{66}^2 c_{44}) ~~~~&(3D)\\
 J_cB \approx (8\pi)^{-1/2} W /(r_p d c_{66}) .          ~~~~&(2D)
\end{eqnarray}
Here $W = n_p \langle f_{p,i}^2\rangle_{\rm pins}$ is the average
pinning force squared, with $n_p$ the volume density of pins
and $f_{p,i}$ the actual force exerted by the $i$th pin on the FLL,
and  $r_p\approx \xi$ is the range of the pinning forces.
 A detailed calculation of $W$ and $r_p$ as functions of $B/B_{c2}$
for various types of random pins was performed from GL theory
by Brandt (1986d).
The main idea of LO is that the random pins destroy the long-range
order of the FLL (Larkin 1970) such that with 3D pinning
the short-range order only is preserved in a cigar-shaped volume of
radius $R_{3c} = 8\pi r_p^2 c_{44}^{1/2} c_{66}^{3/2}/W$
and length (along $B$) $L_{c} = (c_{44}/c_{66})^{1/2} R_{3c} =
8\pi r_p^2 c_{44}c_{66}/W \gg R_{3c}$.
This correlated volume $V_c$ is defined
as the volume within which the pinning-caused mean-square vortex
displacement  $g({\bf r}) = \langle |{\bf u(r) - u}(0)|^2\rangle$
is smaller than the pinning force range squared,
$g({\bf r}) \le r_p^2 $. This condition yields an
elliptical volume $V_c= (4\pi/3)R_{3c}^2 L_{c}$.
For films of thickness $d< L_{c}$ and radius $R$,
pinning becomes 2D and the
correlated volume is a cylinder of height $d$ and radius
$R_{2c} = r_p c_{66} [8\pi d/(W \ln(R/R_{2c}))]^{1/2}$, thus
$V_c = \pi R_{2c}^2 d$. LO then assume that within the volume $V_c$
the pinning forces are uncorrelated and therefore their squares add.
This idea yields for the average pinning force the estimate
 $J_cB = (W/V_c)^{1/2}$, or explicitly (8.19) and (8.20).
In the above LO expressions corrections from nonlocal elasticity
are omitted. With increasing pinning strength these nonlocal
corrections (Brandt 1986b, W\"ordenweber and Kes 1987)
become important when $R_{3c} \le \lambda$. When
pinning is so strong that $R_{3c}$ becomes comparable to the
flux-line spacing $a$, the collective pinning of bundles goes
over to the collective pinning of individual vortices, which
was considered by numerous authors, see below.
When pinning becomes even stronger (or the FLL softer) such that
$L_c$ is reduced to $r_p$ or $\xi$, then the collective pinning
result goes over into the direct summation, $J_c B \approx
(n_p W)^{1/2} = n_p \langle f_{p,i}^2\rangle_{\rm pins}^{1/2}$.
   For {\it dilute} pins, with spacing
$n_p^{1/3}$ exceding the flux-line spacing $(\phi_0/B)^{1/2}$,
this may be written as $J_cB \approx n_p f_p$ since all pins
exert their maximum force $f_p$.
   For {\it dense} pins, not all pins are active and one gets in the
direct summation limit $J_cB \approx (n^{1/3} B/\phi_0)f_p$.

 The 2D LO result (8.20) is borne out {\it quantitatively} by 2D
computer simulations which artificially suppress plastic
deformation by assuming a spring potential between nearest neighbours
and obtain a $J_cB$ which is $\approx 1.1$ times the LO expression
(8.20) (Brandt, unpublished). A nice experimental verification of
(8.20) was obtained with weak pinning films of amorphous MoGe by
Kes and Tsuei (1981, 1983), White et al.\ (1993), and
Garten et al.\ (1995). The 3D LO result (8.19) does not always
 apply to real superconductors since in an infinite 3D FLL even
weak pins may induce {\it plastic deformation}, which softens the FLL
and enhances $J_c$ (Mullok and Evetts 1985). But recently the 3D LO
result (8.19) was confirmed by small-angle neutron scattering from
2H-NbSe$_2$ single crystals (Yaron et al.\ 1994).
Very recently Giamarchi and Le Doussal (1994, 1995) by a powerful
``functional variational method'' confirmed the 3D and 2D  LO
results for $J_cB$ and found the remarkable result that the 2D and 3D
pinned FLL (or any elastic medium in a random pinning potential)
does not have to contain dislocations to lower its strain and energy.

  The  LO  collective pinning theory was extended to films of finite
thickness  by Brandt (1986b) and W\"ordenweber and Kes (1987).
Nice physical ideas on the pinned FLL are given by
Bhattacharya and Higgins (1993, 1994) and by
Yamafuji et al.\ (1994, 1995).
Computer simulations of the pinned FLL
were performed by Brandt (1983a, b),  Jensen et al.\ (1989, 1990),
Shi and Berlinsky (1991),
Pla and Nori (1991), Kato et al.\ (1991), Enomoto et al.\ (1993),
Machida and Kaburaki (1993, 1995), Reefman and Brom (1993),
Wang et al.\ (1995),
and particularly nicely by Koshelev (1992c). Koshelev and Vinokur
(1994) introduce the notion of ``dynamic melting''
of the FLL by defining an effective temperature of the FLL caused
by its jerky motion over random pins during flux flow, which like
thermal fluctuations may be described by a Langevin equation.
Wang et al.\ (1995) compute the complex resistivity of the randomly
pinned FLL by computer simulations.

  The long-range distortion of the pinned FLL, which crucially enters
the pinning summation (Larkin 1970) and which was also discussed
in the context of decoration experiments (Grier et al.\ 1991,
Chudnovsky 1990), was derived from scaling arguments by
Feigel'man et al.\ (1989), who found for these transverse
fluctuations power laws of the type
$g({\bf r}) = \langle |{\bf u(r) -u(}0)|^2 \rangle \sim r^\zeta$,
where $\zeta$ depends on the dimensionalities $d$ and $n$ of the
elastic manifold and of the displacement field ${\bf u}$
(e.g., $d=3$, $n=2$ for the FLL; $d=1$, $n=2$ for a single flux line),
see Blatter et al.\ (1994a) for detailed discussion. Recently this
distortion was calculated by a Gaussian variational method using
the replica method (Bouchaud et al.\ 1992), a replica density
functional approach (Menon and Dasgupta 1994),
and a random phase model (Batrouni and Hwa 1994).

  In my opinion the fundamental problem of an elastic medium in a
random potential in thermal equilibrium, was solved definitively
by Giamarchi and Le Doussal (1994, 1995)
by the Gaussian variational method and the renormalization group;
Giamarchi and Le Doussal find that previous theories strongly
overestimate the pinning-caused displacements of the FLL and that
the correlation function
 $g({\bf r}) = \langle |{\bf u(r) -u(}0)|^2 \rangle$ at large
distances (where $g^{1/2}$ exceeds the lattice spacing $a$)
increases only slowly with the logarithm of $r$, since a pinning
center essentially displaces only the nearest lattice point
(or flux line).
For more work on the random field $xy$-model as a model for
vortex pinning see Le Doussal and Giamarchi (1995).
By computer simulations  \v{S}\'a\v{s}ik et al.\ (1995)
 show that in a 2D superconductor the
vortex-liquid transition is not accompanied by a simultaneous
divergence of the correlation length of the phase of the order
parameter, and that even in the solid FLL phase the phase-angle
coherence-length remains short ranged, of the order of the
magnetic penetration depth $\lambda$.

The problem of a single vortex line depinning from a random potential
has been studied extensively, see e.g.\ Huse et al.\ (1985),
Kardar (1985, 1987), Brandt (1986c), Ivlev and Mel'nikov (1987),
Nattermann et al.\ (1992), Parisi (1992), Dong et al.\ (1993),
Erta\c{s} and Kardar (1994), and the simulation of depinning noise
of one moving flux line by Tang et al.\ (1994) commented by
Brandt (1995b). A detailed discussion is
given in the review paper by Blatter et al. (1994a).
The softening of the randomly pinned FLL and the fluctuation of the
elastic moduli are calculated by a replica method by Ni and Gu (1994).
Collective pinning of Josephson vortices in a long disordered
Josephson junction (e.g.\ between the superconducting Cu-O planes in
HTSC's) was calculated by Vinokur and Koshelev (1990) and in a
comprehensive nice work by Fehrenbacher et al.\ (1992), see also the
recent experiments by Itzler and Tinkham (1995), the theory by
Balents and Simon (1995), and the book on Josephson junctions
and their critical currents by Barone and Paterno (1982).
%
         \\[0.7 cm]
{\bf 9.~ Thermally activated depinning of flux lines}
         \setcounter{subsection}{9}  \setcounter{equation}{0}
         \\*[0.5 cm]
{\it 9.1.~ Pinning force versus pinning energy}
         \\*[0.2 cm]
The ideal pinning picture of section 8  striktly spoken applies
only at zero temperature. At $T>0$, thermally activated
depinning of flux lines causes a non-vanishing resistivity
even below $J_c$ (Anderson 1962, Kim and Anderson
1964). In {\it conventional superconductors} this effect is
observed only close  to the transition temperature $T_c$
as {\it flux creep} (Beasley et al.\ 1969, Rossel et al.\ 1990,
Berghuis and Kes 1990, Suenaga et al.\ 1991,
Svedlindh et al.\ (1991), Schmidt et al.\ 1993).
Flux creep occurs in the critical state after the applied magnetic
field is increased or decreased and then held constant.
The field gradient and the persistent currents and magnetization
then slowly decrease with an approximately logarithmic time law.
  As mentioned in section 8.6 and section 9.6 below,
flux creep is caused by a highly nonlinear,
current-dependent flux-flow resistivity $\rho(J)$, which may be
modeled,  e.g.\  by  $\rho\propto (J/J_c)^n$ ($n\gg 1$),
$\rho\propto\exp(J/J_1)$, or $\rho\propto \exp[-(J_2/J)^\alpha]$.
Initially, the persistent currents in a ring feel a
large $\rho$, but as the current decays,  $\rho$ decreases
rapidly and so does the decay rate  $-\dot{J}(t) \propto\rho(J)$.

   In HTSC's thermal depinning is observed in a  large temperature
interval below T$_c$. This ``giant flux creep'' (Dew-Hughes 1988,
Yeshurun and Malozemoff 1988, Kes et al.\ 1989, Inui et al.\ 1989)
leads to reversible behaviour of HTSC's above an
``irreversibility line'' or ``depinning line'' in the $B$-$T$ plane,
 see e.g.\ Yazyi et al.\ (1991), Schilling et al.\ (1992, 1993),
 and Deak et al.\ (1994).
Thermal depinning occurs mainly because (a) the superconducting coherence
length $\xi$ ($\approx$ vortex core radius) is small, (b)
the magnetic penetration depth $\lambda$ is large, and (c) these
materials are strongly anisotropic (section 3); here the notation
$\xi =\xi_{ab}$, $\lambda =\lambda_{ab}$,
and $\kappa = \lambda_{ab}/\xi_{ab}$ is used.
All three properties decrease the pinning {\it energy}
but tend to increase the pinning {\it force}:

  \underline{Small $\xi$} means that  the elementary pinning
{\it energy} $U_p$ of small pins (e.g.\ oxygen vacancies or
clusters thereof) is small, of the order of
 $ (B_c^2/\mu_0)\xi^3 = (\Phi_0^2/8\pi^2\mu_0 \lambda^2)\xi$.
The elementary pinning {\it force} $U_p /\xi$, however, is
{\it independent} of $\xi$ in this estimate and is thus not
necessarily small in HTSC's, cf.\ section 8.2.3.

 \underline{Large $\lambda$} means that the stiffness of the FLL with
respect to shear deformation and to short-wavelength tilt
(nonlocal elasticity, section 4.2.) is small. Therefore, the flux lines
can better adjust to the randomly positioned pins. According to the
statistical summation theories for random pins
(Labusch 1969b, Campbell and Evetts 1972, Larkin and Ovchinnikov 1979),
this flexibility increases the average pinning {\it force} density.
Therefore, the naive argument that a soft FLL or a
flux-line liquid with vanishing shear stiffness cannot be pinned
since it may flow around the pins, does not apply to the realistic
situation where there are many more pins than flux lines.
However, in HTSC's the decrease of the pinning {\it energy} enhances
the rate of thermally activated depinning and thus effectively
reduces pinning at {\it finite} temperature.

   \underline{Large material anisotropy}, like the large penetrations
depth, effectively softens the FLL and
{\it increases} the average pinning {\it force} but {\it decreases}
the pinning {\it energy}. This will be particularly clear with the
 columnar pins discussed in section 9.2.

A useful quantity which shows in which superconductors thermal
fluctuations and thermal depinning become important, is the
Ginzburg number $Gi$ (section 5.1.). Ginzburg (1960) showed that in
phase transitions of the second order the temperature range
$T_f \le T\le T_c$ where the relative fluctuations of the
order parameter $\psi$ are large,
$| \delta \psi/\psi |^2 (T_f) \approx 1$
is determined by a constant $Gi = 1- T_f/T_c$. For anisotropic
superconductors this Ginzburg number is (in SI units)
\begin{eqnarray}   
Gi = \frac{1}{2} \Big(\frac{k_B T_c }{\xi^3(0) } ~
                 \frac{\Gamma \mu_0}{4\pi B_c^2(0)} \Big)^2
   = \frac{1}{2} \Big(\frac{2\pi\mu_0 k_B}{\phi_0^2}  ~
                 \frac{T_c \lambda(0)^2 \Gamma}{\xi(0)} \Big)^2 \,.
\end{eqnarray}
  This means that fluctuations are large when $T_c$, $\lambda$,
and the anisotropy $\Gamma$ are large, and when $\xi$ is small.
         \\[0.4 cm]
{\it 9.2.~ Linear defects as pinning centres. The ``Bose glass''.}
         \\*[0.2 cm]
Long columnar pins of damaged atomic lattice were generated
by high-energy heavy-ion irradiation perpendicular to the
Cu-O planes in YBCO (Roas et al.\ 1990a, Civale et al.\ 1991,
Budhani et al.\ 1992, Konczykowski et al.\ 1993, Klein et al.\ 1993a, c,
Schuster et al.\ 1993b, 1994c, 1995b, Leghissa et al.\ 1993)
or BSCCO (Gerh\"auser et al.\ 1992, Leghissa et al.\ 1992,
Thompson et al.\ 1992, Schuster et al.\ 1992a, Hardy et al.\ (1993),
Klein et al.\ 1993b, c, Kummeth et al.\ 1994, Kraus et al.\ 1994).
Clearly these linear pins are most effective when the flux lines
are parallel to the pins since then large sections of the
flux line are pinned simultaneously, which means a large pinning
energy and a reduction of thermally activated depinning.
However, this desired effect is observed only in the less
anisotropic YBCO. In the very anisotropic BSCCO the flux lines
have very low line tension (section 4.3.) and thus easily break
into short segments or point vortices which then depin
individually with very small activation energy
(Gerh\"auser et al.\ 1992, Brandt 1992c).
 As a consequence, BSCCO tapes are good superconductors only
at low temperatures, say at $T=4\,$K, where they may be used to built
coils for extremely large magnetic fields.
 In principle, if $B$ could be
kept strictly parallel to the layers, large $J_c$ and weak thermal
depinning could be achieved even at $T=77\,$K, but this geometric
condition is difficult to meet.

  The decrease of the pinning energy with  increasing anisotropy
 $\Gamma = \lambda_c/\lambda_{ab}$  and the eventual depinning of
independent  point vortices from linear pins oriented along the
$c$-axis can be understood as follows. The flux-line tension (4.14)
is $P\approx (\phi_0^2/4\pi \mu_0 \lambda^2_{ab} \Gamma^2)
  \ln(\Gamma L /\xi_{ab})$, where $L=1/k_z < \lambda_{ab}$
is the typical length of the deformation; the pinning energy
per unit length of a linear defect with radius $r_p > \xi_{ab}$
is $U_p = \epsilon (\phi_0^2/4\pi \mu_0 \lambda_{ab}^2)$ with
$\epsilon \le 0.5 +\ln(r_p/\xi_{ab}) \approx 1$;
the equal sign applies for cylindrical holes, cf.\ equation (1.1)
and section 4.5. Now consider the depinning nucleus when a
flux line switches from one line pin to a neighbouring line pin at
a distance $a_p$, and back to the original line pin, {\bf figure} 21.
The energy $U$ and minimum width $2L$ of this double kink is
obtained by minimizing the sum of the depinning energy $2LU_p$
and the tilt energy $(P/2) \int\! u'(z)^2 {\rm d}z$
where $u(z)$ is the flux-line displacement. The result is
a trapezoidal nucleus with
\begin{eqnarray}   
   L = (P/2U)^{1/2} a_p = [2\ln(L\Gamma /\xi_{ab})
        /\epsilon]^{1/2}a_p/ \Gamma \,,~~~
   U = (8PU_p)^{1/2} a_p = 4LU_p \sim 1/\Gamma .
\end{eqnarray}
  Both the width $2L$ and energy $U$ of the depinning nucleus
decrease with increasing anisotropy $\Gamma$ as $1/\Gamma$.
When $L$ has shrunk to the layer separation $s$ at
$\Gamma \approx 2a_p/s$, the anisotropic London theory
ceases to be valid, and $L$ saturates to $L\approx s$.
  A similar picture results when the distance $a_p$ between the
linear pins is so large, or the applied current density $J$
so high, that depinning is triggered by the current. The critical
nucleus for depinning is now a parabola of height $U_p/(\phi_0 J)$,
width $2L = (8U_p P)^{1/2} /(\phi_0 J) \sim 1/\Gamma$,
and energy $U=(4/3)LU_p = (4/3) (2U_p^3 P)^{1/2} /(\phi_0 J)
\sim 1/\Gamma$.
  The general expression for arbitrary $a_p$ and $J$ is given by
Brandt (1992c). Note that the pinning {\it force} per unit length
$F_p \approx U_p/\xi_{ab}$ and the critical current density
$J_c = F_p/\phi_0 \approx \epsilon \phi_0 /(4\pi \mu_0 \lambda_{ab}^2
\xi_{ab})$ do not depend on the anisotropy in this geometry.

  Pinning of flux lines by randomly positioned parallel columnar pins
is an interesting statistical problem (Lyuksyutov 1992). Nelson
and Vinokur (1992, 1993) show that in the model of a
simplified flux-line interaction this system is equivalent to
a ``Bose glass'', which has similar properties as the ``vortex glass''
of section 9.5. While  W\"oltgens et al.\ (1993a) do not find
Bose-glass behaviour in thin films, Krusin-Elbaum et al.\ (1994)
and Jiang et al.\ (1994) in YBCO crystals with columnar defects
observe Bose-glass melting, and Konczykowski (1994) in YBCO and BSCCO,
Konczykowski et al.\ (1995) in BSCCO crystals, Miu et al.\ (1995) in
BSCCO films, and Budhani et al.\ (1994) in a Tl-HTSC with
columnar defects find evidence for Bose-glass behaviour.
In a nice paper Krusin-Elbaum et al.\ (1994), with YBCO crystals
bombarded by 1 GeV Au ions, confirm various predictions of the
Bose-glass melting transition; this transition is assumed to occur
 along the irreversibility line
$H_{\rm irr}(T) \sim (1- T/T_c)^\alpha$
determined from the maximum of the
dissipative component $\chi''$ of the ac susceptibility at 1 MHz,
see Krusin-Elbaum et al.\ (1991) and Civale et al.\ (1992) for
a discussion of this ac method. In this experiment,
with increasing defect density  the effective power exponent
$\alpha$ grows from $\approx 4/3$ to $\approx 2$, in quantitative
agreement with the predicted (Nelson and Vinokur 1992, 1993) shift
of the Bose-glass transition field to higher temperatures.
Van der Beek et al.\ (1995) investigate the vortex dynamics in
BSCCO with columnar defects using an ac-shielding method and
find good agreement with the Bose-glass theory.

  The state when flux lines are locked-in at the columnar defects
was observed by dc measurements (Budhani et al.\ 1992) and by
microwave absorption (Lofland et al.\ 1995).
Ryu et al.\ (1993) performed Monte-Carlo simulations of this system.
The influence of additional (``competing'') disorder by point defects
is calculated by Hwa et al.\ (1993a) and Balents and Kardar (1994).
The effect of non-parallel (``splayed'') columnar pins is considered
by Hwa et al.\ (1993b). This splay of columnar defects reduces the
vortex motion (Civale et al.\ 1994). Thermal depinning and quantum
depinning from a single line defect is considered  by
Kuli\'c et al.\ (1992),  Kr\"amer and Kuli\'c (1993, 1994),
Khalfin and Shapiro (1993), Sonin and Horovitz (1995), Sonin (1995),
and Chudnovsky et al.\ (1995).  The phase diagram of a thin
superconducting film with a hole in an axial magnetic field
is presented by Bezryadin et al.\ (1995) as a model for a
columnar pin.  Thermal depinning from a
single twin plane is calculated by Khalfin et al.\ (1995).
The low-temperature dynamics of flux lines pinned by parallel
twin planes is found to be dominated by Mott variable-range hopping
of superkinks (Marchetti and Vinokur 1994, 1995).

By a decoration method Leghissa et al.\ (1993) observe the
destruction of translational order in the FLL by these linear
defects. Doyle et al.\ (1995) measure the effect of columnar
defects on the elasticty  of the FLL.
Dai et al.\ (1994b) in BSCCO simultaneously observe columnar defects
by etching and flux lines by decoration and thus can see where
the flux lines are pinned.  By scanning tunnelling microscopy
Behler et al.\ (1994b) see the columnar defects and the
pinned flux lines on the same NbSe$_2$ sample.
 In YBCO films irradiated at various angles by heavy ions
Holzapfel et al.\ (1993) observe sharp peaks in the angle-dependent
critical current $J_c(\Theta)$ when $B$ is parallel to the
columnar pins, see also Klein et al.\ (1993a, b).
In a similar HTSC with oblique and crossed linear defects
Schuster et al.\ (1994c, 1995b)
observe anisotropic pinning by magneto-optics.
Becker et al.\ (1995) measure magnetic relaxation in DyBCO
and interpret these data by a distribution of activation energies.
 Klein et al.\ (1993b, c) in BSCCO with radiation-induced
columnar defects inclined at 45$^o$ with respect to the $c$-axis
observe that at $T=40\,$K pinning is isotropic and thus the
point vortices are independent, but at $T=60\,$K anisotropic
pinning indicates that the pinned objects have line character
and thus the point vortices are aligned and form flux lines.
Kraus et al.\ (1994a) give a review on such tailored linear defects.
Lowndes et al.\ (1995) introduce columnar defects by growing
YBCO films on miscut, mosaic LaAlO substrates, and observe strong
asymmetric pinning.
         \\[0.4 cm]
{\it 9.3.~ The Kim-Anderson model}
         \\*[0.2 cm]
In HTSC's at sufficiently high temperatures, a linear (Ohmic)
resistivity $\rho$ is observed at small current densities $J\ll J_c$
due to thermally assisted flux flow (TAFF, Kes et al.\ 1989).
Both effects, flux creep at $J\approx J_c$ and TAFF at
$J\ll J_c$, are limiting cases of the general expression of
Anderson (1962) and Anderson and Kim (1964)
for the electric field  $E(B,T,J)$  caused by
thermally activated flux jumps out of pinning centres,
   \begin{equation}   
E(J) = 2\rho_c J_c\,\exp(-U/k_BT)\sinh(JU/J_c k_BT)~.
   \end{equation}
In (9.3) enter the {\it phenomenological parameters}
$J_c(B)$ (critical current density at $T=0$),
$\rho_c(B,T)$ (resistivity at $J=J_c$),  and
$U(B,T)$ (activation energy for flux jumps).
 The physical idea behind equation (9.3) is that the
Lorentz force density ${\bf J \times B}$ acting on the
FLL {\it increases} the rate of thermally
activated jumps of flux lines or flux-line bundels
along the force,
$\nu_0   \exp[-(U-W)/k_BT]$, and {\it reduces} the jump rate
for backward jumps, $\nu_0\exp[-(U+W)/k_BT]$.
Here $U(B,T)$ is an activation energy, $W=JBVl$
the energy gain during a jump, $V$  the jumping
volume,    $l$ the jump width, and $\nu_0$ is an attempt
frequency. All these quantities depend on the microscopic
model, which is still controversial, but by
defining a critical current density $J_c=JU/W=U/BVl$, only
measurable quantities enter. Subtracting the two jump
rates to give an effective rate $\nu$
and then writing the drift velocity $v=\nu l$ and the
electric field $E= vB = \rho J$ one arrives at (9.3).

For large currents $J\approx J_c$ one has $W\approx U\gg k_BT$ and
thus $E\propto \exp(J/J_1)$ with $J_1=J_ck_BT/U$. For small
currents $J\ll J_1$ one may linearize the $\sinh(W/k_BT)$ in (9.3)
and gets {\it Ohmic} behaviour
with a thermally activated linear resistivity $\rho_{\rm TAFF}
\propto \exp(-U/k_BT)$.  Combining this with the usual
flux-flow resistivity $\rho_{\rm FF}$ valid at $J\gg J_c$
(Qiu and Tachiki 1993)
or with the square-root result [p=2 in (8.1)] for a  particle
moving viscously across a one-dimensional sinusoidal potential
(Schmid and Hauger 1973) one gets ({\bf figure} 22)
     \setcounter{equation}{4}
     \begin{eqnarray}   
                                            \nonumber
 ~~\rho~= & (2\rho_c U/k_BT)\exp(-U/k_BT) ~=~ \rho_{\rm TAFF}
     &\mbox{~for~~$J\ll J_1$~~  (TAFF)~~~} \;~~~~~~~(9.4a)\\
                                            \nonumber
 ~~\rho~= & \rho_c \exp[(J/J_c-1)U/k_BT] \propto \exp(J/J_1)
     &\mbox{~for~~$J\,\approx J_c$~~ (flux creep)~~~} \,~~~(9.4b) \\
                                            \nonumber
 ~~\rho~= & \rho_{\rm FF} (1 - J_c^2 /J^2)^{1/2} ~\approx \rho_{\rm FF}
        \approx \rho_n B/B_{c2}(T)
     &\mbox{~for~~$J\gg J_c$~~  (flux flow).~~}  ~~~~~(9.4c)
     \end{eqnarray}
The linear TAFF regime (9.4a) is observed if $B$ and $T$ are
sufficiently large (Palstra et al.\ 1990b, Shi et al.\ 1991,
Worthington et al.\ 1991, 1992, Seng et al.\ 1992,
van der Beek et al.\ 1992a, Wen and Zhao 1994).
At lower $T$ and $B$ nonlinear resistivity is observed
(Koch et al.\ 1989, Gammel et al.\ 1991, Safar et al.\ 1992a,
Yeh et al.\ 1992, 1993a, b, Dekker et al.\ 1992a,
Worthington et al.\ 1992, Seng et al.\ 1992,
van der Beek et al.\ 1992a, Sandvold and Rossel 1992,
White et al.\ 1993, 1994, W\"oltgens et al.\ 1993a, b,
Sun et al.\ 1993, Jiang et al.\ 1994, Wagner et al.\ 1994, 1995,
Deak et al.\ 1994, Heinzel et al.\ 1994),
which often can be fitted by the vortex-glass
expressions, section 9.5.  It appears that in the TAFF regime
the  FLL is in a ``liquid'' state  (Feigel'man and Vinokur 1990,
Chakravarty et al.\ 1990, Vinokur et al.\ 1990a, 1991b,
Koshelev 1992a), i.e.\  it has no shear stiffness;
therefore,  elastic deformations of the FLL at different points
are not correlated. This assumption leads to an activation
energy $U$ which does not depend on the current density.
Another situation where an extended Kim-Anderson theory fits
the measured $E(J)$ data well is when there are two types of
pinning centers (Adamopoulos et al.\ 1995).
         \\[0.4 cm]
{\it 9.4.~ Theory of Collective Creep}
         \\*[0.2 cm]
 Theories of collective creep go beyond the Anderson model (9.3)
and predict that the thermally jumping volume $V$ of the FLL
depends on the current density $J$ and becomes infinitely large
for $J\to 0$. Therefore, also the activation energy diverges,
e.g.\  as $U\propto V\propto 1/J^\alpha$ with
$\alpha >0$. As a consequence, the thermally activated resistivity
$\rho(J,T) \sim \exp[-U(J)/k_BT\,]$ becomes {\it truly zero}
for $J\to 0$. This result follows if weak random pinning acts on
a FLL that is treated as an ideally elastic medium which may be
isotropic (Feigel'man et al.\ 1989, Natterman 1990, Fischer and
Nattermann 1991,  Feigel'man et al.\ 1991),  anisotropic
(Pokrovsky et al.\ 1992,  Blatter and Geshkenbein 1993),
or layered (Koshelev and Kes 1993).
Qualitatively the same result is obtained by the vortex-glass
 picture, which starts from complete disorder, section 9.5.
Interestingly, into theories of collective creep all three moduli
for compression, shear, and tilt of the FLL enter, whereas into
the collective pinning theory  of Larkin and Ovchinnikov (1979)
(valid at zero temperature) enter only the shear and tilt moduli.
 This is so because the activated (jumping) volume of the FLL
depends also on the compression caused by the driving current, but
the correlated volume $V_c$ of the pinning-caused distortions
of the FLL (section 8.7.) depends only on the combination
$c_{11}^{-p} +c_{66}^{-p} \approx c_{66}^{-p}$ ($p=1$ or 2
for dimensionality $D=2$ or 3).

 An activation energy which diverges when $J\to 0$ is also obtained
in theories that assume the depinning to proceed via a kink
mechanism which allows
the flux lines to climb out from the space between the Cu-O layers
(Ivlev and Kopnin 1990a, b, 1991, Chakravarty et al.\ 1990)
or from columnar pins (Nelson and Vinokur 1992, 1993, Brandt 1992c,
Lyuksyutov 1992, Kr\"amer and Kuli\'c 1993, 1994) (section 9.2.).
         \\[0.4 cm]
{\it 9.5.~ Vortex glass scaling}
         \\*[0.2 cm]
 Similar current--voltage curves as in the collective creep
theory follow from the ``vortex-glass'' picture
(Fisher 1989, Fisher, Fisher, and Huse 1991).
The basic idea of the glassy behaviour is that if there is a
second order phase transition to this vortex glass
similar to that in theories of spin glasses, then both a
characteristic correlation length $\xi_g$ (the size of the
jumping volume) and the relaxation time $\tau_g$ of the fluctuations
of the glassy order parameter {\it diverge} at the glass-transition
temperature $T_g$,
\begin{equation}      
\xi_g (T,B) = \xi_g(B) | 1-T/T_g |^{-\nu}\,, ~~~~
\tau_g (T,B) \simeq \tau_g(B) | 1-T/T_g |^{-\nu z} .
\end{equation}
The vortex-glass picture predicts scaling laws, e.g.\  the
electric field should scale as $E \xi_g^{z-1} =
f_{\pm}(J\xi_g^{D-1})$ where $z\approx 4$, $D$ is the spatial
dimension, and $f_\pm (x)$ are scaling functions for
the regions above and below $T_g$.
For $x\to 0$ one has $f_+(x)$ = const  and
$f_-(x) \to \exp(-x^{-\mu})$. At $T_g$, a power-law current--voltage
curve is expected, $E\propto J^{(z+1)/(D-1)}$, thus
     \begin{eqnarray}   
 \rho~\propto & J^{(z+1)/(D-1) -1}    &\mbox{~for~~$T=T_g$~}\\
 \rho~\propto & \exp[-(J_2/J)^\alpha] &\mbox{~for~~$T<T_g$~}.
     \end{eqnarray}
In the theory of collective creep there is no
explicit glass temperature, but the picture is similar since
collective creep occurs only below a ``melting temperature''
$T_m$ above which the FLL looses its elastic stiffness.
Thus $T_m$ has a similar meaning as $T_g$. A vortex glass
state should not occur in 2D flux-line lattices
(Feigel'man et al.\ 1990, Vinokur et al.\ 1990b, Dekker et al.\ 1992b,
Bokil and Young 1995).
More details may be found in the reviews by Blatter et al.\ (1994a)
and K.\ H.\ Fischer (1995a).

   Experiments which measure the magnetization decay
or the voltage drop with high sensitivity, appear to
confirm this scaling law in various HTSC's in an appropriate
range of $B$ and $T$. For example, by plotting
$T$-dependent creep rates $\dot M = dM/dt$ in reduced form,
$(1/J) |1-T/T_g |^{-\nu (z-1)} \dot M$ versus $J|1-T/T_g|^{-2\nu}$,
 van der Beek at al.\ (1992a) in BSCCO measured  $T_g= 13.3\,$K,
$z=5.8 \pm1$, and $\nu = 1.7 \pm 0.15$, {\bf figure} 23.
 Vortex-glass scaling of $\rho(J,T)$ was observed, e.g., in YBCO
films (Koch et al.\ 1989, Olsson et al.\ 1991, Dekker et al.\ 1992b,
W\"oltgens et al.\ 1993a), in YBCO crystals (Yeh et al.\ 1992,
1993b), in BSCCO films (Yamasaki et al.\ 1994),
and in Tl$_2$Ba$_2$CaCu$_2$O$_8$ films (Li, Li et al.\ 1995).
  Very detailed data $\rho(J)$  for three YBCO samples
with different pinning (without and with irradiation by
protons or Au ions)  are presented by Worthington et al.\ (1992)
for various inductions $B$  with $T$ as parameter.
In the sample with intermediate pinning two transitions are
seen in $\rho(J)$, a ``melting transition'' at $T_m$
(e.g.\  $T_m \approx 91.5\,$K at $B=0.2\,$T,
$J=10^5\,$Am$^{-2}$)  and a ``glass transition'' at
$T_m \approx 90\,$K (above case) or $T_m = 84.92\,$K
(strong pinning sample, very sharp transition at
$B= 4\,$T, $J \le 4\times 10^5\,$Am$^{-2}$).
The FLL phase in between these two transitions was
termed ``vortex slush''.
 Krusin-Elbaum  et al.\ (1992) in YBCO crystals observe a crossover
between single-vortex and collective-pinning regimes by plotting
lines of $J_c(B,T) = $const, and Qiang Li et al.\ (1994) observe a
vortex-glass-to-liquid transition in a BSCCO/Ag tape.
Wagner et al.\ (1995) see a vortex-liquid  vortex-glass transition
in BSCCO films.
Ando et al.\ (1992, 1993) in narrow YBCO strips observe a size
effect on the vortex-glass transition, see also the observation of
collective pinning and fluctuation effects of strongly pinned
flux lines in ultrathin NbTi wires by Ling et al.\ (1995).
   Metlushko et al.\ (1994a) identify different pinning regimes in
BSCCO single crystals, and Moshchalkov et al.\ (1994a) see a
crossover from point pinning to columnar pinning when BSCCO
crystals are irradiated;  these data may also be explained
by the crossover of two $B(T)$ lines corresponding, respectively, to
FLL melting or depinning and to the vanishing of the edge-shape
barrier (section 8.5.) as suggested by Indenbom et al.\ (1994a, b).
Kim, Li and Raffy (1995) find two different scaling behaviours of the
critical current in thin BSCCO films below and above $T_0\approx 25$ K.
A nice comprehensive study of the $E$-$J$-$B$ surface of HTSC's was
performed by Caplin et al.\ (1994) and Cohen et al.\ (1994).

   A different method to check the vortex-glass scaling predictions
is to measure the linear ac response. This has been achieved over
the remarkable frequency range from 3 mHz to 50 MHz (over ten decades)
on the {\it same} specimen by K\"otzler et al.\ (1994a, b), see also
Olsson et al.\ (1991),  Reed et al.\ (1994, 1995),
Wu, Ong, and Li (1993), and Ando et al.\ (1994).
The complex resistivity of a circular YBCO film was obtained
by measuring the ac susceptibility and then applying an inversion
scheme based on the integral equation for the sheet current in
circular disks (Brandt 1993, 1994c, see also section 8.6.),
 which yields the magnetic susceptibility $\chi(\omega)$ as a function
of the complex resistivity $\rho(\omega)$ (section 9.5.).
This contact-free method allows a very precise
determination of both the modulus and the phase of $\rho(\omega)$.
Within the glass model, the magnitude and phase angle of the linear
complex conductivity $\sigma(\omega) = 1/\rho(\omega)$ are predicted
to have the following forms, characteristic for a second-order
physe transition (Fisher, Fisher, and Huse 1991, Dorsey 1991)
\begin{equation}     
 | \sigma(\omega) | = (\tau_g/\xi_g) S_{\pm} (\omega \tau)\,, ~~~~
 \arctan (\sigma''/\sigma') = P_{\pm} (\omega \tau)\,,
\end{equation}
where $S_{\pm} (x)$ and $P_{\pm} (x)$ are homogeneous
scaling functions above
and below $T_g$, which describe the change from pure Ohmic behaviour
$(\sigma ''=0)$ for $T \gg T_g$ to pure screening ($\sigma '=0$)
at $T \ll T_g$. While their limiting behaviours are known, the
detailed shapes of these scaling functions have not yet been worked
out. With their contact-free method K\"otzler et al.\ (1994b) found
that in YBCO films the phase angle of $\sigma(\omega)$  monotonically
increases with $\omega$ at temperatures $T>T_g$ and decreases at
$T<T_g$, but at $T=T_g$ the {\it phase remains frequency-independent
over ten decades in $\,\omega$  at a non-trivial value} $(1-1/z)\pi/2$
as predicted by Dorsey (1991) and Dorsey et al.\ (1992),
 with $z=5.5 \pm 0.5$, {\bf figure} 24. This constancy of the phase
angle follows from (9.5) and (9.8) by applying the Kramers-Kronig
relation between magnitude and phase angle of the linear
conductivity $\sigma(\omega)$. Physically it means
 that at $T_g$ there coexist islands of ideally pinned
``glassy'' FLL and of mobile FLL. For these data both
 the modulus and the phase of $\sigma(\omega)$ {\it scale}
perfectly over more than 14 decades  in the scaled frequency
 $\omega\tau_g$ if $\nu = 1.7 \pm 0.1$ is used, {\bf figure} 25.
         \\[0.4 cm]
{\it 9.6.~ Linear ac response}
         \\*[0.2 cm]
{\it 9.6.1. Thermally activated flux flow.~}
Within the TAFF model (9.4a) the linear complex ac resistivity
$\rho(\omega)$  and ac susceptibility $\chi(\omega)$ of the FLL
are obtained
as follows. Consider a superconducting half space in which a
constant external field generates a uniform FLL with induction $B$.
If the flux lines are rigidly pinned, an additional applied small
ac magnetic field $H_a(t)=H_0 \exp(i\omega t)$ will
penetrate exponentially to the London depth $\lambda$. If the
flux lines are pinned elastically with  force constant $\alpha_L$
(Labusch parameter) such that a small uniform displacement
$u$ causes a restoring force density $-\alpha_L u$, then $H_a(t)$
penetrates exponentially to the Campbell penetration depth
$\lambda_{\rm C} = (B^2/\mu_0\alpha_L)^{1/2}$ (Campbell 1971,
Campbell and Evetts 1972). More precisely, the penetration depth
is now $(\lambda^2 +\lambda_{\rm C}^2)^{1/2}$ (Brandt 1991d).
If the flux lines are parallel to the planar surface
this penetration proceeds by compressional waves in the FLL;
if they are perpendicular to the surface the penetration occurs
by tilt waves. In both cases the same penetration depth
results since the moduli for long-wavelength compression and
tilt are approximately equal, $c_{11}\approx c_{44} =B^2/\mu_0$
(4.6). The viscous drag force density on the FLL,
$-\eta \,\partial u/\partial t$,  may be accounted for by replacing
$\alpha_L$ by $\alpha_L + i\omega\eta$. This means that viscous
damping dominates at frequencies $\omega \gg \tau_0^{-1}$
where $\tau_0 =\eta/\alpha_L$.  At lower frequencies the
ac response of the superconductor with elastically pinned flux lines
is that of the Meissner state with increased penetration depth.

At finite temperatures one has to account for thermal depinning.
This was achieved in three different ways, which all yield
essentially the same result: A model of periodic pins was
considered by Coffey and Clem (1991a); a relaxing Labusch parameter
was introduced by Brandt (1991d); a visco-elastic force-balance
equation was solved following a suggestion of A.\ M.\ Campbell
(Brandt 1992a). If the FLL is displaced homogeneously and then
stopped, the elastic restoring force {\it relaxes} due to
thermally activated (or tunnelling-caused) depinning. This means
the Labusch parameter decreases with time. It ``seems natural''
to assume an exponential decay,
$\alpha_L(t) =\alpha_L(0)\exp(-t/\tau)$. This assumption
in fact reproduces the results obtained by the two other methods,
but not all experimental results confirm this TAFF picture, see
 section 9.5.\ (K\"otzler et al.\ 1994b) and
 section 9.6.3.\ (Behr et al.\ 1992), and also
Koshelev and Vinokur (1991) and van der Beek et al.\ (1993).
The relaxation time may be expressed as
$\tau=B^2/(\rho_{\rm TAFF} \alpha_L) \approx
\tau_0 \exp(-U/k_BT)$ with $\rho_{\rm TAFF}=\rho_{\rm FF}
\exp(U/k_BT)$  the TAFF resistivity from (9.4a)
and $\tau_0 = \eta/\alpha_L = B^2/(\rho_{\rm FF} \alpha_L)$ with
$\rho_{\rm FF}=B^2/\eta$ (5.6) the flux-flow resistivity.
Formally this relaxation replaces $\alpha_L$
by $\alpha_L i\omega/(1+i\omega \tau)$.
The resulting  complex ac penetration depth
$\lambda_{ac}(\omega)$ is thus
   \begin{equation}   
  \lambda_{ac}^2(\omega) = \lambda^2 + \frac{B^2}{\mu_0} \big(
  \frac{\alpha_L} {1-i/\omega\tau} + i\omega \eta \big)^{-1}
   \approx \lambda^2 + \lambda_{\rm C}^2~ \frac{1-i/\omega\tau}
   {1+i\omega\tau_0}.
   \end{equation}
In the second version of (9.9) $\tau = \tau_0\exp(U/kT) \gg \tau_0$
was assumed. From $\lambda_{ac}$ the complex surface
impedance is obtained as $Z_s = i\omega \mu_0\lambda_{ac}(\omega)$.
Notice that the result (9.9) does not specify the physical nature of
the creep and thus in principle applies also to quantum creep where
$\tau = \tau_0 \exp(S_E /\hbar)$ with $S_E$ an Euklidian action,
 section 9.9.

  From the complex ac penetration depth (9.9) and the Maxwell
equations one finds that for the half space the electric field
$E(x)$ has the same spatial dependence $\sim \exp(-x/\lambda_{ac})$
as the current density $J(x)$. Thus the ratio
$E(x)/J(x) = \rho_{ac}$ is spatially constant. This means that the
usual {\it local} relation $E = \rho_{ac} J$ applies although
the penetration of $J$ and $E$ is mediated by coherent flux lines,
see discussion in section 6.4. With (9.9) the complex ac resistivity
$\rho_{ac}(\omega) = E/J = i\omega\mu_0\lambda_{ac}^2$ becomes
    \begin{equation}  
  \rho_{ac}(\omega) \approx i\omega\mu_0\lambda^2 \,+\rho_{\rm TAFF} \cdot
  (1+i\omega\tau)/ (1 +i \omega \tau_0) .
    \end{equation}
For {\it low and high} frequencies, this resistivity is
{\it real} and independent of $\omega$; the flux motion in
this case is {\it diffusive} with diffusivity $D= \rho_{ac}/\mu_0$.
The ac penetration depth is then the usual {\it skin depth}
$\delta =(2D/\omega)^{1/2}$ as in normal conducting metals.
For {\it intermediate} frequencies, $\rho_{ac}$ is imaginary and
proportional to $\omega$; the surface currents are then nearly
loss free due to strong elastic pinning, and the screening
in this case is like in the Meissner state but with a
larger  penetration depth $(\lambda^2 + \lambda_C^2)^{1/2}$.

  From $\lambda_{ac}$ (9.9) or $\rho_{ac}$ (9.10) one in principle
obtains the linear complex susceptibility of superconductors of
arbitrary shape. For example, for a slab with $|x| \le a$ and for
a cylinder with radius $a$ one has the penetrating parallel magnetic
ac fields
    \begin{equation}  
  H_1(|x| \le a) =  H_0 \exp(i\omega t)\,
  \frac{\cosh(x/\lambda_{ac})}{\cosh(a/\lambda_{ac})}\,, ~~~~
   H_1(r\le a) =  H_0 \exp(i\omega t)\,
  \frac{I_0(r/\lambda_{ac})}{I_0(a/\lambda_{ac})}
   \end{equation}
where $I_0(u)$ is a modified Bessel function. Averaging $H_1$
over the specimen cross section one gets the linear complex  ac
susceptibility $\mu(\omega) = 1 +\chi(\omega)
 = \langle H_1\rangle / H_0 $ for slabs and cylinders
(Clem et al.\ 1976) in a parallel ac field,
    \begin{equation}  
  \mu_{\rm slab}(\omega)=  \frac{\tanh(u)}{u}, ~~~~
  \mu_{\rm cyl} (\omega)=  \frac{2\, {I}_1(u)}{u\,{I}_0(u) }
    ~~~\,{\rm with}~~u=\frac{a}{\lambda_{ac}}
    \end{equation}
and $I_1(u) = {\rm d}I_0/{\rm d}u$.
In the Ohmic (diffusive) cases at low and large frequencies,
the known skin-effect behaviour results, with a complex argument
    $u = (1+i)(\omega\, d^2/8D)^{1/2} $ containing the
flux diffusivities $D= \rho_{\rm TAFF}/\mu_0~~ =
\lambda_{\rm C}^2 /\tau$ for $\omega \ll 1/\tau$, and
$D=\rho_{\rm FF}/\mu_0=\lambda_{\rm C}^2 /\tau_0$ for
$1/\tau_0 \ll \omega$. Explicitly one then gets for the Ohmic slab
(Kes et al.\ 1989, Geshkenbein et al.\ 1991, Brandt 1991e, 1992b)
   \begin{equation}  
 \mu(\omega) = \mu'- i\mu'' =\frac{(\sinh v +\sin v)
         -i(\sinh v -\sin v)} {v\,(\cosh v + \cos v)}
   \end{equation}
with     $v = (\omega d^2/ 2D)^{1/2} = d/\delta$, {\bf figure} 26.
This gives  $\mu(\omega) = 1$ for $\omega =0$ and $\mu(\omega) \approx
(1-i)/v$  for $|v| \gg 1$. The dissipative part $\mu''(\omega)$
has a maximum $\mu''_{\rm max} = 0.41723$ at $v=2.2542$
corresponding to $\omega \tau_d = 1.030 \approx 1$ with $\tau_d =
d^2/\pi^2 D$, or to $d/2\delta =1.127 \approx 1$ with
$\delta = (2D/\omega)^{1/2}$ the skin depth.

{\it 9.6.2. General susceptibilities.~}
The linear susceptibilities of rectangular bars in a parallel ac field,
and of thin strips and disks in perpendicular ac field for arbitrary
complex resistivity $\rho_{ac}$ are given by Brandt (1994b, c)
({\bf figure} 27), see also Dorsey (1995). In general, these linear
 permeabilities $\mu$ and susceptibilities $\chi=\mu-1$
may be expressed as infinite sums over poles of first order in the
complex $\omega$ plane (Brandt 1994c),
  \begin{equation}   
  \mu(\omega) = \! \sum_n \frac{c_n}{\Lambda_n + w}, ~~~
  \chi(\omega) = -w\! \sum_n \frac{c_n/\Lambda_n}{\Lambda_n + w} \,.
  \end{equation}
  The variable $w$ in (9.14) is proportional to
$i\omega/\rho_{ac}(\omega)$; for a slab or cylinder in
longitudinal ac field one has
$w=u^2 =(a/\lambda_{ac})^2 = i\omega a^2 \mu_0/\rho_{ac}$,
and for the slab and disk in perpendicular field,
$w= ad/(2\pi \lambda_{ac}^2) = i\omega a d\mu_0/(2\pi \rho_{ac})$.
The real and positive constants $\Lambda_n$ and $c_n$
($n=0 \dots \infty$) follow from an eigenvalue problem, namely,
a differential or integral equation describing local or nonlocal
diffusion of longitudinal flux or of sheet current in parallel
or perpendicular geometry, cf.\ equations (8.4) and (8.18). The pole
positions $\Lambda_n$ are the eigenvalues and the amplitudes $c_n$
are squares of integrals over the eigenfunctions (matrix elements).
 For example, for the strip this eigenvalue problem reads
    \setcounter{equation}{13} {\renewcommand{\theequation}{9.14a}
  \begin{equation}   
  f_n(x) = -\Lambda_n \int_0^1 \! \ln \Big| \frac{x-x'}{x+x'}\Big|
           \, f_n(x') \, {\rm d}x'  \,,
 \end{equation} }
yielding $\Lambda_0= 0.63852$, $\Lambda_n \approx \Lambda_0 +n$,
and $c_n =$ 0.5092, 0.159, 0.091, 0.062, \dots, obtained from
$c_n= 4\Lambda_n^2 b_n^2$ where
$b_n = \int_0^1  f_n(x)\, x\, {\rm d}x$. In particular, for an
Ohmic strip with length $2a$, thickness $d$, and resistivity $\rho$,
the lowest eigenvalue $\Lambda_0$ determines the relaxation time
$\tau_0 =ad\mu_0 /(2\pi \rho \Lambda_0)$ with which the sheet
current decays exponentially after a jump in the applied magnetic
field and which yields the position of the maximum in the
dissipative part of the ac susceptibility
$\chi(\omega) = \chi'-i\chi''$, $\chi''_{\rm max} =
\mu''_{\rm max}= 0.4488$ at $\omega \tau_0 = 1.108 \approx 1$.

For practical purposes the infinite sums (9.14) may be approximated
by their first few terms. Tables of the $c_n$ and $\Lambda_n$
for the slab, cylindre, strip, and disk are given by
Brandt (1994c); the coefficients for the square and for rectangles
will be published. By inverting the formulae (9.14),
K\"otzler et al.\ (1994b) extracted the linear complex ac conductivity
of YBCO and BSCCO disks in parallel and perpendicular dc field
from ac susceptibility measurements; they confirmed the
predictions of vortex-glass scaling (section 9.5.) in three of
 these geometries, but for BSCCO with the dc field along the
$c$-axis this scaling did not work.

Skvortsov and Geshkenbein (1994) obtain the linear susceptibility
of anisotropic ellipsoids. The linear screening by an
infinite type-II superconducting film between two ac coils was
calculated by Hebard and Fiory (1982), Fiory et al.\ (1988),
Jeanneret et al.\ (1989), Clem and Coffey (1992), Pippard (1994),
and Klupsch (1995).  Nonlinear screening is considered by
Nurgaliev (1995) and by Gilchrist and Konczykowski (1990, 1993),
who showed that screening of coils by a superconductor can be
described by a series of inductive loops with given mutual- and
self-inductances and with nonlinear resistivity; see
Gilchrist and van der Beek (1994) for applications
and F\`abrega et al.\ (1994a) for experiments.

{\it 9.6.3. Algebraic relaxation.~}
In ac experiments on ceramic BSCCO by Behr et al.\ (1992)
in the large frequency range 2\,Hz $\leq \omega/2\pi \leq 2\,$MHz
both the real and imaginary parts of the susceptibility
$\mu(\omega)$  (9.12) for cylinders below
$T=T_l \approx 70\,$K could be fitted very well by replacing the
above exponential decay of the Labusch parameter,
$\alpha_L(t) = \alpha_L \exp(-t/\tau)$ yielding
$ \alpha_L(\omega) = \alpha_L i\omega \tau /(1 +i\omega\tau)$,
by an algebraic decay of the form
$\alpha_L(t) = \alpha_L /(1+ t/\tau)^{\beta}$
yielding for $\omega\tau \ll 1$
 \begin{equation}  
\alpha_L(\omega) =\alpha_L (i\omega \tau)^\beta\,\Gamma(1-\beta)\,,~~~
\lambda_{ac}^2 = \lambda_C^2 /[\, (i\omega \tau)^\beta\,
                             \Gamma(1-\beta)\, ]
    \end{equation}
where $\Gamma(1-\beta)$ is Euler's gamma function.
Interestingly, in these experiments the effect of temperature on
the dynamics is entirely
determined by that of the exponent $\beta(T) \approx 1/(1+U/k_BT)$,
see equation (9.16) below, while
$\tau \approx 4\cdot 10^{-12}\,$s (for $B=1\,$T)
is nearly independent of $T$. Moreover,
$\beta(T)$ overlaps with exponents determined from the algebraic
decay of the time dependent magnetization of this sample at the
same field (Spirgatis et al.\ 1993).  Behr et al.\ (1992) find a
depinning limit ($\beta =1$) for
$T_l \approx 70\,$K $\leq T \leq T_c\approx 90\,$K and a glass
transition with  $\beta \approx 0.07$ at $T_g\approx 24\,$K.

{\it 9.6.4. Further work.~}
Further, more microscopic theories of the linear ac susceptibility
of the FLL in layered superconductors are given by Koshelev and
Vinokur (1991), Ivanchenko (1993), and Artemenko et al.\ (1994),
see also Hsu (1993), Tachiki et al.\ (1994), and
Bulaevskii et al.\ (1995a) for the low-frequency magneto-optical
properties.
Sonin and Traito (1994) consider the surface impedance of the
arbitrarily inclined FLL taking into account also a possible surface
barrier (section 8.5.) and the thin surface layer over which the
flux lines bend in order to hit the surface at a right angle
(Hocquet et al.\ 1992, Mathieu et al.\ 1993,
Pla\c{c}ais et al.\  1993, 1994, Brandt 1993a).
Parks et al.\ (1995) measure vortex dynamics in the Terahertz
domain by phase-sensitive spectroscopy and find disagreement with
the usual flux-flow picture.
  Anomalous vortex mobility vanishing logarithmically in the limit
of small frequencies in an ideal 2D superconductor made from a
triangular array of Josephson junctions, was observed by
Th\'eron et al.\ (1993, 1994) and is explained by Beck (1994) and
Korshunov (1994), see also the MC simulations by Yu and Stroud (1994).

A discussion of the linear and nonlinear ac response of HTSC's is
given by van der Beek et al.\ (1993), see also
Matsushita et al.\ (1991) and the discussion below equation (8.13).
Note that a nonlinear frequency-dependent (and time-independent)
resistivity $\rho = E/J$ cannot be defined since in a nonlinear
conductor the time dependences of $E$ and $J$ in general are
different and do not cancel.

Numerous research groups investigate pinning and flux flow by
microwave absorption, see  e.g.\  Blazey et al.\ (1988),
Portis et al.\ (1988), Shridar et al.\ (1989),
Giura et al.\ (1992), Kessler et al.\ (1994), Gough and Exon (1994),
Blackstead et al.\ (1994), Lofland (1995), Golosovsky (1991, 1992,
1993, 1994, 1995), and the reviews by Piel and M\"uller (1991),
 Portis (1992), and Dumas et al.\ (1995).
Unexpected sharp magnetoabsorption resonances with cyclotronic and
anticyclotronic polarisation were observed in BSCCO at 30 and 47 MHz
in fields up to 7 T by Tsui et al.\ (1994)
and are explained by Kopnin et al.\ (1995).
Several groups perform microwave precision measurements of the
penetration depth $\lambda(T)$ in the absence of flux lines in order
to check microscopic theories of superconductivity (e.g., s- or d-wave
pairing), see e.g.\  Anlage and Wu (1992), Hardy et al.\ (1993),
Wu et al.\ (1993), Klein et al.\ (1993d),
Lee et al.\ (1994), Kamal et al.\ (1994), and Xu et al.\ (1995).
By mutual inductance of coaxial coils $\lambda(T)$ was measured, e.g.,
by Fiory et al.\ (1988), Jeanneret et al.\ (1989), Ulm et al.\ (1995),
see also section 9.6.2.
  However, caution has to be taken in interpreting the measured
$\lambda(T)$ as indicative of s-wave or d-wave pairing since
$\lambda(T)$ may be influenced by the presence of vortices,
by fluctuations of the phase $\phi$ of the order parameter
(Coffey 1994b, Roddick and Stroud 1995), by remnants of a
Kosterlitz-Thouless transition (Gingras 1995), or by the layered
structure (Klemm and Liu 1995, Adrian et al.\ 1995).

Careful measurements of $\lambda_{ab}(T)$ and $\lambda_c(T)$
in a suspension of aligned ultrafine YBCO particles smaller than
 $\lambda_{ab}$ were performed by Neminsky and Nikolaev (1993),
who found a power-law of the form $\lambda_{ab}(T)/\lambda_{ab}(0)
\approx \lambda_c(T)/\lambda_c(0) \approx 1 + 0.34 (T/T_c)^{2.5}$,
which they interpret as due to a $T$-dependent inelastic
 scattering length $l_{in}(T)$ inserted in
 $\lambda(T) / \lambda(0) = 1 + \xi_0 /2 l_{in}$.
Buzdin, Neminsky et al.\ (1994) calculate and measure the
quasi-2D fluctuation contribution to the anisotropic penetration
depth in YBCO and find excellent agreement.
%
         \\[0.4 cm]
{\it 9.7.~ Flux creep and activation energy}
         \\*[0.2 cm]
  The decay of screening currents, or of the magnetization,
in principle is completely determined by the geometry and
by the resistivity $\rho(T,B,J)$. In superconducting rings
and hollow cylinders  in axially oriented magnetic
field one simply has $\dot{M} \propto \dot{J} \propto
\rho(T,B,J)$. Within the Kim-Anderson model (9.3) the
decay of persistent currents in this geometry was calculated
analytically for all times (Zhukov 1992, Schnack et al.\ 1992).
Twelve decades of the electric field $E=10^{-13}$ to $10^{-1}\,$V/m
were measured by Sandvold and Rossel (1992) combining
current--voltage curves with highly sensitive measurements of
decaying currents in rings of YBCO films of 3 mm diameter,
 0.1 mm width, and 200 nm thickness.
In a nice work Kunchur et al.\ (1990) show that by the superposition
of decaying flux distributions, usual flux creep can lead to a
{\it memory effect}, which had been observed by Rossel et al.\ (1989)
but then was interpreted as glass-like behaviour.
Dislocation-mediated (``plastic'') flux creep is considered by
van der Beek and Kes (1991a), and  Mee at al.\ (1991) investigate
creep in a granular HTSC with a distribution of coupling strengths.

The current dependent activation energy $U(J)$ may be extracted
from creep experiments by the method of Maley et al.\ (1990),
{\bf figure} 28, see e.g.\ van der Beek et al.\ (1992a),
Kung et al.\ (1993), Sengupta et al.\ (1993), and Rui et al.\ (1995)
(for a Hg-based HTSC). The Kim-Anderson model (9.3) originally
means an effective activation energy $\propto J_c - J$,
cf.\ equation (9.4b), and actually corresponds to a zig-zag shaped
pinning potential, which has a barrier height proportional to the
current-caused tilt of the total potential. As shown by Beasley
et al.\ (1969), a more realistic
rounded potential yields $U \propto (J_c - J)^{3/2}$.
Collective creep theory yields $U \propto 1/J^\alpha$
with $\alpha >0$ depending on $B$ and on the pinning strength.
For single-vortex pinning one predicts $\alpha = 1/7$;
$\alpha$ crosses over from $5/2$ to 1 to $7/9$ as the vortex bundles
increase from small ($a < R_\perp$, $R_\| \approx L_c <\lambda$)
to intermediate ($a < R_\perp <\lambda <R_\| \approx L_c$)
to large ($\lambda < R_\perp, R_\| \approx L_c$) size.  Here
$R_\perp$, $R_\|$, and $L_c$ denote the transverse or parallel
(to the flux motion) and $L_c$ the longitudinal (to the field)
dimensions of the bundle (section 8.7.), and $a$ is the
flux-line spacing (Feigel'man et al.\ 1989, Nattermann 1990,
Fischer and Nattermann 1991, Blatter et al.\ 1994a).
For the more general (uniaxial) anisotropy situation, the result is
more complicated and may be found in Blatter et al.\ (1994a).
   These various exponents $\alpha$ are nicely seen in proton
irradiated YBCO by Thompson et al.\ (1994).
Dekker et al.\ (1992a) find $\alpha =0.19$ at low, and
$\alpha = 0.94$ at high current densities.
  Many experiments suggest a logarithmic dependence $U \sim \ln(J_2/J)$
(Ries et al.\ 1993), corresponding to the limit $\alpha \to 0$
or to a logarithmic interaction between flux lines
(Zeldov et al.\ 1990) or point vortices (Zhukov et al.\ 1994). For
$U \sim \ln(J_2/J)$  the creep rate can be calculated analytically
(Vinokur et al.\ 1991a) and a power-law current--voltage curve
$E\sim J^n$ holds which leads to the nonlinear diffusion equations
(8.4a) (Rhyner 1993, Gilchrist and van der Beek 1994)
 and to self-organized criticality.

   Combining the collective creep result
$U(J) = U_c (J_c/J)^\alpha$ for $J \ll J_c$
with the Kim-Anderson formula $U(J) = U_c (1-J/J_c)$
for $J \approx J_c$ one obtains for the decaying current density
in rings and cylinders the interpolation formula
  \begin{equation}  
J(t) \approx J_c \left[ 1+ \alpha \frac{k_B T}{U_c}
 \ln \frac{t}{t_0} \right]^{-1/\alpha} ~{\rm or}~~
J(t) \approx J_c (t/t_0)^{-k_B T/U_c}
      ~~~\mbox{~for~$\alpha \ll 1$~~}
  \end{equation}
where $t_0$ is an integration constant. Formula (9.16) is confirmed
nicely, e.g.\ with YBCO by Thompson et al.\ (1991). From (9.16) one
obtains a temperature dependent $J(T)$ and the normalized creep rate
 $S$ (Malozemoff and Fisher 1990, Feigel'man et al.\ 1991)
  \begin{equation}  
J(T) \approx J_c \exp(-T/T_0) \,,~~~~T_0 \approx U_c/[k_B \ln(t/t_0)]
  \end{equation} \begin{equation}
S = -\frac{d\,\ln J}{d\,\ln t} \approx \frac{k_B T} {U_c +
   \alpha k_B T \ln(1 +t/t_0) } \,.
  \end{equation}
Experiments on HTSC's by Senoussi et al.\ (1988) yield
$T_0 \approx 10\,$K and $U_c/k_B \approx$ 100 to 1000\,K. Note that
the decay rate $S$ (9.18) decreases with increasing time and
saturates at $k_B T > U_c/[\alpha \ln(t/t_0)]$  to a value
$1/[\alpha \ln(t/t_0)]$.
Numerical solutions to the relaxation rates for various
dependencies $U(J)$ are given by Schnack et al.\ (1992).
Caplin, Perkins and Cohen (1995) show that the often measured
value $S=0.03$ reflects the experimental conditions rather than
being a characterisitic feature of the underlying vortex dynamics.
McElfresh et al.\ (1995) measure and explain the influence of a
transport current on flux creep in YBCO films.

  Based on the concept of self-organized criticality or avalanche
dynamics (Bak et al.\ 1988, Tang and Bak 1988, Tang 1993,
Ling et al.\ 1991, Field et al.\ 1995), Wang and Shi (1993)
derived an expression
$J(t) = J_c -J_c [(k_BT/U_0) \ln(1+\beta t/\tau)]^{1/\beta}$
which is similar to (9.16) and for $\beta =1$ coincides with the
Kim-Anderson result, see also the measurements of creep by
micro Hall probes by Koziol et al.\ (1993).
Good fits to relaxation rates are also achieved by fitting
a {\it spectrum} of activation energies (Hagen and Griessen 1989,
Griessen 1990, Gurevich 1990, Niel and Evetts 1991, Martin and
Hebard 1991, Theuss et al.\ 1992, Theuss 1993). The success of
theories which assume only {\it one} activation energy $U$
suggests that at given  values of $B$, $J$, and $T$
essentially one effective $U$ of an entire spectrum
determines the physical process under consideration.
%
         \\[0.4 cm]
{\it 9.8.~ Universality of flux creep}
         \\*[0.2 cm]
   Gurevich and K\"upfer (1993) recently pointed out a new
simplicity of the flux-creep problem: When
the applied field $H_a(t)$ is increased or decreased with constant
ramp rate  $\dot H_a$ and then held constant for times $t\ge 0$,
then after some transition time the electric field becomes
{\it universal} taking the form ${\bf E(r},t)= {\bf F(r}) /(t+\tau)$
with ${\bf F(r)}$ a space-dependent function and
$\tau \propto 1/|\dot H_a|$. This universal behaviour holds if the
current--voltage law $E(J)$ is sufficiently nonlinear, namely,
if one has $\partial E/\partial J \approx E/J_1$ with a constant
$J_1$. This condition is exact if $E(J) = E_0 \exp(J/J_1)$
[Anderson's creep (9.4b)] and it holds to a good approximation
for a wide class of  $E(J)$ curves, e.g., for $E\propto J^n$
with $n\gg 1$ [it was checked numerically that practically the same
function ${\bf F(r)}$ results for all $n\ge 3$ (Gurevich and Brandt
1994)] or for the vortex-glass and collective creep models which yield
 \begin{equation}  
 E(J) =E_c \exp[\, U(J)/k_BT\,], ~~~~U(J)=[\,(J_c/J)^\alpha -1\,]U_0,
 \end{equation}  \begin{equation}  
   \frac{\partial E}{\partial J} = \frac{U_0 \alpha E}{k_BT J_c}
   \Big[ 1 + \frac{k_BT}{U_0} \ln\frac{E_c}{E} \Big]
    ^{1+1/\alpha} \approx \frac{E}{J_1}\,, ~~~~
   J_1 = \frac{k_BT J_c}{\alpha U_0} \,.
 \end{equation}
After a sufficiently long ramp period the superconductor has reached
the critical state corresponding to its $E(J)$ and to the geometry;
in this saturated state ${\bf E(r)}$ is a {\it linear function}
of ${\bf r}$ since ${\bf \nabla \times E} = -\mu_0 {\bf \dot H}_a$,
see also Zhukov et al.\ (1994).
When ramping is stopped at $t=0$, then the linear ${\bf E(r)}$
starts to curve and to decrease until, after a transient period of
duration $\approx \tau$ (Gurevich et al.\ 1991, Gurevich 1995b),
the universal behaviour $E\propto 1/t$ is reached. In particular,
for a slab of width $2a$ ($|x| \le a$) in {\it parallel} field
Gurevich and K\"upfer (1993) obtain
 \begin{equation}  
  E(x,t) = \frac{\mu_0 J_1}{t+\tau} \, x\,(a - |x|/2)\,, ~~~
  \tau \approx J_1 a/|\dot H_a| \,.
 \end{equation}
This universality applies also to transverse geometry, {\bf figure} 29;
for a strip of width $2a$ ($|x| \le a$) and thickness $d\ll a$ in a
{\it perpendicular} field Gurevich and Brandt (1994) obtain
 \begin{equation}  
  E(x,t) = \frac{\mu_0 J_1 ad}{2\pi (t+\tau)}
                  f\Big( \frac{x}{a} \Big)\,, ~~
  f(u) = (1+u)\ln(1+u) -(1-u)\ln(1-u) -2u\ln|u|
 \end{equation}
and a similar expression for the circular disk.

  The current density ${\bf J(r},t)$, magnetic field ${\bf H(r},t)$,
and magnetization $M(t)$ during creep are obtained by inserting
the universal ${\bf E(r},t)$ into the specific relation $E(J)$, e.g.,
$J(x,t)\propto E(x,t)^{1/n}$  for $E\propto J^n$ or
$J(x,t)= J_c -J_1 \ln [\, E_c/E(x,t)\, ]$ for the vortex glass and
collective creep models (9.20). The latter yield in both geometries
 \begin{eqnarray}  
 M(t) = M(0) -M_1 \ln(1 +t/\tau_0) & &  ~~~(0< t\le \tau_0) \\
 M(t) = M_c - M_1 \ln (t/t_0)      & &  ~~~( t \gg \tau_0 )
 \end{eqnarray}
  where for parallel geometry $\tau_0 =0.66 J_1 a/|\dot H_a|$,
$M_1 =a^2 J_1$,  $t_0 = 8a^2 \mu_0 J_1 /e^3 E_c$,
and for perpendicular geometry $\tau_0 =0.41 J_1 d/|\dot H_a|$,
$M_1 =a^2 J_1$,  $t_0 = a d \mu_0 J_1 /4 E_c$.

In a nice work
Cai et al.\ (1994) applied these results to extract the unrelaxed
irreversible magnetization from combined magnetization and flux-creep
measurements, finding that the anomalous ``fish tail'' or
``peak effect'' observed by many groups in the critical current
 density and thus in the irreversible magnetization and torque,
is a static rather than a dynamic effect; this is debated by other
authors, see, e.g., Schnack et al.\ (1992), Yeshurun et al.\ (1994),
van Dalen et al.\ (1995) and Jirsa et al.\ (1995).
A method to extract the ``real''
critical current (not influenced by creep) is developed by
Schnack et al.\ (1993) and Wen et al.\ (1995).
The influence of the viscous motion of flux lines on the magnetic
hysteresis is considered by Tak\'acs and G\"om\"ory (1990)
and by P\accent23ust (1992). As discussed above, creep does
{\it always} occur in superconductors which are not in magnetic
equilibrium, in particular during variation of the applied field;
this fact is described in a unified picture by treating the
superconductor as a {\it nonlinear conductor}, cf.\ equations (8.4)
and (8.18). Therefore, in HTSC's {\it the measured $J_c$ will always
depend on the time and length scales of the experiment} as confirmed
nicely by the ac experiments of F\`abrega et al.\ (1994a).
   For more work on the ``fish-tail effect'' see
Ling and Budnick (1991), Kobayashi et al.\ (1993),
Klupsch et al.\ (1993), Gordeev et al.\ (1994),
Solovjev et al.\ (1994),
Perkins et al.\ (1995), and D'Anna et al.\ (1995).
 A careful analysis of the fish-tail effect in YBCO and TmBCO with
various oxygen content was recently performed by
Zhukov et al.\ (1995). An explanation of the peak effect in terms of
plastic flow of the FLL is given by Bhattacharya and Higgins (1994).

  An interesting difference between  flux creep in these two
geometries is that the universal $E(x)$ is {\it monotonic}
in parallel geometry  but has a {\it maximum} in perpendicular
geometry at $x=a/\sqrt2$ for the strip and $x= 0.652$ for the
disk. In perpendicular geometry after the increase of $H_a(t)$ is
stopped, the creeping magnetic field $H({\bf r},t)$ {\it increases}
near the centre as expected, but it {\it decreases} near the edges
because of demagnetizing effects. The regions of increasing and
decreasing $H$ are separated by a ``neutral line'' at which
$H({\bf r},t) = H_a$ is constant during creep. For the strip
two neutral lines run at $|x| = a/\sqrt2$ where $E(x)$ is
maximum. For the disk a neutral circle appears at $r=0.876 a$
where $rE(r)$ is maximum. For squares and rectangles the neutral
line, defined by $H_z(x,y) = H_a$ in equation (8.14), has a
star-like shape ({\bf figure} 18, 19) which was observed by
magneto-optics by Schuster et al.\ (1995a). For sample shapes differing
from a circular disk or infinite strip, flux motion is a genuine
2D problem and the creep-solution for the electric field
${\bf E}(x,y,t)$ does not factorise into time and space functions.
As a consequence, the neutral line slightly changes its shape
during a long creep time, but probably this effect is difficult
to observe.

  A universality of different type is discussed by
Geshkenbein et al.\ (1991). In the critical state, i.e.\ during
flux creep, the response to a small perturbation $\delta B(x,t) \ll B$
is described by a diffusion equation in which the time is replaced
by the logarithm of the time, see also section 8.3.
 As a consequence, in the case of
complete penetration the relaxation is universal and does not depend
on the magnetic field. The absorption of a small ac field $\delta B
\sim \cos(\omega t)$ is then proportional to the square root of the
waiting time (creep time) $t$,\, $\mu''(\omega) \sim \sqrt{\omega t}$.
This universality of the linear response during flux creep applies
also to perpendicular geometry (Gurevich and Brandt 1995).
         \\[0.4 cm]
{\it 9.9.~ Depinning by tunnelling}
         \\*[0.2 cm]
In numerous experiments the creep rate plotted versus
$T$ appears to tend to a finite value at zero temperature.
Flux creep observed at low temperatures by Lensink et al.\ (1989),
Mota et al.\ (1989, 1991), Griessen et al.\ (1990),
 Fruchter et al.\ (1991), Lairson et al.\ (1991), Liu et al.\ (1992),
Prost et al.\ (1993), and Garc\'{\i}a et al.\ (1994), and in
ultrathin Nb wires by Ling et al.\ (1995),
in principle may be explained by usual thermal activation
from smooth shallow pinning wells (Griessen 1991), but more likely it
indicates ``quantum creep'' or ``quantum tunnelling'' of vortices
out of the pins
as suggested first by Glazman and Fogel' (1984a) and Mitin (1987).
The thermal depinning rate $\propto \exp(-U/k_B T)$ is now replaced
by the tunnelling rate $\propto \exp(-S_E/\hbar)$ where $S_E$ is the
Euklidean action of the considered tunnelling process;
see Seidler et al.\ (1995) for a recent measurement of $S_E$.
Rogacki et al.\ (1995) measure the low-temperature ($T\ge 0.1$ K)
creep rate in DyBCO crystals by the vibrating reed technique.

Tunnelling of vortices differs from tunnelling of particles by the
smallness of the inertial mass of the vortex
(Suhl 1965, Kuprianov and Likharev 1975, Hao and Clem 1991b,
Coffey and Clem 1991b, \v{S}im\'anek 1992, Coffey 1994a,
Blatter and Geshkenbein 1993, Blatter et al.\ 1994).  This means that
the {\it vortex motion is overdamped} and there are no oscillations or
resonances. This overdamped tunnelling out of pins is calculated
by Ivlev et al.\ (1991c) and Blatter, Geshkenbein and Vinokur (1991a).
Griessen et al.\ (1991) show that the dissipative quantum tunnelling
theory of Caldeira and Leggett (1981) with the usual vortex
viscosity $\eta$  inserted reproduces the overdamped results.
A very general theory of collective creep and of vortex-mass
dominated quantum creep in anisotropic and layered superconductors
is given by Blatter and Geshkenbein (1993) and a short review
on flux tunnelling by Fruchter et al.\ (1994).
Since their coherence length is so small, HTSC's may belong to
a class of very clean type-II superconductors in which quantum
tunnelling of vortices is governed by the Hall term
(Feigel'man et al.\ 1993b, Garc\'{\i}a et al.\ 1994,
Ao and Thouless 1994).
Macroscopic tunnelling of a vortex out of a SQUID (a small
superconducting loop with two Josephson junctions) was calculated
in the underdamped case (Ivlev and Ovchinnikov 1987) and in the
overdamped limit (Morais-Smith et al.\ 1994).
%
         \\[0.7 cm]
{\bf 10.~ Concluding remarks}
         \setcounter{subsection}{9}  \setcounter{equation}{0}
         \\*[0.5 cm]
This review did not present general phase diagrams for the FLL
in HTSC's since these might suggest a universality which is not
generally valid. Lines in the $B$-$T$ plane which separate
 ``phases'', e.g.,
the ideal FLL or elastically distorted FLL from the plastically
distorted or amorphous or liquid FLL, are not always sharp and
in general depend on their
definition and on the pinning strength and purity of the material,
but also on the size and shape of the specimen and on the time scale
or frequency of the experiment.

For example, an irreversibility line or depinning line may be defined
as the line $T_{\rm dp}(B)$ at which the imaginary (dissipative) part
of the linear ac  permeability $\mu''(\omega)$ (9.13)
of a HTSC in the TAFF regime is maximum for a given frequency
(Kes et al.\ 1989, van der Beek et al.\ 1992a, Seng et al.\ 1992,
Karkut et al.\ 1993). This condition coincides
with the maximum attenuation of a vibrating HTSC. Superconductors
performing tilt vibrations in a magnetic field have been used
extensively to study elastic pinning (Esquinazi 1991,
Brandt et al.\ 1986a,b, Esquinazi et al.\ 1986, Brandt 1987)
and depinning lines (Gammel et al.\ 1988,
Kober et al.\ 1991b, Brandt 1992a, Gupta et al.\ 1993,
Ziese et al.\ 1994a, b, c, Drobinin et al.\ 1994, Shi et al.\ 1994).
   A typically sharp maximum in the attenuation occurs when the ac
(or vibrational) frequency $\omega/2\pi$ coincides with a reciprocal
relaxation time ($\omega \tau_d =1$),
or the skin depths equals  half the specimen thickness
($d/2\delta =1$), or when the thermally activated resistivity
$\rho_{\rm TAFF}(B,T) \approx \rho_{\rm FF} \exp[U(B,T)/k_BT]$
goes through the value $\rho_{\rm dp} = \omega d^2 \mu_0 /\pi^2$. From
these three equivalent conditions it is clear that such a depinning
line in the $B$-$T$ plane {\it depends on the frequency and on the size
of the specimen}, at least logarithmically. In certain geometries the
attenuation should exhibit even two peaks, which may belong to two
different relaxation modes (Brandt 1992b, Ziese et al.\ 1994b),
but also other reasons for the occurrence of two or more peaks are
possible (Dur\'an et al.\ 1991,  Xu et al.\ 1992,
Arrib\'ere et al.\ 1993, Jos\'{e} and Ramirez-Santiago 1993,
D'Anna et al.\ 1992, 1994c, Giapintzakis et al.\ 1994).

  Such peaks, therefore, do not necessarily indicate a melting (or other)
transition in the FLL (Brandt et al.\ 1989, Kleimann et al.\ 1989,
Geshkenbein et al.\ 1991, Steel and Graybeal 1992,
Ziese et al.\ 1994a, b, c). Moreover, an observed ``melting line'' may
also correspond to the vanishing of a geometric edge-shape barrier
(section  8.5.) as shown by Indenbom et al.\ (1994a, b) and
Grover et al.\ (1990).
 In extremely weak pinning YBCO
specimens, however, the observed sharp peak in the attenuation
of slow torsional oscillations probably indicates a  melting
transition in the FLL (Farrell et al.\ 1991, Beck et al.\ 1992),
see the discussion in section 5.6.
Notice also that the glass transition detected in YBCO and BSCCO by
K\"otzler et al.\ (1994b) by measuring the complex ac susceptibility
$\chi$ of a disk over ten decades of the frequency, is not seen
directly in the raw data $\chi(B,T,\omega)$ ({\bf figure} 24) but is
evident in the phase angle of the complex resistivity
$\rho(B,T,\omega)$ evaluated from these data, section 9.5.

Several topical features of the FLL which give further insight into
the properties of the pinned or moving FLL, for brevity could not be
discussed here in detail, e.g., the Hall effect (section 7.5.),
the thermo-effects (section 7.6.), and the flux noise (section 7.7.)
reviewed by Clem (1981b). Another such topical problem
is the enhancement of pinning by an applied
electric field (Mannhardt et al.\ 1991, Mannhardt 1992,
Burlachkov et al.\ 1993, Jenks and Testardi 1993, Lysenko et al.\ 1994,
Frey et al.\ 1995, Ghinovker et al.\ 1995).
Also omitted from this review were the rich properties of magnetic
superconductors (Krey 1972, 1973), which may contain self-induced
and spiral flux lines, see the reviews by Tachiki (1982) and
Fulde and Keller (1982), of heavy-fermion superconductors like UPt$_3$
(Kleiman et al.\ 1992), and of possible ``exotic'' superconductors
with more-component order-parameter or with d-wave
(or ``s+id''-wave) pairing; see, e.g., Abrikosov (1995).
Recent theories of the vortex structure in  d-wave superconductors
are given by Ren et al.\ (1995),
Berlinsky et al.\ (1995), and Schopohl and Maki (1995).

This overview was intended to show that the discovery of
high-$T_c$ superconductivity has revived the interest in the
flux-line lattice. The intriguing electromagnetic properties of the
HTSC's inspired numerous novel ideas related to the layered structure
and short coherence length of these cuprates, which both favour
large thermal  fluctuations and thermal depinning of the FLL.
At the same time, old classical work on the FLL was rediscovered
and improved or adapted to these highly anisotropic and
fluctuating superconductors. This review only could present
a selection from the wast amount of work performed in a
very active field of research.
%
        \\[1.0 cm]
{\bf Acknowledgments}
         \\*[0.5 cm]
During the time in which this review was written many helpful
conversations with colleagues contributed to the clarification of
the numerous issues. I particularly want to thank Misha Indenbom for
numerous suggestions and
 Aleksander Gurevich, Alexei Koshelev,
Anatolii Larkin, Yurii Ovchinnikov, Gianni Blatter, Dima Geshkenbein,
Alexei Abrikosov, Lev Bulaevskii, Harald Weber,
Uwe Essmann, Lothar Schimmele,
Thomas Schuster, Michael Koblischka, Holger Kuhn, Alexander Forkl,
Horst Theuss, Helmut Kronm\"uller, Joachim K\"otzler, Asle Sudb{\o},
John Clem, Ted Forgan, Jan Evetts, Archie Campbell, Peter Kes,
Joe Vinen, Colin Gough, Mike Moore,  Boris Ivlev,
Denis Feinberg, John Gilchrist, Ronald Griessen,
H\'el\`ene Raffy, Sadok Senoussi, Ian Campbell,
Kazuo Kadowaki, Yoichi Ando,
Richard Klemm, George Crabtree, Valeri Vinokur,
Vladimir Kogan, Marty Maley, Igor Maksimov, Aleksander Grishin,
Aleksander Zhukov, Victor Moshchalkov, Miodrag Kuli\'{c},
Gertrud Zwicknagl, Roland Zeyher, Roman Mints, Eli Zeldov, Terry Doyle,
Edouard Sonin, Yurii Genenko, Klim Kugel, Leonid Fisher,
Georg Saemann-Ischenko, Roland Busch, Martino Leghissa,
Florian Steinmeyer,
Rudolph Huebener, Rudolf Gross, Konrad Fischer,
Pablo Esquinazi, Dierk Rainer, Werner Pesch, and B{\aa}rd Thrane.
Thanks are also due to Uwe Essmann, Yasuhiro Iye, Martino Leghissa,
Thomas Schuster, Gerd Nakielski, and Kees van der Beek
for figures 4, 11, 13, 20, 24, 25, and 28.
Part of this work was performed at the Institute for Scientific
Interchange in Torino, Italy; this is gratefully acknowledged.
%
   \newpage
{\bf References}
  \begin{description} {\small   \itemsep -6.05  pt 

\item Abrikosov A A 1957 {\it Zh. Eksp. Teor. Fiz.} {\bf 32}
           1442 (1957 {\it Sov. Phys. JETP} {\bf 5} 1174)
\item Abrikosov A A 1964 {\it Zh. Eksp. Teor. Fiz.} {\bf 47} 
           720 (1965 {\it Sov. Phys. JETP} {\bf 20} 480)    
\item Abrikosov A A 1988 {\it Fundamentals of the
      Theory of Metals} (Amsterdam: North Holland)

\item Abrikosov A A, Buzdin A I, Kuli\'c M L and Kuptsov D A 1989
      {\it Supercond.\ Sci.\ Technol.} {\bf 1} 260

\item Abrikosov A A 1995 {\it Phys.\ Rev.\ B} {\bf 51} 11955;
      {\it Physica C} {\bf 244} 243 

\item Adamopoulos N, Soylu B, Yan Y and Evetts J E 1995
      {\it Physica C} {\bf 242} 68 %

\item Adrian S D, Reeves M E, Wolf S A and Krezin V Z 1995
      {\it Phys.\ Rev.\ B} {\bf 51} 6800  

\item Alfimov G L and Silin V P 1994
      {\it Zh.\ Eksp.\ Teor.\ Fiz.} {\bf 106} 671 
      (1994 {\it Sov.\ Phys.\ JETP} {\bf 79} 369)

\item Aliev Yu M and Silin V P 1993    
      {\it Phys.\ Lett.\ A} {\bf 177} 259  

\item Almasan C C, de Andrade M C, Dalichaouch Y, Neumeier J J,
      Seaman C L, Maple M B, Guertin R P, Kuric M V and Garland J C
      1992 {\it Phys.\ Rev.\ B} {\bf 69} 3812

\item Alves S, Boekema C, Halim C, Lam J C, Whang E, Cooke D W
      and Leon M {\it Phys.\ Rev.\ B} {\bf 49} 12396   

\item Ambegaokar V and Baratoff A 1963
      {\it Phys.\ Rev.\ Lett.} {\bf 10} 486; {\bf 11} 104 (erratum) 
\item Ambegaokar V and Halperin B I 1969
      {\it Phys.\ Rev.\ Lett.} {\bf 22} 1364  

\item Ambj{\o}rn J and Olesen P 1989 {\it Phys. Lett. B} {\bf 218} 67

\item Amrein T, Schultz L, Kabius B and Urban K 1995  
      {\it Phys.\ Rev.\ B} {\bf 51} 6792      

\item Anderson P W 1962 {\it Phys.\ Rev.\ Lett.} {\bf 9} 309
\item Anderson P W and Kim Y B 1964 {\it Rev.\ Mod.\ Phys.} {\bf 36} 39

\item Ando Y, Motohira N, Kitazawa K, Takeya J and Akita S 1991
      {\it Phys.\ Rev.\ Lett.} {\bf 67} 2737
\item Ando Y, Kubota H  and Tanaka S 1992
      {\it Phys.\ Rev.\ Lett.} {\bf 69} 2851 
\item Ando Y, Kubota H  and Tanaka S 1993    
      {\it Phys.\ Rev.\ B} {\bf 48} 7716
\item Ando Y, Kubota H, Sato Y and Terasaki I 1994
      {\it Phys.\ Rev.\ B} {\bf 50} 9680

\item Andr{\'e} M-O, D'Anna G, Indenbom M V and Benoit W 1994
      {\it Physica C} {\bf 235-240} 2627 

\item Anlage S M and Wu D-H 1992
      {\it J.\ Supercond.} {\bf 5} 395

\item Antognazza L, Berkowitz S J, Geballe T H and Char K 1995
      {\it Phys.\ Rev.\ B} {\bf 51} 8560 

\item * Ao P 1995 {\it J.\ Supercond.} {\bf 8} ??? 

\item Ao P and Thouless D J 1993 {\it Phys.\ Rev.\ Lett.} {\bf 70} 2158
\item Ao P and Thouless D J 1994 {\it Phys.\ Rev.\ Lett.} {\bf 72} 132
\item Ao P and Zhu X-M 1995 {\it Phys.\ Rev.\ Lett.} {\bf 74} 4718

\item Aranson I, Gitterman M and Shapiro B Ya 1995 
      {\it Phys.\ Rev.\ B} {\bf 51}  3092            

\item Aronov A G and Rapoport A B 1992  
      {\it Mod.\ Phys.\ Lett.\ B} {\bf 6} 1083

\item Arrib\'ere A, Pastoriza H, Goffman M F, de la Cruz F, Mitzi D B
      and Kapitulnik A 1993
      {\it Phys.\ Rev. B} {\bf 48} 7486

\item Artemenko S N, Kr\"amer C and Wonneberger W 1994
      {\it Physica C} {\bf 220} 101

\item Artemenko S N and Kruglov A N 1990
      {\it Phys. Lett. A} {\bf 143} 485

\item Artemenko S N and Latyshev Yu I 1992 
      {\it Mod.\ Phys.\ Lett.\ B} {\bf 6} 367

\item Ashkenazy V D, Jung G, Khalfin I B and Shapiro B Ya 1994
      {\it Phys.\ Rev.\ B} {\bf 50} 13679        
\item Ashkenazy V D, Jung G, Khalfin I B and Shapiro B Ya 1995a
      {\it Phys.\ Rev.\ B} {\bf 51} 1236         
\item Ashkenazy V D, Jung G, and Shapiro B Ya 1995b
      {\it Phys.\ Rev.\ B} {\bf 51} 9052, 9118   

\item Aslamazov L G and Larkin A I 1968
      {\it Phys. Lett. A} {\bf 26} 238

\item Atsarkin V A, Vasneva G A and Zybtsev S G 1994
      {\it Supercond.\ Sci.\ Technol.} {\bf 7} 423  

\item Auer J and Ullmaier H 1973 {\it Phys. Rev.} {\bf 7} 136 

\item Baert M, Metlushko V V, Jonckheere R, Moshchalkov V V and
      Bruynseraede Y 1995 {\it Europhys.\ Lett.} {\bf 29} 157;
      {\it Phys.\ Rev.\ Lett.} {\bf 74} 3269  

\item Bak P, Tang C, and Wiesenfeld K,
      {\it Phys.\ Rev.\ A} {\bf 38} 364 

\item Balatski\u{\i} A V, Burlachkov L I and Gor'kov L P 1986
      {\it Zh.\ Eksp.\ Teor.\ Fiz.} {\bf 90} 1478
      (1986 {\it Sov.\ Phys.\ JETP} {\bf 63} 866)

\item Balents L and Kardar M 1994
      {\it Phys.\ Rev.\ B} {\bf 49} 13030   

\item Balents L and Nelson D R 1994
      {\it Phys.\ Rev.\ Lett.} {\bf 73} 2618   

\item Balents L and Simon S H 1995
      {\it Phys.\ Rev.\ Lett.} {\bf 51} 6515 

\item Balestrino G, Marinelli M, Milani E, Varlamov A A and Yu L 1993
      {\it Phys.\ Rev.\ B} {\bf 47} 6037 

\item Balestrino G, Crisan A, Livanov D V, Milani E, Montuori M and
      Varlamov A A 1995                  
      {\it Phys.\ Rev.\ B} {\bf 51} 9100

\item Bardeen J, Cooper L N and Schrieffer J R 1957
      {\it Phys.\ Rev.} {\bf 108} 1175 
\item Bardeen J 1978 {\it Phys.\ Rev.\ B} {\bf 17} 1472 
\item Bardeen J and Stephen M J 1965
      {\it Phys. Rev.} {\bf 140} A1197
\item Bardeen J and Sherman R D 1975                  
      {\it Phys.\ Rev.\ B} {\bf 12} 2634

\item * Baraduc C, Buzdin A, Henry J-Y, Brison J-P and Puech L 1995
      {\it Physica C} {\bf 247} ??? (in print) 

\item Barone A and Paterno G 1982 {\it Physics and Applications of the
      Josephson effect} (New York: Wiley)

\item Barone A, Larkin A I and Ovchinnikov Yu N 1990
      {\it J.\ Supercond.} {\bf 3} 155  

\item Batrouni G G and Hwa T 1994             
      {\it Phys.\ Rev.\ Lett.} {\bf 72} 4133

\item Barford W and Gunn M 1988 {\it Physica C} {\bf 156} 515 

\item Barford W and Harrison M 1994 {\it Phys.\ Rev.\ B} {\bf 50} 13748 

\item Barowski H, Sattler K M and Schoepe W 1993
      {\it J.\ Low Temp.\ Phys.} {\bf 93} 85 

\item Baym G 1995 
      {\it Phys.\ Rev.\ B} {\bf 51} 11697

\item Bean C P 1964 {\it Rev.\ Mod.\ Phys.} {\bf 36} 31
\item Bean C P 1970 {\it J.\ Appl.\ Phys.} {\bf 41} 2482
\item Bean C P and Livingston J D 1964 {\it Phys.\ Rev.\ Lett.} {\bf 12} 14

\item Beasley M R, Labusch R and Webb W W  1969
      {\it Phys. Rev.} {\bf 181} 682

\item Beck H 1994 {\it Phys.\ Rev.\ B} {\bf 49} 6153

\item Beck R G, Farrell D E, Rice J P, Ginsberg D M and Kogan V G 1992
      {\it Phys.\ Rev.\ Lett.} {\bf 68} 1594

\item Becker Th, Theuss H, Schuster Th, Kronm\"uller H, Kraus M and
      Saemann-Ischenko G 1995
      {\it Physica C} {\bf 245} 273    

\item Bednorz J G and M\"uller K A 1986 
      {\it Z.\ Phys.\ B} {\bf 64} 189

\item Behler S, Pan S H, Jess P, Baratoff A, G\"untherodt H-J,
      L\'evy F, Wirth G and Wiesner J 1994a
      {\it Phys.\ Rev.\ Lett.} {\bf 72} 1750  
\item Behler S, Jess P, Baratoff A and G\"untherodt H-J 1994b
      {\it Physica C} {\bf 235-240} 2703    

\item Behr R, K\"otzler J, Spirgatis A and Ziese M 1992
      {\it Physica A} {\bf 191} 464   

\item Bending S J, van Klitzing K and Ploog K 1990
      {\it Phys.\ Rev.\ B} {\bf 42} 9859

\item Benkraouda M and Ledvij M 1995   
      {\it Phys.\ Rev.\ B} {\bf 51} 6123

\item Berezinskii V L 1970 {\it Zh.\ Eksp.\ Teor.\ Fiz.} {\bf 59} 907
                 (1971 {\it Sov.\ Phys.\ JETP} {\bf 34} 610)

\item Berghuis P and Kes P H  1990
      {\it Physica B} {\bf 165\&166} 1169    
\item Berghuis P, van der Slot A L F and Kes P H  1990
      {\it Phys.\ Rev.\ Lett.} {\bf 65} 2583   

\item Bergk W and Tewordt L 1970
      {\it Z.\ Phys.} {\bf 230} 178  

\item Berkovits R and Shapiro B Ya 1995
      {\it Phys.\ Rev.\ Lett.} {\bf 70} 2617 

\item * Berlinsky A J, Fetter A L, Franz M, Kallin C and Soininen P I 1995
      (preprint: GL Theory of Vortices in d-Wave Superconductors)

\item Bezryadin A, Buzdin A and  and Pannetier B 1995
      {\it Phys.\ Rev.\ B} {\bf 51} 3718  

\item Bezryadin A and Pannetier B 1995
      {\it J.\ Low Temp.\ Phys.} {\bf 98} 251    

\item Bhagwat K V and Chaddah P 1992
      {\it Physica C} {\bf 190} 444

\item Bhattacharya S and Higgins M J 1993
      {\it Phys.\ Rev.\ Lett.} {\bf 70} 2617   
\item Bhattacharya S and Higgins M J 1994
      {\it Phys.\ Rev.\ B} {\bf 49} 10005      
\item Bhattacharya S, Higgins M J and Ramakrishnan T V 1994
      {\it Phys.\ Rev.\ Lett.} {\bf 73} 1699   

\item Bishop D J, Gammel P L, Huse D A and Murray C A 1992
      {\it Science} {\bf 255} 165

\item Blackstead H A, Pulling D B, Horwitz J S and Chrisey D B 1994
      {\it Phys.\ Rev.\ B} {\bf 49} 15335   

\item Blamire M G and Evetts J E 1985         
      {\it Appl.\ Phys.\ Lett.} {\bf 46} 1181

\item Blatter G, Feigel'man M V, Geshkenbein V B, Larkin A I,
      and Vinokur V M 1994a {\it Rev.\ Mod.\ Phys.} {\bf 66} 1125-1388

\item Blatter G and Geshkenbein V B 1993
      {\it Phys.\ Rev.\ B} {\bf 47} 130

\item Blatter G, Geshkenbein V B and Larkin A I 1992  
      {\it Phys.\ Rev.\ Lett.} {\bf 68} 875
\item Blatter G, Geshkenbein V B and Larkin A I 1993
      {\it Phys.\ Rev.\ Lett.} {\bf 71} C302 

\item Blatter G, Geshkenbein V B and  Vinokur V M 1991a   
      {\it Phys.\ Rev.\ Lett.} {\bf 66} 3297
\item Blatter G, Ivlev B and Rhyner J 1991b
      {\it Phys.\ Rev.\ Lett.} {\bf 66} 2392       
\item Blatter G, Rhyner J and  Vinokur V M 1991c 
      {\it Phys.\ Rev.\ B} {\bf 43} 7826

\item Blatter G and Ivlev B 1993
      {\it Phys.\ Rev.\ Lett.} {\bf 70} 2621   
\item Blatter G and Ivlev B 1994
      {\it Phys.\ Rev.\ B} {\bf 50} 10272      
\item * Blatter G and Ivlev B 1995
      {\it Phys.\ Rev.\ B} {\bf 52} (in print) (Helicon modes)

\item Blatter G, Ivlev B, Kagan Yu, Theunissen M, Volotkin Ya
      and Kes P 1994b {\it Phys.\ Rev.\ B} {\bf 50} 13013  

\item Blatter G, Ivlev B and Nordborg H 1993
      {\it Phys.\ Rev.\ B} {\bf 48} 10448   

\item Blazey K W, Portis A M, M\"uller K A and Holtzberg F 1988
      {\it Europhys.\ Lett.} {\bf 462} 1988   

\item Blum T and Moore M A 1995
      {\it Phys.\ Rev.\ B} {\bf 51} 15359 

\item Bokil H S and Young A P 1995
      {\it Phys.\ Rev.\ Lett.} {\bf 74} 3021 

\item Bolle C A, Gammel P L, Grier D G, Murray C A, Bishop D J, Mitzi D B,
      and Kapitulnik A 1991 {\it Phys.\ Rev.\ Lett.} {\bf 66}  112

\item Bonevich J E, Harada K, Kasai H, Matsuda T, Yoshida T, Pozzi G
      and Tonomura A 1994a {\it Phys.\ Rev.\ B} {\bf 49} 6800 

\item Bonevich J E, Harada K, Kasai H, Matsuda T, Yoshida T 
      and Tonomura A 1994b {\it Phys.\ Rev.\ B} {\bf 50} 567  

\item Bontemps N, Davidov D, Monod P, Even R 1991
      {\it Phys.\ Rev.\ B} {\bf 43} 11512  

\item * Borovitskaya E S and Genkin V M 1995 
      {\it Physica C} (in print)  

\item Bouchaud J-P, M\'ezard M and Yedida S 1992
      {\it Phys.\ Rev.\ B} {\bf 46} 14686  

\item Brandt E H 1969a {\it phys.\ stat.\ sol.\ (b)} {\bf 35} 1027;
                                          {\bf 36} 371
\item Brandt E H 1969b {\it phys.\ stat.\ sol.\ (b)} {\bf 36} 381, 393

\item Brandt E H 1972 {\it phys.\ stat.\ sol.\ (b)} {\bf 51} 345
\item Brandt E H 1973 {\it Phys. Lett. A} {\bf 43} 539

\item Brandt E H 1974 {\it phys.\ stat.\ sol.\ (b)} {\bf 64} 257, 
                                              467, {\bf 65} 469   
\item Brandt E H 1975 {\it phys.\ stat.\ sol.\ (b)} {\bf 71} 277 

\item Brandt E H 1976a {\it phys.\ stat.\ sol. (b)} {\bf 77} 105
\item Brandt E H 1976b {\it J.\ Low Temp.\ Phys.} {\bf 24} 409

\item Brandt E H 1977a {\it J.\ Low Temp.\ Phys.} {\bf 26} 709, 735
\item Brandt E H 1977b {\it J.\ Low Temp.\ Phys.} {\bf 28} 263, 291
\item Brandt E H 1977c {\it phys.\ stat.\ sol. (b)} {\bf 84} 269, 637

\item Brandt E H 1978  {\it Phil.\ Mag B} {\bf 37} 293  
\item Brandt E H 1980a {\it Phys.\ Lett.} {\bf A77} 484 
\item Brandt E H 1980b {\it J.\ Low Temp.\ Phys.} {\bf 39} 41 
\item Brandt E H 1980c {\it Phys.\ Lett.} {\bf A79} 207 

\item Brandt E H 1981a {\it J.\ Low Temp.\ Phys.} {\bf 42} 557 
\item Brandt E H 1981b {\it J.\ Low Temp.\ Phys.} {\bf 44} 33, 59 

\item Brandt E H 1982 {\it Phys.\ Rev.\ B} {\bf 25} 5756 
\item Brandt E H 1983a {\it Phys.\ Rev.\ Lett.} {\bf 50} 1599    
\item Brandt E H 1983b {\it J.\ Low Temp.\ Phys.} {\bf 53} 41, 71 

\item Brandt E H 1986a {\it Phys.\ Rev.\ B} {\bf 34} 6514  
\item Brandt E H 1986b {\it J.\ Low Temp.\ Phys.} {\bf 64} 375 
\item Brandt E H 1986c {\it Phys.\ Rev.\ Lett.} {\bf 56} 1381 
\item Brandt E H 1986d {\it Phys.\ Rev.\ Lett.} {\bf 57} 1347 
\item Brandt E H 1987 {\it J.\ de Physique} {\bf C8} 31 

\item Brandt E H 1988a {\it Phys.\ Rev.\ B} {\bf 37} 2349         
\item Brandt E H 1988b {\it J.\ Low Temp.\ Phys.} {\bf 73} 355  
\item Brandt E H 1988c {\it Appl.\ Phys.\ Lett.} {\bf 53} 1554  

\item Brandt E H 1989a {\it Phys.\ Rev.\ Lett.} {\bf 63} 1106  
\item Brandt E H 1989b {\it Physica C} {\bf  162-164} 1167   
\item Brandt E H 1989c {\it Science} {\bf 243} 349           
\item Brandt E H 1990a {\it Am.\ J.\ Phys.} {\bf 58} 43      
\item Brandt E H 1990b {\it Z.\ Phys.\ B} {\bf 80} 167       
\item Brandt E H 1990c {\it Physica C} {\bf 165\&166} 1129   

\item Brandt E H 1991a {\it Physica B} {\bf 169} 91          
\item Brandt E H 1991b {\it Int. J. Mod. Phys. B} {\bf 5} 751
\item Brandt E H 1991c {\it Phys.\ Rev.\ Lett.} {\bf 66} 3213  
\item Brandt E H 1991d {\it Phys.\ Rev.\ Lett.} {\bf 67} 2219  
\item Brandt E H 1991e {\it Physica C} {\bf 185-189} 270  

\item Brandt E H 1992a {\it Physica C} {\bf 195} 1
\item Brandt E H 1992b {\it Phys.\ Rev.\ Lett.} {\bf 68} 3769  
\item Brandt E H 1992c {\it Phys.\ Rev.\ Lett.} {\bf 69} 1105; 
                       {\it Europhys.\ Lett.} {\bf 18} 635
\item Brandt E H 1992d {\it Phys.\ Rev.\ B} {\bf 46} 8628      

\item Brandt E H 1993a {\it Phys.\ Rev.\ B} {\bf 48} 6699      
\item Brandt E H 1993b {\it Phys.\ Rev.\ Lett.} {\bf 71} 2821  

\item Brandt E H 1994a {\it Phys.\ Rev.\ B} {\bf 49} 9024      
\item Brandt E H 1994b {\it Phys.\ Rev.\ B} {\bf 50} 4034      
\item Brandt E H 1994c {\it Phys.\ Rev.\ B} {\bf 50} 13833     %
\item Brandt E H 1994d {\it Physica C} {\bf 235-240} 2939    

\item Brandt E H 1995a {\it Phys.\ Rev.\ Lett.} {\bf 74} 3025 
\item Brandt E H 1995b {\it Phys.\ Rev.\ Lett.} {\bf 74} C3499 
\item Brandt E H 1995c {\it Phys.\ Rev.\ B} (submitted) 

\item Brandt E H, Clem J R and Walmsley D G 1979
      {\it J.\ Low Temp.\ Phys.} {\bf 37} 43    

\item Brandt E H and Essmann U 1975 {\it Phys.\ Lett.} {\bf 51A} 45 
\item Brandt E H and Essmann U 1987 {\it phys.\ stat.\ sol.\ (b)}
            {\bf 144} 13            

\item Brandt E H, Esquinazi P, Neckel H and Weiss G 1986a
      {\it Phys.\ Rev.\ Lett.} {\bf 56} 89             
\item Brandt E H, Esquinazi P and Neckel H 1986b
      {\it J.\ Low Temp.\ Phys.} {\bf 63} 187        
\item Brandt E H, Esquinazi P and Weiss G 1989
      {\it Phys.\ Rev.\ Lett.} {\bf 62} C2330           

\item Brandt E H and Evetts J 1992 {\it Concise Encyclopedia
            of Magnetic \& Superconducting Materials} ed J Evetts
            (Oxford: Pergamon) p 150
\item Brandt E H and Indenbom M 1993
      {\it Phys.\ Rev.\ B} {\bf 48} 12893       
\item Brandt E H, Indenbom M and Forkl A 1993
      {\it Europhys.\ Lett.} {\bf 22} 735     

\item Brandt E H and Seeger A 1986 {\it Adv.\ Physics} {\bf 35} 189

\item Brandt E H and Sudb{\o} A 1991 {\it Physica C} {\bf 180} 426

\item Brandt U, Pesch W and Tewordt 1967 
      {\it Z.\ Physik} {\bf 201} 209

\item Brass A and Jensen H J 1989         
      {\it Phys.\ Rev.\ B} {\bf 39} 9587

\item Brawner D A and Ong N P 1993  
      {\it J.\ Appl.\ Phys.} {\bf 73} 3890

\item Brawner D A, Ong N P and Wang Z Z 1993  
      {\it Phys.\ Rev.\ B} {\bf 47} 1156, {\bf 48} 13188

\item Brechet Y J M, Dou\c{c}ot B, Jensen H J and Shi A-C 1990
      {\it Phys.\ Rev.\ B} {\bf 42} 2116 

\item Brice\~no G, Crommie M F and Zettl A   
      1991 {\it Phys.\ Rev.\ Lett.} {\bf 66} 2164
\item Brice\~no G, Crommie M F and Zettl A   
      1993 {\it Physica C} {\bf 204} 389

\item Brom H B and Alloul H 1991 {\it Physica C} {\bf 177} 297

\item Brongersma S H, Verweij E, Koema N J, de Groot D G, Griessen R
      and Ivlev B I 1993 {\it Phys.\ Rev.\ Lett.} {\bf 71} 2319

\item Br\"ull P, Kirchg\"assner D and Leiderer P 1991
      {\it Physica C} {\bf 182} 339

\item Brunner O, Antognazza L, Triscone J-M, Mi\'eville L and
      Fisher {\O} 1991 {\it Phys.\ Rev.\ Lett.} {\bf 67} 1354 

\item Bryksin V V and Dorogovtsev S N 1993 
      {\it Physica C} {\bf 215} 173

\item Buck W, Selig K-P and Parisi J 1981 
      {\it J.\ Low Temp.\ Phys.} {\bf 45} 21

\item Buckel W 1990 {\it Superconductivity} (New York: VCH Publishers)

\item Budhani R C, Holstein W L and Suenaga M 1994
      {\it Phys.\ Rev.\ Lett.} {\bf 72} 566  
\item Budhani R C, Suenaga M and Liou S H 1992
      {\it Phys.\ Rev.\ Lett.} {\bf 69} 3816   
\item Budhani R C, Liou S H and Cai Z X 1993
      {\it Phys.\ Rev.\ Lett.} {\bf 71}  621  

\item Bulaevskii L N 1972 {\it Zh.\ Eksp.\ Teor.\ Fiz.} {\bf 64} 2241
      (1973 {\it Sov.\ Phys.\ JETP} {\bf 37} 1133)
\item Bulaevskii L N 1990 {\it Int. J. Mod. Phys. B} {\bf 4} 1849  
\item Bulaevskii L N 1991 {\it Phys.\ Rev.\ B} {\bf 44} 910  

\item Bulaevskii L N and Maley M P 1993
      {\it Phys.\ Rev.\ Lett.} {\bf 71}  3541   

\item Bulaevskii L N and Clem J R 1991
      {\it Phys.\ Rev.\ B} {\bf 44} 10234  

\item Bulaevskii L N, Cho J H, Maley M P, Li Q, Suenaga M and Ledvij M 1994a
      {\it Phys.\ Rev.\ B} {\bf 50} 3507  
\item Bulaevskii L N, Maley M P and Schegolev I F 1994b
      {\it Physica B} {\bf 197} 506     
\item Bulaevskii L N, Maley M P and Cho J H 1994c
      {\it Physica C} {\bf 235-240} 87  
\item Bulaevskii L N, Maley M P and Tachiki 1995a
      {\it Phys.\ Rev.\ Lett.} {\bf 74} 801 

\item Bulaevskii L N, Hammel P C and Vinokur V M 1995b
      {\it Phys.\ Rev.\ B} {\bf 51} 15355 

\item Bulaevskii L N, Daemen L L, Maley M P and Coulter J Y 1993a
      {\it Phys.\ Rev.\ B} {\bf 48}  13798  

\item Bulaevskii L N, Kolesnikov N N, Schegolev I F and Vyaselev O M 1993b
      {\it Phys.\ Rev.\ Lett.} {\bf 71} 1891                  

\item Bulaevskii L N, Ledvij M and Kogan V G 1992a
      {\it Phys.\ Rev.\ B} {\bf 46} 366                  
\item Bulaevskii L N, Ledvij M and Kogan V G 1992b
      {\it Phys.\ Rev.\ B} {\bf 46} 11807                
\item Bulaevskii L N, Ledvij M and Kogan V G 1992c
      {\it Phys.\ Rev.\ Lett.} {\bf 68} 3773             
\item Bulaevskii L N, Clem J R, Glazman L and Malozemoff A P 1992d
      {\it Phys.\ Rev.\ B} {\bf 45}  2545              
\item Bulaevskii L N, Clem J R, and Glazman L 1992e 
      {\it Phys.\ Rev.\ B} {\bf 45}  2545

\item Bulaevskii L N, Meshkov S V and Feinberg D 1991
      {\it Phys.\ Rev.\ B} {\bf 43} 3728                 

\item Burlachkov L 1989 {\it Europhys. Lett.} {\bf 8} 673
\item Burlachkov L 1993 {\it Phys.\ Rev.\ B} {\bf 47} 8056
\item Burlachkov L, Khalfin I B and Shapiro B Ya 1993
      {\it Phys.\ Rev.\ B} {\bf 48} 1156    
\item Burlachkov L and Vinokur V M 1994
      {\it Physica C} {\bf 194-196} 1819 
\item Burlachkov L, Geshkenbein V B, Koshelev A E, Larkin A I and
      Vinokur V M 1994 {\it Phys.\ Rev.\ B} {\bf 50} 16770  

\item Busch R, Ries G, Werthner H, Kreiselmeyer G and Saemann-Ischenko G
      1992 {\it Phys.\ Rev.\ Lett.} {\bf 69} 522  

\item Buzdin A I 1993
      {\it Phys.\ Rev.\ B} {\bf 47} 11416 

\item Buzdin A I and Feinberg D 1990 {\it J. Phys. France} {\bf 51} 1971
\item Buzdin A I and Feinberg D 1992
      {\it Phys. Lett. A} {\bf 165} 281, {\bf 167} 89 
\item Buzdin A I and Feinberg D 1994a
      {\it Physica C} {\bf 220} 74 
\item Buzdin A I and Feinberg D 1994b
      {\it Physica C} {\bf 235-240} 2755 

\item Buzdin A I, Krotov S S and Kuptsov D A 1991
      {\it Physica C} {\bf 175} 42


\item Buzdin A I, Neminsky A, Nikolaev P and Baraduc C 1994
      {\it Physica C} {\bf 227} 365 

\item Buzdin A I and Simonov A Yu 1990 {\it Physica C} {\bf 168} 421

\item Buzdin A I and Simonov A Yu 1991 {\it Physica C} {\bf 175} 143

\item Cai X Y, Gurevich A, Larbalestier D C, Kelley R J, Onellion M,
      Beger H and Margaritondo G 1994       
      {\it Phys.\ Rev.\ B} {\bf 50} 16774

\item Caldeira A O and  Leggett A J 1981
      {\it Phys.\ Rev.\ Lett.} {\bf 46} 211

\item Calzona V, Cimberle M R, Ferdeghini C, Grasso G, Iorio A, Licci F,
      Putti M, Rizzuto C and Siri A S 1993
      {\it Supercond.\ Sci.\ Technol.} {\bf 6} 46

\item Camerlingo C, Monaco R, Ruggiero B, Russo M and Testa G 1995
      {\it Phys.\ Rev.\ B} {\bf 51} 6493 

\item Campbell A M 1971 {\it J. Phys. C} {\bf 4} 3186    
\item Campbell A M 1978 {\it Phil.\ Mag.\ B} {\bf 37} 149, 169 
\item Campbell A M 1980 {\it Helvetica Physica Acta} {\bf 53} 404 
\item Campbell A M 1982 {\it Cryogenics} {\bf 82} 3 
\item Campbell A M 1991 {\it IEEE Trans.\ Magn.} {\bf 27} 1660 

\item Campbell A M and Evetts J E 1972 {\it Adv.\ Phys.} {\bf 72} 199

\item Campbell L J, Doria M M  and Kogan V G 1988 {\it Phys.\ Rev.\ B}
      {\bf 38} 2439

\item Caplin A D, Angadi M A, Laverty J R and de Oliveira A L 1992
      {\it Supercond.\ Sci.\ Technol.} {\bf 5} S186 

\item Caplin A D, Cohen L F, Perkin G K and Zhukov A A 1994
      {\it Supercond.\ Sci.\ Technol.} {\bf 7} 412

\item Caplin A D, Perkin G K and Cohen L F 1995 
      {\it Supercond.\ Sci.\ Technol.} {\bf 8} 366

\item Carneiro G 1992 {\it Phys.\ Rev.\ B} {\bf 45} 2391, 2403 
\item Carneiro G 1994 {\it Phys.\ Rev.\ B} {\bf 50} 6982 

\item Carneiro G, Doria M M and de Andrade S C B 1992
      {\it Physica C} {\bf 203} 167
\item Carneiro G, Cavalcanti R and Gartner A 1993
      {\it Phys.\ Rev.\ B} {\bf 47} 5263 

\item Caroli C, DeGennes P G and Matricon J 1964
      {\it Phys.\ Lett.} {\bf 9} 307              

\item Carraro C and Fisher D S 1995
      {\it Phys.\ Rev.\ B} {\bf 51} 534 

\item Carretta P 1992 {\it Phys.\ Rev.\ B} {\bf 45} 5760 
\item Carretta P 1993 {\it Phys.\ Rev.\ B} {\bf 48} 528  

\item Castro H and Rinderer L 1995       
      {\it Physica C} {\bf 235-240} 2915 

\item Cave J R and Evetts J E 1978
      {\it Phil.\ Mag.\ B} {\bf 37} 111  

\item Chakravarty S, Ivlev B I and Ovchinnikov Yu N 1990
      {\it Phys.\ Rev.\ Lett.} {\bf 64} 3187,
      {\it Phys.\ Rev.\ B} {\bf 42} 2143

\item Chan W C, Jwo D S, Lin Y F and Huang Y 1994
      {\it Physica C} {\bf 230} 349  

\item Charalambous M, Chaussy J and Lejay P 1992
      {\it Phys.\ Rev. B} {\bf 45} 5091  
\item Charalambous M, Chaussy J, Lejay P and Vinokur V 1993
      {\it Phys.\ Rev. Lett.} {\bf 71} 436  

\item Chattopadhyay B and Shenoy S R 1994   
      {\it Phys.\ Rev.\ Lett.} {\bf 72} 400 

\item Chen D-X and Goldfarb R B 1989
      {\it J.\ Appl.\ Phys.} {\bf 66} 2489  

\item Chen D-X, Goldfarb R B, Cross R W and Sanchez A 1993
      {\it Phys.\ Rev.\ B} {\bf 48} 6426  

\item Chen S Q, Skocpol W J, DeObaldia E, O'Malley M and Mankiewich P M
      1993 {\it Phys.\ Rev.\ B} {\bf 47} 2936  

\item Chen T and Teitel S 1994 {\it Phys.\ Rev.\ Lett.} {\bf 72} 2085
\item Chen T and Teitel S 1995 {\it Phys.\ Rev.\ Lett.} {\bf 74} 2792
\item Chen J L and Yang T J 1994 {\it Phys.\ Rev.\ B} {\bf 50} 4064 

\item Chien T R, Jing T W, Ong N P and Wang Z Z 1991 
      {\it Phys.\ Rev.\ Lett.} {\bf 66} 3075

\item Chikumoto N, Konczykowski M, Motohiro N and Malozemoff A P 1992
      {\it Phys.\ Rev.\ Lett.} {\bf 69} 1260  

\item * Cho J H, Maley M P, Safar H, Coulter J Y and Bulaevskii L N 1995
      {\it Phys.\ Rev.\ Lett.} {\bf 74} (submitted) 

\item Choi E-J, Lihn H-T S, Drew H D and Hsu T C 1994
      {\it Phys.\ Rev.\ B} {\bf 49} 13271  

\item Choi M-S and Lee S-I 1995
      {\it Phys.\ Rev.\ B} {\bf 51} 6680 

\item Chudnovsky E 1990 {\it Phys.\ Rev.\ Lett.} {\bf 65} 3060 
\item Chudnovsky E 1995 {\it Phys.\ Rev.\ B} {\bf 51} 1181,  
                                               15351  
\item Chudnovsky E, Ferrera A and Vilenkin A 1995
      {\it Phys.\ Rev.\ B} {\bf 51} 1181  

\item Civale L, Krusin-Elbaum L, Thompson J R, Wheeler R, Marwick A D,
      Kirk M A, Sun Y R, Holtzberg F and Feild C 1994
      {\it Phys.\ Rev.\ B} {\bf 50} 4102                 

\item Civale L, Marwick A D, Worthington T K, Kirk M A, Thompson J R,
      Krusin-Elbaum L, Sun Y, Clem J R and Holtzberg F 1991
      {\it Phys.\ Rev.\ B} {\bf 67} 648

\item Civale L, Worthington T K, Krusin-Elbaum L, Sun Y, and Holtzberg F
      1992 {\it Magnetic Susceptibility of Superconductors and Other Spin
      Systems} ed R A Hein, T L Francavilla and D H Liebenberg
      (New York: Plenum Press) 

\item Clem J R 1974a {\it Proc.\ 13th Conf.\ an Low Temp.\ Phys.}
      ed K D Timmerhaus, W J O'Sullivan and E F Hammel
      (New York: Plenum) vol {\bf 3} p 102   
\item Clem J R 1974b {\it Phys.\ Rev.\ B} {\bf 9} 898    

\item Clem J R 1975a {\it J.\ Low Temp.\ Phys.} {\bf 18} 427   
\item Clem J R 1975b {\it Phys.\ Rev.\ B} {\bf 12} 1742  
\item Clem J R 1977 {\it Phys.\ Rev.\ Lett.} {\bf 24} 1425  
\item Clem J R 1979 {\it J.\ Appl. Phys.} {\bf 50} 3518       
\item Clem J R 1980 {\it J.\ Low Temp.\ Phys.} {\bf 38} 353   

\item Clem J R 1981a {\it Physica B} {\bf 107} 453    
\item Clem J R 1981b {\it Phys.\ Rep.} {\bf 75} 1     
\item Clem J R 1988 {\it Physica C} {\bf 153-155} 50  
\item Clem J R 1991 {\it Phys.\ Rev.\ B} {\bf 43} 7837  

\item Clem J R 1992 {\it Concise Encyclopedia of    
      Magnetic \& Superconducting Materials} ed J Evetts
      (Oxford: Pergamon) p 612

\item Clem J R, Bumble B, Raider S I, Gallagher W J and Shih Y C 1987
      {\it Phys.\ Rev.\ B} {\bf 35} 6637  

\item Clem J R, Huebener R P and Gallus D E 1973
      {\it J.\ Low Temp.\ Phys.} {\bf 12} 449   

\item Clem J R, Kerchner  H R and Sekula T S 1976
      {\it Phys.\ Rev.\ B} {\bf 14} 1893        

\item Clem J R and Coffey M W 1990 {\it Phys.\ Rev.\ B} {\bf 42} 6209
\item Clem J R and Coffey M W 1992 {\it Phys.\ Rev.\ B} {\bf 46} 14662 

\item Clem J R and Hao Z 1993 {\it Phys.\ Rev.\ B} {\bf 48} 13774 
\item Clem J R and Sanchez A 1994 {\it Phys.\ Rev.\ B} {\bf 50} 9355  

\item Coffey M W 1994a {\it Phys.\ Rev.\ B} {\bf 49} 9774     
\item Coffey M W 1994b {\it Physica C} {\bf 235-240} 1961  

\item * Coffey M W 1995      
      {\it Phys.\ Rev.\ B} {\bf 52} (in print)

\item Coffey M and Clem J R 1991a  
      {\it Phys.\ Rev.\ Lett.} {\bf 67} 386
\item Coffey M and Clem J R 1991b  
      {\it Phys.\ Rev.\ B} {\bf 44} 6903

\item Cohen L F, Zhukov A A, Perkins G, Jensen H J, Klestov S A,
      Voronkova V I, Abell S, K\"upfer H, Wolf T and Caplin A D 1994
      {\it Physica C} {\bf 230} 1  

\item Conen E and Schmid A 1974                   
      {\it J.\ Low Temp.\ Phys.} {\bf 17} 331

\item Conner L W and Malozemoff A P 1991 {\it Phys.\ Rev.\ B} {\bf 43} 402

\item Cooley L D, Lee P J and Larbalestier D C 1994
      {\it Appl.\ Phys.\ Lett.} {\bf 64} 1298

\item Cooley L D and Grishin A M 1995         
      {\it Phys.\ Rev.\ Lett.} {\bf 74} 2788  

\item Cooper J R, Loram J W and Wade J M 1995
      {\it Phys.\ Rev.\ B} {\bf 51} 6179  

\item Crabtree G W, Kwok W K, Welp U, Downey J, Fleshler S and
      K G Vandervoort 1991 {\it Physica C} {\bf 185-189} 282  

\item Crabtree G W, Fendrich J, Kwok W K and Glagola B G 1994
      {\it Physica C} {\bf 235-240} 2543 

\item Cribier D, Jacrot B, Rao L M and Farnoux B 1964
      {\it Phys. Lett.} {\bf 9} 106

\item de la Cruz F, Pastoriza H, L\'opez D, Goffmann M F, Arrib\'ere A and
      Nieva G 1994a {\it Physica C} {\bf 235-240} 83 
\item de la Cruz F, L\'opez D and Nieva G 1994b
      {\it Phil.\ Mag.\ B} {\bf 70} 773 

\item Cubitt R, Forgan E M, Warden M, Lee S L, Zimmermann P, Keller H,
      Savi\'c I M, Wenk P, Zech D, Kes P H, Li T W, Menovsky A A and
      Tarnawski Z 1993a {\it Physica C} {\bf 213} 126 

\item Cubitt R, Forgan E M, Yang G, Lee S L, Paul D McK, Mook H A,
      Yethiraj M, Kes P H, Li T W, Menovsky A A, Tarnawski Z
      and Mortensen K 1993b {\it Nature} {\bf 365} 407

\item D'Anna G, Benoit W, Sadowski W and Walker E 1992
      {\it Europhys.\ Lett.} {\bf 20} 167    

\item D'Anna G, Indenbom M V, Andr\`e M-O, Benoit W and Walker E 1994a
      {\it Europhys.\ Lett.} {\bf 25} 225    

\item D'Anna G, Andr\`e M-O, Indenbom M V and Benoit W 1994b
      {\it Physica C} {\bf 230} 115     

\item D'Anna G, Andr\`e M-O,  and Benoit W 1994c  
      {\it Europhys.\ Lett.} {\bf 25} 539

\item * D'Anna G, Gammel P L, Safar H, Alers G B and Bishop D J 1995
      {\it Phys.\ Rev.\ Lett.} (submitted) 

\item Deak J, McElfresh M, Clem J R, Hao Zh, Konczykowski M,
      Muenchausen R, Foltyn S and Dye R 1994
      {\it Phys.\ Rev.\ B} {\bf 49} 6270  

\item Daemen L L, Campbell L J, and Kogan V G 1992
                  {\it Phys.\ Rev.\ B} {\bf 46} 3631 
\item Daemen L L, Bulaevskii L N, Maley M P and Coulter J Y  1993a
                  {\it Phys.\ Rev.\ Lett.} {\bf 70} 1167,
                  {\it Phys.\ Rev.\ B} {\bf 47} 11291
\item Daemen L L, Campbell L J, Simonov A Yu and Kogan V G 1993b
                  {\it Phys.\ Rev.\ Lett.} {\bf 70} 2948

\item Dai H, Liu J and Lieber C M 1994a     
      {\it Phys.\ Rev.\ Lett.} {\bf 72} 748
\item Dai H, Yoon S, Liu J, Budhani C and Lieber C M 1994b
      {\it Science} {\bf 265} 1552 

\item D\"aumling M and Larbalestier D C 1989
      {\it Phys.\ Rev.\ B} {\bf 40} 9350

\item D\"aumling M, Walker E and Fl\"ukiger R 1994
      {\it Phys.\ Rev.\ B} {\bf 50} 13024

\item de Andrade R and de Lima O F 1995
      {\it Phys.\ Rev.\ B} {\bf 51} 9383 

\item DeGennes P G 1966 {\it Superconductivity of Metals and
      Alloys} (New York: Benjamin)

\item Dekker C, Eidelloth W and Koch R H 1992a   
      {\it Phys.\ Rev.\ Lett.} {\bf 68} 3347
\item Dekker C, W\"oltgens P J M, Koch R H, Hussey B W and Gupta A 1992b
      {\it Phys.\ Rev.\ Lett.} {\bf 69} 2717       

\item Delrieu J M 1972 {\it J.\ Low Temp.\ Phys.} {\bf 6} 197
\item Delrieu J M 1973 {\it J.\ Phys.\ F: Metal Physics} {\bf 3} 893
\item Delrieu J M 1974 {\it State Thesis, Universit\'e Paris-Sud}

\item Deltour R and Tinkham M 1966
      {\it Phys. Lett.} {\bf 23} 183  
\item Deltour R and Tinkham M 1967
      {\it Phys.\ Rev.\ Lett.} {\bf 19} 125 
\item Deltour R and Tinkham M 1968
      {\it Phys. Rev.} {\bf 174} 478  

\item Denniston C and Tang C 1995
      {\it Phys.\ Rev.\ B} {\bf 51} 8457 

\item Dersch H and Blatter G 1988
      {\it Phys.\ Rev.\ B} {\bf 38} 11391  

\item De Soto S M, Slichter C P, Wang H H, Geiser U and Williams J M
      1993 {\it Phys.\ Rev.\ Lett.} {\bf 70} 2956 

\item Deutscher G 1991 {\it Physica C} {\bf 185-189} 216 

\item Deutscher G and DeGennes P G 1969 {\it Superconductivity}
      ed R D Parks (New York: Marcel Dekker) vol II 1005

\item Deutscher G and Entin-Wohlman O 1978 
      {\it Phys.\ Rev.\ B} {\bf 17} 1249 

\item Deutscher G, Imry Y and Gunther L 1974
      {\it Phys.\ Rev.\ B} {\bf 10} 4598 

\item Dew-Hughes D 1987 {\it Phil.\ Mag.\ B} {\bf 55} 459 
\item Dew-Hughes D 1988 {\it Cryogenics} {\bf 28} 674     

\item Dimos D, Chaudhari P, Mannhardt J and LeGoues F K 1988
      {\it Phys.\ Rev.\ B} {\bf 61} 219 

\item Ding S Y, Wang G Q, Yao X X, Peng H T, Peng Q Y and Zhou S H 1995
      {\it Phys.\ Rev.\ B} {\bf 51} 9107 

\item Dobrosavljevi\'{c}-Gruji\'{c} L and Radovi\'{c} Z 1993
      {\it Supercond.\ Sci.\ Technol.} {\bf 6} 537  

\item Dodgson M J W and Moore M A 1995 
      {\it Phys.\ Rev.\ B} {\bf 51} 11887

\item Doettinger S G, Huebener R P, Gerdemann R, K\"uhle A, Anders S,
      Tr\"auble T G and Vill\'egier J C 1994
      {\it Phys.\ Rev.\ Lett.} {\bf 73} 1691 

\item Dolan G J 1974
      {\it J. \ Low Temp.\ Phys.} {\bf 15} 133 

\item Dolan G J, Chandrashekar G V, Dinger T R, Feild C and Holtzberg F
      1989a {\it Phys.\ Rev.\ Lett.} {\bf 62} 827    
\item Dolan G J, Holtzberg F, Feild C and Dinger T R
      1989b {\it Phys.\ Rev.\ Lett.} {\bf 62} 827    

\item Dom\'{\i}nguez D, Bulaevskii L N, Ivlev B, Maley M and Bishop A R
      1995 {\it Phys.\ Rev.\ Lett.} {\bf 74} 2579;  
           {\it Phys.\ Rev.\ B} {\bf 51} 15649  

\item Dong M, Marchetti M C, Middleton A A and Vinokur V M 1993
      {\it Phys.\ Rev.\ Lett.} {\bf 70} 662  

\item Dong S Y and Kwok H S 1993
      {\it Phys.\ Rev.\ B} {\bf 48} 6488  

\item Doniach S 1990 {\it High-Temperature Superconductivity}
      ed K S Bedell, D Coffey, D E Meltzer, D Pines and J R Schrieffer
      (New York: Addison-Wesley) p 406

\item Doria M M, Gubernatis J E and Rainer D 1989
      {\it Phys.\ Rev.\ B} {\bf 39} 9573               

\item Doria M M and de Oliveira I G 1994a
      {\it Phys.\ Rev.\ B} {\bf 49} 6205 

\item Doria M M and de Oliveira I G 1994b
      {\it Physica C} {\bf 235-240} 2577 

\item Dorin V V, Klemm R A, Varlamov A A, Buzdin A I and Livanov D V
      1993  {\it Phys.\ Rev.\ B} {\bf 48} 12951  

\item Dorosinskii L A, Indenbom M V, Nikitenko V I, Ossip'yan Yu A,
      Polyanskii A A and Vlasko-Vlasov V K 1992
      {\it Physica C} {\bf 203}  149  

\item Dorosinskii L A, Nikitenko V I and Polyanskii A A 1994
      {\it Phys.\ Rev.\ B} {\bf 50}  501  

\item Dorsey A T 1991 {\it Phys.\ Rev.\ B} {\bf 43} 7575 
\item Dorsey A T 1992 {\it Phys.\ Rev.\ B} {\bf 46} 8376
\item Dorsey A T 1995 {\it Phys.\ Rev.\ B} {\bf 51} 15329 
\item Dorsey A T, Huang M and Fisher M P A 1992
      {\it Phys.\ Rev.\ B} {\bf 45} 523

\item Doyle R A, Campbell A M and Somekh R E 1993
      {\it Phys.\ Rev.\ Lett.} {\bf 71} 4241         

\item Doyle R A, Seow W S, Johnson J D, Campbell A M, Berghuis P,
      Somekh R E, Evetts J E, Wirth G and Wiesner J  1995
      {\it Phys.\ Rev.\ B} {\bf 51} 12763   

\item * Doyle T B and Labusch R 1995               
      {\it Proc.\ Intnl.\ Cryogenics Materials Conf.\ 1994.
      Critical State in Superconductors} (Oct.\ 1994 in Honolulu)

\item Drobinin A V, Tsipenyuk Yu M and Kamentsev K E 1994
      (preprint for LT20, unpublished)

\item * Duarte A, Fernandez Righi E, Bolle C A, de la Cruz F, Gammel P L,
      Oglesby C S, Batlogg B and Bishop D J 1995 
      {\it Phys.\ Rev.\ Lett.} (submitted) 

\item Dumas J, Neminsky A, Thrane B P and Schlenker C 1994
      {\it Molecular Phys.\ Rep.} {\bf 8} 51 

\item Dur\'an C, Yazyi J, de la Cruz F, Bishop D J, Mitzi D B
      and Kapitulnik A 1991
      {\it Phys.\ Rev. B} {\bf 44} 7737

\item Ebner C and Stroud D 1985
      {\it Phys.\ Rev.\ B} {\bf 31} 165  

\item Efetov K B 1979 {\it Zh. Eksp. Teor. Fiz.} {\bf 76} 1781
                (1979 {\it Sov. Phys. JETP} {\bf 49} 905)

\item Eilenberger G 1964 {\it Z.\ Physik} {\bf 180} 32    
\item Eilenberger G 1966 {\it Phys.\ Rev.} {\bf 153} 584  
\item Eilenberger G 1967 {\it Phys. Rev.} {\bf 164} 628   
\item Eilenberger G 1968 {\it Z.\ Physik} {\bf 214} 195   

\item Ekin J W, Serin B and Clem J R 1974
      {\it Phys.\ Rev.\ B} {\bf 9} 912    
\item Ekin J W and Clem J R 1975
      {\it Phys.\ Rev.\ B} {\bf 12} 1753  

\item Eltsev Yu, Holm W and Rapp \"O 1994 
      {\it Phys.\ Rev.\ B} {\bf 49} 12333; {\it Physica C} {\bf 235-240} 2605

\item Eltsev Yu and Rapp \"O 1995 
      {\it Phys.\ Rev.\ B} {\bf 51} 9419

\item Enomoto Y, Katsumi K and Maekawa S 1993 {\it Physica C} {\bf 215} 51 

\item Erta\c{s} D and Kardar M 1994
      {\it Phys.\ Rev.\ Lett.} {\bf 73} 1703 

\item Essmann U 1971 {\it Superconductivity} F Chilton ed
      (Amsterdam: North Holland) p 38

\item Essmann U and Tr\"auble H 1967 {\it Phys. Lett. A} {\bf 24} 526

\item Essmann U and Tr\"auble H 1971 {\it Sci. Am.} {\bf 224} 75

\item Esquinazi P 1991 {\it J.\ Low Temp.\ Phys.} {\bf 85} 139 
\item Esquinazi P, Neckel H, Brandt E H and Weiss G 1986
      {\it J.\ Low Temp.\ Phys.} {\bf 64} 1        

\item Evetts J E and Glowacki B A 1988
      {\it Cryogenics} {\bf 28} 641   

\item Ezaki T and Irie F 1976
      {\it J.\ Phys.\ Soc.\ Japan} {\bf 40} 682   

\item F\`abrega L, Fontcuberta J, Civale L and Pi\~nol 1994a
     {\it Phys.\ Rev.\ B} {\bf 50} 1199  

\item F\`abrega L, Fontcuberta J, Mart\'{\i}nez B and Pi\~nol 1994b
     {\it Phys.\ Rev.\ B} {\bf 50} 3256  

\item F\`abrega L, Fontcuberta J and Pi\~nol 1995
     {\it Physica C} {\bf 245} 325  

\item F\"ahnle M 1977  {\it phys.\ stat.\ sol.\ (b)} {\bf 83} 433 
\item F\"ahnle M 1979  {\it phys.\ stat.\ sol.\ (b)} {\bf 92} K127 
\item F\"ahnle M 1981
      {\it J.\ Low Temp.\ Phys.} {\bf 46} 3;  
      {\it IEE Trans.\ Magn.} {\bf Mag-17} 1707 

\item F\"ahnle M and Kronm\"uller H 1978
      {\it J.\ Nucl.\ Mater.} {\bf 72} 249 

\item Farrell D E 1994 {\it Physical Properties of High Temperature
      Superconductors IV} ed D M Ginsberg (Singapore: World Scientific)

\item Farrell D E, Williams C M, Wolf S A, Bansal N P and Kogan V G 1988
     {\it Phys.\ Rev.\ Lett.} {\bf 61} 2805       

\item Farrell D E, Bonham S, Foster J, Chang Y C, Jiang P Z,
      Vandervoort K G, Lam D J and Kogan V G 1989
     {\it Phys.\ Rev.\ Lett.} {\bf 63} 782        

\item Farrell D E, Rice J P and Ginsberg D M 1991 
     {\it Phys.\ Rev.\ Lett.} {\bf 67} 1165

\item Farrell D E, Rice J P, Ginsberg D M and Liu Z D 1990
     {\it Phys.\ Rev.\ Lett.} {\bf 64} 1573

\item Farrell D E, Kwok W K, Welp U, Fendrich J and Crabtree G W 1995
     {\it Phys.\ Rev.\ B} {\bf 51} 9148  

\item Fehrenbacher R, Geshkenbein V B and Blatter G 1992
     {\it Phys.\ Rev.\ B} {\bf 45} 5450  

\item Feigel'man M V and Vinokur V M 1990
      {\it Phys.\ Rev.\ B} {\bf 41} 8986      

\item Feigel'man M V, Geshkenbein V B and Larkin A I 1990
      {\it Physica C} {\bf 167} 177         

\item Feigel'man M V, Geshkenbein V B, Larkin A I and Vinokur V M 1989
      {\it Phys.\ Rev.\ Lett.} {\bf 63} 2303   

\item Feigel'man M V, Geshkenbein V B, Larkin A I and Vinokur V M 1994
      {\it Physica C} {\bf 235-240} 3127    

\item * Feigel'man M V, Geshkenbein V B, Larkin A I and Vinokur V M 1995
      {\it Phys.\ Rev.\ Lett.} (submitted)

\item Feigel'man M V, Geshkenbein V B and Vinokur V M 1991
      {\it Phys.\ Rev. B} {\bf 43} 6263     

\item Feigel'man M V, Geshkenbein V B, Ioffe L B and Larkin A I 1993a
      {\it Phys.\ Rev.\ B} {\bf 48} 16641   

\item Feigel'man M V, Geshkenbein V B, Larkin A I and Levit S 1993b
      {\it Pis'ma Zh.\ Eksp.\ Teor.\ Fiz.} {\bf 57} 699 
      (1993 {\it Sov.\ Phys.\ JETP Lett.}    {\bf 57} 711)

\item Feinberg D 1992 {\it Physica C} {\bf 194} 126            
\item Feinberg D 1994 {\it J.\ Phys.\ III France} {\bf 4} 169  

\item Feinberg D and Ettouhami A M 1993a
      {\it Physica Scripta} {\bf T49} 159      
\item Feinberg D and Ettouhami A M 1993b
      {\it Int. J. Mod. Phys. B} {\bf 7} 2085  
\item Feinberg D, Theodorakis S and Ettouhami A M 1994
      {\it Phys.\ Rev.\ B} {\bf 49} 6285       

\item Feinberg D and Villard C 1990 {\it Mod. Phys. Lett. B} {\bf 4} 9

\item Fendrich J A, Kwok W K, Giapintzakis J, van der Beek C J, Vinokur V M,
      Fleshler S, Welp U, Viswanathan H K and Crabtree G W 1995
      {\it Phys.\ Rev.\ Lett.} {\bf 74} 1210 

\item Ferrari M J, Wellstood F C, Kingston J J and Clarke J 1991
      {\it Phys.\ Rev.\ Lett.} {\bf 67} 1346

\item Ferrari M J, Johnson M, Wellstood F C, Kingston J J, Shaw T J,
      and Clarke J 1994  
      {\it J.\ Low Temp.\ Phys.} {\bf 94} 15

\item Ferrell R A 1992 {\it Phys. Rev.} {\bf 68} 2524

\item Fetter A L 1967 {\it Phys. Rev.} {\bf 163} 390 
\item Fetter A L 1980 {\it Phys.\ Rev.\ B} {\bf 22} 1200 
\item Fetter A L 1994 {\it Phys.\ Rev.\ B} {\bf 50} 13695

\item Fetter A L and Hohenberg P C 1967 
      {\it Phys.\ Rev.} {\bf 159} 330

\item Fetter A L and Hohenberg P C 1969 {\it Superconductivity}
      ed R D Parks (New York: Marcel Dekker) vol II p 817

\item Field S, Witt J, Nori F and Ling Xinsheng 1995
      {\it Phys.\ Rev.\ Lett.} {\bf 74} 1206    

\item Fink H J and Haley S B 1991
      {\it Phys.\ Rev.\ Lett.} {\bf 66} 216     

\item Fink H J, Lopez A and Maynard R 1985
      {\it Phys.\ Rev.\ B} {\bf 25} 5237     

\item Fink H J, Rodriguez D and Lopez A 1988 
      {\it Phys.\ Rev.\ B} {\bf 38} 8767

\item Fink H J, Buisson O and Pannetier B 1991 
      {\it Phys.\ Rev.\ B} {\bf 43} 10144

\item Fiory A T 1971 {\it Phys.\ Rev.\ Lett.} {\bf 27} 501
\item Fiory A T 1973 {\it Phys.\ Rev.\ B} {\bf 7} 1881

\item Fiory A T, Hebard A F, Mankiewich P M and Howard R E 1988
      {\it Appl.\ Phys.\ Lett.} {\bf 52} 2165
\item Fiory A T, Hebard A F and Somekh  S 1978
      {\it Appl.\ Phys.\ Lett.} {\bf 32} 73  

\item Fischer K and Teichler H 1976 {\it Phys. Lett. A} {\bf 58} 402

\item Fischer K H 1991 {\it Physica C} {\bf 178} 161

\item Fischer K H 1992a {\it Physica C} {\bf 193} 401
\item Fischer K H 1992b {\it Physica C} {\bf 200} 23   

\item Fischer K H 1995a      
      {\it Superconductivity Review} {\bf 1} 153-206

\item * Fischer K H 1995b (preprint: Influence of disorder on KT transition)
      (PRL ?) 

\item Fischer K H and Nattermann T 1991
      {\it Phys.\ Rev.\ B} {\bf 43} 10372

\item Fischer {\O}, Brunner O, Antognazza L, Mi\'eville L and Triscone L-M
      1992 {\it Phys.\ Scr.\ T} {42} 46  

\item Fisher D S 1980 {\it Phys.\ Rev.\ B} {\bf 22} 1190

\item Fisher D S 1992 {\it Phenomenology and Application of
      High-Temperature Superconductors} ed K S Bedell, M Inui,
      D E Meltzer, J R Schrieffer and S Doniach
      (New York: Addison-Wesley) p 287

\item Fisher D S, Fisher M P A and Huse D A 1991
      {\it Phys.\ Rev.\ B} {\bf 43} 130

\item Fisher L M, Il'in N V, Podlevskikh N A, and Zakharchenko S I 1990
      {\it Solid State Commun.} {\bf 73} 687  

\item Fisher L M, Il'in N V, Voloshin I F,
      Makarov N M, Yampol'skii V A, Perez Rodriguez F and Snyder R L 1993
      {\it Physica C} {\bf 206} 195 

\item Fisher L M, Voloshin I F, Gorbachev V S, Savel'ev S E and
      Yampol'skii V A 1995           
      {\it Physica C} {\bf 245} 231

\item Fisher M P A 1989 {\it Phys.\ Rev.\ Lett.} {\bf 62} 1415

\item Fleshler S, Kwok W-K, Welp U, Vinokur V M, Smith M K, Downey J
      and Crabtree G W 1993 {\it Phys.\ Rev.\ B} {\bf 47} 14448 

\item Flippen R B, Askew T R, Fendrich J A and van der Beek C J 1995
      {\it Phys.\ Rev.\ Lett.} (submitted) 

\item Forgan E M 1973 {\it J.\ Phys. F: Metal Physics} {\bf 3} 1429 

\item Forgan E M, Cubitt R, Yethiraj M, Christen D K, Paul D M, Lee S L
      and Gammel P 1995 {\it Phys.\ Rev.\ Lett.} {\bf 74} C1697 

\item Forgan E M and Gough C E 1968
      {\it Phys. Lett.} {\bf 26} 602  

\item Forgan E M, Paul D McK, Mook H A, Timmins P A, Keller H, Sutton S
      and J S Abell 1990
      {\it Nature} {\bf 343 } 735  

\item Forkl A, Dragon T and Kronm\"uller H 1991a  
            {\it J.\ Appl.\ Phys.} {\bf 67} 3047
\item Forkl A, Habermeier H-U, Leibold B, Dragon T and
            Kronm\"uller H 1991b {\it Physica C} {\bf 180} 155
\item Forkl A 1993 {\it Physica Scripta T} {\bf 49} 148
\item Forkl A and Kronm\"uller H 1994 {\it Physica C} {\bf 228} 1

\item Fortini A and Paumier E 1972 {\it Phys.\ Rev.\ B} {\bf 5} 1850 
\item Fortini A and Paumier E 1976 {\it Phys.\ Rev.\ B} {\bf 14} 55  
\item Fortini A, Haire A and Paumier E 1980
      {\it Phys.\ Rev.\ B} {\bf 21} 5065

\item Frankel D J 1979 {\it J.\ Appl.\ Phys.} {\bf 50} 5402

\item Franz M and Teitel S 1994 {\it Phys.\ Rev.\ Lett.} {\bf 73} 480 
\item Franz M and Teitel S 1995 {\it Phys.\ Rev.\ B} {\bf 51} 6551
\item Freimut A, Hohn C and Galffy M 1991
      {\it Phys.\ Rev.\ B} {\bf 44} 10396      

\item Frey E, Nelson D R and Fisher D S 1994
      {\it Phys.\ Rev.\ B} {\bf 49} 9723       

\item Frey T, Mannhardt J, Bednorz J G and Williams E J 1995
      {\it Phys.\ Rev.\ B} {\bf 51} 3257  

\item Friedel J 1989
      {\it J.\ Phys.\ Cond.\ Mat.} {\bf 1} 7757 

\item Friesen M 1995   
      {\it Phys.\ Rev.\ B} {\bf 51} 632, 12786

\item Fruchter A, Malozemoff A P, Campbell I A, Sanchez J, Konczykowski M,
      Griessen R and Holtzberg F 1991 {\it Phys.\ Rev.\ B} {\bf 43} 8709

\item Fruchter A, Prost D and Campbell I A 1994
      {\it Physica C} {\bf 235-240} 249 

\item Fu C M, Moshchalkov V V, Boon W, Temst K, Bruynseraede Y,
      Jakob G, Hahn T and Adrian H 1993a
      {\it Physica C} {\bf 205} 111  

\item Fu C M, Moshchalkov V V, Rosseel E, Baert M, Boon W and
      Bruynseraede Y 1993b
      {\it Physica C} {\bf 206} 110  

\item Fulde P and Keller J 1982              
      {\it Superconductivity in Ternary Compounds II: Superconductivity
      and Magnetism} ed M B Maple and {\O} Fischer (Berlin: Springer) p 249

\item Fujita A and Hikami S 1995
      {\it Phys.\ Rev.\ B} {\bf 51} 16259 

\item Gaitan F 1995 {\it Phys.\ Rev.\ B} {\bf 51} 9061 

\item Gal'perin Yu M and Sonin E B 1976        
      {\it Fiz.\ Tverd.\ Tela} {\bf 18} 3034
      (1976 {\it Sov.\ Phys.\ Solid State} {\bf 18} 1768)

\item Gammel P L, Bishop D J, Dolan G J, Kwo J R, Murray C A,
      Schneemeyer L F and Wacsczk J V 1987
     {\it Phys.\ Rev.\ Lett.} {\bf 59} 2592    
\item Gammel P L, Bishop D J, Rice J P and Ginsberg D M 1992a
     {\it Phys.\ Rev.\ Lett.} {\bf 68} 3343    
\item Gammel P L, Dur\'an C A, Bishop D J, Rice J P, Kogan V G, Ledvij M,
      Simonov A Yu, Rice J P and Ginsberg D M 1992b
      {\it Phys.\ Rev.\ Lett.} {\bf 69} 3808    
\item Gammel P L, Schneemeyer L F, Waszczak J V and Bishop D J 1988
     {\it Phys.\ Rev.\ Lett.} {\bf 61} 1666    
\item Gammel P L, Schneemeyer L F and Bishop D J 1991
     {\it Phys.\ Rev.\ Lett.} {\bf 66} 953     

\item Gao L, Xue Y Y, Chen F, Xiong Q, Meng R L, Ramirez D, Chu C W,
      Eggert J H and Mao H K 1994 {\it Phys.\ Rev.\ B} {\bf 50} 4260 

\item Gao Z X, Osquiguil, Maenhoudt M, Wuyts B, Libbrecht S and
      Bruynseraede Y 1993 {\it Phys.\ Rev.\ Lett.} {\bf 66} 953 

\item Garc\'{\i}a A, Zhang X X, Tejada J, Manzel M and Bruchlos H 1994
      {\it Phys.\ Rev.\ B} {\bf 50} 9439  

\item Garten F, White W R and Beasley M R 1995
      {\it Phys.\ Rev.\ B} {\bf 51} 1318  

\item Gerber A, Li J N, Tarnawski Z,  Franse J J M and Menovsky A A 1993
      {\it Phys.\ Rev.\ B} {\bf 47} 6047  

\item Genenko Yu A 1992
      {\it Sverkhprovodimost'} {\bf 5} 1402      
      (1992 {\it Superconductivity} {\bf 5} 1376)
\item Genenko Yu A 1993 {\it Physica C} {\bf 215} 343 

\item Genenko Yu A 1994 
      {\it Pis'ma Zh.\ Eksp.\ Teor.\ Fiz.} {\bf 59} 807
      (1994 {\it Sov.\ Phys.\ JETP Lett.}    {\bf 59} 841);
      {\it Phys.\ Rev.\ B} {\bf 49} 6950  

\item Genenko Yu A 1995                  
      {\it Phys.\ Rev.\ B} {\bf 51} 3686 

\item Genenko Yu A and Medvedev Yu V  1992 
      {\it Phys.\ Lett.\ A} {\bf 167} 427

\item Genenko Yu A, Medvedev Yu V and Shuster G V 1993
      {\it Pis'ma Zh.\ Eksp.\ Teor.\ Fiz.} {\bf 57} 705
      (1993 {\it Sov.\ Phys.\ JETP Lett.} {\bf 57} 717) 

\item Genoud J-Y, Triscone G, Junod A, Tsuakamoto T and M\"uller J 1995
      {\it Physica C} {\bf 242} 143 

\item Gerh\"auser W, Ries G, Neum\"uller H-W, Schmidt W, Eibl O,
      Saemann-Ischenko G and Klaum\"unzer S 1992
      {\it Phys.\ Rev.\ Lett.} {\bf 68} 879          

\item Geshkenbein V B, Feigel'man M V and Vinokur V M 1991
      {\it Physica C} {\bf 185-189} 2511 

\item Geshkenbein V B, Ioffe L B and Larkin A I 1993
      {\it Phys.\ Rev.\ B} {\bf 48} 9917  

\item Geshkenbein V B and Larkin A I 1994
      {\it Phys.\ Rev.\ Lett.} {\bf 73} C609  

\item Geshkenbein V B, Vinokur V M and Fehrenbacher R 1991
      {\it Phys.\ Rev.\ B} {\bf 43} 3748

\item Ghinovker M, Sandomirsky V B and Shapiro B Ya 1995
      {\it Phys.\ Rev.\ B} {\bf 51} 8404  

\item Giaever I 1966 {\it Phys.\ Rev.\ Lett.} {\bf 16} 460 

\item Giura M, Marcon R, Fastampa R and Silva E 1992 
      {\it Phys.\ Rev.\ B} {\bf 45} 7387

\item Giamarchi T and Le Doussal P 1994         
      {\it Phys.\ Rev.\ Lett.} {\bf 72} 1530    
\item * Giamarchi T and Le Doussal P 1995
      {\it Phys.\ Rev.\ B} {\bf 52} (in print)  

\item Giapintzakis J, Neiman R L, Ginsberg D M and Kirk M A    
      1994 {\it Phys.\ Rev.\ B} {\bf 50} 16001 

\item Gilchrist J 1972 {\it J.\ Phys.\ D: Appl.\ Phys.} {\bf 5} 2252
\item Gilchrist J 1990a {\it Supercond.\ Sci.\ Technol.} {\bf 3} 93
\item Gilchrist J 1990b {\it Solid State Commun.} {\bf 75} 225 
\item Gilchrist J 1994a {\it Physica C} {\bf 219} 67   
\item Gilchrist J 1994b {\it Supercond.\ Sci.\ Technol.} {\bf 7} 849
\item Gilchrist J and Dombre T 1994 
      {\it Phys.\ Rev.\ B} {\bf 49} 1466
\item Gilchrist J and Konczykowski M 1990 {\it Physica C} {\bf 168} 123
\item Gilchrist J and Konczykowski M 1993 {\it Physica C} {\bf 212} 43
\item Gilchrist J and van der Beek C J 1994
      {\it Physica C} {\bf 231} 147  

\item * Gingras M J P 1995 {\it preprint} Finite-scale remnants of a
      KT transition                               
\item Ginsberg D M and Manson J T 1995
      {\it Phys.\ Rev.\ B} {\bf 51} 515 

\item Ginzburg V L 1960 {\it Fiz.\ Tverd.\ Tela} {\bf 2} 2031
      (1961 {\it Sov.\ Phys.\ Sol.\ State} {\bf 2} 1824) 

\item Ginzburg V L and Landau L D 1950
      {\it Zh.\ Eksp.\ Teor.\ Fiz.} {\bf 20} 1064 [English translation
      in {\it Men of Physics: L D Landau} ed D ter Haar
      (New York: Pergamon 1965)] vol 1 pp 138-167

\item Giroud M, Buisson O, Wang Y Y and Pannetier B 1992
      {\it J.\ Low Temp.\ Phys.} {\bf 87} 683 

\item Glazman L and Fogel' N Ya 1984a {\it Fiz.\ Nizk.\ Temp.} {\bf 10} 95
      (1984 {\it Sov.\ J.\ Low Temp.\ Phys.} {\bf 10} 51)  
\item Glazman L and Fogel' N Ya 1984b {\it Fiz.\ Tverd.\ Tela} {\bf 26} 800
      (1984 {\it Sov.\ Phys.\ Sol.\ State} {\bf 26} 482)

\item Glazman L and Koshelev A 1990
      {\it Zh.\ Eksp.\ Teor.\ Fiz.} {\bf 97} 1371 
      (1990 {\it Sov.\ Phys.\ JETP} {\bf 70} 774) 
\item Glazman L and Koshelev A 1991a {\it Phys.\ Rev.\ B} {\bf 43} 2835
\item Glazman L and Koshelev A 1991b {\it Physica C} {\bf 173} 180

\item Glyde H R, Moleko L K and Findeisen P 1992a
      {\it Phys.\ Rev.\ B} {\bf 45} 2409
\item Glyde H R, Lesage F and Findeisen P 1992b
      {\it Phys.\ Rev.\ B} {\bf 46} 9175

\item Goldfarb R B, Lelental M, Thompson C A 1991
    {\it Magnetic Susceptibility of Superconductors and Other Spin Systems}
      ed R A Hein, T L Francavilla and D H Liebenberg
      (New York: Plenum Press) p 49  

\item Golosovsky M, Naveh Y and Davidov D 1991  
      {\it Physica C} {\bf 180} 164
\item Golosovsky M, Naveh Y and Davidov D 1991  
      1992 {\it Phys.\ Rev.\ B} {\bf 45} 7495
\item Golosovsky M, Ginodman V, Shaltiel D, Gerh\"auser W  
      and Fischer P 1993 {\it Phys.\ Rev.\ B} {\bf 47} 9010
\item Golosovsky M, Tsindlekht M, Chayet H and Davidov D 1994
      {\it Phys.\ Rev.\ B} {\bf 50} 470        
                             (erratum {\bf 51} 12062)
\item Golosovsky M, Snortland H J and Beasley M R 1995
      {\it Phys.\ Rev.\ B} {\bf 51} 6462 

\item Golubov A A and Hartmann U 1994
      {\it Phys.\ Rev.\ Lett.} {\bf 72} 3602  

\item Golubov A A and Logvenov G Yu 1995
      {\it Phys.\ Rev.\ B} {\bf 51} 3696 

\item G\"om\"ory F and Tak\'acs S  1993
      {\it Physica C} {\bf 212} 297

\item Goodman B B 1966 {\it Rep.\ Prog.\ Phys.} {\bf 29} 445

\item Gordeev S N, Jahn W, Zhukov A A, K\"upfer H and Wolf T 1994
      {\it Phys.\ Rev.\ B} {\bf 49} 15420

\item Gor'kov L P 1959a
      {\it Zh.\ Eksp.\ Teor.\ Fiz.} {\bf 36} 1918
      (1959 {\it Sov.\ Phys.\ JETP } {\bf 9} 1364)     
\item Gor'kov L P 1959b
      {\it Zh.\ Eksp.\ Teor.\ Fiz.} {\bf 37} 833
      (1960 {\it Sov.\ Phys.\ JETP } {\bf 10} 593)     

\item Gor'kov L P and Kopnin N B 1975
       {\it Usp.\ Fiz.\ Nauk} {\bf 116} 413
      (1976 {\it Sov.\ Phys.\ - Uspechi} {\bf 18} 496)

\item Gorlova I G and Latyshev Yu I 1990
      {\it Pis'ma Zh.\ Eksp.\ Teor.\ Fiz.} {\bf 51} 197
      (1990 {\it Sov.\ Phys.\ JETP Lett.} {\bf 51} 224)     

\item Gorter C J and Casimir H B G 1934  
      {\it Z.\ Tech.\ Phys.} {\bf 15} 539

\item Gough C E and Exon N J 1994        
      {\it Phys.\ Rev.\ B} {\bf 50} 488

\item Grant P D, Denhoff M W, Xing W, Brown P, Govorkov S, Irwin J C,
      Heinrich B, Zhou H, Fife A A and Cragg A R 1994
      {\it Physica C} {\bf 229} 289           

\item Gray K E, Kampwirth R T and Farrell D E 1990
      {\it Phys.\ Rev.\ B} {\bf 41} 819
\item Gray K E, Kampwirth R T, Miller D J, Murduck J M, Hampshire D,
      Herzog R and Weber H W 1991 {\it Physica C} {\bf 174} 340 
\item Gray K E, Kim D H, Veal B W, Seidler G T, Rosenbaum T F and
      Farrell D E 1992 {\it Phys.\ Rev.\ B} {\bf 45} 10071  

\item Graybeal J M, Luo J and White W R 1994
      {\it Phys.\ Rev.\ B} {\bf 49} 12923  

\item Grier D G, Murray C A, Bolle C A, Gammel P L, Bishop D J, Mitzi D B
      and Kapitulnik A 1991 {\it Phys.\ Rev.\ Lett.} {\bf 66} 2270
\item Griessen R 1990 {\it Phys.\ Rev.\ Lett.} {\bf 64} 1674
\item Griessen R 1991 {\it Physica C} {\bf 172} 441

\item Griessen R, Lensink J G and Schnack H G 1991
      {\it Physica C} {\bf 185-189} 337                   

\item Griessen R, Lensink J G, Schr\"oder T A M and Dam B 1990
      {\it Cryogenics} {\bf 30} 536

\item Griessen R, Wen H-H, van Dalen A J J , Dam B, Rector J,
      Schnack H G, Libbrecht S, Osquiguil E and Bruynseraede Y 1994
      {\it Phys.\ Rev.\ Lett.} {\bf 72} 1910 

\item Grigorieva I V 1994
      {\it Supercond.\ Sci.\ Technol.} {\bf 7} 161  

\item Grigorieva I V, Steeds J W and Sasaki K 1993
      {\it Phys.\ Rev.\ B} {\bf 48} 16865 

\item Grigorieva I V, Steeds J W, Balakrishnan G  and Paul D M 1995
      {\it Phys.\ Rev.\ B} {\bf 51} 3765 

\item Grishin A M, Martynovich A Yu and Yampol'skiy S V 1990
      {\it Zh.\ Eksp.\ Teor.\ Fiz.} {\bf 97} 1930
      (1990 {\it Sov.\ Phys.\ JETP} {\bf 70} 1089)

\item Grishin A M, Martynovich A Yu and Yampol'skiy S V 1992
      {\it Zh.\ Eksp.\ Teor.\ Fiz.} {\bf 101} 649
      (1992 {\it Sov.\ Phys.\ JETP} {\bf 74} 345)

\item Grover A K, Kumar R, Chaddah P, Subramanian C K and 
      Sankaranarayanan Y 1990 {\it Physica C} {\bf 170} 431

\item Grover A K, Ramakrishnan S, Kumar R, Paulose P L, Chandra G,
      Malik S K and Chaddah P 1992    
      Sankaranarayanan Y 1992 {\it Physica C} {\bf 192} 372

\item Gross R and Koelle D 1994 
      {\it Rep.\ Prog.\ Phys.} {\bf 57} 651-741          

\item Gugan D 1994 {\it Physica C} {\bf 233} 165 

\item Gugan D and Stoppard O 1993 {\it Physica C} {\bf 213} 109  

\item Gupta A, Kopelevich Y, Ziese M, Esquinazi P, Fischer P,
      Schulz F I and Braun H F 1993
      {\it Phys.\ Rev.\ B} {\bf 48} 6359                

\item Guinea F 1990 {\it Phys.\ Rev.\ B} {\bf 42} 6244 
\item Guinea F and Pogorelov Yu 1995    
      {\it Phys.\ Rev.\ Lett.} {\bf 74} 462

\item Gurevich A 1990a {\it Phys.\ Rev.\ Lett.} {\bf 65} 3197 
\item Gurevich A 1990b {\it Phys.\ Rev.\ B} {\bf 42} 4852 
\item Gurevich A 1992a {\it Phys.\ Rev.\ B} {\bf 46} 3187, E14329 
\item Gurevich A 1992b {\it Phys.\ Rev.\ B} {\bf 46} 3638;
      {\it Supercond.\ Sci.\ Technol.} {\bf 5} 383 
\item Gurevich A 1993 {\it Phys.\ Rev.\ B} {\bf 48} 12857 
\item Gurevich A 1995a {\it Physica C} {\bf 243} 191 
\item Gurevich A 1995b
      {\it Int.\ J.\ Mod.\ Phys.\ B} (in print) 

\item Gurevich A and Cooley L D 1994 {\it Phys.\ Rev.\ B} {\bf 50} 13563
\item Gurevich A, K\"upfer H, Runtsch B and Meier-Hirmer R 1991
      {\it Phys.\ Rev.\ B} {\bf 44} 12090   
\item Gurevich A and K\"upfer H 1993 {\it Phys.\ Rev.\ B} {\bf 48} 6477
\item Gurevich A and Brandt E H 1994 {\it Phys.\ Rev.\ Lett.} {\bf 73} 178
\item Gurevich A and Brandt E H 1995 (unpublished) 

\item Gutlyanskii E D 1994
      {\it Pis'ma Zh.\ Eksp.\ Teor.\ Fiz.} {\bf 59} 459 
      (1994 {\it Sov.\ Phys.\ JETP Lett.}    {\bf 59} 459)

\item Gygi F and Schl\"uter M 1991      
      {\it Phys.\ Rev.\ B} {\bf 43} 7609

\item Gyorgy E M, van Dover R B, Jackson K A, Schneemeyer L F and
      Waszczak J V 1989  {\it Appl.\ Phys.\ Lett.} {\bf 55} 283   

\item Habermeier H-U and Kronm\"uller H 1977 
      {\it Appl.\ Phys.} {\bf 12} 297

\item Hackenbroich G and Scheidl S 1991 {\it Physica C} {\bf 181} 163

\item Hagen C W and Griessen R 1989
      {\it Phys.\ Rev.\ Lett.} {\bf 62} 2857  

\item Hagen S J, Smith A W, Rajeswari M, Peng J L, Li Y Z,
      Greene R L, Mao S N, Xi X X, Bhattacharya S, Li Q and Lobb C J
      1993  {\it Phys.\ Rev.\ B} {\bf 47} 1064  

\item Halbritter J 1992  {\it Phys.\ Rev.\ B} {\bf 46} 14861; 
      {\it J.\ Appl.\ Phys.} {\bf 71} 339                     
\item Halbritter J 1993  {\it Phys.\ Rev.\ B} {\bf 48} 9735   

\item Haley S B 1995  
      {\it Phys.\ Rev.\ Lett.} {\bf 74} 3261

\item Halperin B I and Nelson D R 1979 {\it Phys.\ Rev.\ B} {\bf 19} 2457

\item Hao Z 1993 {\it Phys.\ Rev.\ B} {\bf 48} 16822  

\item Hao Z and Clem J R 1991a {\it IEEE Trans. Magn.} {\bf 27} 1086
\item Hao Z and Clem J R 1991b {\it Phys.\ Rev.\ B} {\bf 43} 7622 
\item Hao Z and Clem J R 1992 {\it Phys.\ Rev.\ B} {\bf 46} 5853 %
\item Hao Z and Clem J R 1993 {\it Phys.\ Rev.\ Lett.} {\bf 71} C301 

\item Hao Z, Hu C-R and Ting C S 1995    
      {\it Phys.\ Rev.\ B} {\bf 51} 9387 

\item Hao Z, Clem J R, McElfresh M W, Civale L, Malozemoff A P
            and Holtzberg F 1991 {\it Phys.\ Rev.\ B} {\bf 43} 2844 

\item Hao Z and Hu C-R 1993 {\it Phys.\ Rev.\ B} {\bf 48} 16818 

\item Harada K, Matsuda T, Kasai H, Bonevich J E, Yoshida T, Kawabe U
      and Tonomura A 1993
      {\it Phys.\ Rev.\ Lett.} {\bf 71} 3371 

\item Hardy V, Simon Ch, Provost J and Groult D 1993
      {\it Physica C} {\bf 205} 371    

\item Hardy W N, Bonn D A, Morgan D C, Liang R and Zhang K 1993
      {\it Phys.\ Rev.\ Lett.} {\bf 70} 3999  

\item Harris J M, Ong N P and Yan Y F 1993
      {\it Phys.\ Rev.\ Lett.} {\bf 71} 1455 

\item Harris J M, Ong N P, Gagnon R and Taillefer J 1995a
      {\it Phys.\ Rev.\ Lett.} {\bf 74} 3684 

\item Harris J M, Ong N P, Matl P, Gagnon R, Taillefer J, Kimura T
      and Kitazawa K 1995b
      {\it Phys.\ Rev.\ Lett.} {\bf 74} 3684 

\item Harris J M, Yan Y F, Tsui O K C, Matsuda Y and Ong N P 1994
      {\it Phys.\ Rev.\ Lett.} {\bf 73} 1711    

\item Harshman D R, Schneemeyer L F, Waszczak J V, Aeppli G, Cava R J,
      Batlogg B, Rupp L W, Ansaldo E J and Williams D Ll 1989
      {\it Phys.\ Rev.\ B} {\bf 39} 851  

\item Harshman D R, Kleiman R N, Inui M, Espinosa G P, Mitzi B,
      Kapitulnik A, Pfiz T and Williams D Ll  1991
      {\it Phys.\ Rev.\ Lett.} {\bf 67} 3152  

\item Harshman D R, Brandt E H, Fiory A T, Inui M, Mitzi B,
      Schneemeyer L F and Waszczak J V 1993
      {\it Phys.\ Rev.\ B} {\bf 47} 2905  

\item Hasegawa S and Kitazawa K 1990
      {\it Jap.\ J. Appl.\ Phys.} {\bf 29} 434  

\item Hasegawa S, Matsuda T, Endo J, Osakabe N, Igarashi M,
      Kobayashi T, Naito M, Tonomura A and Aoki R 1991
      {\it Phys.\ Rev.\ B} {\bf 43} 7631  

\item Hebard A F and Fiory A T 1982
      {\it Physica B} {\bf 109\&110}  1637  

\item Heinzel Ch, H\"aring J, Yiao Z L and Ziemann P 1994
      {\it Physica C} {\bf 235-240} 2679  

\item Helfand E and Werthamer N R 1966
      {\it Phys. Rev.} {\bf 147} 288     

\item Hellerquist M C, Ryu S, Lombardo L W and Kapitulnik A 1994
      {\it Physica C} {\bf 230} 170      

\item Hensel B, Grivel J-C, Jeremie A, Perin A, Pollini A and Fl\"ukiger R
      1993 {\it Physica C} {\bf 205} 329 

\item Hensel B, Grasso G and Fl\"ukiger R 1995
      {\it Phys.\ Rev.\ B} {\bf 51} 15456   

\item Hergt R, Andr\"a W and Fischer K 1991  
      {\it Physica C} {\bf 185-189} 2197

\item Herlach D, Majer G, Rosenkranz J, Schmolz M, Schwarz W,
      Seeger A, Templ W, Brandt E H, Essmann U, F\"urderer K
      and Gladisch M 1990 {\it Hyperfine Interactions} {\bf 63-65} 41

\item Hess H F, Robinson R B and Waszczak J V  1990
      {\it Phys.\ Rev.\ Lett.} {\bf 64} 2711
\item Hess H F, Murray C A and Waszczak J V  1992
      {\it Phys.\ Rev.\ Lett.} {\bf 69} 2138
\item Hess H F, Murray C A and Waszczak J V  1994
      {\it Phys.\ Rev.\ B} {\bf 50} 16528            

\item Hetzel R E, Sudb{\o} A and Huse D A {\it Phys.\ Rev.\ Lett.} {\bf 69} 518

\item Hikami S and Larkin A I 1988    
      {\it Mod.\ Phys.\ Lett.\ B} {\bf 2} 693
\item Hikami S, Fujita A and Larkin A I 1991
      {\it Phys.\ Rev.\ B} {\bf 44} 10400

\item Hilzinger H R 1977 {\it Phil.\ Mag.} {\bf 36} 225

\item Hindmarsh M B and Kibble T W B 1995
      {\it Rep.\ Prog.\ Phys.} {\bf 58} 477

\item Hochmut H and Lorenz M 1994 {\it Physica C} {\bf 220} 209 

\item Hocquet T, Mathieu P and Simon Y 1992
      {\it Phys.\ Rev.\ B} {\bf 46} 1061  

\item Hofmann S and K\"ummel R 1993
      {\it Phys.\ Rev.\ Lett.} {\bf 70} 1319  

\item Hohn C, Galffy M, Dascoulidou A, Freimuth A, Soltner H and
      Poppe U 1991  {\it Z.\ Phys.\ B} {\bf 85} 161 

\item Holzapfel B, Kreiselmeyer G, Kraus M, Saemann-Ischenko G, Bouffard S,
      Klaum\"unzer S and Schultz L 1993
      {\it Phys.\ Rev.\ B} {\bf 48} 600  
\item Hopfeng\"artner R, Hensel B and Saemann-Ischenko G 1991
      {\it Phys.\ Rev.\ B} {\bf 44} 741  
\item Hopfeng\"artner R, Leghissa M, Kreiselmeyer G, Holzapfel B,
      Schmitt P and Saemann-Ischenko G 1993
      {\it Phys.\ Rev.\ B} {\bf 47} 5992 

\item Horovitz B 1991 {\it Phys.\ Rev.\ Lett.} {\bf 67} 378   
\item Horovitz B 1992 {\it Phys.\ Rev.\ B} {\bf 45} 12632
\item Horovitz B 1993 {\it Phys.\ Rev.\ B} {\bf 47} 5947, 5964
\item Horovitz B 1994 {\it Phys.\ Rev.\ Lett.} {\bf 72} C1569 
\item Horovitz B 1995 {\it Phys.\ Rev.\ B} {\bf 51} 3989
\item Houghton A and Moore M A 1988
      {\it Phys.\ Rev.\ Lett.} {\bf 60} 1207     
\item Houghton A, Pelcovits R A and Sudb{\o} A 1989
      {\it Phys.\ Rev.\ B} {\bf 40} 6763         
\item Houghton A, Pelcovits R A and Sudb{\o} A 1990
      {\it Phys.\ Rev.\ B} {\bf 42} 906 

\item Houlrik J, Jonsson A and Minnhagen P 1994 
      {\it Phys.\ Rev.\ B} {\bf 50}  3953

\item Hsu T C 1993 {\it Physica C} {\bf 213} 305 

\item Hsu J W P and Kapitulnik A 1992 {\it Phys.\ Rev.\ B} {\bf 45} 4819 

\item Hu C-R and Thompson R S 1972 {\it Phys.\ Rev.\ B} {\bf 6} 110

\item Hu J and MacDonald A H 1993 {\it Phys.\ Rev.\ Lett.} {\bf 71} 432

\item Huang C Y, Shapira Y, McNiff E J, Peters P N, Schwartz B B,
      Wu M K, Shull R D and Chiang C K    1988  
      {\it Mod.\ Phys.\ Lett.\ B} {\bf 2} 869, 1027

\item Huang Y, Lin Y F, Wang M J and Wu M K 1993
      {\it J.\ Appl.\ Phys.} {\bf 73} 3580

\item Hubermann B A and Doniach S 1979
      {\it Phys.\ Rev.\ Lett.} {\bf 43} 950

\item Hubert A  1967 {\it phys.\ stat.\ sol.} {\bf 24} 669
\item Hubert A  1972 {\it phys.\ stat.\ sol.\ (b)} {\bf 53} 147

\item Huebener R P 1979 {\it Magnetic Flux Structures in Superconductors}
         (Berlin: Springer)            
\item Huebener R P 1984   
      {\it Rep.\ Prog.\ Phys.} {\bf 47} 175

\item Huebener R P 1995              
      {\it Supercond.\ Sci.\ Technol.} {\bf 8} 189

\item Huebener R P and Brandt E H 1992   
      {\it Physica C} {\bf 202} 321
\item Huebener R P, Kober F, Ri H-C, Knorr K, Tsuei C C, Chi C C and
      Scheuermann M R 1991 {\it Physica C} {\bf 181} 345 
\item Huebener R P, Ri H-C, Gross R and Gollnik F 1993
      {\it Physica C} {\bf 209} 27  
\item Huebener R P, Ustinov A and Kaplunenko V 1990 
      {\it Phys.\ Rev.\ B} {\bf 42} 4831

\item Hug H J, Moser A, Parashikov I, Stiefel B, Fritz O, G\"untherodt H-J
      and Thomas H 1994        
      {\it Physica C} {\bf 235-240} 2695

\item Hulm J K and Matthias B T 1980 
      {\it Science} {\bf 208} 881

\item H\"unneckes C, Bohn H G, Schilling W and Schulz H 1994
      {\it Phys.\ Rev.\ Lett.} {\bf 72} 2271 

\item Huse D A 1992
      {\it Phys.\ Rev.\ B} {\bf 46} 12230 

\item Huse D A, Henley C L and Fisher D S 1985
      {\it Phys.\ Rev.\ Lett.} {\bf 55} 2924

\item Huse D A and Majumdar S N  1993
      {\it Phys.\ Rev.\ Lett.} {\bf 71} 2473  

\item Hussey N E, Carrington A, Cooper J R and Sinclair D C 1994
      {\it Physica C} {\bf 235-240}  2711 

\item Hwa T, Nelson D and Vinokur V M 1993a
      {\it Phys.\ Rev.\ B} {\bf 48} 1167     
\item Hwa T, Le Doussal P, Nelson D and Vinokur V M 1993b
      {\it Phys.\ Rev.\ B} {\bf 48} 1167     

\item Hyun O B, Clem J R and Finnemore D K 1989  
      {\it Phys.\ Rev.\ B} {\bf 40} 175

\item Ichioka M 1995 {\it Phys.\ Rev.\ B} {\bf 51} 9423 

\item Iga F, Grover A K, Yamaguchi Y, Nishihara Y, Goyal N and Bhat S V 1995
      {\it Phys.\ Rev.\ B} {\bf 51} 8521 

\item Ihle D 1971 {\it phys.\ stat.\ sol.\ (b)} {\bf 47} 423

\item Ikeda R 1992a {\it Physica C} {\bf 201} 386 
\item Ikeda R 1992b {\it Phys.\ Rev.\ B} {\bf 46} 14842

\item Ikeda R, Ohmi T and Tsuneto T 1990
      {\it J.\ Phys.\ Soc.\ Japan} {\bf 59} 1740  
\item Ikeda R, Ohmi T and Tsuneto T 1991a
      {\it Physica C} {\bf 185-189} 1563  
\item Ikeda R, Ohmi T and Tsuneto T 1991b
      {\it J.\ Phys.\ Soc.\ Japan} {\bf 60} 1051  
\item Ikeda R, Ohmi T and Tsuneto T 1992
      {\it J.\ Phys.\ Soc.\ Japan} {\bf 61} 254   

\item Ikuta H, Hirota N, Nakayama Y, Kishio K and Kitazawa K 1993
      {\it Phys.\ Rev.\ Lett.} {\bf 70} 2166  

\item Ikuta H, Hirota N, Kishio K and Kitazawa K 1994
      {\it Physica C} {\bf 235-240} 237 

\item Indenbom M V and Brandt E H 1994        
      {\it Phys.\ Rev.\ Lett.} {\bf 73} C1731

\item Indenbom M V, D'Anna G D, Andr\'e M-O, Benoit W, Kronm\"uller H,
      Li T W and Kes P H 1994a
      {\it Critical Currents in Superconductors} ed H W Weber  
      (Singapore: World Scientific) p 327  

\item Indenbom M V, D'Anna G D, Andr\'e M-O, Kabanov V V and Benoit W
      1994b {\it Physica C} {\bf 235-240}  201 

\item Indenbom M V, Forkl A, Habermeier H-U and Kronm\"uller H 1993
      {\it J.\ Alloys and Compounds} {\bf 195} 499  

\item Indenbom M V, Forkl A, Ludescher B, Kronm\"uller H,
      Habermeier H-U, Leibold B, D'Anna G, Li T W, Kes P H and
      Menovsky A A 1994c {\it Physica C} {\bf 226} 325 

\item Indenbom M V, Kronm\"uller H, Li T W, Kes P H and Menovsky A A 1994d
      {\it Physica C} {\bf 222} 203                   

\item Indenbom M V, Schuster Th, Kuhn H, Kronm\"uller H, Li T W and
      Menovsky A A 1995
      {\it Phys.\ Rev.\ B} {\bf 51} 15484  

\item Inui M and Harshman D R 1993
      {\it Phys.\ Rev.\ B} {\bf 47} 12205 

\item Inui M, Littlewood P B and Coppersmith S N 1989
      {\it Phys.\ Rev.\ Lett.} {\bf 63} 2421

\item Ioffe L B, Larkin A I, Varlamov A A and Yu L 1993
      {\it Phys.\ Rev.\ B} {\bf 47} 8936 

\item Irie F, Matsushita T, Otabe S, Matsuno T and Yamafuji K 1989
      {\it Cryogenics} {\bf 29} 317  

\item Isaac I, Jung J, Murakami M, Tanaka S, Mohamed M A-K and Friedrich L
      1995 {\it Phys.\ Rev.\ B} {\bf 51} 11806  

\item Itzler M A, Danner G M, Bojko R and Chaikin P M 1994
      {\it Phys.\ Rev.\ B} {\bf 49} 6815          

\item Itzler M A and Tinkham M 1995
      {\it Phys.\ Rev.\ B} {\bf 51} 435, 9411 

\item Ivanchenko Yu M 1993
      {\it Phys.\ Rev.\ B} {\bf 48} 15966 

\item Ivanchenko Yu M, Belevtsov L V, Genenko Yu A and Medvedev Yu V 1992
      {\it Physica C} {\bf 193} 291

\item Ivanchenko Yu M, Khirny{\u{\i}} V F and Mikheenko P N 1979
      {\it Zh.\ Eksp.\ Teor.\ Fiz.} {\bf 77} 952
      (1979 {\it Sov.\ Phys.\ JETP} {\bf 50} 479)

\item Ivlev B I and Mel'nikov V I 1987   
      {\it Phys.\ Rev.\ B} {\bf 36} 6889   

\item Ivlev B I and Campbell L J 1993    
      {\it Phys.\ Rev.\ B} {\bf 47} 14514

\item Ivlev B I and Kopnin N B 1990a
      {\it J. Low Temp. Phys.} {\bf 80} 161
\item Ivlev B I and Kopnin N B 1990b
      {\it Phys.\ Rev.\ Lett.} {\bf 64} 1828;
      {\it Phys.\ Rev.\ B} {\bf 42} 10052
\item Ivlev B I and Kopnin N B 1991
      {\it Phys.\ Rev.\ B} {\bf 44} 2747

\item Ivlev B I, Kopnin N B and Pokrovsky V L 1990
      {\it J. Low Temp. Phys.} {\bf 80} 187

\item Ivlev B I, Kopnin N B and Salomaa M M  1991a
      {\it Phys.\ Rev.\ B} {\bf 43} 2896        

\item Ivlev B I and Ovchinnikov Yu N 1987     
      {\it Zh.\ Eksp.\ Teor.\ Fiz.} {\bf 93} 668
      (1987 {\it Sov.\ Phys.\ JETP} {\bf 66} 378)

\item Ivlev B I, Ovchinnikov Yu N and  Pokrovsky V L 1991b
   {\it Mod.\ Phys.\ Lett.\ B} {\bf 5} 73     

\item Ivlev B I, Ovchinnikov Yu N and  Thompson R S 1991c
   {\it Phys.\ Rev.\ B} {\bf 44} 7023           

\item Iye Y, Nakamura S, Tamegai T, Terashima T, Yamamoto K
      and Bando Y 1990a {\it Physica C} {\bf 166} 62  
\item Iye Y, Watanabe A, Nakamura S, Tamegai T, Terashima T, Yamamoto K
      and Bando Y 1990b {\it Physica C} {\bf 167} 278 
\item Iye Y, Terashima T and Bando Y 1991a
      {\it Physica C} {\bf 184} 362                   
\item Iye Y, Fukushima A, Tamegai T, Terashima T and Bando Y 1991b
      {\it Physica C} {\bf 185-189} 297       
\item Iye Y, Oguro I, Tamegai T, Datars W R, Motohira N and
      Kitazawa K 1992 {\it Physica C} {\bf 199} 154   

\item Janossy B, de Graaf A, Kes P H, Kopylov V N and Togonidze T G 1995
      {\rm Physica C} {\bf 246} 277 

\item Jeanneret B, Gavilano J L, Racine G A, Leemann Ch and Martinoli P
      1989 {\rm Appl. Phys. Lett.} {\bf 55} 2336

\item Jenks W G and Testardi L R 1993
      {\it Phys.\ Rev.\ B} {\bf 48} 12993

\item Jensen H J 1990   
      {\it Phys.\ Rev.\ Lett.} {\bf 64} 3103

\item Jensen H J, Brass A, Shi A-Ch and Berlinski A J 1990
      {\it Phys.\ Rev.\ B} {\bf 41} 6394
\item Jensen H J, Brechet Y and Brass A 1989
      {\it J.\ Low Temp.\ Phys.} {\bf 74} 293

\item Jensen H J and Minnhagen P 1991
      {\it Phys.\ Rev.\ Lett.} {\bf 66} 1630

\item Jensen H J, Minnhagen P, Sonin E and Weber H 1992
      {\it Europhys. Lett.} {\bf 20} 463

\item Jiang W, Yeh N-C, Reed D S, Kriplani U, Beam D A,
      Konczykowski M, Tombrello T A and Holtzberg F 1994
      {\it Phys.\ Rev.\ Lett.} {\bf 72} 550    

\item Jiang W, Yeh N-C, Reed D S, Kriplani U, and Holtzberg F 1995
      {\it Phys.\ Rev.\ Lett.} {\bf 74} 1438  

\item Jiang X, Connolly P J, Hagen S J and Lobb C J 1994
      {\it Phys.\ Rev.\ B} {\bf 49} 9244   

\item Jin R, Schilling A and Ott H R 1994
      {\it Phys.\ Rev.\ B} {\bf 49} 9218   

\item * Jirsa M, Koblischka M R and van Dalen A J J  1995
      {\it Phys.\ Rev.\ B} {\bf 52} (submitted) 

\item Jonsson A and Minnhagen P 1994 
      {\it Phys.\ Rev.\ Lett.} {\bf 73}  3576

\item Jos\'{e} J V and Ramirez-Santiago G 1993
      {\it J.\ Phys.\ A: Math.\ Gen.} {\bf 26} L535

\item Josephson B D 1962 {\it Phys.\ Lett.} {\bf 1} 251 

\item Jung J, Isaac I and Mohammed M A-K 1993
      {\it Phys.\ Rev.\ B} {\bf 48} 7526
\item Jung J, Mohammed M A-K, Isaak I and Friedrich L 1994
      {\it Phys.\ Rev.\ B} {\bf 49} 12188   

\item Kadowaki K, Songliu Y and Kitazawa K 1994
      {\it Supercond.\ Sci.\ Technol.} {\bf 7} 519

\item Kamal S, Bonn D A, Goldenfeld N, Hirschfeld P J, Liang R and
      Hardy W N 1994 {\it Phys.\ Rev.\ Lett.} {\bf 73} 1845 

\item Kamerlingh Onnes H 1911 {\it Leiden Comm.} {\bf 120b, 122b, 124c}

\item Kammerer U 1969 {\it Z.\ Phys.} {\bf 227} 125

\item Kaplunenko V K, Moskvin S I and Shmidt V V 1985
      {\it Fiz.\ Nizk.\ Temp.} {\bf 11} 846
      (1985 {\it Sov.\ J.\ Low Temp.\ Phys.} {\bf 11} 464)  

\item Kardar M 1985 {\it Phys.\ Rev.\ Lett.} {\bf 55} 2923 
\item Kardar M 1987 {\it Nucl.\ Phys.\ B} {\bf 290} 582    

\item Karkut M G, Slaski M, Heill L K, Sagdahl L T and Fossheim K 1993
      {\it Physica C} {\bf 215} 19   

\item Kato Y, Enomoto Y and Maekawa S 1991  
      {\it Phys.\ Rev.\ B} {\bf 44} 6916

\item Kato Y and Nagaosa N 1993             
      {\it Phys.\ Rev.\ B} {\bf 47} 2932; {\bf 48} 7383

\item Keimer B, Shih W Y, Erwin R W, Lynn J W, Dogan F and Aksay I A 1994
      {\it Phys.\ Rev.\ Lett.} {\bf 73} 3459  

\item Kerchner R 1981 {\it J.\ Low Temp.\ Phys.} {\bf 46} 205
\item Kerchner R 1983 {\it J.\ Low Temp.\ Phys.} {\bf 50} 337

\item Kes P H 1992a {\it Phenomenology and Application of
      High-Temperature Superconductors} ed K S Bedell, M Inui,
      D E Meltzer, J R Schrieffer and S Doniach
      (New York: Addison-Wesley) p 390
\item Kes P H 1992b {\it Concise Encyclopedia     
            of Magnetic \& Superconducting Materials} ed J Evetts
            (Oxford: Pergamon) p 88, p 163           
\item Kes P H, Aarts J, van den Berg J, van der Beek C J,
      Mydosh J A 1989 {\it Supercond.\ Sci.\ Technol.} {\bf 1} 242

\item Kes P H, Aarts J, Vinokur V M and van der Beek C J 1990
      {\it Phys.\ Rev.\ Lett.} {\bf 64} 1063  

\item Kes P H, van der Beek C J, Maley M P, McHenry M E, Huse D A,
      Menken M J V and Menovsky A A 1991
      {\it Phys.\ Rev.\ Lett.} {\bf 67} 2383  

\item Kes P H and Tsuei C C 1981
      {\it Phys.\ Rev.\ Lett.} {\bf 47} 1930  
\item Kes P H and Tsuei C C 1983
      {\it Phys.\ Rev.\ B} {\bf 28} 5126      

\item Kessler C, Nebendahl B, Peligrad D-N, Dul\v{c}ic A, Habermeier H-U
      and Mehring M 1994 {\it Physica C} {\bf 219} 233 

\item Khalfin I B and Shapiro B Ya 1993
      {\it Physica C} {\bf 207} 359       
\item Khalfin I B, Malomed B A and Shapiro B Ya 1994
      {\it Phys.\ Rev.\ B} {\bf 50} 9351  
\item Khalfin I B, Dorosinskii L A and Shapiro B Ya 1995
      {\it Phys.\ Rev.\ B} {\bf 51} 1245  

\item Khalmakhelidze Yu G and Mints R G 1991
      {\it Cryogenics} {\bf 29} 1041 

\item Khasanov R I, Talanov Yu I, Vashakidze Yu M and Teitelbaum G B 1995
      {\it Physica C} {\bf 242} 333 

\item Khaykovich B, Zeldov E, Konczykowski M, Majer D, Larkin A I and 
      Clem J R 1994  {\it Physica C} {\bf 235-240} 2757 

\item Kim S F, Li Z Z and Raffy H 1995
      {\it Physica C} {\bf 244} 78 

\item Kim D H, Gray K, Kampwirth R T, and McKay D M 1990a
      {\it Phys.\ Rev.\ B} {\bf 41} 11642  
\item Kim D H, Gray K, Kampwirth R T, Woo K C, McKay D M and Stein J 1990b
      {\it Phys.\ Rev.\ B} {\bf 42} 6249   

\item Kim D H, Kang W N, Kim Y H, Park J H, Lee J J, Yi G H, Hahn T S
      and Choi S S  1995 {\it Physica C} {\bf 246} 235 

\item Kim Y B, Hempstead C F and Strnad A R 1963
      {\it Phys.\ Rev.} {\bf 129} 528    
\item Kim Y B, Hempstead C F and Strnad A R 1965
      {\it Phys.\ Rev.} {\bf 139} A1163  

\item Kim Y B and Stephen M J 1969 {\it Superconductivity}
      ed R D Parks (New York: Marcel Dekker) vol II 1107  

\item Kim Y C, Thompson J R, Ossandon J G, Christen D K and Paranthaman M
      1995 {\it Phys.\ Rev.\ B} {\bf 51} 11767 

\item Kitaguchi H, Takada J, Oda K, Osaka A, Miura Y, Tomii Y, Mazaki H and
      Takano M 1989 {\it Physica C} {\bf 157} 267 

\item Kleiman R N, Gammel P L, Schneemeyer L F, Waszczak J V and Bishop D J
      1989 {\it Phys.\ Rev.} {\bf 62} R2331 

\item Kleiman R N, Broholm C, Aeppli G, Bucher E, St\"ucheli N, Bishop D J,
      Clausen K N, Mortensen K, Pedersen J S and Howard B 1992
      {\it Phys.\ Rev.\ Lett.} {\bf 69} 3120

\item Klein L, White W R, Beasley M R and Kapitulnik A 1995
      {\it Phys.\ Rev.\ B} {\bf 51} 6796 

\item Klein L, Yacoby E R, Wolfus Y, Yeshurun Y, Burlachkov L,
      Shapiro B Ya, Konczykowski M and Holtzberg F 1993a 
      {\it Phys.\ Rev.\ B} {\bf 47} 12349  

\item Klein L, Yacoby E R, Yeshurun Y, Konczykowski M and Kishio K 1993b
      {\it Phys.\ Rev.\ B} {\bf 48} 3523  

\item Klein L, Yacoby E R, Yeshurun Y, Konczykowski M, Holtzberg F
      and Kishio K 1993c {\it Physica C} {\bf 209} 251  

\item Klein N, Tellmann N, Schulz H, Urban K, Wolf S A and Krezin V Z
      1993d {\it Phys.\ Rev.\ Lett.} {\bf 71} 3355 

\item Klein U 1987  {\it J.\ Low Temp.\ Phys.} {\bf 69} 1   
\item Klein U 1990  {\it Phys.\ Rev.\ B} {\bf 41} 4819      

\item Klein U and P\"ottinger 1991    
      {\it Phys.\ Rev.\ B} {\bf 44} 7704
\item Klein U, Rammer J and Pesch W 1987
      {\it J.\ Low Temp.\ Phys.} {\bf 66} 55     

\item Kleiner R, M\"uller P, Kohlstedt H, Pedersen N F and Sakai S 1994
      {\it Phys.\ Rev.\ B} {\bf 50} 3942       
\item Kleiner R, Steinmeyer F, Kunkel G and M\"uller P 1992
      {\it Phys.\ Rev.\ Lett.} {\bf 68} 2394   

\item Kleiner W H, Roth L M and Autler S W 1964
      {\it Phys.\ Rev.} {\bf 133} A1226            

\item Klemm R A 1988  {\it Phys.\ Rev.\ B} {\bf 38} 6641
\item Klemm R A 1990  {\it Phys.\ Rev.\ B} {\bf 41} 117   
\item Klemm R A 1993  {\it Phys.\ Rev.\ B} {\bf 47} 14630 
\item * Klemm R A 1995  {\it SIAM J.\ Appl.\ Math.}  (August 1995)
\item Klemm R A 1996  {\it Layered Superconductors} (New York: Oxford)
                      (to appear)
\item Klemm R A and Liu S H 1995 
      {\it Phys.\ Rev.\ Lett.} {\bf 74} 2343

\item Klemm R A, Luther A and Beasley M R 1975
      {\it Phys.\ Rev.\ B} {\bf 12} 877

\item Klimesch P and Pesch W 1978
      {\it J.\ Low Temp.\ Phys.} {\bf 32} 869 

\item Klupsch Th 1995 {\it Physica C} {\bf 244} 165 

\item Klupsch Th, Hergt R and Hiergeist R 1993 
      {\it Physica C} {\bf 211} 65

\item Knorpp R, Forkl A, Habermeier H-U and Kronm\"uller H 1994
      {\it Physica C} {\bf 230} 128

\item Kobayashi T, Nakayama Y, Kishio K, Kimura T, Kitazawa K and
      Yamafuji K 1993 {\it Appl.\ Phys.\ Lett.} {\bf 62} 1830  

\item Kober F, Ri H-C, Gross R, Koelle D, Huebener R P and Gupta A 1991a
      {\it Phys.\ Rev.\ B} {\bf 44} 11951  

\item Kober J, Gupta A, Esquinazi P, Braun H F and Brandt E H 1991b
      {\it Phys.\ Rev.\ Lett.} {\bf 66} 2507            

\item Koblischka M R and Wijngaarden R J 1995 
      {\it Supercond.\ Sci.\ Technol.} {\bf 8} 199

\item Koblischka M R, Schuster Th and Kronm\"uller H 1993
      {\it Physica C} {\bf 211} 263
\item Koblischka M R, Schuster Th and Kronm\"uller H 1994
      {\it Physica C} {\bf 219} 205

\item Koch R H, Foglietti V, Gallagher W J, Koren G, Gupta A and
      Fisher M P A 1989 {\it Phys.\ Rev.\ Lett.} {\bf 63} 1511

\item Kogan V G 1988 {\it Phys.\ Rev.\ B} {\bf 38} 7049        
\item Kogan V G 1990 {\it Phys.\ Rev.\ Lett.} {\bf 64} 2192 
\item Kogan V G 1994 {\it Phys.\ Rev.\ B} {\bf 49} 15874       

\item Kogan V G, Bulaevskii L N, Miranovi\'c P and Dobrosavlevi\'c-Gruji\'c L
      1995 {\it Phys.\ Rev.\ B} {\bf 51} 15344

\item Kogan V G and Campbell L J 1989 {\it Phys.\ Rev.\ Lett.} {\bf 62} 1552

\item Kogan V G and Clem J R 1981 {\it Phys.\ Rev.\ B} {\bf 24} 2497

\item Kogan V G , Fang M M and Mitra S 1988
      {\it Phys.\ Rev.\ B} {\bf 38} 11958    

\item Kogan V G, Ledvij M and Bulaevskii L N 1992  
      {\it Phys.\ Rev.\ Lett.} {\bf 70} 1870

\item Kogan V G, Simonov A Yu and Ledvij M 1993a 
      {\it Phys.\ Rev.\ B} {\bf 48} 392          

\item Kogan V G, Ledvij M, Simonov A Yu, Cho J H and Johnston D C 1993b
      {\it Phys.\ Rev.\ Lett.} {\bf 70} 1870

\item Kogan V G, Nakagawa N, Thiemann S L 1990
      {\it Phys.\ Rev.\ B} {\bf 42} 2631

\item Konczykowski M 1994        
      {\it Physica C} {\bf 235-240} 197

\item Konczykowski M, Burlachkov L I, Yeshurun Y and Holtzberg F 1991
      {\it Phys.\ Rev.\ B} {\bf 43} 13707  

\item Konczykowski M, Chikumoto N, Vinokur V M and Feigel'man M V
      1995 {\it Phys.\ Rev.\ B} {\bf 51} 3957 

\item Konczykowski M, Vinokur V M, Rullier-Albenque F, Yeshurun Y and
      Holtzberg F 1993 {\it Phys.\ Rev.\ B} {\bf 47} 5531 

\item Kopelevich Y, Esquinazi P and Ziese M 1994 
      {\it Mod.\ Phys.\ Lett.\ B} {\bf 8} 1529   

\item Kopnin N B and Kravtsov V E 1976  
      {\it Pis'ma Zh.\ Eksp.\ Teor.\ Fiz.} {\bf 23} 631
      (1976 {\it Sov.\ Phys.\ JETP Lett.}    {\bf 23} 578);
      {\it  Zh.\ Eksp.\ Teor.\ Fiz.} {\bf 71} 1644
      (1976 {\it Sov.\ Phys.\ JETP}    {\bf 44} 861)

\item Kopnin N B and Lopatin A V 1995 
      {\it Phys.\ Rev.\ B} {\bf 51} 15291

\item Kopnin N B, Lopatin A V, Sonin E B and Traito K B 1995
      {\it Phys.\ Rev.\ Lett.} {\bf 74} 4527 

\item Kopnin N B and Salomaa M M 1991       
      {\it Phys.\ Rev.\ B} {\bf 44} 9667

\item Kopnin N B, Ivlev B I and Kalatsky V A 1993
      {\it J. Low Temp. Phys.} {\bf 90} 1

\item Kopylov V N, Koshelev A E, Schegolev I F and Togonidze 1990
      {\it Physica C} {\bf 170} 291     

\item Korevaar P, Kes P H, Koshelev A E and Aarts J 1994
      {\it Phys.\ Rev.\ Lett.} {\bf 72} 3250

\item Korshunov S E 1990 {\it Europhys.\ Lett.} {\bf 11} 757
\item Korshunov S E 1993 {\it Phys.\ Rev.\ B} {\bf 48} 3969
\item Korshunov S E 1994 {\it Phys.\ Rev.\ B} {\bf 50} 13616 
\item Korshunov S E and Larkin A I 1992  
      {\it Phys.\ Rev.\ B} {\bf 46} 6395

\item Koshelev A E 1992a {\it Phys.\ Rev.\ B} {\bf 45} 12936 
\item Koshelev A E 1992b {\it Physica C} {\bf 191} 219     
\item Koshelev A E 1992c {\it Physica C} {\bf 198} 371     
\item Koshelev A E 1993  {\it Phys.\ Rev.\ B} {\bf 48} 1180  
\item Koshelev A E 1994a {\it Physica C} {\bf 223} 276     
\item Koshelev A E 1994b {\it Phys.\ Rev.\ B} {\bf 50} 506 
                             (erratum {\bf 51} 12063)
\item * Koshelev A E, Glazman L I and Larkin A I 1995
      {\it Phys.\ Rev.\ B} (submitted) 

\item Koshelev A E and Kes P H 1993 {\it Phys.\ Rev.\ B} {\bf 48} 6539

\item Koshelev A E and Vinokur V M 1991 {\it Physica C} {\bf 173} 465 

\item Koshelev A E and Vinokur V M 1994
      {\it Phys.\ Rev.\ Lett.} {\bf 73} 3580 

\item Koshelev A E and Vinokur V M 1995
      {\it Phys.\ Rev.\ B} {\bf 51} 11678 

\item Kosterlitz J M and Thouless D J 1973
      {\it J.\ Phys.\ C} {\bf 6} 1181

\item K\"otzler J, Kaufmann M, Nakielski G and Behr R 1994a
      {\it Phys.\ Rev.\ Lett.} {\bf 72} 2081

\item K\"otzler J, Nakielski G, Baumann M, Behr R, Goerke F and
      Brandt E H 1994b {\it Phys.\ Rev.\ B} {\bf 50} 3384 

\item Koyama T and Tachiki M 1992 
      {\it Physica C} {\bf 193} 163
\item Koyama T and Tachiki M 1993 
      {\it Physica C} {\bf 210} 509

\item Koziol Z, de Ch\^atel P, Franse J J M, Tarnavski Z and Menovsky A A
      1993  {\it Physica C} {\bf 212} 133

\item Kr\"ageloh U 1970 {\it phys.\ stat.\ sol.} {\bf 42} 559 

\item Kr\"ageloh U, Kumpf U and Seeger A 1971
      {\it Proc.\ 12th Intnl.\ Conf.\ of Low Temperature Physics LT12}
      ed E Kanda (Kyoto: Academic Press of Japan)  p 473 

\item Kramer E J 1973  {\it J.\ Appl.\ Phys.} {\bf 44} 1360  
\item Kramer E J 1978  {\it J.\ Nuclear Mater.} {\bf 72} 5   

\item Kramer L 1971 {\it Phys.\ Rev.\ B} {\bf 11} 3821

\item Kramer L and Pesch W 1974 {\it Z.\ Phys.} {\bf 269} 59

\item Kr\"amer A and Kuli\'c M L 1993 {\it Phys.\ Rev.\ B} {\bf 48} 9673
\item Kr\"amer A and Kuli\'c M L 1994 {\it Phys.\ Rev.\ B} {\bf 50} 9484

\item Krasnov V M, Pedersen N F and Golubov A A 1994
      {\it Physica C} {\bf 209} 579    

\item Kraus M, Kreiselmeyer G, Daniel J, Leghissa M,  
      Saemann-Ischenko G, Holzapfel B, Kummeth P, Scholz R
      and Vinnikov L Ya 1994a {\it Nuclear Instr.\ Methods B} {\bf 89} 307

\item Kraus M, Leghissa M and Saemann-Ischenko G 1994b
      {\it Phys.\ Bl.} {\bf 50} 333

\item Krey U 1972 {\it Intl.\ J.\ Magnetism} {\bf 3} 65  
\item Krey U 1973 {\it Intl.\ J.\ Magnetism} {\bf 4} 153 

\item Krivoruchko V N 1993 {\it J.\ Low Temp.\ Phys.} {\bf 92} 217 

\item Kronm\"uller H 1970
     {\it phys.\ stat.\ sol.} {\bf 40} 295  
\item Kronm\"uller H 1974  {\it Internat. Discussion Meeting on
      Flux Pinning in Superconductors} ed P Haasen and H C Freyhardt
      (G\"ottingen: Akademie der Wissenschaften) p 1 

\item Kronm\"uller H and Riedel H 1970
     {\it phys.\ stat.\ sol.} {\bf 38} 403 
\item Kronm\"uller H and Riedel H 1976
     {\it phys.\ stat.\ sol.} {\bf 77} 581 

\item Kruithof G H, van Son P C and Klapwijk T M 1991
      {\it Phys.\ Rev.\ Lett.} {\bf 67} 2725  

\item Krusin-Elbaum L, Civale L, Holtzberg F, Malozemoff A P, and Field C
      1991 {\it Phys.\ Rev.\ Lett.} {\bf 67} 3156  
\item Krusin-Elbaum L, Civale L, Vinokur V M and Holtzberg F
      1992 {\it Phys.\ Rev.\ Lett.} {\bf 69} 2280  
\item Krusin-Elbaum L, Civale L, Blatter G, Marwick A D, Holtzberg F
      and Feild C 1994 {\it Phys.\ Rev.\ Lett.} {\bf 72} 1914 

\item Kugel K I, Lisovskaya T Ya and Mints R G 1991
      {\it Superconductivity} {\bf 4} 2166   

\item Kugel K I and Rakhmanov  A L 1992   
      {\it Physica C} {\bf 196} 17

\item Kuklov A, Bulatov A and Birman J L 1994
      {\it Phys.\ Rev.\ Lett.} {\bf 72} 2274 

\item Kuli\'{c} M L, Kr\"amer A and Schotte K D 1992
      {\it Solid State Commun.} {\bf 82} 541    

\item Kuli\'{c} M L and Rys F S 1989
      {\it J.\ Low Temp.\ Phys.} {\bf 76} 167    

\item Kulik I O 1968       
      {\it Zh. Eksp. Teor. Fiz.} {\bf 55} 889
      (1969 {\it Sov. Phys. JETP} {\bf 28} 461)

\item K\"ummel R, Sch\"ussler U, Gunsenheimer U and Plehn H 1991
      {\it Physica C} {\bf 185-189} 221 

\item Kummeth P, Struller C, Neum\"uller H-W and Saemann-Ischenko G 1994
      {\it Appl.\ Phys.\ Lett.} {\bf 65} 1302 

\item Kunchur M N, Christen D K and Phillips J M 1993
      {\it Phys.\ Rev.\ Lett.} {\bf 70} 998
\item Kunchur M N, Christen D K, Klabunde C E and Phillips J M 1994
      {\it Phys.\ Rev.\ Lett.} {\bf 72} 752, 2259
\item Kunchur M N, Poon S J and Subramanian M A 1990
      {\it Phys.\ Rev.\ B} {\bf 41} 4089  

\item Kung P J, Maley M P, McHenry M E, Willis J O, Murakami M
      and Tanaka S 1993 {\it Phys.\ Rev.\ B} {\bf 48} 13922

\item K\"upfer H, Apfelstedt I, Fl\"ukiger R, Keller C, Meier-Hirmer R,
      Runtsch B, Turowski A, Wiech U and Wolf T 1988
      {\it Cryogenics} {\bf 28} 650  

\item Kuprianov M Yu and Likharev K K 1974
      {\it Fiz.\ Tverd.\ Tela} {\bf 16} 2829
      (1975 {\it Sov.\ Phys.\ Solid State} {\bf 16} 1835)

\item Kuprianov M Yu and Likharev K K 1975
      {\it Zh.\ Eksp.\ Teor.\ Fiz.} {\bf 68} 1506
      (1975 {\it Sov.\ Phys.\ JETP} {\bf 41} 755)

\item Kwok W K, Welp U, Crabtree G W, Vandervoort K, Hulscher G and
      Liu J Z 1990 {\it Phys.\ Rev.\ Lett.} {\bf 64} 966    
\item Kwok W K, Welp U, Vinokur V M, Fleshler S, Downey J and
    Crabtree G W 1991 {\it Phys.\ Rev.\ Lett.} {\bf 67} 390 
\item Kwok W K, Fleshler S, Welp U, Vinokur V M, Downey J and
    Crabtree G W 1992 {\it Phys.\ Rev.\ Lett.} {\bf 69} 3370 
\item Kwok W K, Fendrich J, Welp U, Fleshler S, Downey J and
    Crabtree G W 1994a {\it Phys.\ Rev.\ Lett.} {\bf 72} 1088, 1092 
\item Kwok W K, Fendrich J, van der Beek C J and Crabtree G W 1994b
    {\it Phys.\ Rev.\ Lett.} {\bf 73} 2614 

\item Labusch R 1966 {\it Phys.\ Lett.} {\bf 22} 9  
\item Labusch R 1968 {\it Phys.\ Rev.} {\bf 170} 470 
\item Labusch R 1969a {\it phys.\ stat.\ sol.} {\bf 32} 439
\item Labusch R 1969b {\it Crystal Lattice Defects} {\bf 1} 1

\item Labusch R and Doyle T B 1995                     
      {\it Proc.\ Intnl.\ Cryogenics Materials Conf.\ 1994.
      Critical State in Superconductors} (Oct.\ 1994 in Honolulu)

\item Labdi S, Raffy H, Laborde O and Monceau P 1992
      {\it Physica C} {\bf 197} 274  

\item Lairson M, Sun J Z, Geballe T H, Beasley M R and Bravman J C 1991
      {\it Phys.\ Rev.\ B} {\bf 43} 10405

\item Landau L D and Lifshitz E M 1960 {\it Electrodynamics
            of Continuous Media} (New York: Pergamon) p 178

\item Lang W, Heine G, Kula W and Sobolewski R 1995
      {\it Phys.\ Rev.\ B} {\bf 51} 9180

\item Larbalestier D C 1991                 
      {\it Physics Today} {\bf 44}(6) 74

\item Larkin A I 1970 {\it Zh.\ Eksp.\ Teor.\ Fiz.} {\bf 58} 1466
            (1970 {\it Sov.\ Phys.\ JETP} {\bf 31} 784)

\item Larkin A I and Ovchinnikov Yu N 1971
            {\it Zh.\ Eksp.\ Teor.\ Fiz.} {\bf 61} 1221
            (1972 {\it Sov.\ Phys.\ JETP} {\bf 34} 651)

\item Larkin A I and Ovchinnikov Yu N 1973
            {\it Zh.\ Eksp.\ Teor.\ Fiz.} {\bf 65} 1704
            (1974 {\it Sov.\ Phys.\ JETP} {\bf 38} 854)
\item Larkin A I and Ovchinnikov Yu N 1975
            {\it Zh.\ Eksp.\ Teor.\ Fiz.} {\bf 68} 1915
            (1976 {\it Sov.\ Phys.\ JETP} {\bf 41} 960)

\item Larkin A I and Ovchinnikov Yu N 1979 {\it J. Low Temp. Phys.}
            {\bf 73} 109
\item Larkin A I and Ovchinnikov Yu N 1986 {\it Nonequilibrium
      Superconductivity}  ed D N Langenberg and A I Larkin
      (Amsterdam: Elsevier) p 493
\item Larkin A I and Ovchinnikov Yu N 1995
      {\it Phys.\ Rev.\ B} {\bf 51} 5965    

\item Lawrence W E and Doniach S 1971
      {\it Proc.\ 12th Intnl.\ Conf.\ of Low Temperature Physics LT12}
      ed E Kanda (Kyoto: Academic Press of Japan)  p 361

\item LeBlanc M A R, Celebi S, Wang S X and Plech\'achek V 1993
      {\it Phys.\ Rev.\ Lett.} {\bf 71} 3367  

\item Le Doussal P and Giamarchi T 1995
      {\it Phys.\ Rev.\ Lett.} {\bf 74} 606  

\item Lee C Y and Kao Y H 1995
      {\it Physica C} {\bf 241} 167 

\item Lee H H and Moore M A 1994 {\it Phys.\ Rev.\ B} {\bf 49} 9240 

\item Lee J Y, Paget K M, Lemberger T R, Foltyn S R and Wu X 1994
      {\it Phys.\ Rev.\ B} {\bf 50} 3337  

\item Lee P A and Rice T M 1979 {\it Phys.\ Rev.\ B} {\bf 19} 3970

\item * Lee S L, Warden M, Keller H, Schneider J W, Zimmermann P, Cubitt R,
      Forgan E M, Wylie M T, Kes P H, Li T W, Menovsky A A and
      Tarnavski Z 1995                         
      {\it Phys.\ Rev.\ Lett.} {\bf ??} (submitted)

\item Lee S L, Zimmermann P, Keller H, Warden M, Savi{\'c} I M,
      Schauwecker R, Zech D, Cubitt R, Forgan E M, Kes P H,
      Li T W, Menovsky A A and Tarnavski Z 1993
      {\it Phys.\ Rev.\ Lett.} {\bf 71} 3862

\item Lee Th S, Missert N, Sagdahl L T, Clarke J, Clem J R, Char K,
      Eckstein J N, Fork D K, Lombardo L, Kapitulnik A, Schneemeyer L F,
      Waszczak J V and van Dover R B 1995
      {\it Phys.\ Rev.\ Lett.} {\bf 74} 2796 

\item Leghissa M, Gurevich L A, Kraus M, Saemann-Ischenko G and
      Vinnikov L Ya 1993 {\it Phys.\ Rev.\ B} {\bf 48} 31341

\item Leghissa M, Schuster Th, Gerh\"auser W, Klaum\"unzer S,
      Koblischka M R, Kronm\"uller H, Kuhn H, Neum\"uller H-W and
      Saemann-Ischenko G 1992 {\it Europhys.\ Lett.} {\bf 19} 323 

\item Leiderer P, Boneberg J, Br\"ull P, Bujok V and Herminghaus S 1993
     {\it Phys.\ Rev.\ Lett.} {\bf 71} 2646  

\item Lemmens P, Fr\"onig P, Ewert S, Pankert J, Marbach G and
      Comberg A  1991 {\it Physica C} {\bf 174} 289  

\item Lensink J G,  Flipse C F J, Roobeek J, Griessen R and Dam B 1989
      {\it Physica C} {\bf 162-164} 663

\item Levitov L S 1991 {\it Phys.\ Rev.\ Lett.} {\bf 66} 224

\item Li H C, Li J W, Wang R L, Yin B, Zhao Z X, Yan S L, Fang L
      and Si M S 1995 {\it Physica C} {\bf 246} 330 

\item Li Y-H and Teitel S 1991 {\it Phys.\ Rev.\ Lett.} {\bf 66} 3301
\item Li Y-H and Teitel S 1992 {\it Phys.\ Rev.\ B} {\bf 45} 5718
\item Li Y-H and Teitel S 1993 {\it Phys.\ Rev.\ B} {\bf 47} 359
\item Li Y-H and Teitel S 1994 {\it Phys.\ Rev.\ B} {\bf 49} 4136

\item Li Qi, Kwon C, Xi X X, Bhattacharya S, Walkenhorst A,
      Venkatesan T, Hagen S J, Jiang W and Green R L 1992
      {\it Phys.\ Rev.\ Lett.} {\bf 69} 2713

\item Li Qiang, Suenaga M, Hikata T and Sato K 1992
         {\it Phys.\ Rev.\ B} {\bf 46} 5857
\item Li Qiang, Suenaga M, Bulaevskii L N, Hikata T and Sato K 1993
         {\it Phys.\ Rev.\ B} {\bf 48} 13865
\item Li Qiang, Wiesmann H J, Suenaga M, Motowidlow L and Haldar P 1994
         {\it Phys.\ Rev.\ B} {\bf 50} 4256

\item Liang R, Dosanjh P, Bonn D A, Hardy W N and Berlinsky A J 1994
         {\it Phys.\ Rev.\ B} {\bf 50} 4212

\item Likharev K K 1971 {\it Iz.\ Vuzov Radiofiz.} {\bf 14} 909, 919
         (1971 {\it Radiophys.\ Quantum Electron.} {\bf 14} 714, 722)

\item Likharev K K 1979
      {\it Rev. Mod. Phys.} {\bf 51} 101 

\item Ling X S and Budnick J I 1991
    {\it Magnetic Susceptibility of Superconductors and Other Spin Systems}
      ed R A Hein, T L Francavilla and D H Liebenberg
      (New York: Plenum) p 377  

\item Ling X S, Shi D and Budnick J I 1991
      {\it Physica C} {\bf 185-189} 2181   

\item Ling X S, McCambridge J D, Rizzo N D, Sleight J W, Prober D E,
      Motowidlo L R and Zeitlin B A  1995 
      {\it Phys.\ Rev.\ Lett.} {\bf 74} 805

\item Lischke B and Rodewald W 1974
      {\it phys.\ stat.\ sol.\ (b)} {\bf 63} 97 

\item Liu C and Narahari Achar B N 1990
      {\it Physica C} {\bf 168} 515      

\item Liu Ch J, Schlenker C, Schubert J and Stritzker B 1993
      {\it Phys.\ Rev.\ B} {\bf 48} 13911 

\item Liu J Z, Jia Y X, Shelton R N and Fluss M J 1991
      {\it Phys.\ Rev.\ Lett.} {\bf 66} 1354               

\item * Liu Wu, Clinton T W and Lobb C J 1995
      {\it Phys.\ Rev.\ B} {\bf 52} ???  

\item Liu Y, Haviland D B, Glazman L I and A M Goldman 1992
      {\it Phys.\ Rev.\ Lett.} {\bf 68} 2224        

\item Livingston J D and DeSorbo W 1969 {\it Superconductivity} 
      ed R D Parks (New York: Marcel Dekker) vol II p 1235

\item Lofland S E, Bhagat S M, Rajeswari M, Venkatesan T, Kanjilal D,
      Senapati L and Mehta G K 1995
      {\it Phys.\ Rev.\ B} {\bf 51} 8489 

\item Logvenov G Yu, Ryazanov V V, Ustinov A V and Huebener R P 1991
      {\it Physica C} {\bf 175} 179     

\item L\'opez D, Nieva G, de la Cruz F, Jensen H J and O'Kane D 1994  
      {\it Phys.\ Rev.\ B} {\bf 50} 9684 

\item Lowell J 1972 {\it J. Phys. F} {\bf 2} 547 

\item Lowndes D H, Christen D K, Klabunde C E, Wang Z I, Kroeger D M,
      Budai J D, Zhu S and Norton D P 1995   
      {\it Phys.\ Rev.\ Lett.} {\bf 74} 2355 

\item Luo S, Yang G and Gough C E 1995
      {\it Phys.\ Rev.\ B} {\bf 51} 6655  

\item Lykov A N 1993 {\it Adv.\ Phys.} {\bf 42} 263 

\item Lynn J W, Rosov N, Grigereit T E, Zhang H and Clinton T W 1994
      {\it Phys.\ Rev.\ Lett.} {\bf 72} 3413

\item Lyuksyutov I F 1992 {\it Europhys. Lett.} {\bf 20} 273

\item Lysenko V S, Gomeniuk Y V, Lozovski V Z, Tyagulski I P  
      and Varukhin V N 1994  {\it Physica C} {\bf 235-240} 2631

\item Ma H and Chui S T 1991 {\it Phys.\ Rev.\ Lett.} {\bf 67} 505
\item Ma H and Chui S T 1992 {\it Phys.\ Rev.\ Lett.} {\bf 68} 2528

\item Machida M and Kaburaki H 1993
      {\it Phys.\ Rev.\ Lett.} {\bf 71} 3206
\item Machida M and Kaburaki H 1995
      {\it Phys.\ Rev.\ Lett.} {\bf 74} 1434 

\item Maki K 1964 {\it Physics} {\bf 1} 21          
\item Maki K 1969a {\it Superconductivity} 
      ed R D Parks (New York: Marcel Dekker) vol II p 1035
\item Maki K 1969b {\it J.\ Low Temp.\ Phys.} {\bf 1} 45    
\item Maki K 1982  {\it Progr.\ Theor.\ Phys.} {\bf 41} 902 
\item Maki K and Takayama H 1971
      {\it Progr.\ Theor.\ Phys.} {\bf 46} 1651 
\item Maki K and Thompson R S 1989
      {\it Phys.\ Rev.\ B} {\bf 39} 2767 

\item Maksimov I L 1988 {\it Phys. Lett. A} {\bf 128} 289 
\item Maksimov I L 1994 {\it Physica C} {\bf 235-240} 3017 

\item Maksimov I L and Elistratov A A 1995
      {\it Pis'ma Zh.\ Eksp.\ Teor.\ Fiz.} {\bf 61} 204   
      (1995 {\it Sov.\ Phys.\ JETP Lett.} {\bf 61} 208) 

\item Maley M P, Willis J O, Lessure H and McHenry M E 1990
      {\it Phys.\ Rev.\ B} {\bf 42} 2639    

\item Malozemoff A P 1991 {\it Physica C} {\bf 185-189} 264 

\item Malozemoff A P and Fisher M P A 1990
      {\it Phys.\ Rev.\ B} {\bf 42} 6784

\item Mannhardt J 1992 {\it Mod.\ Phys.\ Lett.\ B} {\bf 6} 555 

\item Mannhardt J, Bosch J, Gross R and Huebener R P 1987
      {\it Phys.\ Rev.\ B} {\bf 35} 5267  

\item Mannhardt J and Tsuei C C 1989
      {\it Z.\ Phys.\ B} {\bf 77} 53  

\item Mannhardt J, Scholm D G, Bednorz J G and M\"uller K A 1991
      {\it Phys.\ Rev.\ Lett.} {\bf 67} 2099                   

\item Mansky P, Chaikin P M and Haddon R C 1994  
      {\it Phys.\ Rev.\ Lett.} {\bf 70} 1323

\item Marchetti M C 1991 {\it Phys.\ Rev.\ B} {\bf 43} 8012
\item Marchetti M C 1992 {\it Physica C} {\bf 200} 155     

\item Marchetti M C and Nelson D R 1990 {\it Phys.\ Rev.\ B} {\bf 41} 1910;
                                                           {\bf 42} 9938
\item Marchetti M C and Nelson D R 1991 {\it Physica C} {\bf 174} 40
\item Marchetti M C and Nelson D R 1993 {\it Phys.\ Rev.\ B} {\bf 47} 12214
\item * Marchetti M C and Nelson D R 1995 {\it Phys.\ Rev.\ B} {\bf 52} ??
                        Theory of Double Sided Flux Decorations 

\item Marchetti M C and Vinokur M V 1994
      {\it Phys.\ Rev.\ Lett.} {\bf 72} 3409 

\item Marchetti M C and Vinokur M V 1995
      {\it Phys.\ Rev.\ B} {\bf 51} 16276 

\item * Marchevsky M, Gurevich L A, Kes P H and Aarts J 1995
      {\it Phys.\ Rev.\ Lett.} (submitted) 

\item Marcon R, Silva E, Fastampa R and Giura M 1992
      {\it Phys.\ Rev.\ B} {\bf 46} 3612           
\item Marcon R, Silva E, Fastampa R, Giura M and Sarti S 1994
      {\it Phys.\ Rev.\ B} {\bf 50} 13684          

\item Markiewicz R S 1990a
      {\it J.\ Phys.: Cond.\ Matt.} {\bf 2} 4005  
\item Markiewicz R S 1990b
      {\it Physica C} {\bf 171} 479 

\item Marley A C, Higgins M J and Bhattacharya S 1995
      {\it Phys.\ Rev.\ Lett.} {\bf 74} 3029 

\item Marsh G E 1994 {\it Phys.\ Rev.\ B} {\bf 49} 450; {\bf 50} 571

\item Martin S and Hebard A F 1991 {\it Phys.\ Rev.\ B} {\bf 43} 6253

\item Martinez J C, Brongersma S H, Koshelev A, Ivlev B, Kes P H,
      Griessen R P, de Groot D G, Tarnavski Z and Menovsky A A
      1992 {\it Phys.\ Rev.\ Lett.} {\bf 69} 2276

\item Martinez J C, van der Linden P J E M, Bulaevskii L N, Brongersma S H,
      Koshelev A, Perenboom J A A J, Menovsky A A and Kes P H 1994
      {\it Phys.\ Rev.\ Lett.} {\bf 72} 3614 

\item Martinoli P 1978     
      {\it Phys.\ Rev.\ B} {\bf 17}  1175
\item Martinoli P, Olsen J L and Clem J R 1978 
      {\it J.\ Less Comm.\ Met.} {\bf 62} 315

\item Martynovich A Yu 1993 {\it Physica C} {\bf 218} 19 

\item Martynovich A Yu 1994 {\it Physica C} {\bf 227} 22 

\item Marx A, Fath U, Ludwig W, Gross R and Amrein T 1995
      {\it Phys.\ Rev.\ B} {\bf 51} 6735  

\item Mathieu P, Pla\c{c}ais B and Simon Y 1993
      {\it Phys.\ Rev.\ B} {\bf 48} 7376      

\item Matsuda Y, Ong N P, Yan Y F, Harris J M and Peterson J B 1994
      {\it Phys.\ Rev.\ B} {\bf 49} 4380      

\item Matsushita T 1989
      {\it Physica C} {\bf 160} 328      
\item Matsushita T 1994
      {\it Physica C} {\bf 220} 1172     

\item Matsushita T and Irie F 1985
      {\it J.\ Phys.\ Soc.\ Japan} {\bf 54} 1066  

\item Matsushita T, Otabe E S and Ni B 1991
      {\it Physica C} {\bf 182} 95      
\item Matsushita T and Yamafuji K 1979  
      {\it J.\ Phys.\ Soc.\ Jpn.} {\bf 47} 1426

\item Mawatari Y and Yamafuji K 1994
      {\it Physica C} {\bf 228} 336 

\item McElfresh M, Zeldov E, Clem J R, Darwin M, Deak J and Hou L 1995
      {\it Phys.\ Rev.\ B} {\bf 51} 9111  

\item McHenry M E, Lessure H S, Maley M P, Coulter J Y, Tanaka I and
      Kojima H 1992 {\it Physica C} {\bf 190} 403 

\item Mee C, Rae A I M, Vinen W F and Gough C E 1991
      {\it Phys.\ Rev.\ B} {\bf 43} 2946      

\item Mehring M 1990 {\it Earlier and Recent Aspects of
      Superconductivity} ed J G Bednorz and K A M\"uller
      (Berlin: Springer) p 467 

\item Mehring M, Hentsch F, Mattausch Hj and Simon A 1990
      {\it Solid State Commun.} {\bf 75} 753  

\item Meilikhov E A 1993               
      {\it Physica C} {\bf 209} 566

\item Meilikhov E A and Farzetdinova R M 1993 
      {\it Physica C} {\bf 210} 473

\item Meilikhov E A and Farzetdinova R M 1994 
      {\it Physica C} {\bf 221} 27; {\it J.\ Supercond.} {\bf 7} 897

\item Meissner W and Ochsenfeld R 1933 {\it Naturwissenschaften}
            {\bf 21} 787
\item Menon G I and Dasgupta C 1994 
      {\it Phys.\ Rev.\ Lett.} {\bf 73} 1023

\item Metlushko V V, G\"untherodt G, Moshchalkov V V and Bruynseraede Y 1994a
      {\it Europhys.\ Lett.} {\bf 26} 371 

\item Metlushko V V, Baert M, Jonckheere R, Moshchalkov V V and
      Bruynseraede Y 1994b      
      {\it Solid State Commun.} {\bf 91} 331

\item Miesenb\"ock H M 1984               
      {\it J.\ Low Temp.\ Phys.} {\bf 56} 1

\item Mikheenko P N, Genenko Yu A, Medvedev Yu V, Usoskin A I
      and Chukanova I N 1993 {\it Physica C} {\bf 212} 332

\item Mikheenko P N and Kuzovlev 1993 {\it Physica C} {\bf 204} 229

\item Mikheev L V and Kolomeisky E B 1991 
      {\it Phys.\ Rev.\ B} {\bf 43} 10431

\item Mikitik G P 1995  
      {\it Physica C} {\bf 245} 287

\item Minenko E V and Kulik I O 1979 {\it Fiz.\ Nizk.\ Temp.} {\bf 5} 1237
      (1979 {\it Sov.\ J.\ Low Temp.\ Phys.} {\bf 5} 583)  

\item Minnhagen P 1991 {\it Phys.\ Rev.\ B} {\bf 44} 7546

\item Minnhagen P and Olsson P 1991        
      {\it Phys.\ Rev.\ Lett.} {\bf 67} 1039
\item Minnhagen P and Olsson P 1992 {\it Phys.\ Rev.\ B} {\bf 45} 5722;
        {\it Physica Scripta} {\bf T42} 29 

\item Mints R G and Rakhmanov A L 1981
      {\it Rev.\ Mod.\ Phys.} {\bf 53} 551      

\item Mints R G and Snapiro I B  1993
      {\it Phys.\ Rev.\ B} {\bf 47} 3273;         
      {\it Europhys.\ Lett.} {\bf 21} 611       
\item Mints R G and Snapiro I B  1994
      {\it Phys.\ Rev.\ B} {\bf 49} 6188          
\item Mints R G and Snapiro I B  1995           
      {\it Phys.\ Rev.\ B} {\bf 51} 3054 

\item Mitin A V 1987 {\it Zh. Eksp. Teor. Fiz.} {\bf 93} 590
      (1987 {\it Sov. Phys. JETP} {\bf 66} 335)

\item Miu L, Wagner P, Hadish A, Hillmer F, Adrian H, Wiesner J and Wirth G
      1995 {\it Phys.\ Rev.\ B} {\bf 51} 3953 

\item Mkrtchyan G S and Schmidt V V 1971   
      {\it Zh.\ Eksp.\ Teor.\ Fiz.} {\bf 61} 367
      (1972 {\it Sov.\ Phys.\ JETP} {\bf 34} 195)

\item Monceau P, Saint-James D and Waysand G 1975
      {\it Phys.\ Rev.\ B} {\bf 12} 3673 

\item Moon F C, Weng K C and Cheng P Z 1989
      {\it J.\ Appl.\ Phys.} {\bf 66} 5643
\item Moon F C, Yanoviak M M and Ware R 1988
      {\it Appl.\ Phys.\ Lett.} {\bf 52} 1534

\item Moonen J T and Brom H B 1995 
      {\it Physica C} {\bf 244} 1, 10 

\item Moore M A 1989 {\it Phys.\ Rev.\ B} {\bf 39} 136  
\item Moore M A 1992 {\it Phys.\ Rev.\ B} {\bf 45} 7336 
\item Moore M A and Wilkin N K 1994                     
      {\it Phys.\ Rev.\ B} {\bf 50} 10294

\item Morais-Smith C, Ivlev B and Blatter G 1994  
      {\it Phys.\ Rev.\ B} {\bf 49} 4033  
\item * Morais-Smith C, Ivlev B and Blatter G 1995  
      {\it Phys.\ Rev.\ B}  {\bf 52} (in print)

\item Moser A, Hug H J, Parashikov I, Stiefel B, Fritz O, Thomas H,
      Baratoff A, G\"untherodt H-J and Chaudhari P 1995
      {\it Phys.\ Rev.\ Lett.} {\bf 74} 1847 

\item Moser E, Seidl E and Weber H W  1982
      {\it J.\ Low Temp.\ Phys.} {\bf 49} 585 

\item Moshchalkov V V, Gielen L, Dhall{\'e} M, Van Haesendonck C and
      Bruynseraede Y 1993 {\it Nature} {\bf 361} 617 

\item Moshchalkov V V, Metlushko V V, G\"untherodt G, Goncharov I N,
      Didyk A Yu and Bruynseraede Y 1994a
      {\it Phys.\ Rev.\ B} {\bf 50} 639  

\item Moshchalkov V V, Gielen L, Baert M, Metlushko V, Neuttiens G,
      Strunk C, Bruyndoncx V, Qiu X, Dhall{\'e} M, Temst K, Potter C,
      Jonckheere R, Stockman L, Van Bael M, Van Haesendonck C
      and Bruynseraede Y 1994b {\it Physica Scripta} {\bf T55} 168

\item Mota A C, Juri G, Visani P and Pollini A 1989
       {\it Physica C} {\bf 162-164} 1152
\item Mota A C, Juri G, Visani P, Pollini A, Teruzzi T and Aupke K 1991
      {\it Physica C} {\bf 185-189} 343

\item Mou Chung-Yu, Wortis R, Dorsey A T and Huse D A 1995
      {\it Phys.\ Rev.\ B} {\bf 51} 6575  

\item M\"uller K H 1989 {\it Physica C} {\bf 159} 717     
\item M\"uller K H 1990 {\it Physica C} {\bf 168} 585
\item M\"uller K H 1991 {\it Physica C} {\bf 185-189} 1609 

\item M\"uller K H and Andrikidis C 1994
      {\it Phys.\ Rev.\ B} {\bf 49} 1294    

\item M\"uller P 1994                       
      {\it Physica C} {\bf 235-240} 289

\item Mullok S J and Evetts J E 1985
      {\it J.\ Appl.\ Phys.} {\bf 57} 2588  

\item Murakami M 1990 {\it Mod.\ Phys.\ Lett.\ B} {\bf 4} 163 
\item Murakami M, Oyama T, Fujimoto H, Taguchi T, Gotoh S, Shiohara Y,
      Koshizuga N and Tanaka S 1990
      {\it Jpn.\ J.\ Appl.\ Phys.} {\bf 29} L1991 

\item Muzikar P 1991 {\it Phys.\ Rev.\ B} {\bf 43} 10201 

\item Nakamura N, Gu G D and Koshizuka N 1993
      {\it Phys.\ Rev.\ Lett.} {\bf 71} 915

\item Narahari Achar B N 1991 {\it Solid State Commun.} {\bf 77} 689

\item Natterman T 1990 {\it Phys.\ Rev.\ Lett.} {\bf 64} 2454
\item Natterman T, Stepanov S, Tang L-H and Leshorn H 1992
      {\it J.\ Phys.\ II France} {\bf 2} 1483 

\item Neerinck D, Temst K, Baert M, Osquiguil E, Van Hasendonck C,
      Bruynseraede Y, Gilabert A and Schuller I K 1991
      {\it Phys.\ Rev.\ Lett.} {\bf 67} 2577             

\item Nelson D R 1988 {\it Phys.\ Rev.\ Lett.} {\bf 60} 1973
\item Nelson D R 1989 {\it J. Stat. Phys.} {\bf 57} 511
\item Nelson D R and Le Doussal 1990 {\it Phys.\ Rev.\ B} {\bf 42} 10113
\item Nelson D R and Seung H S 1989 {\it Phys.\ Rev.\ B} {\bf 39} 9153

\item Nelson D R and Vinokur V M 1992 {\it Phys.\ Rev.\ Lett.} {\bf 68} 2398
\item Nelson D R and Vinokur V M 1993 {\it Phys.\ Rev.\ B} {\bf 48} 13060

\item Neminsky A M, Dumas J, Thrane B P, Schlenker C, Karl H, and
      Stritzker B 1994 {\it Phys.\ Rev.\ B} {\bf 50} 3307

\item Neminsky A M and Nikolaev P N  1993
      {\it Physica C} {\bf 212} 389 

\item Nemoshkalenko V V, Ivanov M A, Nikitin B G, Pogorelov Yu G and
      Klimenko G A 1990  {\it Solid State Commun.} {\bf 74} 637 

\item Nemoshkalenko V V, Ivanov M A, Nikitin B G and Pogorelov Yu G 1992
      {\it High-$T_c$ Superconductivity} ed A S Davydov and V M Loktev
      (Berlin: Springer) p 173-193  

\item Neumann L and Tewordt L 1966 {\it Z.\ Physik} {\bf 189} 55, 
                                                    {\bf 191} 73  
\item Ni Jun and Gu Binglin 1994        
      {\it Phys.\ Rev.\ B} {\bf 49} 15276

\item Nieber S and Kronm\"uller H 1993a {\it Physica C} {\bf 210} 188
\item Nieber S and Kronm\"uller H 1993b {\it Physica C} {\bf 213} 43

\item Niel L and Evetts J 1991 {\it Europhys.\ Lett.} {\bf 15} 543

\item Nielsen H B and Olesen P 1979 {\it Nucl. Phys. B} {\bf 160} 380

\item Nikolsky R 1989 {\it Cryogenics} {\bf 29} 388 

\item Niu Q, Ao P and Thouless D J 1994
      {\it Phys.\ Rev.\ Lett.} {\bf 72} 1706

\item Norris W T 1970 {\it J.\ Phys.\ D: Appl.\ Phys.} {\bf 3} 489

\item Noshima T, Kinoshita M, Nakano S and Kuwasawa Y 1993
      {\it Physica C} {\bf 206} 387,       

\item Nozi\`ers P and Vinen W F 1966 {\it Phil.\ Mag.} {\bf 14} 667

\item * Nurgaliev T 1995 {\it Physica C} (in print) 

\item Obara H, Andersson M, F\`abrega L, Fivat P, Triscone J-M,
      Decroux M and Fischer {\O} 1995
      {\it Phys.\ Rev.\ Lett.} {\bf 74} 3041

\item Obst B 1971 {\it phys. stat. sol.} {\bf 45} 467

\item Obst B and Brandt E H 1978 {\it Phys. Lett. A} {\bf 64} 460

\item Obukhov S P and Rubinstein M 1990 {\it Phys.\ Rev.\ Lett.} {\bf 65} 1279

\item Okuda K, Kawamata S, Noguchi S, Itoh N, Kadowaki K  1991
      {\it J. Phys. Soc. Jan.} {\bf 60} 3226

\item Olsson H K, Koch R H, Eidelloth W and Robertazzi R P 1991
      {\it Phys.\ Rev.\ Lett.} {\bf 66} 2661  

\item O'Neill J A and Moore M A 1992 {\it Phys.\ Rev.\ Lett.} {\bf 69} 2582
\item O'Neill J A and Moore M A 1993 {\it Phys.\ Rev.\ B} {\bf 48} 374

\item Ota T, Tsukada I, Terasaki I and Uchinokura K 1994
      {\it Phys.\ Rev.\ B} {\bf 50} 3363  

\item Oussena M, Gagnon R, Wang Y and Aubin M 1992
      {\it Phys.\ Rev.\ B} {\bf 46} 528       
\item Oussena M, de Groot P A J, Gagnon R and Taillefer L 1994a
      {\it Phys.\ Rev.\ B} {\bf 49} 9222   
\item Oussena M, de Groot P A J, Gagnon R and Taillefer L 1994b
      {\it Phys.\ Rev.\ Lett.} {\bf 72} 3606  
\item Oussena M, de Groot P A J, Porter S J, Gagnon R and Taillefer L 1995
      {\it Phys.\ Rev.\ B} {\bf 51} 1389  

\item Ovchinnikov Yu N 1977                 
            {\it Z. Physik B} {\bf 27} 239
\item Ovchinnikov Yu N 1980
            {\it Zh. Eksp. Teor. Fiz.} {\bf 79}  1496
            (1980 {\it Sov. Phys. JETP} {\bf 52} 755) 
\item Ovchinnikov Yu N 1982
            {\it Zh. Eksp. Teor. Fiz.} {\bf 82} 2020
            (1982 {\it Sov. Phys. JETP} {\bf 55} 1162)   
\item Ovchinnikov Yu N 1983                              
            {\it Zh. Eksp. Teor. Fiz.} {\bf 84} 237
            (1983 {\it Sov. Phys. JETP} {\bf 57} 136)
\item Ovchinnikov Yu N and Brandt E H 1975               
            {\it phys.\ stat.\ sol.} {\bf 67} 301
\item * Ovchinnikov Yu N and Kresin V Z 1995  
            {\it Phys.\ Rev.\ B} {\bf 52} (in print)

\item Palstra T T M, Battlog B, Schneemeyer L F and Waszczak J V 1990a
      {\it Phys.\ Rev.\ Lett.} {\bf 64} 3090   
\item Palstra T T M, Battlog B, van Dover R B, Schneemeyer L F and
      Waszczak J V 1990b  {\it Phys.\ Rev.\ B} {\bf 41} 6621  

\item Pan V M 1993 {\it J.\ Alloys and Compounds} {\bf 195} 387 

\item Pankert J, Marbach G, Comberg A, Lemmens P, Fr\"onig P
      and Ewert S 1990 {\it Phys.\ Rev.\ Lett.} {\bf 56} 3052  

\item Pannetier B 1991 {\it Quantum Coherence in Mesoscopic Systems}
      ed B Kramer (New York: Plenum) p 457  

\item Park S J and Kouvel J S 1993  
      {\it Phys.\ Rev.\ B} {\bf 48} 13995 

\item Parks B, Spielman S, Orenstein J, Nemeth D T, Ludwig F, Clarke J,
      Merchant P and Lew D J 1995
      {\it Phys.\ Rev.\ Lett.} {\bf 74} 3265 

\item Pashitski A E, Polyanskii A, Gurevich A, Parrell J A and
      Larbalestier D C  1991
      {\it Physica C} {\bf 246} 133 

\item Pastoriza H, Goffmann M F, Arrib\'ere A and de la Cruz F 1994
      {\it Phys.\ Rev.\ Lett.} {\bf 72} 2951 

\item Pearl J 1964 {\it J. Appl. Phys. Lett.} {\bf 5} 65 
\item Pearl J 1966 {\it J. Appl. Phys.} {\bf 37} 4139

\item Perez-Gonzalez A and Clem J R 1985  
      {\it Phys.\ Rev.\ B} {\bf 31} 7048;
      {\it J.\ Appl.\ Phys.} {\bf 58} 4326
\item Perez-Gonzalez A and Clem J R 1991  
      {\it Phys.\ Rev.\ B} {\bf 43} 7792

\item * Perkins G K, Cohen L F, Zhukov A A and Caplin A D 1995
      {\it Phys.\ Rev.\ B} {\bf 52} ???  

\item Pereyra P and Kunold A 1991        
      {\it Phys.\ Rev.\ B} {\bf 51} 3820 

\item Pesch W 1975 {\it Z.\ Physik B} {\bf 21} 263  

\item Pesch W and Kramer L 1974
      {\it J.\ Low Temp.\ Phys.} {\bf 15} 367
\item Pesch W, Watts-Tobin R and Kramer L 1974 
      {\it Z.\ Physik} {\bf 269} 253

\item Peters P N, Sisk R C, Urban E W, Huang C Y and Wu M K 1988
      {\it Appl.\ Phys.\ Lett.} {\bf 52} 2066 

\item Petersen R L and Ekin J W 1988  
      {\it Phys.\ Rev.\ B} {\bf 37} 9848

\item Petzinger K G and Warren G A 1990 {\it Phys.\ Rev.\ B} {\bf 42} 2023
\item Petzinger K G and Tuttle B 1993 {\it Phys.\ Rev.\ B} {\bf 47} 2909

\item Piel H and M\"uller G 1991      
      {\it Trans.\ Mag.} {\bf 27} 85

\item Pierson S W 1994 {\it Phys.\ Rev.\ Lett.} {\bf 73} 2496

\item Pierson S W 1995 {\it Phys.\ Rev.\ B} {\bf 51} 6663;
      {\it Phys.\ Rev.\ Lett.} {\bf 74} 2359  

\item Pippard A B 1969 {\it Phil. Mag.} {\bf 19} 220
\item Pippard A B 1969 {\it Phil. Mag.} {\bf 19} 220
\item Pippard A B 1994 {\it Supercond.\ Sci.\ Technol.} {\bf 7} 696

\item Pla O and Nori F 1991 {\it Phys.\ Rev.\ Lett.} {\bf 67} 919

\item Pla\c{c}ais B, Mathieu P and Simon Y 1993
      {\it Phys.\ Rev.\ Lett.} {\bf 60} 1973      
\item Pla\c{c}ais B, Mathieu P and Simon Y 1994
      {\it Phys.\ Rev.\ B} {\bf 49} 1581      

\item Pokrovsky V L, Lyuksyutov I and Nattermann T 1992
      {\it Phys.\ Rev.\ B} {\bf 46} 3071        

\item Porter S J, Daniel G J, Oussena M and de Groot P A J 1994
      {\it Physica C} {\bf 235-240} 3099 

\item Portis A M 1992 {\it J.\ Supercond.} {\bf 5} 319 
\item Portis A M, Blazey K W and Waldner F 1988
      {\it Physica C} {\bf 153-155} 308                

\item Preosti G and Muzikar P 1993
      {\it Phys.\ Rev.\ B} {\bf 48} 8831, 9921   

\item Proki\'{c} V, Davidovi\'{c} D and Dobrosavljevi\'{c}-Gruji\'{c} L
      1995 {\it Phys.\ Rev.\ B} {\bf 51} 1270 

\item Provost J, Paumier E and Fortini A 1974 {\it J. Phys. F} {\bf 4} 439

\item Prost D, Fruchter L, Campbell I A, Motohira N and
      Konczykowski M 1993 {\it Phys.\ Rev.\ B} {\bf 47} 3457 

\item Pudikov V 1993
      {\it Physica C} {\bf 212} 

\item P\"umpin B, Keller H, K\"undig W, Odermatt W, Patterson D,
      Schneider J W, Simmler H, Connell S, M\"uller K A, Bednorz J G,
      Blazey K W, Morgenstern I, Rossel C and Savi\'c M
             1988 {\it Z.\ Phys.\ B} {\bf 72} 175
\item P\"umpin B, Keller H, K\"undig W, Odermatt W, Savi\'c I M,
      Schneider J W, Simmler H, Zimmermann P, Kaldis E, Rusiecki S,
      Maeno Y and Rossel C 1990 {\it Phys.\ Rev. B} {\bf 42} 8019  

\item P\accent23ust L 1992
      {\it Supercond.\ Sci.\ Technol.} {\bf 5} 497

\item Qiu X G and Tachiki M 1993 {\it Physica C} {\bf 207} 255
\item Qiu X G, Zhao B R, Guo S Q, Zhang J L, Li L and Tachiki M 1993
      {\it Phys.\ Rev.\ B} {\bf 48} 16180

\item Radzihovsky L and Frey E 1993 {\it Phys.\ Rev.\ B} {\bf 48} 10357

\item Raffy H, Renard J C and Guyon E 1972  
      {\it Solid State Commun.} {\bf 11} 1679
\item Raffy H, Renard J C and Guyon E 1974  
      {\it Solid State Commun.} {\bf 14} 427, 431

\item Raffy H, Labdi S, Laborde O and Monceau P 1991
      {\it Phys.\ Rev.\ Lett.} {\bf 66} 2515, 
      {\it Physica C} {\bf 184} 159                             
\item Raffy H, Labdi S, Li Z Z, Rifi H, Kim S F, Megtert S, Laborde O
      and Monceau P 1994
      {\it Physica C} {\bf 235-240} 182 

\item Rammer J 1988 {\it J.\ Low Temp.\ Phys.} {\bf 71} 323
\item Rammer J 1991a {\it Physica C} {\bf 177} 421  
\item Rammer J 1991b {\it Physica C} {\bf 181} 99   
\item Rammer J, Pesch W and Kramer L 1987
      {\it Z.\ Phys.\ B} {\bf 68} 49                

\item Raveau B 1992
      {\it Physics Today} {\bf 45} 53        

\item Reed D S, Yeh N-C, Jiang W, Kriplani U, Beam D A and Holtzberg F
      1994 {\it Phys.\ Rev.\ B} {\bf 49} 4384   

\item Reed D S, Yeh N-C, Konczykowski M, Samoilov A V and Holtzberg F
      1995 {\it Phys.\ Rev.\ B} {\bf 51} 16448 

\item Reefman D and Brom H B 1993 {\it Physica C} {\bf 213} 229;
         {\it Phys.\ Rev.\ B} {\bf 48} 3567

\item Reinel D, Dieterich W, Majhofer A and Wolf T 1995
      {\it Physica C} {\bf 245} 193 

\item Reittu H J and Laiho R 1992  
      {\it Supercond.\ Sci.\ Technol.} {\bf 5} 448

\item Ren Y, Xu J-H and Ting C S 1995  
      {\it Phys.\ Rev.\ Lett.} {\bf 74} 3680

\item Renner Ch, Kent A D, Niedermann Ph and Fisher O 1991
      {\it Phys.\ Rev.\ Lett.} {\bf 67} 1650

\item Rhyner J 1993
      {\it Physica C} {\bf 212} 292 

\item Rhyner J and Blatter G 1989        
      {\it Phys.\ Rev.\ B} {\bf 40} 829  

\item Ri H-C, Gross R, Gollnik F, Beck A, Huebener R P, Wagner P and
      Adrian H 1994 {\it Phys.\ Rev.\ B} {\bf 50} 3312  

\item Ri H-C, Kober F, Beck A, Alff L, Gross R and Huebener R P 1993
      {\it Phys.\ Rev.\ B} {\bf 47} 12312

\item Richardson R A, Pla O and Nori F 1994
      {\it Phys.\ Rev.\ Lett.} {\bf 72} 1268

\item Riek C T, Scharnberg K and Schopohl N 1991  
      {\it J.\ Low Temp.\ Phys.} {\bf 84} 381

\item Ries G, Neum\"uller H-W, Busch R, Kummeth P, Leghissa M, Schmitt P
      and Saemann-Ischenko G 1993    
      {\it J. Alloys and Comp.} {\bf 72} 1268

\item * Riseman T M, Brewer J H, Chow K H, Hardy W N, Kiefl R F,
      Kreitzman S R, Liang R, MacFarlane W A, Mendels P, Morris G D,
      Rammer J, Schneider J W, Niedermayer C and Lee S L 1995
      {\it Phys.\ Rev.\ B} (submitted) 

\item Roas B, Hensel B, Henke S, Klaum\"unzer S, Kabius B, Watanabe W,
      Saemann-Ischenko G, Schultz L and Urban K 1990a
      {\it Europhys.\ Lett.} {\bf 11} 669   

\item Roas B, Schultz L and Saemann-Ischenko G 1990b
      {\it Phys.\ Rev.\ Lett.} {\bf 64} 479

\item Roest W and Rekveldt M Th 1993    
      {\it Phys.\ Rev.\ B} {\bf 48} 6420

\item Roddick E and Stroud D 1995        
      {\it Phys.\ Rev.\ Lett.} {\bf 74} 1430

\item Rogacki K, Esquinazi P, Faulhaber E and Sadowski W 1995
      {\it Physica C} {\bf 246} 123 

\item Roitburd A, Swartzendruber L J, Kaiser D L, Gayle F W and
      Bennett L H 1990 {\it Phys.\ Rev.\ Lett.} {\bf 64} C2962 

\item Rossel C, Maeno Y and Morgenstern I 1989  
      {\it Phys.\ Rev.\ Lett.} {\bf 62} 681

\item Rossel C, Sandvold E, Sergent M, Chevrel R and Potel M 1990
      {\it Physica C} {\bf 165} 233

\item Roth B J, Sepulveda N G and Wiskwo J P 1989 
      {\it J.\ Appl.\ Phys.} {\bf 65} 361   

\item Rui Y, Ji H L, Xu X N, Shao H M, Qin M J, Jin X, Yao X X, Rong X S,
      Ying B and Zhao Z X 1995          
      {\it Phys.\ Rev.\ B} {\bf 51} 9161

\item Runge K and Pannetier B 1993
      {\it J.\ Phys.\ I France} {\bf 3} 389

\item Ryu S, Doniach S, Deutscher G and Kapitulnik A 1992
      {\it Phys.\ Rev.\ Lett.} {\bf 68} 710
\item Ryu S, Kapitulnik A and Doniach S 1993
      {\it Phys.\ Rev.\ Lett.} {\bf 71} 4245

\item Safar H, Gammel P L, Bishop D J, Mitzi D B and Kapitulnik A
      1992a {\it Phys.\ Rev.\ Lett.} {\bf 68} 2672    
\item Safar H, Gammel P L, Huse D A, Bishop D J, Rice J P and Ginsberg D M
      1992b {\it Phys.\ Rev.\ Lett.} {\bf 69} 824     
\item Safar H, Gammel P L, Huse D A, Majumdar S N, Schneemeyer L F,
      Bishop D J, L\'opez D, Nieva G and de la Cruz F 1994
      {\it Phys.\ Rev.\ Lett.} {\bf 72} 1272          
\item Safar H, Rodr\'{\i}guez E, de la Cruz F, Gammel P L, Schneemeyer L F
      and Bishop D J 1992c
      {\it Phys.\ Rev.\ B} {\bf 46} 14238  

\item Saint-James D, Sarma G and Thomas E J 1969
      {\it Type-II Superconductivity} (New York: Benjamin)

\item Saint-James D and DeGennes P G 1963
      {\it Phys.\ Lett.} {\bf 7} 306       

\item Salamon M B and Shi Jing 1992        
      {\it Phys.\ Rev.\ Lett.} {\bf 69} C1622

\item Samoilov A V 1993 {\it Phys.\ Rev.\ Lett.} {\bf 71} 617
\item Samoilov A V, Ivanov Z G and Johansson L-G 1994
      {\it Phys.\ Rev.\ B} {\bf 49} 3667
\item Samoilov A V, Legris A, Rullier-Albenque F, Lejay P, Bouffard S,
      Ivanov Z G and Johansson L-G 1995
      {\it Phys.\ Rev.\ B} {\bf 51} 2351 

\item Sandvold E and Rossel C 1992 {\it Physica C} {\bf 190} 309

\item Sardella E 1991 {\it Phys.\ Rev.\ B} {\bf 44} 5209
\item Sardella E 1992 {\it Phys.\ Rev.\ B} {\bf 45} 3141

\item Sardella E and Moore M A 1993    
      {\it Phys.\ Rev.\ B} {\bf 48} 9664

\item Sarma N V 1968 {\it Phil.\ Mag.} {\bf 18} 171 

\item Sarti S, Silva E, Fastampa R, Giura M and Marcon R 1994
      {\it Phys.\ Rev.\ B} {\bf 49} 556

\item \v{S}\'a\v{s}ik R and Stroud D  1993     
      {\it Phys.\ Rev.\ B}     {\bf 48} 9938
\item \v{S}\'a\v{s}ik R and Stroud D  1994a    
      {\it Phys.\ Rev.\ Lett.} {\bf 72} 2462
\item \v{S}\'a\v{s}ik R and Stroud D  1994b    
      {\it Phys.\ Rev.\ B} {\bf 49} 16074, {\bf 50} 3294
\item * \v{S}\'a\v{s}ik R and Stroud D  1995
      {\it Phys.\ Rev.\ B} {\bf 52} ????         

\item \v{S}\'a\v{s}ik R, Stroud D  and Te\v{s}anovich 1995
      {\it Phys.\ Rev.\ B} {\bf 51} 3042  

\item Sauerzopf F M, Seidl E and Weber H W 1982
      {\it J.\ Low Temp.\ Phys.} {\bf 49} 249   

\item Sauerzopf F M, Wiesinger H P, Weber H W and Crabtree G W 1995
      {\it Phys.\ Rev.\ B} {\bf 51} 6002 

\item Schalk R M, Weber H W, Barber Z H, Przylupsky P and Evetts J E
      {\it Physica C} {\bf 199} 311

\item Scheidl S and Hackenbroich G 1992 {\it Phys.\ Rev.\ B} {\bf 46} 14010

\item Schelten J, Ullmaier H and Lippmann G 1975
      {\it Phys.\ Rev.\ B} {\bf 12} 1772     

\item Schilling A, Jin R, Guo J D and Ott H R 1993
      {\it Phys.\ Rev.\ Lett.} {\bf 71} 1899     

\item Schilling A, Ott H R and Wolf Th 1992
      {\it Phys.\ Rev.\ B} {\bf 46} 14253        

\item Schimmele L, Kronm\"uller H and Teichler H 1988
      {\it phys.\ stat.\ sol.\ b} {\bf 147} 361

\item Schindler G, Seebacher B, Kleiner R, M\"uller P and Andres K 1992
      {\it Physica C} {\bf 196} 1  

\item Schmid A 1966 {\it Phys. Kond. Materie} {\bf 5} 302   
\item Schmid A 1982 {\it J.\ Low Temp.\ Phys.} {\bf 49} 609 

\item Schmid A and Hauger W 1973 {\it J.\ Low Temp.\ Phys.} {\bf 11} 667

\item Schmidt H 1968              
      {\it Z. Physik} {\bf 216} 336

\item Schmidt M F, Israeloff N E and Goldmann A M 1993
      {\it Phys.\ Rev.\ Lett.} {\bf 70} 2162  

\item Schmitt P, Kummeth P, Schultz L and Saemann-Ischenko G 1991
      {\it Phys.\ Rev.\ Lett.} {\bf 67} 267                 
\item Schmucker R 1977 {\it Phil.\ Mag.} {\bf 35} 431, 453

\item Schmucker R and Brandt E H 1977
       {\it phys.\ stat.\ sol.\ (b)} {\bf 79} 479
\item Schmucker R and Kronm\"uller H 1974
       {\it phys.\ stat.\ sol.\ (b)} {\bf 61} 181

\item Schnack H G, Griessen R, Lensink J G, van der Beek C J and
      Kes P H 1992 {\it Physica C} {\bf 197} 337

\item Schnack H G, Griessen R, Lensink J G, and Wen Hai-hu 1993
      {\it Phys.\ Rev.\ B} {\bf 48} 13178 

\item Schneider E and  Kronm\"uller H 1976
       {\it phys.\ stat.\ sol.\ (b)} {\bf 74} 261

\item Schoenes J, Kaldis E and Karpinski J 1993
      {\it Phys.\ Rev.\ B} {\bf 48} 16869        

\item Sch\"onenberger A M, Geshkenbein V B and Blatter G 1993
      {\it Phys.\ Rev.\ B} {\bf 48} 15914

\item * Sch\"onenberger A M, Geshkenbein V B and Blatter G 1995
      {\it Phys.\ Rev.\ Lett.} {\bf 74} (submitted) 

\item Schopohl N and Baratoff A 1988 {\it Physica C} {\bf 153-155} 689
\item * Schopohl N and Maki K 1995 {\it Europhys.\ Lett.} ???

\item Schuster Th, Indenbom M V, Koblischka M R, Kuhn H
      and Kronm\"uller H 1994a {\it Phys.\ Rev.\ B} {\bf 49} 3443

\item Schuster Th, Indenbom M V, Kuhn H, Brandt E H
      and Konczykowski M 1994b {\it Phys.\ Rev.\ Lett.} {\bf 73} 1424

\item Schuster Th, Indenbom M V, Kuhn H, Kronm\"uller H, Leghissa M,
      and Kreiselmeyer G 1994c        
      {\it Phys.\ Rev.\ B} {\bf 50} 9499

\item Schuster Th, Koblischka M R, Kuhn H, Gl\"ucker M, Ludescher B 
      and Kronm\"uller H 1993a {\it J.\ Appl.\ Phys.} {\bf 74} 3307

\item Schuster Th, Kuhn H, Koblischka M R, Theuss H, Kronm\"uller H,
      Leghissa M, Kraus M and Saemann-Ischenko G 1993b
      {\it Phys.\ Rev.\ B} {\bf 47} 373                 

\item Schuster Th, Koblischka M R, Kuhn H, Kronm\"uller H, Friedl G,
      Roas B and Schultz L 1993c    
      {\it Appl.\ Phys.\ Lett.} {\bf 62} 768

\item Schuster Th, Koblischka M R, Kuhn H, Kronm\"uller H, Leghissa M,
      Gerh\"auser W, Saemann-Ischenko G, Neum\"uller H-W and 
      Klaum\"unzer S 1992a {\it Phys.\ Rev.\ B} {\bf 46} 8496

\item Schuster Th, Koblischka M R, Reininger T, Ludescher B, Henes R
      and Kronm\"uller H 1992b  
      {\it Supercond.\ Sci.\ Technol.} {\bf 5} 614

\item Schuster Th, Koblischka M R, Ludescher B, Moser N   
      and Kronm\"uller H 1991 {\it Cryogenics} {\bf 31} 811

\item Schuster Th, Kuhn H, Brandt E H, Indenbom M V, Koblischka M R
      and Konczykowski M 1994d {\it Phys.\ Rev.\ B} {\bf 50} 16684 

\item Schuster Th, Kuhn H and Brandt E H 1995a
            {\it Phys.\ Rev.\ B} {\bf 51} 697  

\item Schuster Th, Kuhn H, Indenbom M V, Leghissa M, Kraus M and
      Konczykowski M  1995b           
      {\it Phys.\ Rev.\ B} {\bf 51} 16358

\item Seeger A 1955 {\it Theorie der Gitterfehlstellen.
      Handbuch der Physik} vol 7 part 1
      ed S Fl\"ugge (Berlin: Springer) p 383
\item Seeger A 1958 {\it Kristallplastizit\"at.
      Handbuch der Physik} vol 7 part 2
      ed S Fl\"ugge (Berlin: Springer) p 1

\item Seeger A 1970 {\it Metallurg.\ Trans.} {\bf 1} 2987  
\item Seeger A 1979    
      {\it Phys.\ Lett.} {\bf 77A} 259
\item Seeger A and Kronm\"uller H 1968       
      {\it phys.\ stat.\ sol.} {\bf 27} 371

\item Seidler G T, Rosenbaum T F, Beauchamp K M, Jaeger H M, Crabtree G W,
      Welp U and Vinokur V M  1995           
      {\it Phys.\ Rev.\ Lett.} {\bf 74} 1442

\item Seng Ph, Gross R, Baier U, Rupp M, Koelle D, Huebener R P,
      Schmitt P and Saemann-Ischenko G 1992
      {\it Physica C} {\bf 192} 403          

\item Sengupta S, Dasgupta C, Krishnamurty H R, Menon G I,
      Ramakrishnan T V 1991 {\it Phys.\ Rev.\ Lett.} {\bf 67} 3444 
\item Sengupta S, Shi D, Wang Z, Smith M E and McGinn P J 1993
      {\it Phys.\ Rev.\ B} {\bf 47} 5165  

\item Senoussi S 1992 {\it J. Physique III (Paris)} {\bf 2} 1041 
\item Senoussi S, Aguillon C and Hadjoudi S 1991
      {\it Physica C} {\bf 175} 215
\item Senoussi S, Hadjoudi S, Maury R and Fert A 1990
      {\it Physica C} {\bf 165} 364  
\item Senoussi S, Mosbah F, Sarrhini O and Hammond S 1993
      {\it Physica C} {\bf 211} 288  
\item Senoussi S, Ouss\'ena M, Collin G and Campbell I A 1988
      {\it Phys.\ Rev. B} {\bf 37} 9792

\item Seow W S, Doyle R A, Johnson J D, Kumar D, Somekh R, Walker D J C,
      and Campbell A M 1995
      {\it Physica C} {\bf 241} 71   

\item Serene J W and Rainer D 1983   
      {\it Physics Reports} {\bf 101} 221  

\item Shapira Y and Neuringer L J 1967
      {\it Phys.\ Rev.} {\bf 154} 375  

\item Shapoval E A 1995
      {\it Pis'ma Zh.\ Eksp.\ Teor.\ Fiz.} {\bf 61} 124 
      (1995 {\it Sov.\ Phys.\ JETP Lett.}  {\bf 61} 131)

\item Shatz S, Shaulov A and Yeshurun Y 1993 
      {\it Phys.\ Rev.\ B} {\bf 48} 13871

\item Shehata L N 1981
      {\it phys.\ stat.\ sol.\ (b)} {\bf 105} 77 

\item Shenoy S R and Chattopadhyay B 1995 
      {\it Phys.\ Rev.\ B} {\bf 51} 9129  

\item Shi An-Chang and Berlinsky A J 1991
      {\it Phys.\ Rev.\ Lett.} {\bf 67} 1926 

\item Shi Donglu, Kourous H E, Xu Ming and Kim D H 1991
      {\it Phys.\ Rev. B} {\bf 43} 514  

\item Shi X D, Chaikin P M, Ong N P and Wang Z Z 1994
      {\it Phys.\ Rev. B} {\bf 50} 13845  

\item Shimshoni E 1995
      {\it Phys.\ Rev.\ B} {\bf 51} 9415 

\item Shmidt V V and Mkrtchyan G S                      
       1974 {\it Usp.\ Fiz.\ Nauk} {\bf 112} 459
      (1974 {\it Sov.\ Phys.\ Usp.} {\bf 17} 170)

\item Shridar S, Wu D-H and Kennedy W 1989
      {\it Phys.\ Rev.\ Lett.} {\bf 63} 1873   

\item \v{S}im\'anek E 1992 {\it Phys.\ Rev.\ B} {\bf 46} 14054

\item Simon Y and Thorel P 1971  
      {\it Phys. Lett.} {\bf 35A} 450

\item Skvortsov M A and Geshkenbein V B 1994
      {\it Zh.\ Eksp.\ Teor.\ Fiz.} {\bf 105} 1379
      (1994 {\it Sov.\ Phys.\ JETP} {\bf 78} 743)

\item Smith A W, Clinton T W, Tsuei C C and Lobb C J 1994
      {\it Phys.\ Rev.\ B} {\bf 49} 12927   

\item Soininen P I, Kallin C and Berlinsky A J 1994
      {\it Physica C} {\bf 235-240} 2593  

\item Solovjov V F, Pan V M and Freyhardt H C 1994
      {\it Phys.\ Rev.\ B} {\bf 50} 13724   

\item Song Y-Q 1995 {\it Physica C} {\bf 241} 187 

\item Song Y-Q, Lee M, Halperin W P, Tonge L M and Marks T J 1992
      {\it Phys.\ Rev.\ B} {\bf 45} 4945  
\item Song Y-Q, Halperin W P, Tonge L, Marks T J, Ledvij M, Kogan V G
      and Bulaevskii L N 1993 {\it Phys.\ Rev.\ Lett.} {\bf 70} 3127  
\item Song Y-Q, Tripp S, Halperin W P, Tonge L and Marks T J 
      and Bulaevskii L N 1994 {\it Phys.\ Rev.\ B} {\bf 50} 16570

\item Sonier J E, Kiefl R F, Brewer J H, Bonn D A, Carolan J F,
      Chow K H, Dosanjh P, Hardy W N, MacFarlane W A, Mendels P,
      Morris G D, Riseman T M and Schneider J W 1994
      {\it Phys.\ Rev.\ Lett.} {\bf 72} 744  

\item * Sonin E B 1995
      {\it Physica B} {\bf ??} ?? preprint, 31.10.94 vortices in He and scs

\item Sonin E B and Horovitz B 1995  
      {\it Phys.\ Rev.\ B} {\bf 51} 6526

\item Sonin E B and Krusius M 1994 {\it  The Vortex State} 1994
      ed N Bontemps, Y Bruynseraede, G Deutscher and A Kapitulnik
      (Dordrecht: Kluwer) p 193

\item Sonin E B and Traito K B 1994   
      {\it Phys.\ Rev.\ B} {\bf 50} 13547

\item Soulen R J, Francavilla T L, Fuller-Mora W W, Miller MM, Joshi C H,
      Carter W L, Rodenbush A J, Manlief M D and Aized D 1994
      {\it Phys.\ Rev.\ B} {\bf 50} 478 

\item Spirgatis A, Trox R, K\"otzler J and Bock J 1992
      {\it Cryogenics} {\bf 33} 138

\item Spivak B and Zhou F 1995 
      {\it Phys.\ Rev.\ Lett.} {\bf 74} 2800

\item Stamp P C E, Forro L and Ayache C 1988
      {\it Phys.\ Rev.\ B} {\bf 38} 2847  

\item Steegmans A, Provoost R, Moshchalkov V V, Silverans R E,
      Libbrecht S, Buekenhoudt A and Bruynseraede Y 1993
      {\it Physica C} {\bf 218} 295

\item Steel D G and Graybeal J M 1992
      {\it Phys.\ Rev.\ B} {\bf 45} 12643

\item Steel D G, White W R and Graybeal J M 1993
      {\it Phys.\ Rev.\ Lett.} {\bf 71} 161

\item Steji{\'c} G, Gurevich A, Kadyrov E, Christen D, Joynt R and
      Larbalestier D C 1994 {\it Phys.\ Rev.\ B} {\bf 49} 1274

\item Steinmeyer F, Kleiner R, M\"uller P, M\"uller H and Winzer K 1994a
      {\it Europhys. Lett.} {\bf 25} 459
\item Steinmeyer F, Kleiner R, M\"uller P and Winzer K 1994b
      {\it Physica B} {\bf 194-196} 2401

\item Stephen M J 1994 {\it Phys.\ Rev.\ Lett.} {\bf 72} 1534

\item Stoddart S T, Bending S J, Geim A K and Henini M 1993
      {\it Phys.\ Rev.\ Lett.} {\bf 71} 3854

\item Stoppard O and Guggan D 1995
      {\it Physica C} {\bf 241} 375  

\item Stucki F, Rhyner J and Blatter G 1991
      {\it Physica C} {\bf 181} 385  

\item Sudb{\o} A 1992 {\it Physica C} {\bf 201} 369

\item Sudb{\o} A and Brandt E H 1991a {\it Phys.\ Rev.\ Lett.} {\bf 66} 1781

\item Sudb{\o} A and Brandt E H 1991b {\it Phys.\ Rev.\ B} {\bf 43} 10482
\item Sudb{\o} A and Brandt E H 1991c {\it Phys.\ Rev.\ Lett.} {\bf 67} 3176

\item Sudb{\o} A, Brandt E H and Huse D A 1993
      {\it Phys.\ Rev.\ Lett.} {\bf 71} 1451

\item Suenaga M, Gosh A K, Xu Y and Welsh D O 1991
      {\it Phys.\ Rev.\ Lett.} {\bf 66} 1777

\item Suh B J, Torgeson D R and Borsa F 1993
      {\it Phys.\ Rev.\ Lett.} {\bf 71} 3011

\item Suhl H 1965 {\it Phys.\ Rev.\ Lett.} {\bf 14} 226

\item Sun J Z, Gallagher W J and Koch R H 1994 
      {\it Phys.\ Rev.\ B} {\bf 50} 13664

\item Sun Y R, Thompson J R, Chen Y J, Christen D K and Goyal A 1993
      {\it Phys.\ Rev.\ B} {\bf 47} 14481

\item Sun Y R, Thompson J R, Schwartz J, Christen D K, Kim Y C and
      Paranthaman M 1995
      {\it Phys.\ Rev.\ B} {\bf 47} 14481

\item Suzuki T and Seeger A 1971
      {\it Comments Solid State Phys.} {\bf 3} 128, 173

\item Suzuki T, Iwai T and Takanaka K 1995 
      {\it Physica C} {\bf 242} 90

\item Svedlindh P, Niskanen K, Norling P, Nordblad P, Lundgren L, Rossel C,
      Sergent M, Chevrel R and Potel M 1991  
      {\it Phys.\ Rev.\ B} {\bf 43} 2735

\item Svensmark H and Falikov L M 1990 
      {\it Phys.\ Rev.\ B} {\bf 42} 9957

\item Swartzendruber L J, Roitburd A, Kaiser D L, Gayle F W and  
      Bennett L H 1990 {\it Phys.\ Rev.\ Lett.} {\bf 64} 483  

\item Tachiki M 1982   
      {\it Physica B} {\bf 109\&110} 1699

\item Tachiki M, Koyama T and Takahashi S 1991
      {\it Physica C} {\bf 185-189} 303  

\item Tachiki M, Koyama T and Takahashi S 1994
      {\it Phys.\ Rev.\ B} {\bf 50} 7065  

\item Tachiki M and Takahashi S 1989 {\it Solid State Commun.}
      {\bf 70} 2991
\item Tachiki M and Takahashi S 1991 {\it Physica B} {\bf 169} 121

\item Tachiki M, Takahashi S and Sunaga K 1991 
      {\it Phys.\ Rev.\ B} {\bf 47} 6095

\item Tak\'acs S 1982 {\it phys. stat. sol. (a)} {\bf 74} 437

\item Tak\'acs S and G\"om\"ory F 1990
      {\it Supercond.\ Sci.\ Technol.} {\bf 3} 94

\item Takanaka K 1975 {\it phys. stat. sol. (b)} {\bf 68} 623

\item Takayama H and Maki K 1973             
      {\it J.\ Low Temp.\ Phys.} {\bf 12} 195

\item Takezawa N, Koyama T and Tachiki M 1993 
      {\it Physica C} {\bf 207} 231

\item Tamegai T, Krusin-Elbaum L, Civale L, Santhanam P, Brady M J,
      Masselink W T, Holtzberg F and Feild C 1992 
      {\it Phys.\ Rev.\ B} {\bf 45} 8201          

\item Tang C 1993 {\it Physica A} {\bf 194} 315
\item Tang C and Bak P 1988 {\it Phys.\ Rev.\ Lett.} {\bf 60} 2347
\item Tang C, Feng S and Golubovic L 1994
      {\it Phys.\ Rev.\ Lett.} {\bf 72} 1264

\item Teichler H 1975 {\it phys. stat. sol. (b)} {\bf 72} 211

\item Ternovski\u{\i} F F and Shehata L N 1972  
      {\it Zh.\ Eksp.\ Teor.\ Fiz.} {\bf 62} 2297
      (1972 {\it Sov.\ Phys.\ JETP} {\bf 35} 1202)

\item Te\v{s}anovi\'{c} Z  1991 {\it Phys.\ Rev.\ B} {\bf 44} 12635
\item Te\v{s}anovi\'{c} Z  1995 {\it Phys.\ Rev.\ B} {\bf 51} 16204

\item Te\v{s}anovi\'{c} Z and Xing L 1991
      {\it Phys.\ Rev.\ Lett.} {\bf 67} 2729
\item Te\v{s}anovi\'{c} Z, Xing L, Bulaevskii L, Li Q and Suenaga M 1992
      {\it Phys.\ Rev.\ Lett.} {\bf 69} 3563

\item Th\'eron R, Simond J-B, Leemann Ch, Beck H, Martinoli P and
      Minnhagen P 1993
      {\it Phys.\ Rev.\ Lett.} {\bf 71} 1246 
\item Th\'eron R, Korshunov S E, Simond J-B, Leemann Ch and Martinoli P 1994
      {\it Phys.\ Rev.\ Lett.} {\bf 72} 562  

\item Theodorakis S 1990 {\it Phys.\ Rev.\ B} {\bf 42} 10172 

\item Theodorakis S and Etthouami A M 1995
      {\it Phys.\ Rev.\ B} {\bf 51} 11664 

\item Theuss H 1993 {\it Physica C} {\bf 208} 155

\item Theuss H, Becker T and Kronm\"uller H 1994
      {\it Physica C} {\bf 233} 179
\item Theuss H, Forkl A and Kronm\"uller H 1992a
      {\it Physica C} {\bf 190} 345
\item Theuss H, Reininger T and Kronm\"uller H 1992b
      {\it J.\ Appl.\ Phys.} {\bf 72} 1936

\item Thiemann S, Radovi\'c Z and Kogan V G 189 {\it Phys.\ Rev.\ B}
      {\bf 39} 11406

\item Thompson J R, Ossandon J G, Christen D K, Chakoumakos B C,
      Sun Y R, Paranthaman M and Brynestad J 1993
      {\it Phys.\ Rev.\ B} {\bf 48} 14031 

\item Thompson J R, Sun Y R, and Holtzberg F 1991 
      {\it Phys.\ Rev.\ B} {\bf 44} 458

\item Thompson J R, Sun Y R, Kerchner H R, Christen D K, Sales B C,
      Chakoumakos B C, Marwick A D, Civale L and Thomson J O 1992
      {\it Appl.\ Phys.\ Lett.} {\bf 60} 2306

\item Thompson J R, Sun Y R, Christen D K, Marwick A D and Holtzberg F
      1994 {\it Phys.\ Rev.\ B} {\bf 49} 13287

\item Thompson R S 1975                        
      {\it Zh. Eksp. Teor. Fiz.} {\bf 69} 2249
      (1976 {\it Sov. Phys. JETP} {\bf 42} 1144)

\item Thompson R S and Hu C-R 1971 {\it Phys.\ Rev.\ Lett.} {\bf 20} 1352
\item Thompson R S and Hu C-R 1974 {\it Proc.\ 13th Intnl.\ Conf.\ on
      Low Temp.\ Physics} ed K D Timmerhaus, W J O'Sullivan
      and E F Hammel (New York: Plenum) p 6763

\item Thorel P, Kahn P, Simon Y and Cribier C 1973
      {\it J.\ Physique} {\bf 34} 447

\item Thuneberg E V 1984 {\it J.\ Low Temp.\ Phys.} {\bf 57} 415 
\item Thuneberg E V 1986 {\it J.\ Low Temp.\ Phys.} {\bf 62} 27
\item Thuneberg E V 1989 {\it Cryogenics} {\bf 29} 236

\item Thuneberg E V, Kurkij\"arvi J and Rainer D 1984
      {\it Phys.\ Rev.\ B} {\bf 29} 3913
\item Tilley D R and Tilley J 1974 {\it Superfluidity and Superconductivity}
      (New York: van Nostrand Reinhold)

\item Timms W E and Walmsley D G 1976  
      {\it J.\ Phys.\ F: Metal Phys.} {\bf 6} 2107

\item Tinkham M 1963 {\it Phys. Rev.} {\bf 129} 2413     
\item Tinkham M 1964 {\it Phys.\ Rev.\ Lett.} {\bf 13} 804
\item Tinkham M 1975 {\it Introduction to Superconductivity}
                     (McGraw-Hill: New York)
\item Tinkham M 1988 {\it Phys.\ Rev.\ Lett.} {\bf 61} 1658
\item Tinkham M and Lobb C J 1989 {\it Solid State Physics} {\bf 42} 91
\item Tinkham M and Skocpol W J 1975
      {\it Rep.\ Prog.\ Phys.} {\bf 38} 1049

\item Tkachenko V K 1969 {\it Zh.\ Eksp.\ Teor.\ Fiz.} {\bf 56} 1763
      (1969 {\it Sov.\ Phys.\ JETP} {\bf 29} 945)   

\item Tomlinson E J, Przylupsky P and Evetts J E 1993
      {\it Cryogenics} {\bf 33} 28 

\item Triscone J-M, Fischer {\O}, Brunner O, Antognazza L, Kent A D and
      Karkut M G 1990
      {\it Phys.\ Rev.\ Lett.} {\bf 64} 804 

\item Triscone J-M, Fivat P, Andersson M, Decroux M and Fischer {\O} 1994
      {\it Phys.\ Rev.\ B} {\bf 50} 1229   

\item Troy R J and Dorsey A T 1993 {\it Phys.\ Rev.\ B} {\bf 47} 2715

\item Troy R J and Dorsey A T 1995       
      {\it Phys.\ Rev.\ B} {\bf 51} 11728

\item Tsui O K C, Ong N P, Matsuda Y, Yan Y F and Petersen J B 1994
      {\it Phys.\ Rev.\ Lett.} {\bf 73} 724 

\item Ullah S and Dorsey A T 1990 {\it Phys.\ Rev.\ Lett.} {\bf 65} 2066
\item Ullah S and Dorsey A T 1991 {\it Phys.\ Rev.\ B} {\bf 44} 262
\item Ullah S, Dorsey A T and Buchholtz L J 1990
      {\it Phys.\ Rev.\ B} {\bf 42} 9950  

\item Ulm E R, Kim J-T, Lemberger T R, Foltyn S R and Wu X 1995
      {\it Phys.\ Rev.\ B} {\bf 51} 9193  

\item Ullmaier H 1975 {\it Irreversible Properties of Type-II
      Superconductors} (Berlin: Springer)

\item Ullrich M, M\"uller D, Heinemann K, Niel L and Freyhardt H C 1993
      {\it Appl.\ Phys.\ Lett.} {\bf 63} 406 

\item Uprety K K and Dom\'{\i}nguez D 1995
      {\it Phys.\ Rev.\ B} {\bf 51} 5955 

\item Urbach J S, Lombardo L W, White W R, Beasley M R and
      Kapitulnik A 1994 {\it Physica C} {\bf 219} 93  

\item Usadel K 1970 {\it Phys.\ Rev.\ Lett.} {\bf 25} 560

\item * van Dalen A J J, Koblischka M R, Griessen R, Jirsa M and
      Ravi Kumar G 1995 {\it Physica C} (submitted)

\item van der Beek C J and Kes P H 1991a 
      {\it Phys.\ Rev.\ B} {\bf 43} 13032
\item van der Beek C J and Kes P H 1991b 
      {\it Physica C} {\bf 185-189} 2241

\item van der Beek C J Kes P H, Maley M P, Menken M J V and Menovsky A A
      1992a {\it Physica C} {\bf 195} 302  
\item van der Beek C J, Nieuwenhuys G J, Kes P H, Schnack H G and Griessen R P
      1992b {\it Physica C} {\bf 197} 320 
\item van der Beek C J, Geshkenbein V B and Vinokur V M 1993
      {\it Phys.\ Rev.\ B} {\bf 48} 3393      

\item van der Beek C J, Konczykowski M, Vinokur V M, Crabtree G W, Li T W
      and Kes P H 1995 {\it Phys.\ Rev.\ B} {\bf 51} 15492
\item van Dover R B 1990
      {\it Appl.\ Phys.\ Lett.} {\bf 56} 2681 

\item * van Otterlo A, Feigel'man M V, Geshkenbein V B and Blatter G 1995
      (preprint on Hall anomaly) 

\item Vecris G and Pelcovits R A 1991 {\it Phys.\ Rev.\ B} {\bf 44} 2767

\item Vinen W F 1969 {\it Superconductivity}
      ed R D Parks (New York: Marcel Dekker) vol II 1167  

\item Vinen W F, Forgan E M, Gough C E and Hood M J 1971
      {\it Physica} {\bf 55} 94  

\item Vinen W F and Warren A C 1967
      {\it Proc.\ Phys.\ Soc.\ London} {\bf 91} 409

\item Vinnikov L Ya, Gurevich L A, Yemel'chenko G A and Ossipyan Yu A
      1988 {\it Solid.\ State Commun.} {\bf 67} 421

\item Vinokur V M and Koshelev A E 1990    
      {\it Zh.\ Eksp.\ Teor.\ Fiz.} {\bf 97} 976
      (1990 {\it Sov.\ Phys.\ JETP} {\bf 70} 547)

\item Vinokur V M, Feigel'man M V, Geshkenbein V B and Larkin A I 1990a
      {\it Phys.\ Rev.\ Lett.} {\bf 65} 259  

\item Vinokur V M, Kes P H and Koshelev A E 1990b
      {\it Physica C} {\bf 168} 28                

\item Vinokur V M, Feigel'man M V and Geshkenbein V B 1991a
      {\it Phys.\ Rev.\ Lett.} {\bf 67} 915        
\item Vinokur V M, Feigel'man M V, Geshkenbein V B and Blatter G 1993
      {\it Phys.\ Rev.\ Lett.} {\bf 71} 1242       

\item Vinokur V M, Geshkenbein V B, Feigel'man M V and Larkin A I 1991b
      {\it Zh. Eksp.\ Teor.\ Fiz.} {\bf 100} 1104
      (1991 {\it Sov.\ Phys.\ JETP} {\bf 73} 610)   

\item Vlasko-Vlasov V K, Indenbom M V, Nikitenko V I,
      Polyanski\u{\i} A A, Prozorov R L, Grekhov I V, Delimova L A,
      Lini\u{\i}chuk I A, Antonov A V and Gusev M Yu 1992
      {\it Sverkhprovodimost'} {\bf 5} 1637      
      (1992 {\it Superconductivity} {\bf 5} 1582)

\item Vlasko-Vlasov V K, Nikitenko V I, Polyanski\u{\i} A A, Crabtree G W,
      Welp G W and Veal B W 1994a {\it Physica C} {\bf 222} 361 

\item Vlasko-Vlasov V K, Dorosinskii L A, Polyanski\u{\i} A A,
      Nikitenko V I, Welp U, Veal B W and  Crabtree G W 1994b
      {\it Phys.\ Rev.\ Lett.} {\bf 72} 3246 

\item Voloshin I F, Fisher L M, Gorbachev V S, Savel'ev S E and
      Yampol'skii V A 1994 
      {\it Pis'ma Zh.\ Eksp.\ Teor.\ Fiz.} {\bf 59} 55
      (1994 {\it Sov.\ Phys.\ JETP Lett.} {\bf 59} 55)  

\item Volovik G E 1993
      {\it Pis'ma Zh.\ Eksp.\ Teor.\ Fiz.} {\bf 58} 457 
      (1993 {\it Sov.\ Phys.\ JETP Lett.}  {\bf 58} 469)

\item Vulcanescu V, Collin G, Kojima H, Tanaka I and Fruchter L 1994
      {\it Phys.\ Rev.\ B} {\bf 50}

\item Wadas A, Fritz O, Hug H J and G\"untherodt H-J  1992
      {\it Z.\ Phys.\ B} {\bf 88} 317 

\item Wagenleithner P 1982
      {\it J.\ Low Temp.\ Phys.} {\bf 48} 25

\item Wagner R, Hillmer F, Frey U and Adrian H 1994
      {\it Phys.\ Rev.\ B} {\bf 49} 13184              
\item Wagner R, Frey U, Hillmer F, and Adrian H 1995
      {\it Phys.\ Rev.\ B} {\bf 51} 1206 

\item Wahl A, Maignan A, Martin C, Hardy V, Provost J and Simon Ch 1995 
      {\it Phys.\ Rev.\ B} {\bf 51} 9123 

\item Walmsley D G 1972         
      {\it J.\ Phys.\ F: Metal Phys.} {\bf 2} 510

\item Walmsley D G and Timms W E 1977  
      {\it J.\ Phys.\ F: Metal Phys.} {\bf 7} 2373

\item Wan Y M, Hebboul S E and Garland J C 1994
      {\it Phys.\ Rev.\ Lett.} {\bf 72} 3867       

\item Wang L, Zhu Y, Zhao H L and Feng S 1990
      {\it Phys.\ Rev.\ Lett.} {\bf 64} 3094       

\item Wang Yu-P and Zhang D-L 1995
      {\it Phys.\ Rev.\ Lett.} {\bf 74} C2838  

\item Wang Z and Shi D 1993       
      {\it Phys.\ Rev.\ B} {\bf 48} 4208, 9782, 16179

\item Wang Z D, Dong J and Ting C S 1994 
      {\it Phys.\ Rev.\ Lett.} {\bf 72} 3875

\item Wang Z D and Ting C S 1992 {\it Phys.\ Rev.\ B} {\bf 46} 284; 
      {\it Phys.\ Rev.\ Lett.} {\bf 69} 1435          

\item Wang Z D, Ho K M, Dong J and Zhu J-X 1995
      {\it Phys.\ Rev.\ B} {\bf 51} 6119 

\item Watson G I and Canright G S 1993 {\it Phys.\ Rev.\ B} {\bf 48} 15950

\item Watts-Tobin R, Kramer L and Pesch W 1974
            {\it J.\ Low Temp.\ Phys.} {\bf 17} 71

\item Weber H and Jensen H J 1991 
      {\it Phys.\ Rev.\ B} {\bf 44} 454

\item Weber H W 1974 {\it J.\ Low Temp.\ Phys.} {\bf 17} 49 
\item Weber H W 1977 {\it Anisotropy Effects in Superconductors}
       (New York: Plenum)
\item Weber H W 1991 {\it Physica C} {\bf 185-189} 309  
\item Weber H W, Botlo M, Sauerzopf F M, Wiesinger H and Klein U 1987
            {\it Jap.\ J.\ Appl.\ Phys.} {\bf 26} 917 
\item Weber H W, Sauerzopf F M and Seidl E 1981
            {\it Physica B} {\bf 107} 429             
\item Weber H W, Schelten J and Lippmann G 1973
            {\it phys.\ stat.\ sol.\ (b)} {\bf 57} 515 
\item Weber H W, Schelten J and Lippmann G 1974
            {\it J.\ Low Temp.\ Phys.} {\bf 16} 367    
\item Weber H W, Seidl E, Botlo M, Laa C, Mayerhofer E, Sauerzopf F M,
            Schalk R M, Wiesinger H P and Rammer J 1989
            {\it Physica C} {\bf 161} 272              

\item Weber M, Amato A, Gygax F N, Schenck A, Maletta H, Duginov V N,
      Grebinnik V G, Lazarev A B, Olshevsky V G, Pomjakushin V Yu,
      Shilov S N, Zhukov V A, Kirillov B F, Pirogov A V, Ponomarev A N,
      Storchak V G, Kapusta S and Bock J 1993      
      {\it Phys.\ Rev.\ B} {\bf 48} 13022

\item Wellstood F C, Ferrari M J, Kingston J J, Shaw T J and Clarke J
      1993  {\it Phys.\ Rev.\ Lett.} {\bf 70} 89

\item Welp U, Fleshler S, Kwok W K, Klemm R A, Vinokur V M, Downey J,
      Veal B and Crabtree G W 1991                   
      {\it Phys.\ Rev.\ Lett.} {\bf 67} 3180

\item Welp U, Gardiner T, Gunther D, Crabtree G W, Vlasko-Vlasov V K and
      Nikitenko V I 1994 {\it Phys.\ Rev.\ Lett.} {\bf 67} 3180 

\item Wen Hai-hu, Schnack H G, Griessen R, Dam B and Rector J 1995
      {\it Physica C} {\bf 241} 353  

\item Wen Hai-hu and Zhao Zhong-xian 1994
      {\it Phys.\ Rev.\ B} {\bf 50} 13853

\item White W R, Kapitulnik A and Beasley M R 1991 
      {\it Phys.\ Rev.\ Lett.} {\bf 66} 2826
\item White W R, Kapitulnik A and Beasley M R 1993 
      {\it Phys.\ Rev.\ Lett.} {\bf 70} 670
\item White W R, Kapitulnik A and Beasley M R 1994 
      {\it Phys.\ Rev.\ B} {\bf 50} 6303                    

\item Wilkin N and Moore M A 1993 {\it Phys.\ Rev.\ B} {\bf 47} 957;
                                                     {\bf 48} 3464 
\item Wipf S L 1967 {\it Phys. Rev.} {\bf 161} 404  
\item Wipf S L 1991 {\it Cryogenics} {\bf 31} 936  

\item W\"oltgens P J M, Dekker C, Sw\"uste J and de Wijn H W 1993a
      {\it Phys.\ Rev.\ B} {\bf 48} 16826       
\item W\"oltgens P J M, Dekker C,  and de Wijn H W 1993b
      {\it Phys.\ Rev.\ Lett.} {\bf 71} 3858    

\item Woo K C, Gray K E, Kampwirth R T, Kang J H, Stein S J, East R,
      and McKay D M 1989 
      {\it Phys.\ Rev.\ Lett.} {\bf 63} 1877

\item W\"ordenweber R 1992 {\it Phys.\ Rev.\ B} {\bf 46} 3076 
\item W\"ordenweber R and Kes P H 1985
      {\it Physica B} {\bf 135} 136 
\item W\"ordenweber R and Kes P H 1986 {\it Phys.\ Rev.\ B} {\bf 34} 494
\item W\"ordenweber R and Kes P H 1987 {\it J. Low Temp. Phys.} {\bf 67} 1

\item Worthington T K, Olsson E, Nichols C S, Shaw T M and Clarke D R 1991
         {\it Phys.\ Rev.\ B} {\bf 43} 10538
\item Worthington T K, Fisher M P A, Huse D A, Toner J, Marwick A D,
         Zabel T, Feild C A and Holtzberg F 1992
         {\it Phys.\ Rev.\ B} {\bf 46} 11854

\item Wu D H, Mao J, Mao S N, Peng J L, Xi X X, Venkatesan T, Greene R L
      and Anlage S 1993 {\it Phys.\ Rev.\ Lett.} {\bf 70} 85   

\item Wu H, Ong N P and Li Y Q 1993           
           {\it Phys.\ Rev.\ Lett.} {\bf 71} 2642

\item Wu M K, Ashburn J R, Torng C J, Hor P H, Meng R L,
           Gao L, Huang Z J, Wang Y Q and Chu C W 1987
           {\it Phys.\ Rev.\ Lett.} {\bf 58} 908
\item Xing L and Te\v{s}anovi\'{c} Z 1990
      {\it Phys.\ Rev.\ Lett.} {\bf 65} 794

\item Xing W, Heinrich B, Zhou Hu, Fife A A and Cragg A R 1994 
      {\it J.\ Appl.\ Phys.} {\bf 76} 4244

\item Xu J H, Miller J H and Ting C S 1995
      {\it Phys.\ Rev.\ B} {\bf 51} 424

\item Xu M, Finnemore D K, Crabtree G W, Vinokur V M, Dabrovski B,
      Hinks D G and Zhang K 1993
      {\it Phys.\ Rev.\ B} {\bf 48} 10630

\item Xu Z, Drulis H, Hou J-C, De Long L E and Brill J W 1992
      {\it Physica C} {\bf 202} 256   

\item Yamafuji K, Fujiyoshi T, Toko K and Matsushita T 1988
      {\it Physica C} {\bf 159} 743 

\item Yamafuji K, Fujiyoshi T, Toko K, Matsuno T, Kobayashi T
      and Kishio K 1994 {\it Physica C} {\bf 226} 133 

\item Yamafuji K and Irie F 1967 {\it Phys.\ Lett.\ A} {\bf 25} 387

\item Yamafuji K, Wakuda T, Fujiyoshi T and Funaki K 1995
      {\it Physica C} {\bf 242} 251 

\item Yamamoto K, Mazaki H and Yasuoka H 1993
      {\it Phys.\ Rev.\ B} {\bf 47} 915

\item Yamasaki H, Endo K, Kosaka S, Umeda M, Yoshida S and Kajimura K
      1994 {\it Phys.\ Rev.\ B} {\bf 50} 12959 

\item Yang G, Shang P, Sutton S D, Jones I P, Abell J S and Gough C E 1993
      {\it Phys.\ Rev.\ B} {\bf 48} 4054  

\item Yang Z J, Johansen T H, Bratsberg H, Bhatnagar A and Skjeltorp A T
      1992 {\it Physica C} {\bf 197} 136     

\item Yao Z, Yoon S, Dai H, Fan S and Lieber C 1994
      {\it Nature} {\bf 371} 777   

\item Yarmchuk E J, Gordon M J V and Packard R E 1979 
      {\it Phys.\ Rev.\ Lett.} {\bf 43} 214

\item Yarmchuk E J and Packard R E 1982           
      {\it J.\ Low Temp.\ Phys.} {\bf 46} 479

\item Yaron U, Kornyushin Y and Felner I 1992 
      {\it Phys.\ Rev.\ B} {\bf 46} 14823

\item Yaron U, Gammel P L, Huse D A, Kleiman R N, Oglesby C S, Bucher E,
      Batlogg B, Bishop D J, Mortensen K, Clausen K N, Bolle C A and
      de la Cruz F 1994 {\it Phys.\ Rev.\ Lett.} {\bf 73} 2748 

\item * Yaron U, Gammel P L, Huse D A, Kleiman R N, Oglesby C S, Bucher E,
      Batlogg B, Bishop D J, Mortensen K and Clausen K N 1995
      {\it Nature} (submitted) 

\item Yazdani A, White W R, Hahn M R, Gabay M, Beasley M R, Kapitulnik A
           1993 {\it Phys.\ Rev.\ Lett.} {\bf 70} 505

\item Yazdani A, Howald C M, White W R, Beasley M R, Kapitulnik A 
           1994 {\it Phys.\ Rev.\ B} {\bf 50} 16117 

\item Yazyi J, Arrib\'ere A, Dur\'an C, de la Cruz F, Mitzi D B and
      Kapitulnik A 1991 {\it Physica C} {\bf 184} 254 

\item Yeh N-C, Reed D S, Jiang W, Kriplani U, Holtzberg F, Gupta A,
      Hunt B D, Vasquez R P, Foote M C and Bajuk L 1992
      {\it Phys.\ Rev.\ B} {\bf 45} 5654               
\item Yeh N-C, Reed D S, Jiang W, Kriplani U, Tsuei C C,
      Chi C C and Holtzberg F 1993a
      {\it Phys.\ Rev.\ Lett.} {\bf 71} 4043           
\item Yeh N-C, Jiang W, Reed D S, Kriplani U and Holtzberg F 1993b
      {\it Phys.\ Rev.\ B} {\bf 47} 6146         

\item Yeshurun Y, Bontemps N, Burlachkov L and Kapitulnik A 1994
      {\it Phys.\ Rev.\ B} {\bf 49} 1548 

\item Yeshurun Y and Malozemoff A P  1988
      {\it Phys.\ Rev.\ Lett.} {\bf 60} 2202

\item Yethiraj M, Mook H A, Wignall G D, Cubitt R, Forgan E M, Paul D McK
      and Armstrong T 1993 {\it Phys.\ Rev.\ Lett.} {\bf 70} 857

\item Ying Q Y and Kwok H S 1990 
      {\it Phys.\ Rev.\ B} {\bf 42} 2242

\item Yu W and Stroud D 1994 {\it Phys.\ Rev.\ B} {\bf 50} 13632 
\item Yu W and Stroud D 1995 {\it Phys.\ Rev.\ B} {\bf 51} 3725 
\item Zavaritski\u{\i} N V 1993
      {\it Pis'ma Zh.\ Eksp.\ Teor.\ Fiz.} {\bf 57} 695 
      (1993 {\it Sov.\ Phys.\ JETP Lett.}    {\bf 57} 707)

\item Zeldov E, Amer N M, Koren G and Gupta A 1990
      {\it Appl.\ Phys.\ Lett.} {\bf 56} 1700

\item Zeldov E, Clem J R, McElfresh M and Darwin M 1994a
      {\it Phys.\ Rev.\ B} {\bf 49} 9802  

\item Zeldov E, Larkin A I, Geshkenbein V B, Konczykowski M, Majer D,
      Khaykovich B, Vinokur V M and Shtrikman H 1994b  
      {\it Phys.\ Rev.\ Lett.} {\bf 73} 1428

\item Zeldov E, Larkin A I, Konczykowski M, Khaykovich B, Majer D,
      Geshkenbein V B and Vinokur V M  1994c        
      {\it Physica C} {\bf 235-240} 2761 

\item Zeldov E, Majer D, Konczykowski M, Geshkenbein V B, Vinokur V M
      and Shtrikman H 1995a 
      {\it Nature} {\bf 375} 373   

\item Zeldov E, Majer D, Konczykowski M, Larkin A I, Vinokur V M,
      Geshkenbein V B, Chikumoto N and Shtrikman H 1995b
      {\it Europhys.\ Lett.} {\bf 30} 367  

\item Zeuner S, Prettl W, Renk K F and Lengfellner H 1994
      {\it Phys.\ Rev.\ B} {\bf 49} 9080

\item Zhao Y, Gu G D, Cochrane J W, Russell G J, Wen J G, Nakamura N,
      Tajima S and Koshizuka N 1995   
      {\it Phys.\ Rev.\ B} {\bf 51} 3806

\item Zhu J, Mester J, Lockhart J and Turneaure J 1993
      {\it Physica C} {\bf 212} 216

\item Zhu Y-D, Zhang F C and Sigrist M 1995 
      {\it Phys.\ Rev.\ B} {\bf 51} 1105

\item Zhukov A A 1992 {\it Solid State Commun.} {\bf 82} 983

\item Zhukov A A, Komarkov D A, Karapetrov G, Gordeev S N and Antonov R I
      1992 {\it Supercond.\ Sci.\ Technol.} {\bf 5} 338 

\item Zhukov A A, K\"upfer H, Rybachuk V A, Ponomarenko L A,
      Murashov V A and Martynkin A Yu 1994
      {\it Physica C} {\bf 219} 99        

\item Zhukov AA and Moshchalkov V V 1991
      {\it Sverkhprovodimost'} {\bf 4} 850      
      (1991 {\it Superconductivity} {\bf 4} 759)

\item Zhukov A A, K\"upfer H, Perkins G, Cohen L F, Caplin A D,
      Klestov S A, Claus H, Voronkova V I, Wolf T and W\"uhl H 1995
      {\it Phys.\ Rev.\ B} {\bf 51} 12704  

\item Ziese M and Esquinazi P 1995  
      {\it Z.\ Phys.\ B} {\bf 94} 265

\item Ziese M, Esquinazi P, Gupta A and Braun H F 1994a
      {\it Phys.\ Rev.\ B} {\bf 50} 9491
\item Ziese M, Esquinazi P, Kopelevich Y and Sherman A B 1994b
      {\it Physica C} {\bf 224} 79
\item Ziese M, Esquinazi P, and Braun H F 1994c
      {\it Supercond. Sci.\ Technol.} {\bf 7} 869  

\item Zuo F, Khizroev S, Jiang X, Peng J L and Greene R L 1994a 
      {\it Phys.\ Rev.\ Lett.} {\bf 72} 1746          
\item Zuo F, Khizroev S, Voss S and Hermann A M 1994b 
      {\it Phys.\ Rev.\ B} {\bf 49} 9252              

\item Zuo F, Vacaru D, Duan H M and Hermann A M 1993
      {\it Phys.\ Rev.\ B} {\bf 47} 8327

\item Zwicknagl G E and Fulde P 1981
      {\it Z.\ Phys.\ B} {\bf 43} 23        

\item Zwicknagl G E and Wilkins J W 1984
      {\it Phys.\ Rev.\ Lett.} {\bf 53} 1276  
    } 
  \end{description}

  \newpage
{
 \baselineskip 14.4 pt  \parskip 2 pt 
\parindent 0pt
\setlength{\parskip}{\baselineskip}
  {\bf Figure Captions} ~~
 (Brandt, The Flux-Line Lattice in Superconductors, 6 June 1995)
                 \\   \\
 {\bf Figure 1.} The flux-line lattice (FLL) in type-II superconductors.
{\it Left:~} FLL with two-dimensional defects (simple edge
dislocation, stacking fault connecting two partial dislocations,
vacancy, interstitial).
{\it Right:~} FLL with a screw dislocation (dash-dotted line).
                 \\  \\
  {\bf Figure 2.}  The periodic magnetic field $B(x,y)$ ({\it left and
right}) and order parameter $|\psi|^2(x,y)$ ({\it right}) of the ideal
flux-line lattice. The flux-line positions correspond to the
maxima in $B(x,y)$ and zeros in $|\psi|^2$.
{\it Top:~} Low induction $B\ll B_{c2}$.
{\it Bottom:~} High induction $B\approx B_{c2}$.
The dashed zig-zag line shows the exotic field profile in clean
superconductors at zero temperature (section 1.3.)

  {\bf Figure 3.}  Reversible (ideal, pin-free) magnetization curves
of long cylinders or strips in parallel magnetic field $B_a$
(demagnetization factor $N=0$, {\it left}\/) and of a sphere ($N=1/3$,
{\it right}\/).
{\it Top:~} Type-I superconductor with Meissner state for
$B_a <(1-N)B_c$, intermediate state for $(1-N)B_c \le B_a \le B_c$,
and normal state for $B_a>B_c$.
The dashed line indicates the
enhancement of the penetration field when the wall energy of the
normal and superconducting domains in the intermediate state is
accounted for (Kronm\"uller and Riedel 1977).
{\it Middle:~} Type-II/1 superconductors with Ginzburg-Landau parameter
$\kappa$ close to $1/\sqrt2$, for $N=0$ exhibit a jump of height $B_0$
in their induction and magnetization at $B_a =B_{c1}$ due to an
attractive interaction between flux lines. For $N>0$, this jump is
stretched over a finite range of $B_a$,
allowing one to observe an intermediate mixed state with domains
of Meissner phase and FLL phase.
{\it Bottom:~} Type-II/2 superconductors with $\kappa \gg 1/\sqrt2$.
Meissner phase for $B_a < (1-N)B_{c1}$, mixed state (FLL) for
$(1-N)B_{c1} \le B_a \le B_{c2}$, and normal state for $B_a > B_{c2}$.

  {\bf Figure 4.}  The flux-line lattice observed at the surface of
type-II superconductors in an electron microscope after decoration
with Fe microcrystallites. (a) PbIn alloy ($T_c=7.5$ K, $\kappa =2$)
at $T=1.2$ K in a remanent magnetic field  of 70 Gauss, yielding
a flux-line spacing $a=0.6\,\mu$m.
(b) High-purity Nb disks 1 mm thick, 4 mm diameter,
of different crystallographic
orientations [110] and [011], at $T=1.2$ K and $B_a = 800$ Gauss
($B_{c1} = 1400$ Gauss). Due to demagetizing effects and the small
$\kappa \approx 0.70$, round islands of Meissner phase are
surrounded by a regular FLL (``intermediate mixed state'').
(c) High-purity Nb foil 0.16 mm thick at $T=1.2$ K and
$B_a = 173$ Gauss. Round islands of FLL embedded in a Meissner phase.
(d) Square disk $5\times 5\times 1$ mm$^3$ of high-purity
polycrystalline Nb at $T=1.2$ K and $B_a = 1100$ Gauss. As $B_a$
is increased, magnetic flux penetrates
from the edges in form of fingers (upper left) which are composed
of FLL shown enlarged in the right picture. The rectangular cross
section of the disk causes an edge barrier (section 8.5.).
 As soon as this  edge barrier is overcome, single flux lines or
droplets of FLL (lower right) pull apart from these fingers and
jump to the center, filling the disk with flux from the center.
 (Courtesy U.\  Essmann).
                     \\   \\
  {\bf Figure 5.}   The periodic magnetic field of a clean superconductor
near $B_{c2}$, equation (1.9), for reduced temperature t=$T/T_c =$
0.99, 0.6, 0.3, and 0. Note the conical maxima and minima at $t=0$.
At $t>0.6$ ($t<0.6$) the saddle point lies between two (three) flux lines.
                 \\   \\
  {\bf Figure 6.}  {\it Left:~} The line integrals yielding the
interaction energy and self energy of curved flux lines,
equations (2.8) and (3.3).
{\it Middle:~} FLL elastically distorted by a point
pinning force ${\bf f}$, equation (4.17).
{\it Right:~} Thermal fluctuation of flux lines.
                 \\  \\
  {\bf Figure 7.} {\it Top left:~} The magnetic field lines of a
point vortex (pancake vortex) in a stack of superconducting layers
indicated by horizontal lines.
{\it Top right:~} Stack of point vortices forming a straight and
a distorted flux line which threads the superconducting Cu-O layers.
{\it Bottom left:~} Josephson currents (vertical arrows) between,
and supercurrents (horizontal arrows) inside the Cu-O layers,
circulating around a Josephson vortex whose core is marked by a dot.
{\it Bottom right:~} Kinked flux lines in a layered superconductor
consist of point vortices connected by Josephson strings.

  {\bf Figure 8.}   Elastic matrix of the FLL defined by (4.2)
({\it left}\/) and elastic moduli (4.3) ({\it right}\/) from isotropic
GL theory.  The depicted matrix elements give the elastic energy of
pure tilt, compression, and shear waves with wavelength $2\pi/k$.
The parabola (dashed line) gives the local approximation to the
compressional and tilt energies valid for $k\ll \lambda^{-1}$.
                 \\    \\
  {\bf Figure 9.}  Six-terminal experiment on platelets of a layered
high-$T_c$ superconductor in a perpendicular magnetic field and
with applied current $I$.
{\it From top to bottom:~} At low temperatures the flux lines
(stacks of Josephson-coupled point vortices) are coherent and
parallel; the resistivity along the flux lines is thus zero, and
therefore the top ($V_1$) and bottom ($V_2$) voltage drops are
equal if the current is not too large ({\it left}\/). (After
 Busch et al.\ 1992 and L\'opez et al.\ 1994).
With increasing temperature the Josephson coupling between
the superconducting layers is reduced by thermal fluctuations; the
flux lines will then break apart if the current is large enough
 ({\it middle}\/, the arrow denotes the vortex velocity {\bf v}).
The resistivity perpendicular to the layers thus becomes larger
and larger, and $V_2$ decreases. At high temperatures the current
flows mainly near the top surface.
{\it Bottom:~} In a flux transformer two films are separated by a very
thin insulating layer and a perpendicular field generates flux lines
which thread both films. When a current is applied to the upper film,
the upper flux lines will move and drag the flux lines in the
lower film, see the field lines of two flux lines depicted in the
middle figure. The moving flux lines generate a voltage drop also in
the lower film, though no current is applied to this film.
This dragging of one FLL by the drifting periodic field of another
FLL ceases when the film separation exceeds the flux-line spacing
divided by $2\pi$ or when the velocity becomes too large.

  {\bf Figure 10.} Visualization of the penetration of a small
additional perpendicular magnetic field into an isotropic type-II
 superconducting strip in a large longitudinal field.
 (a) {\it Left:~}
 If no longitudinal magnetic field is applied, the flux of the
perpendicular field will penetrate from the long edges, taking the
  shortes way.
{\it Middle:~} With a longitudinal flux present, the penetration of
  a perpendicular flux proceeds from the short egdes along the strip.
{\it Left:~}  The stream lines of the screening currents induced
on the two flat surfaces of the strip when a perpendicular field is
applied. These surface currents exert a Lorentz force {\it only on the
ends} of the longitudinal flux lines, where the U-turning current and
the flux lines are at a right angle. Near the long edges the current is
parallel to the flux lines and thus does not exert a force on them.
 (b) The tilt of the longitudinal vortices penetrates from the edges,
driven by the U-turning screening current. After some time (in an
Ohmic strip with flux diffusion) or when the perpendicular field is
increased (in the Bean model with highly non-linear resistivity)
the tilt of the flux lines, and thus the perpendicular flux, have
penetrated completely.
 (c) The penetration of an additional perpendicular field
from the long edges can be described as the cutting and reconnection of
the flux lines belonging to the longitudinal and perpendicular flux
components. The resulting slightly tilted flux lines are then pushed
into the strip by the screening current flowing near the long edges;
however, for small tilt angle the driving force is small compared to
the pinning force, and thus this type of flux penetration is suppressed.
               \\  \\
  {\bf Figure 11.}  (a) The resistivity of a BSCCO platelet ($1\times
2 \times 0.2$ mm$^3$, \,$T=74.36$ K) as a function of the angle $\theta$
of the magnetic field $B_a$ with respect to the $ab$-plane for ({\it from
top to bottom\/}) $B_a$ = 0.1, 0.3, 0.5, 1, 1.5, 2, 2.5, 3, 3.5, 4, 4.5,
 5, 5.5, 6, 6.5, and 7 T. The current is applied in the $ab$ plane
parallel to the axis of rotation.
(b) These curves collaps into one single curve when plotted versus the
perpendicular field component $B_a \sin\theta$. (After Iye et al.\ 1992).

  {\bf Figure 12.} Various configurations of superconducting levitation.
Clockwise from {\it top left:~}  The levitation of a
type-I superconductor  above a magnet usually is unstable.
A type-II superconductor can levitate above, and be suspended below,
a magnet in a stable position and orientation since part of the magnetic
flux penetrates in form of flux lines which are pinned. A ring magnet
magnetized along its axis, exhibits two points on its axis
where the magnetic field is zero. Any diamagnet or superconductor
is attracted to these points and, if light enough, can levitate there
(Kitaguchi et al.\ 1989).
Levitation of a magnet above a flat type-II (type-I) superconductor
is stable (unstable). The levitation above an appropriately curved
``watch glass'' type-I superconductor can be stable.
              \\    \\
  {\bf Figure 13.}  The critical current density $J_c$ as a function
of the angle $\Theta$ between the applied field and the $c$-axis.
(a) YBCO film irradiated by 340 MeV Xe ions at an angle
$\Theta_\Phi = 60^\circ$ until a density of linear defects was
reached that corresponds to a magnetic induction  of $B_\Phi= 1.7$ T
if there is one flux line per defect.
(b) BSCCO film irradiated by 0.9 GeV Pb ions with
$\Theta_\Phi = 30^\circ$ and $B_\Phi = 2.1$ T. Note the sharp maximum of
$J_c$ when $B\perp c$ (intrinsic pinning by the $ab$-planes), and in
YBCO also when $B$ is along the linear defects caused by the
irradiation. In BSCCO this latter maximum is absent, indicating
that point vortices are pinned independently at $T=40$ K and $B=1$ T.
(After Kraus et al.\ 1994).
                 \\  \\
  {\bf Figure 14.} Bean model in a cylinder or slab in a parallel
magnetic field $H_a$. {\it Top left:~} Flux lines pinned by point pins;
the Lorentz force of density $BJ_c$ acts to the right (schematic).
{\it Top right:~} Negative magnetization versus applied field. The
virgin curve, from  equation (8.2), saturates when full penetration
is reached at $H_a= H_c$. The hysteresis loop follows from
equation (8.3).
{\it Middle:~} Profiles of the penetrating ({\it left}\/) and leaving
({\it right}\/) flux. The numbers correspond to the numbers in the
magnetization curve.
{\it Bottom:~} The current density during penetration and exit of
flux and penetration of reverse flux.

  {\bf Figure 15.} Modified Bean model for a superconducting strip
in a perpendicular magnetic field (section 8.4.) which increases
({\it top and middle\/}) and then decreases ({\it bottom\/}).
{\it Top left:~} The saturation of the current density $J$ to its
maximum allowed value $J_c$ near the right strip edge $x=a$.
{\it Top middle:~} Current density $J(x)$ (8.7) for fields
$H_a /H_c =$ 0.5, 1, 1.5, 2, 2.5.\ with $H_c = J_c d/\pi$.
{\it Top right:~} Penetration depth $a-b = a - a/\cosh(H_a/H_c)$.
{\it Middle:~}  Perpendicular magnetic field $H(x)$ (8.8)
for $H_a /H_c=$ 0.5, 1, 1.5, 2, 2.5.
{\it Bottom:~}  Perpendicular magnetic field $H(x)$ ({\it left\/})
and current density $J(x)$ ({\it right\/}) when the applied field
is decreased from $H_a /H_c=$ 1.5 (upper curves) via 1, 0.5, 0
(dashed curves, remanent state), -0.5, -1 to $-1.5$ (lower curve).
The corresponding curves for increasing $H_a$ are obtained by
changing the signs of $H(x)$ and $J(x)$.
                 \\   \\
  {\bf Figure 16.} Visualization of the edge barrier for flux
penetration into a thin type-II superconducting strip or disk with
rectangular cross section
({\it left insert\/}). Plotted is the hysteresis loop of the negative
magnetization. The points A, B, and C on this curve correspond to the
patterns of penetrating flux lines shown in the {\it right insert\/}.
At point B the edge barrier is overcome easily and flux penetrates
directly to the middle of the superconductor (dash-dotted line).
               \\   \\
  {\bf Figure 17.}
  Flux penetration into a disk with radius $a$ and current--voltage
power-law $E(J)= (J/J_c)^{19} E_c$ when the applied field is
increased from zero to $H_a= 0.1,$ 0.2, 0.3, 0.4, 0.5, 0.7, 1
in units of the maximum sheet current $J_{s0} = J_{c0} d$.
This smooth (non-steplike) $E(J)$ smears the vertical slopes of the
sheet current $J(r)d$ and field $H(r)$ (both in units $J_{s0}$) at the
flux front as compared to the Bean model.
{\it Top:~} Constant $J_c = J_{c0}$.
{\it Middle:~} Field dependent $J_c(H)=J_{c0}/(0.8 +|H|)$.
{\it Bottom:~} Ring-shaped disk with a hole of radius $0.4a$, for
constant $J_c=J_{c0}$. Flux penetration into the ring is complete at
$H_a\ge 0.6J_0$.

  {\bf Figure 18.}  Critical state of a rectangular
type-II superconductor film with side ratio $b/a=1.35$ in a
perpendicular field, calculated from equation (8.14).
{\it Left:~} The local magnetization $g(x,y)$ ({\it top\/}) and
the current stream lines, which are lines $g(x,y)=$ const
({\it bottom\/}).
{\it Right:~} The flux density or magnetic fied $H(x,y)$.
Note the sharp maxima of $H$ at the edges and the sharp minima
of $H$ at the discontinuity lines of the current.
The thicker line marks the locus where the self field (8.16) of the
circulating currents is zero and thus $H(x,y)=H_a$. On this
``neutral line''  $H(x,y)$ stays constant during flux creep.

  {\bf Figure 19.}  Penetration of perpendicular magnetic flux into
a rectangular film with side ratio $b/a=1.35$, computed from
equations (8.18) and (8.16) using a grid of $18\times 24$ independent
points on a Personal Computer.
  The film is characterized by $B=\mu_0 H$ and a highly nonlinear
current--voltage curve, $E(J) = (J/J_c)^{79} E_c$, ($E_c=J_c=1$)
which models a type-II superconductor at low temperatures.
Depicted are the current stream lines ({\it top row\/})
and the lines of constant magnetic field $H(x,y)$
({\it bottom row\/}); the thicker line marks $H(x,y)=H_a$.
The applied fields are, from {\it left to right\/},
 $H_a = 0.24$, 0.38, 0.61, and 1.16  in units of $J_c d$. At
$H_a =1.16$ flux has penetrated almost completely, and a critical
state is reached which looks as calculated from (8.14) and depicted
in figure 18.

  {\bf Figure 20.}  Distribution of perpendicular magnetic flux in
a square YBCO film ($d=800\,$nm, $T=20\,$K) and in a rectangular
YBCO single crystal ($d=80\,\mu$m, $T=65\,$K) visualized
magneto-optically by a ferrimagnetic garnet indicator.
For the square the perpendicular field $B_a$ is applied in the
sequence (a) 54 mT, (b) 92 mT, (c) 151 mT, and (d) --24 mT.
 In (a) and (b) the flux penetration starts in a star-like
pattern; in (c) full penetration is almost reached and the
flux density exhibits sharp ridges along the diagonals of the
square; in (d) the
penetration of reversed flux is seen at the edges.
For the rectangle $B_a$ is (e) 149 mT and (f) 213 mT;
the black line indicates the sample edge. At almost complete
penetration (f), current discontinuity lines with the shape of a
double fork are seen as sharp negative ridges of the flux density.
(Courtesy Th.\ Schuster).
                 \\   \\
  {\bf Figure 21.}  Depinning of a flux line from linear defects
visualized as straight cylinders.
{\it Left:~}  By thermal activation and supported by the applied
current, a pinned flux line can extend a parabolic loop (nucleus)
towards a neighbouring linear defect. This loop is then inflated by
the current,
and the flux line gradually slips to the neighbouring defect.
 Kinks in flux lines pinned at two or more linear defects
are driven along the defect by the current, allowing the flux line
to move in the direction of the Lorentz force.
{\it Right:~} Depinning of a flux line from two linear pins in a
layered superconductor. When the Josephson coupling between the layers
is weak (the anisotropy large), individual point vortices can depin
and be trapped by a neighbouring defect.
                 \\   \\
  {\bf Figure 22.}  {\it Left:~} Electric field $E(J)$ induced by moving
flux lines which are driven by a current density $J$, cf.\ equations
(9.4a, b, c). The inset shows a flux line or flux-line bundle hopping
over a periodic pinning potential which is tilted by the Lorentz force
exerted on the flux lines by the current. Thermally activated hopping
becomes more likely at high temperature $T$ and high induction $B$.
{\it Right:} Irreversibility lines (depinning lines) in the $BT$-plane
separate the region where free flux flow or thermally activated
flux flow (TAFF) occurs [both with linear $E(J)$\,] from the flux-creep
region [with highly nonlinear $E(J)$].

  {\bf Figure 23.}  {\it Left:~} Electric field $E$ and resistivity
$\rho =E/J$ versus current density $J$ in a log-log plot for a clean
YBCO crystal in a field of 1 T for various temperatures. Note the
abrupt change of curvature in a narrow temperature interval of 0.1 K
(courtesy T.\ K.\ Worthington).
{\it Right:~} Vortex-glass scaling of the resistivity $\rho(J)$ of a
thick YBCO film. Plotted in this way, data taken at various
temperatures collapse into one curve. (After Dekker et al.\ 1992b).
                 \\   \\
  {\bf Figure 24.} {\it Left:~} Temperature dependence of the real
and imaginary parts of the magnetic susceptibility $\chi$ of a YBCO
film measured at fixed frequencies in ac and dc fields parallel to
the $c$-axis and perpendicular to the film.
 (a) $B_a= 0.4$ T and (b) $B_a = 4$ T.
{\it Right:~} Frequency dependence of (a) the phase angle and
(b) the modulus of the dynamic conductivity $\sigma(\omega)$
evaluated from the dynamic susceptibility $\chi(\omega)$ by
inverting equation (9.14) for the YBCO film data at $B_a= 0.4$ T.
Note that at $T=88.9$ K, interpreted as glass temperature $T_g$, the
phase angle has a frequency-independent non-trivial value.
(After K\"otzler et al.\ 1994b).
                \\  \\
  {\bf Figure 25.} Dynamical (frequency) scaling of the linear
conductivity $\sigma(\omega)$ of the YBCO film of figure 24 at
$B_a = 0.4$ T and $B_a= 4$ T.  Both the phase angle (a) and modulus
(b) of $\sigma$ scale well and exhibit a transition from pure Ohmic
to full screening behaviour. (After K\"otzler et al.\ 1994b).
                 \\ \\
  {\bf Figure 26.}  Complex ac permeability $\mu(\omega) = 1 +
 \chi(\omega) =\mu' - i\mu''$ of a strip of width $2a$, thickness
$d \ll a$, and Ohmic resistiviy $\rho$  in a longitudinal ac field
[solid line, $\mu_\|$ (9.13)\,]  and perpendicular ac field
[dashed line, $\mu_\perp$ (Brandt 1993b, 1994a)\,].
The time scales $\tau_\| = d^2 \mu_0 /16\rho$
and $\tau_\perp = a d \mu_0 /2\pi \rho$ are chosen such that the
slopes of $\mu_\|''$ and $\mu_\perp''$ coincide at $\omega=0$. These
Ohmic $\mu(\omega)$  apply to normal conducting materials and to
superconductors in the free flux-flow or thermally assisted
flux-flow states.

  {\bf Figure 27.} The ac permeabilities $\mu(\omega) = \mu'-i\mu''$
in various longitudinal and transverse geometries for Ohmic
resistivity plotted on a log-log scale.  Time unit is the
relaxation time $\tau_0$ of the slowest relaxation mode in each of
these geometries, e.g.\ $\tau_0 =d^2/D\pi^2$ ($\tau_0 =0.24924 ad/D$)
for strips with cross section $2a \times d$ and with
flux diffusivity $D=\rho/\mu_0$ in longitudinal (perpendicular)
ac field.  For details see Brandt (1994b).
At large $\omega\tau_0$ one has in
longitudinal ac field $\mu' \approx \mu'' \sim 1/\sqrt\omega$,
and in perpendicular ac field
$\mu \sim\ln({\rm const}\cdot\omega)/\omega$, thus $\mu'\sim 1/\omega$
and $\mu'' = (2/\pi) \ln({\rm const}\cdot \omega) \mu'$.
                 \\   \\
  {\bf Figure 28.}  The current-density dependence of the activation
energy $U(J)$ extracted from several experiments on the same BSCCO
single crystal at different temperatures (5 K to 21 K) using the
 method of Maley et al.\ (1990). Plotted in this way, all curves
closely fall on one line, and a nearly temperature-independent
$U(J)$ is obtained. (Courtesy C.\ J.\ van der Beek, see also
  van der Beek et al.\ 1992a.)
                 \\   \\
  {\bf Figure 29.} {\it Upper curves:\/} Universality of the electric
field $E(x,t)$ and $E(r,t)$ approached during flux creep in strips of
half width $a$ ($|x| \le a$) and disks of radius $a$ ($r\le a$).
Shown is the normalized field $E(\eta,t)/E(1,t)$ versus
$\eta = x/a = r/a$ obtained numerically for power-law resistivities
$E(J) = (J/J_c)^n E_c$  for $n= 1,3,$ and $\infty$. The asymptotic
curve $n=\infty$ (bold line) (9.22) is nearly reached at $n\ge 9$.
 {\it Bottom:\/} Relaxation (creep) of $E(x,t)$ for a strip with $n=25$
at times $t = 0, 0.003, 0.01, 0.03, 0.1, 0.3, 1,$ and 3 in time unit
$t_1 =a d \mu_0 J_c /2\pi E_c \approx nt_0$ [cf.\ (9.24), $J_c=J_1 n$].
(After Gurevich and Brandt 1994).
                 \\  \\
  {\bf Figure 30.}
  Flux creep in a disk with nonlinear current-voltage law
$E(J) = (J/J_c)^{19} E_c$ after the increase of $H_a$ was
stopped at $t=0$.  Shown is the relaxation of the sheet current
$J_s(r,t)$ and magnetic field $H(r,t)$ (both in units $J_c d$)
at
 times $t=0.01$, 1, 100, $10^4$, $10^6$, $10^8$, and $10^{12}$ in
units $t_1 = ad \mu_0 J_c /2\pi E_c$. {\it Inset:~} The
magnetization $M(t)$ at $t \gg t_1$ is approximately
$M(t)/M(0) \approx (t_1/t)^{1/18} \approx 1 - (1/18)\ln(t/t_1)$.
  }
 \end{document}